\documentclass[letterpaper,12pt]{book}

\usepackage{mystonybrookthesis}
\usepackage{tikz}
\usepackage{amsmath,amssymb}
\usepackage{graphicx}   
\usepackage{tocbibind}
\usepackage{color}
\usepackage[sort&compress,numbers]{natbib}
\usepackage{notoccite}
\usepackage[colorlinks=true,linkcolor=blue,citecolor=blue, 
urlcolor=blue]{hyperref}   
\usepackage{bm} 

\usepackage[scriptsize]{subfigure}
\usepackage{appendix}

\usepackage{chngcntr}
\usepackage{etoolbox}
\AtBeginEnvironment{subappendices}{%
\addappheadtotoc
\section*{Appendices}
}

\def\snn{{\sqrt{s_{\scriptscriptstyle NN}}}}
\def\Chap#1{Chapter~\ref{#1}}
\title{Fluctuations in ultra-relativistic heavy ion collisions}
\author{Aleksas Mazeliauskas}
\degree{Physics}
\advisor{Derek Teaney}

\def\Eq#1{Eq.~(\ref{#1})}
\def\Eqs#1{Eqs.~(\ref{#1})}
\def\eq#1{(\ref{#1})}
\def\app#1{Appendix~\ref{#1}}

\def\snn{{\sqrt{s_{\scriptscriptstyle NN}}}}
\def\llangle{\left\langle}
\def\rrangle{\right\rangle}
\def\p{\mathbf{p}}

\def\Ref#1{Ref.~\cite{#1}}
\def\x{{\mathbf x}}

\def\p{\mathbf{p}}
\def\x{\mathbf{x}}
\def\k{\mathbf{k}}

\def\kp{\mathbf{k'}}
\def\b{\mathbf{b}}
\newcommand{\ti}{\tau_{\rm init}}
\def\Fig#1{Fig.~\ref{#1}}
\def\Figs#1{Figs.~\ref{#1}}
\def\Sect#1{Section~\ref{#1}}

\def\p{{\bf p}}

\def\Eq#1{Eq.~(\ref{#1})}
\def\Eqs#1{Eqs.~(\ref{#1})}
\def\eq#1{(\ref{#1})}
\def\app#1{Appendix~\ref{#1}}
\def\Fig#1{Fig.~\ref{#1}}
\def\Figs#1{Figs.~\ref{#1}}
\def\Sect#1{Sect.~\ref{#1}}
\def\snn{{\sqrt{s_{\scriptscriptstyle NN}}}}
\def\llangle{\left\langle}
\def\rrangle{\right\rangle}

\def\dd{{d}}
\def\x{{\bm x}}
\def\k{{\bm k}}

\def\p{{\bf p}}

\newcommand\nda{\end{align}}
\def\half{{\textstyle\frac{1}{2}}}
\def\third{{\textstyle\frac{1}{3}}}

\def\Eq#1{Eq.~(\ref{#1})}
\def\app#1{Appendix~\ref{#1}}
\def\Fig#1{Fig.~\ref{#1}}
\def\Sect#1{Sec.~\ref{#1}}
\def\Ref#1{Ref.~\cite{#1}}
\def\llangle{\left\langle}
\def\rrangle{\right\rangle}

\advance\parskip 1.9pt
\advance\voffset -0.2in

\def\Refs{\cite}
\def\element{\in}
\def\parts{{N_{\rm parts}}}

\begin{document}
\maketitle

Fluctuations are one of the main probes of the physics of the new state of hot 
and
dense 
nuclear
matter called the Quark Gluon Plasma (QGP) which is created in the 
ultra-relativistic heavy ion 
collisions. In this dissertation we extend and improve upon the existing 
descriptions of heavy ion collisions in three different directions: we study 
the 
new signatures of initial state fluctuations, the propagation of perturbations 
in the early stages of the collision, and the effect of thermal fluctuations on 
the hydrodynamic expansion of the QGP.
 
First, in \Chap{chap:pca} we study initial state fluctuations  by
examining the 
complete statistical information contained in the
two-particle correlation measurements in hydrodynamic simulations of Pb+Pb 
collisions 
at the CERN Large Hadron Collider ($\snn=2.76\,{\rm TeV}$).  
We use Principal Component Analysis (PCA) to decompose  the spectrum of 
harmonic 
flow, $v_n(p_T)$ for 
$n=0\text{--}5$, into dominant components.
The leading component is identified with the standard 
event plane 
$v_n(p_T)$, while the subleading component describes additional fluctuations in 
the two-particle correlation function.
We find good geometric predictors for the 
orientation and the magnitude of the leading and the subleading flows.
The subleading $v_0$, 
$v_1$, and 
$v_3$ flow harmonics are shown to be a response to the radial 
excitation of the corresponding eccentricity $\varepsilon_n$. In contrast, for 
$v_2$ the 
subleading flow
in {peripheral collisions} is dominated by the nonlinear mixing between 
the leading elliptic flow and radial flow fluctuations. Nonlinear mixing also 
plays a significant role in generating subleading $v_4$ and $v_5$ harmonics.
The PCA gives a systematic way of studying the full information of the 
two-particle correlation matrix and identifying the subleading flows, which  we 
show are
responsible for factorization breaking in 
hydrodynamics.

Second, in \Chap{chap:ekt} we study the thermalization and hydrodynamization 
of fluctuations at the early stages of heavy ion 
collisions.
We use 
leading order effective kinetic theory, accurate at weak coupling, to simulate 
the pre-equilibrium evolution of transverse energy and flow perturbations. 
For the short 
evolution we can use a linear response theory to construct the pre-equilibrium 
Green functions.
Then the energy-momentum tensor at a time when hydrodynamics becomes 
applicable can be expressed as a linear convolution of response functions with 
the initial 
perturbations. 
We propose combining effective kinetic theory   with  
weak coupling initial state models, such as IP-Glasma, to model the 
complete 
pre-thermal evolution from saturated nuclei to hydrodynamics in a 
weak coupling framework.

Last, in \Chap{chap:noise} we consider out-of-equilibrium 
hydrodynamic fluctuations in the expanding QGP. We develop 
a set of kinetic equations for a correlator of thermal fluctuations
which are equivalent to nonlinear hydrodynamics with noise.
We first show that the kinetic
response precisely reproduces the one-loop renormalization of the shear
viscosity
for a static fluid.
We then use the hydro-kinetic equations to analyze thermal fluctuations
for a Bjorken expansion.
The steady state solution to the kinetic equations determine the coefficient of 
the first fractional 
power of the gradient expansion  ($\propto 1/(\tau T)^{3/2}$), which was  
computed here for the first time. 
The formalism of 
hydro\nobreakdash-kinetic 
equations can be applied to 
more general background flows and coupled to existing viscous hydrodynamic 
codes to 
incorporate the physics of hydrodynamic fluctuations.

\clearpage 
\vspace*{1\baselineskip}
\centerline{\bf{Contributions of Authors}}
\vspace*{1\baselineskip}

\Chap{chap:pca} contains published material of the co-authored papers  
\cite{Mazeliauskas:2015vea} and \cite{Mazeliauskas:2015efa}. I, the present 
author,  
contributed significantly to all parts of this work and, in particular, was 
responsible 
for  
performing the PCA and finding the optimal geometric predictors (including 
writing 
the necessary computer code).
I also offered the interpretation of the
subleading elliptic flow as a response to radial size modulations of the 
background 
$\varepsilon_2$ in \Sect{elliptic}.
The 
hydrodynamic simulations of heavy ion collisions used in this publication were 
done by others.

\Chap{chap:ekt} is a partial reproduction of the published article written with 
co-authors~\cite{Keegan:2016cpi}.
I made significant 
contributions in the preparation of this paper. I identified 
the 
causality constrains and constructed the coordinate space Green functions. I 
also 
found the analytic limit of the response to small wavenumber 
perturbations in \Sect{smallk}. I contributed to, but was not the main author 
of the kinetic theory simulation code used for this paper.

\Chap{chap:noise} is a reproduction of the published article with 
collaborators~\cite{Akamatsu:2016llw}.
I contributed to all stages of this work. I reproduced the renormalization 
results 
for the static fluid in \Sect{gravitysec} and
I obtained the final numerical results of finite noise 
contributions for the Bjorken background, which are summarized in 
Table~\ref{tab1}.

The introductory and final chapters of this dissertation 
(Chapters~\ref{chap:foreword}, \ref{chap:intro} and \ref{chap:conclusion}) 
were not published before, except for the figures taken from the referenced 
sources.

\cleardoublepage

\vspace*{10\baselineskip}
\begin{center}
\emph{Mano mamai ir  t\.e\v ciui}\par
\vspace*{1\baselineskip}
\emph{To my mother and father}
\end{center}
\cleardoublepage
\begin{center}
\vspace*{4\baselineskip}
\includegraphics[width=\linewidth]{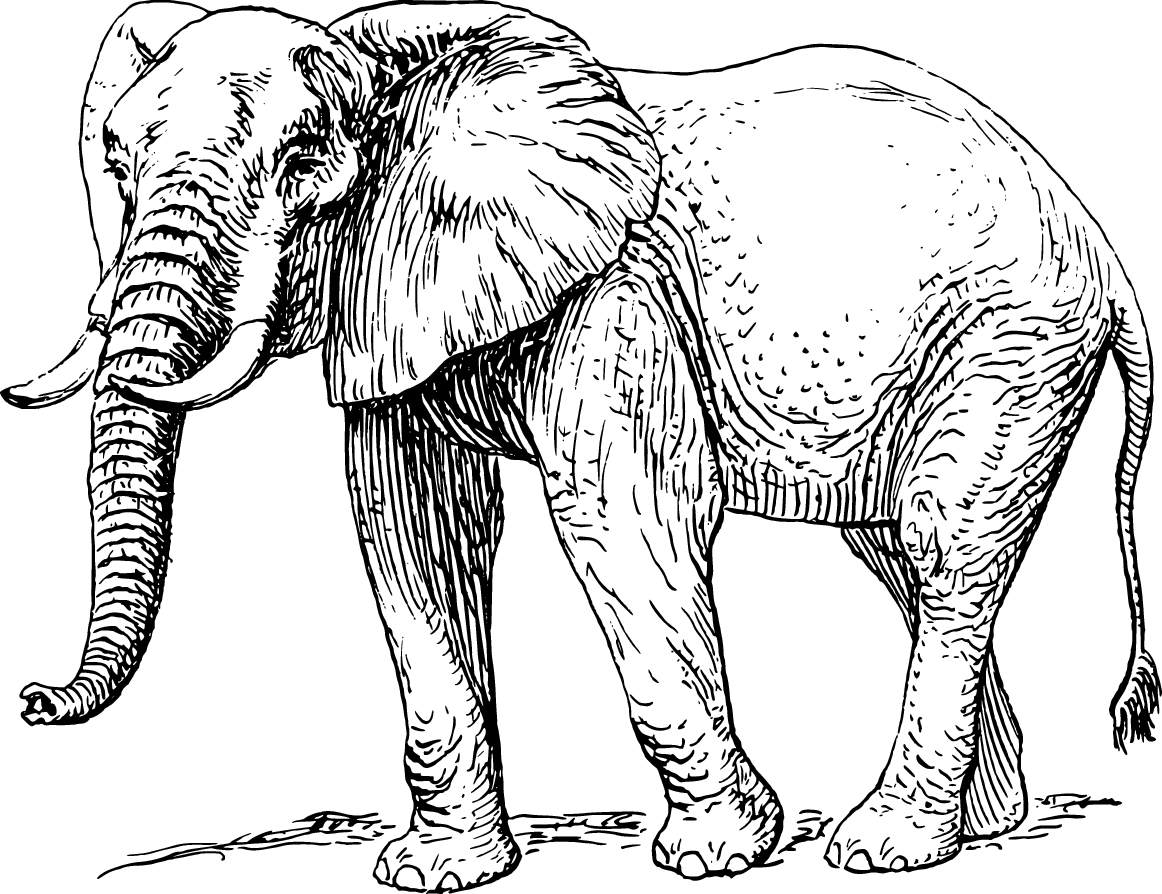}
\vspace*{4\baselineskip}

\emph{We have to remember that what we observe is not nature in itself, but 
nature exposed to our method of questioning} --- Werner 
Heisenberg~\cite{heisenberg1958physics}
\end{center}
\tableofcontents
\listoffigures
\listoftables

\chapter*{Acknowledgments\phantomsection}
\addcontentsline{toc}{chapter}{Acknowledgments}%
\vspace*{4\baselineskip}

First, I would like to thank my advisor Derek Teaney for being an exemplary 
scientist and  
caring mentor.  I 
was privileged to work closely with and learn from someone whose physical 
insight, knowledge, 
hard work, 
and passion for physics is an inspiring example for a young physicist.
His optimism and enthusiastic support were invaluable during my PhD.

I would also like to thank Nuclear Theory Group (NTG) professors 
Thomas Kuo,
Edward Shuryak,
Jacobus Verbaarschot,
Ismail Zahed, and
Dmitri Kharzeev
for welcoming me into the  group, lively lunch discussions and a character 
building seminar experience.  I am thankful to Edward Shuryak for the
NTG generous 
support of  graduate student travel.

I am grateful for the opportunity to meet and work with the numerous NTG 
postdocs and visiting scholars. My special thanks to Jean-Fran\c cois Paquet 
for being a great colleague and friend.

For the work presented in this dissertation, I gratefully acknowledge my 
collaborators Derek Teaney, Yukinao Akamatsu,  
Liam Keegan and Aleksi Kurkela. I am especially thankful to Aleksi Kurkela for 
inviting me to CERN where our collaboration started.

I would like to thank my fellow  nuclear theory graduate students Moshe, 
Rasmus, Mark, and Adith  for keeping me the 
company in C-115 and helping organizing the NTG Friday socials.

There are many other people, who made my 
five year 
stay at Stony Brook both enjoyable and worthwhile. I thank  the  Physics and 
Astronomy department administrative 
staff for the professionalism (especially Sara Lutterbie for the superluminal 
reply speed), 
professors for the 
excellent 
lectures (especially Peter van Nieuwenhuizen and his advanced physics courses), 
graduate students for their friendship 
(%
Abhishodh,
Ahsan,
Andrea,
Andrew and his wife Lizzy,  
Bertus,
Ben,
Choi,
Eric, 
Hari,
JP, 
Naveen,
Mehdi, 
Mingliang,
Saebyeok,
Spencer,
Xinan,
Yiqian, 
and many 
others),  and my host families Marilyn and Harold,  Laima and 
Virginijus, and Rachel and Derek 
for the hospitality.

This dissertation is a culmination of my long held dream of becoming a 
professional physicist. It was an incredible journey, which brought me to many 
wonderful places and helped me to meet many remarkable people. At every step 
along the way I was helped by numerous people. I cherish the education I 
received from all my 
teachers and mentors, financial support of many benefactors, and the 
encouragement  I received from the countless people.
There is no 
way I can repay the debt I owe 
them, but I hope that I have met their expectations.
Finally, I thank my parents  for their love and especially my mother for the
never wavering belief in me.

\cleardoublepage 
\chapter*{Publications}
\vspace*{4\baselineskip}
\begin{itemize}
\item A. Mazeliauskas and D. Teaney, \emph{Subleading harmonic flows in 
hydrodynamic 
simulations of heavy ion collisions}, Phys. Rev. C91, 044902
(2015)~\cite{Mazeliauskas:2015vea}.  Copyright (2015) by the American Physical 
Society
\item A. Mazeliauskas and D. Teaney, \emph{Fluctuations of harmonic and radial
flow in heavy ion collisions with principal components}, Phys. Rev. C93,
024913 (2016)~\cite{Mazeliauskas:2015efa},  Copyright (2016) by the American 
Physical Society
\item L. Keegan, A. Kurkela, A. Mazeliauskas and D. Teaney, \emph{Initial 
conditions for hydrodynamics from weakly coupled pre-equilibrium evolution}
J. High Energ. Phys. 08, 171 (2016)~\cite{Keegan:2016cpi}. Copyright (2016) by 
authors.
\item Y. Akamatsu, A. Mazeliauskas and D. Teaney, \emph{A kinetic regime of 
hydrodynamic fluctuations and long time tails for a Bjorken expansion},
Phys. Rev. C95, 014909 (2017)~\cite{Akamatsu:2016llw}. 
Copyright (2017) by the American Physical Society
\end{itemize}

\cleardoublepage

\pagenumbering{arabic}


\chapter{Foreword}
\label{chap:foreword}
For the most of science history physicists investigated phenomena 
governed by just two fundamental forces---gravity and 
electromagnetism. Gravity holds together the large scale structures 
like planets and galaxies, while electromagnetic interactions bond 
atoms together. Electromagnetic forces are also
responsible for many different 
phases of matter and endless diversity of chemical compounds.

The turn of the twentieth century marked a new era in physics with the
discovery of subatomic particles, e.g. electron and proton. Two new  and short 
ranged forces, i.e. acting 
only at subatomic distances,  were introduced: a weak interaction to 
explain the $\beta$ decay~\cite{fermi} and a strong interaction  to provide a 
necessary bond 
between protons and neutrons in the 
nucleus~\cite{yukawa}. In the following decades huge 
theoretical and experimental advances lead to a detailed description of 
electromagnetic, weak and strong interactions in terms of quantum 
field 
theories~%
\cite{Glashow:1961tr,Salam:1968rm,Weinberg:1967tq,GellMann:1964nj,Zweig:1981pd,Zweig:1964jf}.
This description together with the list of discovered 
particles forms a highly successful particle physics framework known as 
the Standard 
Model~\cite{Burgess,Langacker,Olive:2016xmw,Peskin}. It is 
still an ongoing 
research 
effort to incorporate gravity on the same 
footing as other interactions~\cite{Woodard:2009ns}.

The quantum field theory which 
describes the 
strong interactions between the fundamental constituents of a nucleus ({quarks} 
and 
{gluons}) is called the {Quantum Chromodynamics} 
(QCD)~\cite{Burgess, Langacker,Peskin}. 
Unlike the Quantum Electrodynamics, QCD is 
a 
non-abelian gauge theory and the 
particular  number of quark {flavors} and 
{color} charges  
make QCD look weak at very short distances---a phenomena known as 
{asymptotic freedom} \cite{Yang:1954ek,Politzer:1973fx,Gross:1973id}. Therefore 
certain 
processes at high energy, which probe the small distance limit of the QCD, can 
be 
accurately described by perturbative expansion 
 in the coupling constant $\alpha_s$.  In contrast, at large 
distances (or small energies) the strength of the strong interaction grows and 
new, 
non-perturbative phenomena appear, e.g. color confinement---quarks and gluons 
are hidden in color neutral 
hadrons~\cite{Wilson:1974sk,Mandelstam:1974pi,tHooft:1975krp}. 

Although the fundamental QCD equations of motion are known, most of their 
physical 
consequences remain an unsolved mathematical 
problem~\cite{Jaffe:2000ne}. Thanks to the 
complex nature of the strong force and many body interactions, a hot and dense 
QCD 
medium is a  
fascinating phase of matter with unprecedented 
properties~\cite{Anderson393,Shuryak:1988ck}. 
One very successful approach of studying the strongly interacting QCD 
has been the discretized QCD formulation 
on a 
lattice~\cite{Wilson:1974sk}.
With the 
progress of numerical methods and computational power, Lattice QCD is able to 
reproduce and even predict some properties of the QCD, notably, the hadron 
spectrum 
and the QCD
equation of state~\cite{Fodor:2012gf,Ding:2015ona}. However, transport 
properties of the QCD medium or 
simulations at the
non-zero baryon
chemical potential are very difficult in Lattice QCD~\cite{Ding:2015ona}. 
String 
theory research (which too has its roots in the attempt to 
describe the non-perturbative phenomena of strong 
interactions~\cite{Veneziano1968}) produced a 
new perspective of  
strongly coupled systems 
known as a gauge/gravity duality~\cite{Maldacena:1997re}. This holographic  
duality 
allows to map the non-perturbative regime of certain
gauge theories to classical supergravity in higher dimensions, where 
calculations are tractable~\cite{Chesler:2016vft}. Although no dual of the QCD 
is known, other gauge 
theories 
are 
used to infer generic features of strongly coupled 
systems, e.g. viscosity over entropy ratio $\eta/s$ 
~\cite{Policastro:2002se,Kovtun:2004de}.

Physics is an empirical science and its 
ultimate guidance principle must be the experiment. To create  matter where 
the asymptotic degrees of freedom are 
quarks 
and gluons one has to reach energies and densities exceeding that of nucleus 
core~\cite{Gyulassy:2004zy}. Such state of matter known as the Quark Gluon 
Plasma 
(QGP) is believed to 
have 
existed during the first few microseconds of the Big bang and  possibly exists
at the center of neutron 
stars~\cite{Shuryak:1977ut,Lattimer:2012nd,Weinberg,Wei-Qin1994}. 
On 
Earth the only way of creating such extreme conditions is in the relativistic 
hadron colliders~\cite{Fischer:2014wfa}. Nuclei of heavy elements, e.g. lead 
(Pb) or gold (Au), are 
accelerated to ultra\nobreakdash-relativistic 
energies and collided, momentarily creating a droplet of melted nuclear matter 
which evaporates into a cloud of hadrons in 
less than  $\sim 10^{-22}$ seconds. 
The debris of the collision are then used to 
infer the interesting properties of the ultra\nobreakdash-dense nuclear matter. 
Such events or ``little Big 
bangs" are repeated billions of times,  thus making statistical analysis of  
the 
QGP 
evolution possible.

In the last couple decades the successful hadron collision 
programs  in the US and Europe produced a wealth 
of experimental data, many surprising results, and new 
understanding of the behavior of matter under the extreme conditions.
Heavy ion collisions are a dynamical and multistage process, which requires a 
sophisticated description of every stage of the collision: the initial 
impact, creation and equilibration of  the QGP, the fast QGP 
expansion and cooling down, and the  transition from quarks and gluons into
the observed hadrons. To make the problem more tractable, the different regimes 
of the QCD medium are treated in a number of theoretical limits, which have 
become sizable subfields in their own right (Color Glass Condensate,  
relativistic hydrodynamics, gauge/gravity duality of QCD-like theories, finite 
temperature QCD, etc.~\cite{Wang:2016opj}).  The contrasting pictures of 
different 
stages of the heavy ion collision evolution
is not unlike a famed group 
of blind men touching an elephant---individual descriptions 
might look contradictory, but when combined correctly,  a remarkably 
successful model of the expanding QGP fireball emerges.

In this dissertation I will review my contributions to the physics of 
heavy ion collisions, which, needless to say, depend on the 
many decades of experimental and theoretical work done before. 
 Specifically, together with collaborators I worked on  
the fate of initial and thermal fluctuations in the expanding QGP medium. 
In 
\Chap{chap:intro}, I
give a short introduction to the main components of heavy ion 
physics
and the outline of the three research directions addressed in this manuscript. 
In \Chap{chap:pca}, I present a published 
work on the Principal Component 
Analysis of harmonic flow and factorization 
breaking~\cite{Mazeliauskas:2015vea,Mazeliauskas:2015efa}. 
Chapter~\ref{chap:ekt} covers a published
work on the QGP equilibration in a weak coupling 
framework~\cite{Keegan:2016cpi}.  
Finally in  
\Chap{chap:noise}, I present my work with collaborators on the effects of 
thermal fluctuations on the hydrodynamic expansion of the
QGP~\cite{Akamatsu:2016llw}.  In \Chap{chap:conclusion}, I conclude with a 
short summary of the conducted work 
and an outlook.


\chapter{Introduction}
\label{chap:intro}

\section{Overview}
Ultra-relativistic heavy ion collisions is nearly half a century old 
subject~\cite{Baym:2001in}. Below we  touch only on the recent development 
of the field driven by discoveries at RHIC and LHC  over the past fifteen 
years. 
There is a considerable body of 
terminology specific to this field, but to 
avoid interrupting the discussion  definitions of the main concepts in 
heavy ion collisions are given separately in \Sect{topics}.

The Relativistic Heavy Ion Collider 
(RHIC) at Brookhaven National Laboratory in the United States was built with 
the 
purpose of mass production 
of the new state of matter named the Quark Gluon Plasma 
(QGP)~\cite{Baym:2001in}. Since 
the start of 
operation in 2000, RHIC has been exploring a number of different ion 
systems,
Au-Au, Cu+Cu, U+U, and etc., at various collisions energies 
$\snn=8\text{-}200\,\text{GeV}$~\cite{RHIC_Runs}.
This makes the RHIC a versatile machine and  
particularly well suited for mapping out the phase diagram of the dense QCD 
matter. 
The ongoing Beam Energy Scan program (BES) is searching for the QCD critical 
point---a conjectured end point of  the first order phase transition line 
between the
QGP and the hadronic matter~\cite{Luo:2015doi}. 
In 
2010, the Large Hadron Collider (LHC) at CERN in Switzerland has also started a 
heavy 
ion 
collision program~\cite{Fischer:2014wfa,Muller:2012zq}.
With higher achievable  collision 
energies  ($\snn=2.76\text{-}5.02\,\text{TeV}$ for Pb+Pb), the LHC can create a 
hotter, larger and 
longer lasting  QGP droplets. The larger space\nobreakdash-time volume of QGP 
medium 
allows 
physicists  to 
understand better the bulk properties of hot nuclear matter. In fact, 
thanks to 
the high 
collision energies, measurements of high multiplicity  p+Pb and p+p collisions
suggest the formation of a QGP fireball even in smaller systems, defying all
expectations~\cite{CMS:2012qk,Aaboud:2016yar,Aad:2012gla}.

\begin{figure}
\centering
\includegraphics[width=0.45\linewidth]{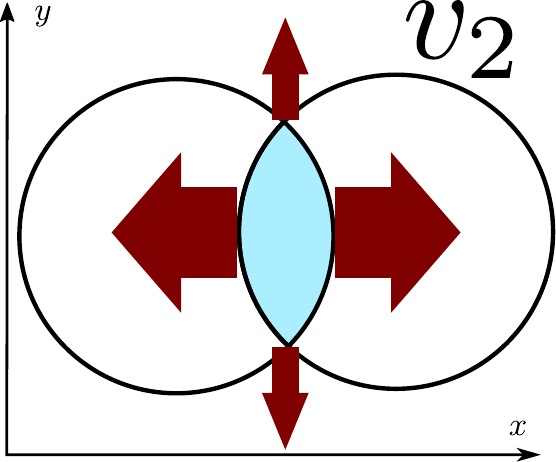}\qquad
\includegraphics[width=0.45\linewidth]{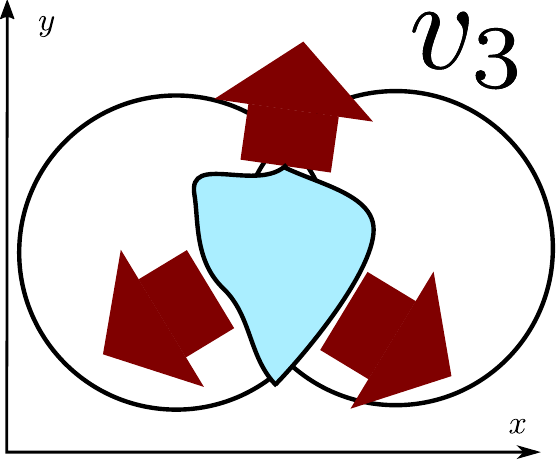}
\caption{Illustration of the overlap region (shaded area) of two colliding 
nuclei (circles). 
The arrows indicate the hydrodynamic flow. 
(a) The event averaged ``almond shaped" overlap of peripheral collisions 
driving the elliptic 
flow $v_2$.
(b)~Triangular deformations due to event-by-event fluctuations causing the 
triangular flow $v_3$.}
\label{fig:flow}
\end{figure}

\begin{figure}
\centering
\includegraphics[width=\linewidth]{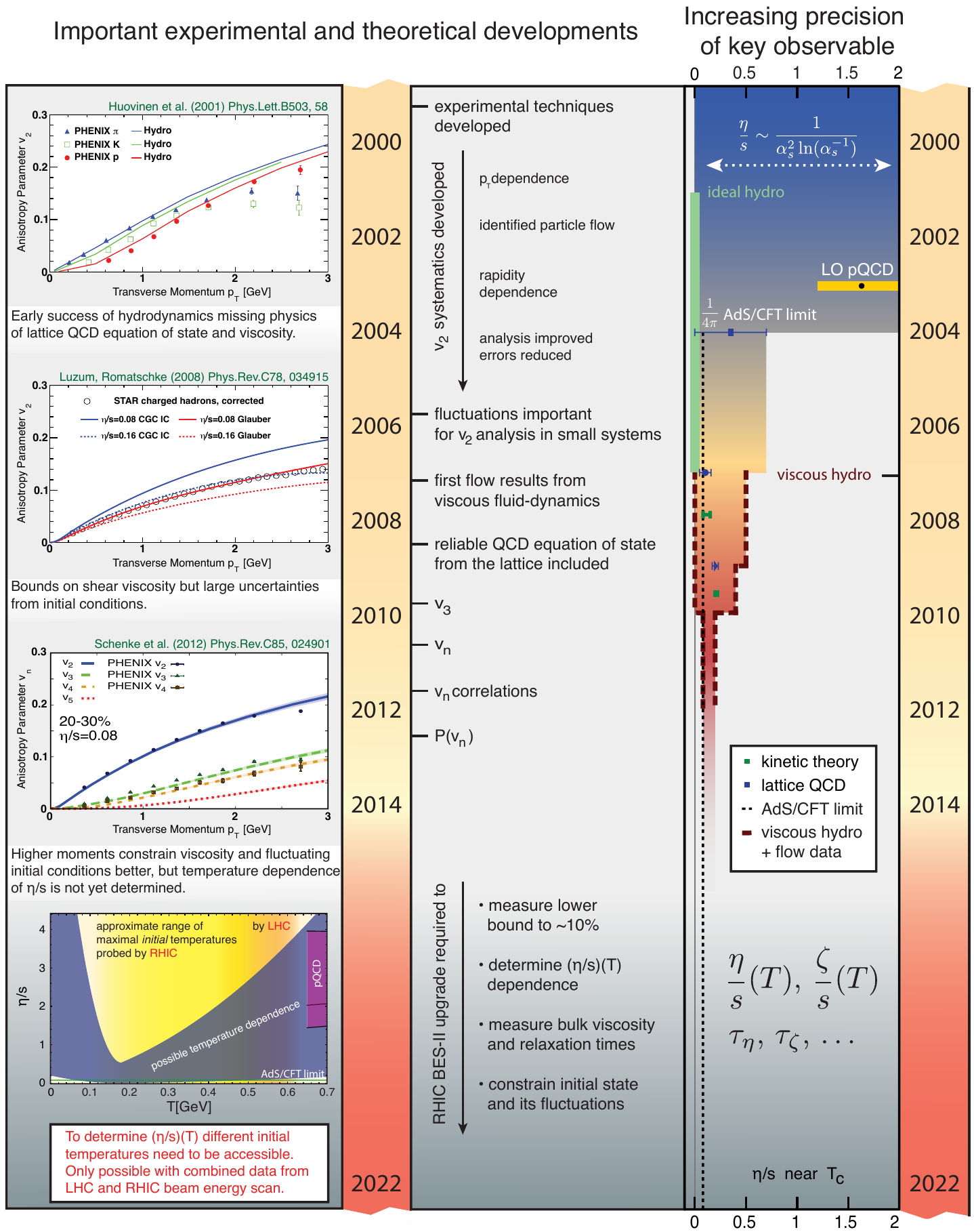}
\caption{The early years of soft heavy ion observables in the RHIC. Figure 
taken 
from \Ref{NSAC}\label{fig:whitepaper-cropped}}
\end{figure}

Since the start of the RHIC over 15 years ago, heavy ion physics has 
progressed  tremendously (see the timeline of RHIC physics in 
\Fig{fig:whitepaper-cropped}). 
The first results of Au+Au collisions at RHIC showed that the 
produced particle 
distribution in the plane 
transverse to the beam has a significant azimuthal anisotropy quantified by the
elliptic flow coefficient $v_2$
\begin{equation}
v_2 = 
\frac{\left<p_x^2\right>-\left<p_y^2\right>}{\left<p_x^2\right>+\left<p_y^2\right>},
\end{equation}
where $\left<p_x^2\right>$ and $\left<p_y^2\right>$ are momentum averages of 
observed particles relative to the event plane (see 
\Fig{fig:flow}(a))~\cite{Ackermann:2000tr}.
 The presence of the elliptic flow was a 
strong evidence that QGP expansion is driven by hydrodynamic pressure 
gradients. In 
peripheral 
collisions  the average overlap of two colliding nuclei is ``almond 
shaped" and the pressure gradients are the 
largest along the short axis of the ellipsoid. Then anisotropically flowing 
medium emits particles boosted in the flow 
direction and therefore the observed particles have a momentum anisotropy 
approximately
proportional to the initial elliptic deformation, or eccentricity, 
$\varepsilon_2$ 
\begin{equation}
\varepsilon_2 
=-\frac{\left<x^2\right>-\left<y^2\right>}{\left<x^2\right>+\left<y^2\right>},\quad
 v_2\propto \varepsilon_2.\label{eq:epssimple}
\end{equation}
Here $\left<x^2\right>$ and $\left<y^2\right>$ are coordinate averages in the 
center of mass frame of the overlap energy density.  
At the early years of RHIC, observed momentum anisotropy was reasonably well 
described 
by  modeling QGP as ideal fluid (see \Fig{fig:whitepaper-cropped} top 
left). Later, the importance of viscous effects was recognized and heavy ion 
collisions became a promising way of determining the transport coefficients of 
high density hot nuclear mater. Relativistic viscous hydrodynamic 
simulations favored viscosity over entropy $\eta/s$ value quite 
close 
to the quantum bound $\eta/s\geq 1/(4\pi)$  obtained for a  certain class of 
supersymmetric gauge theories (\Fig{fig:whitepaper-cropped} center 
left)~\cite{Teaney:2003kp,Romatschke:2007mq,Policastro:2002se,Kovtun:2004de,Danielewicz:1984ww}.
 The accurate  determination of viscosity and other 
transport 
coefficients is an ongoing work, which requires strong constraints on the 
initial conditions and detailed understanding of the many stages of heavy ion 
collisions~\cite{Antinori:2016zxe,Shen:2015msa}.

The elliptic flow associated with eccentricity in initial state
does not vanish even in central collisions. It was later understood that on 
event-by-event basis the initial geometry is not smooth, but fluctuates around 
the 
average geometry due to random 
positions of a finite number of colliding nucleons (see 
\Fig{fig:flow}(b))~\cite{Alver:2010gr}. 
The 
initial state 
fluctuations can produce any order geometric deformations: elliptical 
$\varepsilon_2$, 
triangular $\varepsilon_3$, quadruple $\varepsilon_4$, and other harmonics, 
which in 
turn 
generate momentum anisotropies, 
and  are quantified by flow harmonics 
$v_n$ 
(see 
\Sect{topics} for the definitions).
Higher order 
harmonics are damped by 
the dissipative effects in the QGP~\cite{Staig:2011wj,Schenke:2011bn}, but 
thanks to small viscosity 
harmonics up to 
$n=6$ were measured experimentally at the 
LHC~\cite{Chatrchyan:2013kba,ATLAS:2012at,ALICE:2011ab} and the RHIC 
\cite{Adare:2011tg,Adams:2003zg,Adamczyk:2013waa}, and successfully reproduced 
(and sometimes predicted) by viscous hydrodynamic 
simulations 
(see \Fig{fig:whitepaper-cropped})~\cite{Schenke:2011bn,Gale:2012rq}.

Improvements in experimental data, better description of initial state 
fluctuations, and 
sophistication of 
hydrodynamic evolution enabled physicists to study  detailed 
properties of flow harmonics $v_n$. For example, the event-by-event probability 
distribution of flow magnitudes $p(v_n)$, the nonlinear correlations between 
different harmonic orders, e.g. $v_4$ and $v_2^2$, and flow anisotropy 
dependence on transverse momentum $p_T$ and rapidity 
$\eta$~\cite{Heinz:2013th,Luzum:2013yya,Gale:2013da}. However, despite the 
substantial 
success, the 
heavy ion collision picture is not yet complete. 
One of the main theoretical challenges in heavy ion collision physics is the 
description of the initial moments of the collision. The initial conditions 
is a primary source of uncertainties in hydrodynamic modeling of QGP 
expansion~\cite{Antinori:2016zxe}. In addition, the applicability of 
hydrodynamic description 
at early times raises questions of how 
equilibrates~\cite{Romatschke:2016hle,Kurkela:2016vts,Chesler:2016vft,Gelis:2016rnt}.
 The 
apparent flow signals from small, 
short lived collision 
systems like p+Pb and p+p makes this issue especially 
urgent~\cite{Loizides:2016tew,Stankus:2016usz}.
Finally, in order to improve our understanding of the hot and dense 
nuclear matter, one must consider additional physical effects, e.g. large 
stochastic fluctuations  near the QCD 
critical point~\cite{Stephanov:1998dy}. To address these important issues, we 
pursued several 
different 
directions, outlined in \Sect{outline}, of improving and extending the current 
heavy ion collision picture.

\section{Topical review}
\label{topics}
In this section we briefly review the main concepts of heavy 
ion 
collision physics pertaining to this work. When a concrete collision system is 
needed, we  
 consider 
lead-lead ($^{207}\text{Pb}+^{207}\text{Pb}$) 
collisions at the LHC
center of mass energy $\sqrt{s}=2.76\,\rm{TeV}$\footnote{We use the natural 
units 
 $c=\hbar=k_\text{B}=1$ in which
\begin{alignat*}{3}
&1\,\text{[Length]}\cdot\text{[Energy]} &&= \hbar\cdot c &&= 
 0.1973\,\text{GeV}\cdot\text{fm},\\
&1\,\text{[Length]}\cdot\text{[Time]}^{-1}&&=c&&=2.998\times 
 10^{23}\,\text{fm}\cdot\text{s}^{-1},\\
&1\,\text{[Energy]}\cdot\text{[Temperature]}^{-1}&&=k_B&&=8.617\times10^{-13}\,\text{GeV}\cdot\text{K}^{-1},\\
&1\,\text{[Energy]}\cdot\text{[Mass]}^{-1}&&=c^2&&=5.610\times10^{26}\,\text{GeV}\cdot\text{kg}^{-1}.
\end{alignat*}
 } per nucleon pair (or  
$\sim46\,\mu\text{J}$ for each nucleus).
 For  more complete review please see 
 Refs.~\cite{Heinz:2013th,Teaney:2009qa,Luzum:2013yya,Gale:2013da,Romatschke:2009im}.

\subsection{Longitudinal geometry and Bjorken expansion}
\label{sec:bjorken}

\begin{figure}
\centering
\includegraphics[scale=1.2]{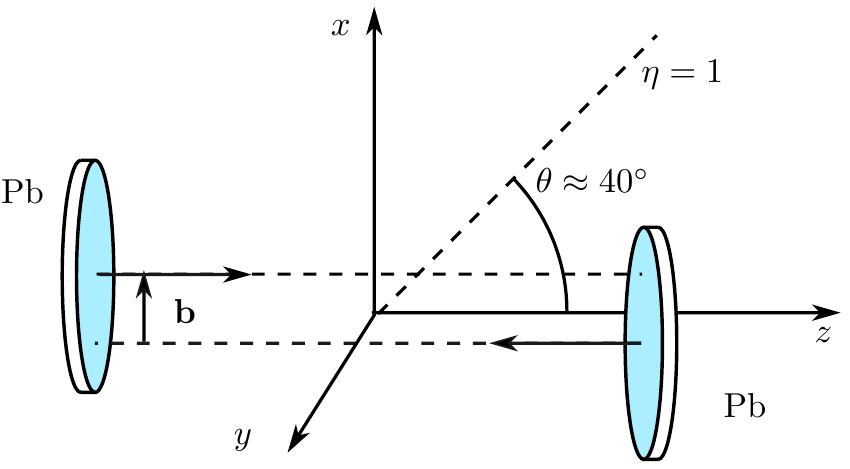}
\caption{The Lorentz contracted 
nuclei traveling along the $z$-axis collide with the impact parameter $\b$, 
which in each event  is 
randomly oriented in 
the transverse $x\text{--}y$ plane. The observed particle 
orientation 
along the 
beam 
axis is given by pseudo-rapidity $\eta$, \Eq{eq:pseudoeta}. 
\label{fig:rapidity}}
\end{figure}

The high center of mass energy of ultra-relativistic ion collisions allows 
considerable simplification of the longitudinal kinematics  (see 
\Fig{fig:rapidity}). In the $z$-axis 
direction (along the beam), the nuclei are traveling at 
$99.99998\%$ of speed of light and are Lorentz contracted by a relativistic 
factor of 
$\gamma \sim 1500$, so in the laboratory frame their passage can be considered 
instantaneous. 

The direction of detected particles along 
the beam axis is given 
in terms of pseudo-rapidity $\eta$
\begin{equation}
	\eta_\text{pseudo}\equiv\frac{1}{2}\ln\frac{|\p|+p_z}{|\p|-p_z}=-\ln \tan 
	\frac{\theta}{2}\label{eq:pseudoeta}
\end{equation}
where $\theta$ is the angle from the beam axis and $\p$ is the momentum of a  
freely moving 
particle. Note 
that  $\eta=1$ corresponds 
to $\theta\approx 40^{\circ}$ and $\eta = 2.5$ to $\theta\approx9.4^{\circ}$ 
(typical acceptance windows in a detector). 
When the mass of a particle can be neglected, $E\approx |\p|$,  
the 
pseudo\nobreakdash-rapidity  
agrees with 
the particle momentum rapidity $y$ defined as
\begin{equation}
	y_\text{particle}\equiv\frac{1}{2}\ln\frac{E+p_z}{E-p_z}.\label{eq:rapidity}
\end{equation}

The heavy ion collision picture known as the Bjorken 
expansion was a result of the important observation that the measured 
inclusive particle production is 
approximately flat in central rapidity $\eta\approx0$ 
region~\cite{Bjorken:1982qr}, (see the current data in \Fig{fig:production}). 
The key assumption of this picture is that for the central rapidity 
region the 
system 
is boost invariant---system evolution and particle production looks the same in 
all
reference frames related by a Lorentz boost in $z$-direction.  {
\begin{figure}
\centering
\includegraphics[height=0.5\linewidth, trim={275 0 0 0}, clip]{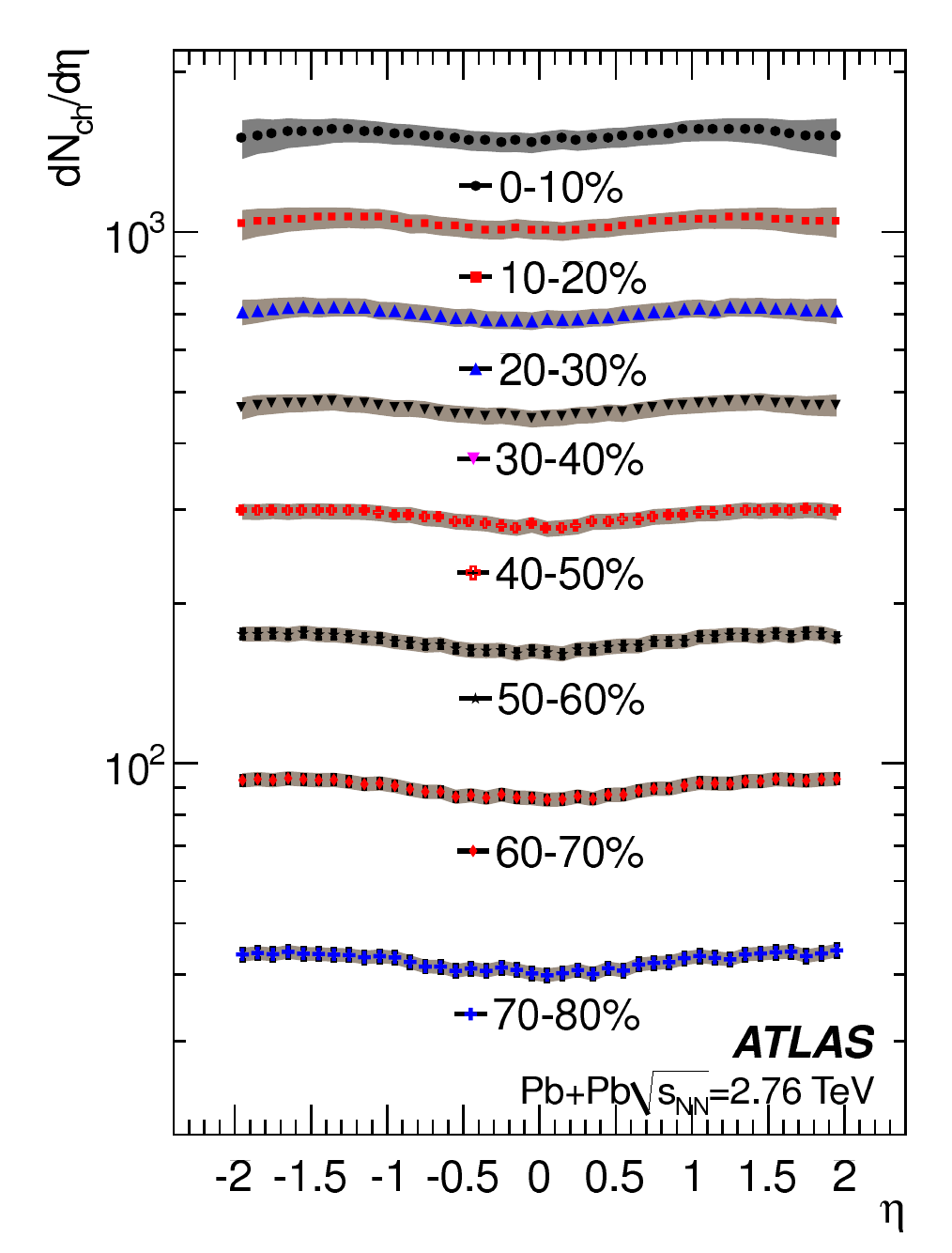}\qquad
\includegraphics[height=0.5\linewidth]{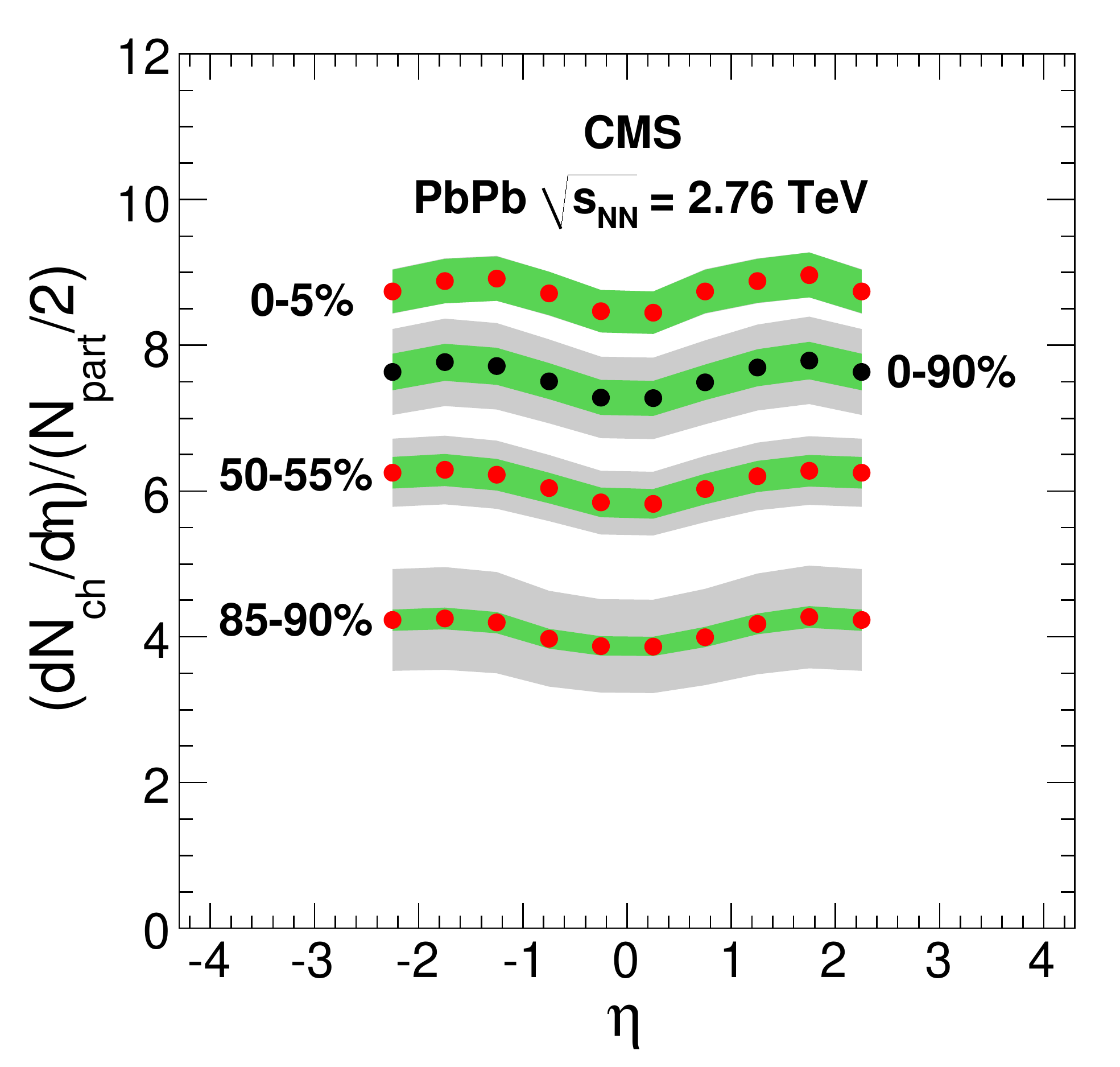}
\caption{Approximately boost invariant measured charged particle multiplicity 
$dN_\text{ch}/d\eta$ in various centrality bins by (a) 
ATLAS~\cite{ATLAS:2011ag} and (b) CMS~\cite{Chatrchyan:2011pb} experiments. 
\label{fig:production}}
\label{fig:fig02atlas}
\end{figure}
}
Such kinematics 
is best 
studied in the coordinates of ``proper time" $\tau=\sqrt{t^2-z^2}$ and 
space-time 
rapidity\footnote{Note 
that $\tau$ is the real proper time  only for observers moving with constant 
velocity 
$v^z=z/t$.}
\begin{equation}
	\eta_\text{space-time}\equiv\frac{1}{2}\ln\frac{t+z}{t-z}.\label{eq:spacetimeeta}
\end{equation}
In $\tau\text{--}\eta$ coordinates, the (mostly positive) metric becomes
\begin{equation}
ds^2=-d\tau^2+dx^2+dy^2+\tau^2 d\eta^2.
\end{equation}
and  boost-invariance in $z$ direction is equivalent to translational 
invariance in $\eta$ coordinate.  
The typical longitudinal momentum of produced particles is much larger than the 
transverse 
momentum  $p^z\gg p_\perp$ and the subsequent expansion does not change the 
particle rapidity significantly. Since $p^z/E=v^z\approx z/t$ 
produced particles with momentum rapidity  $y$ will reach the 
detector  at space-time rapidity $\eta\approx y$.

\Eqs{eq:pseudoeta}, \eq{eq:rapidity}, \eq{eq:spacetimeeta} define three 
different quantities: pseudo, 
particle and space-time rapidities, which are often used interchangeably when 
making estimates in heavy 
ion physics. Experimental results are given in 
pseudo\nobreakdash-rapidity 
$\eta$, which directly relates to the  detector coordinate system. 
Meanwhile simulations are often done using space-time rapidity $\eta$. The 
invariant mass of a pion---the most abundant species of 
detected hadrons---$m_\pi \approx 140\,{\rm MeV}$ is small compared to the 
typically measured energy of pions with transverse momentum $p_\perp \sim 
1\,\text{GeV}$. Therefore massless pion 
approximation is adequate and indeed all three types of rapidities are 
approximately equal~\cite{McLerran:2001sr}
\begin{equation}
\eta_\text{pseudo}\approx y_\pi \approx \eta_\text{space-time}.
\end{equation}
In the subsequent discussion we will not make a distinction between 
different types of rapidity and use the boost invariant picture of heavy ion 
collisions. 

\subsection{Initial conditions}
One of the least understood stages in heavy ion collisions is the period 
between the passing of two nuclei and the formation of approximately 
equilibrated plasma around $\tau\sim 1\,\text{fm}$~\cite{Huovinen:2002rn}. The 
first principle 
QCD 
calculation of heavy ion nucleus wavefunction  is not 
yet possible and a number of different models are used to describe the 
collision process~\cite{Lappi:2015jka}.

\begin{figure}
\centering
\includegraphics[scale=1.2]{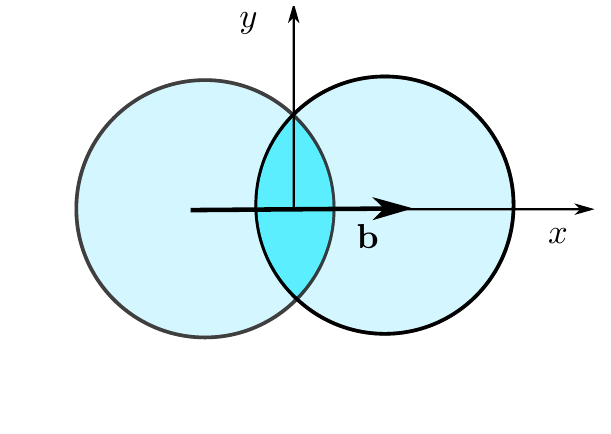}
\quad
\includegraphics[width=0.40\linewidth]{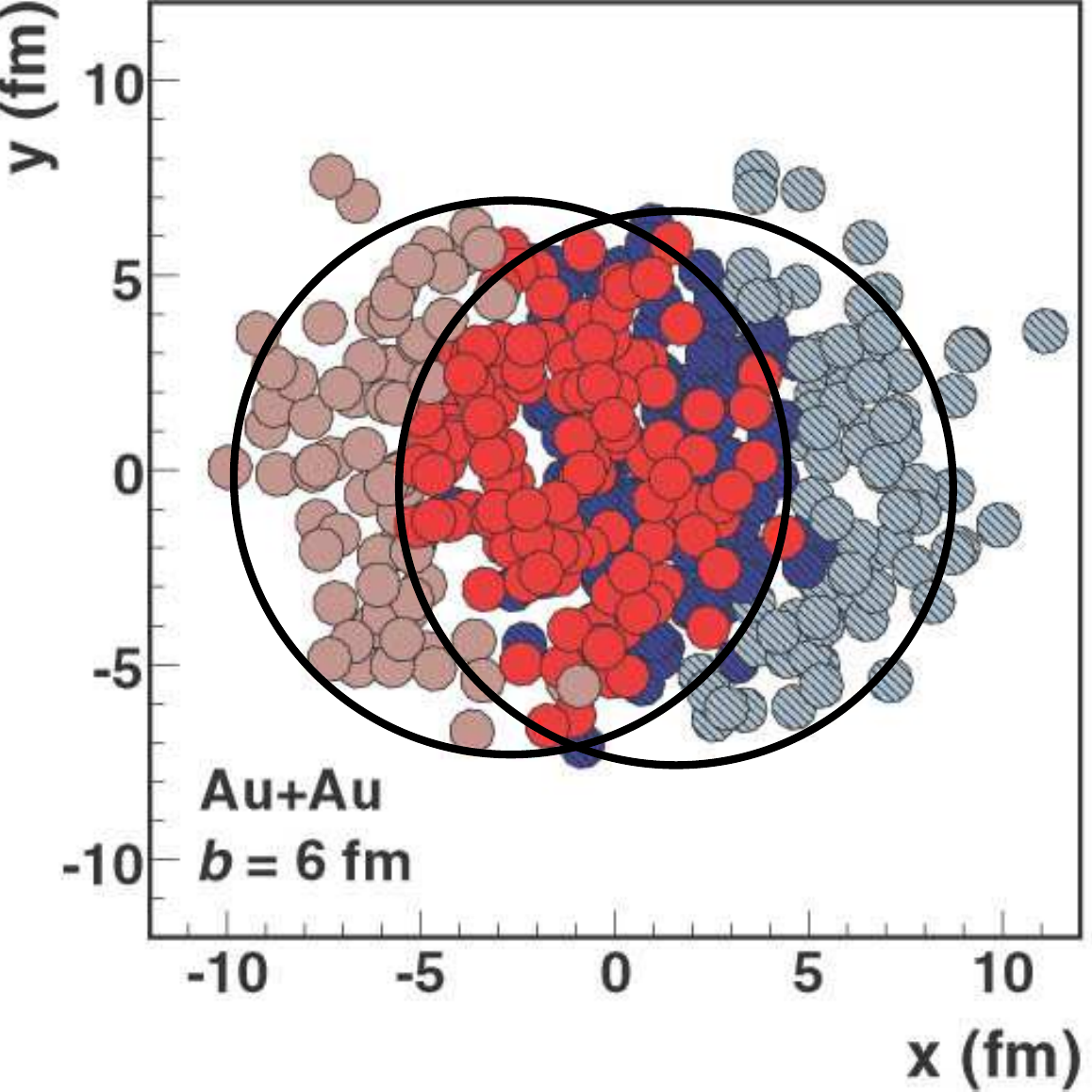}
\caption{Event plane view of the colliding nucleons along the beam axis. (a)  
The ``almond shaped" overlap region (generally impact parameter $\b$ is 
randomly oriented in the transverse plane).
(b) Monte Carlo Glauber sampled nucleon positions of participant 
nucleons (dark colored circles) and spectators (light colored circles) 
(Figure 
adapted from 
\Ref{Miller:2007ri}).\label{fig:eventplane}}
\end{figure}

Two  nuclei collide when they pass each other at distance $|\mathbf{b}|\lesssim
2 R_\text{rms}$, where $R_\text{rms}\sim 
5\,\text{fm}$ is the 
root mean squared charge radius of a Pb nucleus (see  
\Fig{fig:eventplane}(a))~\cite{Angeli201369}.
Current heavy ion collision models are based on a picture of a heavy 
ion 
nucleus composed of nucleons, whose positions are randomly distributed 
according to the 
average 
nuclear charge 
density function 
and subject to  certain physical 
constrains, e.g. nucleons can not overlap due to 
nucleon\nobreakdash-nucleon repulsion (see \Fig{fig:eventplane}(b) for an 
example of nucleon positions sampled according to a popular Monte Carlo Glauber 
(MCG) procedure)~\cite{Miller:2007ri,Blaizot:2014wba,Alver:2008aq}. 
At the LHC energies the two nuclei pass each other almost instantaneously  
$\Delta 
t_\text{passing}\sim 2R_\text{Pb}/(c\gamma) \sim 0.01\,\text{fm}$.  
Furthermore, the longitudinal boost of the nucleus is much 
larger 
than the intrinsic energy scale $\Lambda_\text{QCD}\sim 
200\,\text{MeV}$ inside a 
nucleus and the 
transverse nuclear 
dynamics can be neglected\footnote{$\Lambda_\text{QCD}$ is (somewhat 
ill-defined) energy 
scale at which the 
strong coupling $\alpha_s$ diverges, i.e. it characterizes the non-perturbative 
QCD energy scale~\cite{Olive:2016xmw}.}. Then the nucleons appear frozen in the 
transverse 
plane during the passage and travel on straight lines along the beam 
axis~\cite{Miller:2007ri,McLerran:2001sr}.

\begin{figure}
    \begin{center}
        \includegraphics[width=0.5\textwidth]{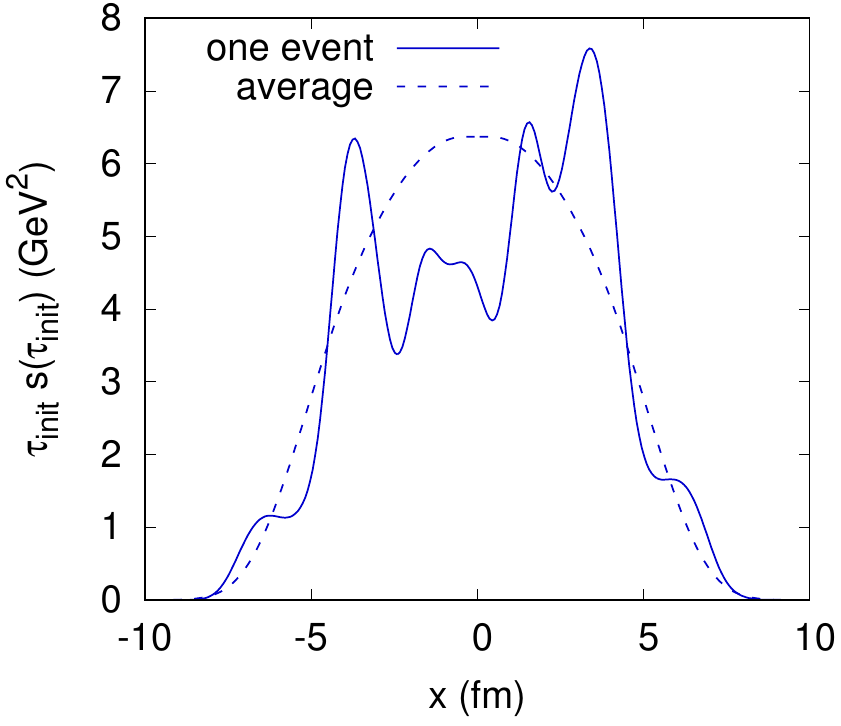}
        \caption{A typical entropy density profile (times 
        ${\ti}\sim0.6\,\text{fm}$) for 
            a single event used 
            as an initial condition in 
            current hydrodynamic simulations at the LHC for a 0-5\% centrality 
            class~\cite{Mazeliauskas:2015vea}.  
            An
        event averaged initial condition is shown by the dashed line. 
    \label{introichydro}}
    \end{center}
\end{figure}

The principal difference between numerous initial state models (MCG, MCKLN, 
MCrcBK, IP-Glasma, EKRT) is in their 
treatment of 
which nucleons collide and how energy is deposited in the collision 
region~\cite{Lappi:2015jka}.
The nucleons which do interact 
are called  wounded 
nucleons or participants, while the remaining undeflected nucleons are known 
as spectators. Interaction between participant nucleons results in the extended 
energy deposition in the rapidity direction, from which the  Quark Gluon Plasma 
is created.
 In 
the popular Monte Carlo Glauber model colliding nucleons 
are 
treated like spherical balls, which collide with one or more nucleons from the 
other nucleus if they are less than one diameter $D$ away in the transverse 
plane
\begin{equation}
D=\sqrt{\sigma_{NN}/\pi}\sim 1.5\,\text{fm}.
\end{equation}
Here $\sigma_{NN}$ is a nucleon-nucleon inelastic cross-section obtained from 
p+p collision experiments~\cite{Miller:2007ri}. When wounded 
nucleons are 
identified, one has to decide on energy or entropy deposition in the event. The 
equilibration process of colliding nucleons is highly 
nontrivial and largely unsolved 
problem~\cite{Gelis:2016rnt,Chesler:2016vft,Kurkela:2016vts}, 
but essential 
features of 
the process can be efficiently 
parametrized by a handful of quantities and then fitted to reproduce the 
measured yields of particles~\cite{Hirano:2009ah,Qiu:2013wca}. According to the 
two-component 
model 
used 
in simulations in \Chap{chap:pca}, a lump amount of entropy density per 
rapidity is 
deposited in the transverse plane for each participant nucleon with adjustments 
for multiple binary collisions. This energy density
distribution (example shown in \Fig{introichydro}) is then passed to the 
subsequent 
hydrodynamic evolution at time 
$\tau_\text{init}\sim 0.6\,\text{fm}$.

 A considerable success of heavy ion 
simulations using MC Glauber initial conditions indicates that, indeed, the 
wounded nucleon distribution in the transverse collision area is the main 
source of variance in the collision~\cite{Alver:2010gr,Qiu:2011iv}.
A more  microscopic description of the early stages 
of hadron 
collisions can be obtained using the  idea of Color Glass Condensate 
(CGC)~\cite{Iancu:2002xk,Iancu:2003xm,Gelis:2010nm, 
Gelis:2007kn, Lappi:2011ju,Albacete:2014fwa}, which is motivated by the 
precision 
data from the 
DESY Hadron-Electron Ring Accelerator 
(HERA)~\cite{Aaron:2009aa}.
At large collision energies the central rapidity region is dominated by the 
interactions of small Bjorken-$x$ partons (mostly gluons)\footnote{Bjorken-$x$ 
is the fraction of the total longitudinal 
nucleus momentum carried by a parton~\cite{Peskin}.}. The small $x$ gluons are 
produced (long before the collision) by highly boosted valence color charges, 
which evolve at much longer 
time scales and appear frozen, i.e. like Color Glass. Gluon density saturates 
at very high values $\rho \sim 1/\alpha_s \gg 1$, when gluon fusion becomes 
as favorable as collinear emission. Such high density gluons can be described 
as classical fields governed by Yang-Mills equations of 
motion~\cite{Krasnitz:1999wc}. 
The classical field evolution describes the energy liberation from the 
color currents, but does not 
lead 
to equilibrium and cannot be smoothly matched to hydrodynamic 
evolution~\cite{Berges:2013eia,Berges:2013fga}. 
In 
\Chap{chap:ekt} we use effective kinetic theory described in \Sect{highT} to 
construct a practical implementation of the ``bottom-up'' thermalization 
scenario matching CGC initial conditions to 
hydrodynamics
~\cite{Baier:2000sb,Kurkela:2016vts}.
However, even without the precise description of equilibration, heavy ion 
collision 
simulations using CGC initial conditions show improvements over MC Glauber 
initial conditions for some observables, e.g. probability distribution of the 
magnitude of flow harmonics~\cite{Gale:2012rq}.  

In \Chap{chap:pca} we use hydrodynamic heavy ion simulations with MC Glauber 
initial conditions, while in \Chap{chap:ekt} we study the equilibration of CGC 
inspired initial state gluon distribution function.

\subsection{Transverse geometry and spatial anisotropies}
The transverse collision geometry is the dominant source of variance in heavy 
ion collisions and its description is of paramount importance for the 
interpretation of many experimental 
observables~\cite{Heinz:2013th,Luzum:2013yya}.
 
It is customary to classify heavy ion collisions by their multiplicity---the 
number of produced charged particles (pions, kaons and protons) in each 
event~\cite{ATLAS:2011ag, 
Chatrchyan:2011pb,Aamodt:2010cz}. For the small impact parameter 
$|\mathbf{b}|$, i.e. central (or head on) collision, the number of interacting 
nucleons is large and such collisions are likely to produce more particles. 
Conversely, the collisions at large $|\mathbf{b}|$ are peripheral and low in 
multiplicity. 
However, centrality does not completely fix the initial geometry 
configuration due 
to fluctuating number of participants even at a fixed impact parameter. 
Nevertheless, comparing central, mid-central and peripheral centrality bins 
allows one to study the effects of average background geometry.

The positions of wounded nucleons (few tens to four hundred) are the most 
import ingredients in determining the initial transverse energy distribution in 
the collision.
 Note that because of the
finite number of wounded nucleons, participant distribution does not need to 
align with the overlap of average nuclear density illustrated in  
\Fig{fig:eventplane}(a) and thus the collision 
geometry 
fluctuates on event-by-event basis as shown in
\Fig{fig:eventplane}(b)~\cite{Alver:2010gr}. These event-by-event fluctuations 
in the transverse geometry 
are typically characterized by a series of eccentricity coefficients 
$\varepsilon_{n,m}$ defined in the center of mass frame  
as~\cite{Heinz:2013th,Luzum:2013yya}
	\begin{align}
		\varepsilon_{n,m} e^{in \Phi_{n,m}} &\equiv   -\frac{[r^m e^{in \phi 
		}]}{ [r^m] }.\label{eq:epsnm}
\end{align}
Here index $n$ is the order of the azimuthal harmonic $e^{in\phi}$, $m$ 
is 
the power of  the
radial weight $r^m$, and phase angle $\Phi_{n,m}$ (defined modulo 
${2\pi}/{n}$) indicates the orientation of the particular azimuthal 
anisotropy relatively to the impact parameter $\mathbf{b}$. For 
example, the elliptic eccentricity given by \Eq{eq:epssimple} 
corresponds to the case $n=m=2$ and $\Phi_{2,2}=0$, i.e. the situation 
illustrated in \Fig{fig:flow}(a).

Eccentricity is not experimentally measurable quantity and the precise way 
of 
computing them differs among 
authors, so brackets $[\ldots]$ can denote averaging with respect to  either
transverse energy or entropy density. Often the normalization factor 
$[r^m]$ is 
computed event-by-event together with the nominator to ensure that 
eccentricities are bounded above, $|\varepsilon_{n,m} e^{in 
\Phi_{n,m}}|<1$. In 
\Chap{chap:pca}, we deviate from the common practice and use 
eccentricities defined in terms of 
transverse entropy distribution with event-class averaged normalization $R_{\rm 
rms}^m$. 

On average, peripheral collisions have an ``almond shaped" 
overlap, therefore  $\varepsilon_{2,2}$, characterizing the elliptical 
deformation of collision geometry (see \Fig{fig:eventplane}(a)), is the 
dominant eccentricity in all but the 
most central collisions. During the QGP 
expansion geometric anisotropies $\varepsilon_n$ are converted to momentum  
anisotropies $v_n$ of 
detected particles. A conventional minus sign in the eccentricity definition, 
\Eq{eq:epsnm}, is chosen for a positive correlation between the two types of 
anisotropies.  Thanks to event-by-event fluctuations, many other 
eccentricities can be excited, but the subsequent 
dissipative evolution of QGP suppresses higher 
harmonics~\cite{Staig:2011wj,Schenke:2011bn}.

In \Chap{chap:pca} we show that standard  eccentricities alone, 
given by \Eq{eq:epsnm}, are not 
sufficient to capture all relevant information about the initial geometry and 
generalized eccentricities have to be considered to improve the correlation 
between the initial geometry and final momentum anisotropies.

\subsection{Hydrodynamics}
\label{hydro}

Significant momentum anisotropies $v_n$ of the observed particle spectrum in 
heavy ion collisions are strongly  suggestive of the collective hydrodynamic 
motion in the expanding QGP fireball~\cite{Ollitrault:1992bk}. Indeed, 
phenomenological models  
have been 
remarkably successful in describing many soft observables and generally 
predicts the hydrodynamic phase to last from $\tau\sim1\,\text{fm}$ to 
$\tau\sim10\,\text{fm}$~\cite{Heinz:2013th,Teaney:2009qa,Luzum:2013yya,Gale:2013da,Romatschke:2009im}.

Hydrodynamics is the long 
wavelength effective theory based on fundamental conservation laws 
of energy and momentum\footnote{There can by additional conserved charges, 
e.g. baryon number with an associated chemical potential 
$\mu_B$.}~\cite{LandauFluids}
\begin{align}
&D_\mu T^{\mu\nu}=0,\label{eq:cons}
\end{align}
where $D_\mu$ is the covariant derivative\footnote{$D_\mu T^{\mu\nu}\equiv 
\partial_\mu 
T^{\mu\nu}+\Gamma^{\mu}_{\mu\rho}T^{\rho\nu}+\Gamma^\nu_{\mu\rho}T^{\mu\rho}$. 
For $\tau\text{--}\eta$ 
coordinates 
discussed in \Sect{sec:bjorken} the non-zero Christoffel symbols are 
$\Gamma_{\eta\eta}^\tau = \tau$ and 
$\Gamma_{\eta\tau}^\eta=\Gamma_{\tau\eta}^\eta=\tau^{-1}$. }.
 The simplifying assumption of hydrodynamics is that $T^{\mu\nu}$ is a function 
 of only four independent fields: the rest-frame energy density 
 $e$ and the  time-like flow velocity $u^\mu$  ($u^\nu u_\nu=-1$), satisfying
\begin{equation}
 T^{\mu\nu}u_\nu =- e u^{\mu}.\label{eq:fields}
\end{equation}
The system of equations, \Eq{eq:cons}, is closed by specifying the constitutive 
equations for the energy momentum tensor 
$T^{\mu\nu}(e(\x),u^\mu(\x))$, which for the smooth background fields  can 
be systematically expanded in gradients
\begin{equation}
T^{\mu\nu}(e,u^\mu) = T^{\mu\nu}_\text{ideal} + T^{\mu\nu}_\text{1st 
order}(\partial e,\partial u^\mu)+T^{\mu\nu}_\text{2nd order}(\partial^2  
e,\partial^2 u^\mu)+\mathcal{O}(\partial^3).
\end{equation}
Then truncation at the $n$th order in gradients defines the $n$th order 
hydrodynamics. 
\subsubsection{Ideal}
The 
zeroth order truncation gives the ideal energy momentum tensor
\begin{align}
 T^{\mu\nu}_\text{ideal}=e u^\mu u^\nu+p\Delta^{\mu\nu},\label{eq:Tideal}
\end{align}
where $\Delta^{\mu\nu}=g^{\mu\nu}+u^\mu u^\nu$ is the projection operator 
orthogonal to $u^\nu$ and $p$  is the pressure given by the equation 
of state 
$p=P(e)$ (see \Sect{sec:eos}). At high temperatures $T\gg T_c\sim 
155\,\text{MeV}$, QGP is often approximated 
as a 
gas of massless particles with constant speed of sound $c_s^2\equiv\partial 
p/\partial e={1}/{3}$, i.e. $p=c_s^2 e$.

Using \Eqs{eq:cons} and  \eq{eq:Tideal}  it is easy to show that the ideal 
equations 
of motion for energy density $e$ and flow velocity $u^\mu$ are
\begin{align}
D e = -(e+p)\theta,\quad
D u^\mu = - \frac{\nabla^\nu p}{(e+p)}\label{eq:eomideal},
\end{align}
with the following covariant notation
\begin{equation}
D \equiv u^\mu D_\mu,\quad \nabla_\mu \equiv 
	\Delta_{\mu}^{\phantom{\mu}\nu} D_\nu,\quad \theta\equiv D_\mu u^\mu.
\end{equation}

An important translationally invariant solution of \Eq{eq:eomideal} in 
$\tau\text{--}\eta$ coordinates is the 
Bjorken flow. 
It corresponds to
a stationary fluid, where $u^\tau=1$ and $\theta\equiv D_\mu 
u^\mu = 
{\tau^{-1}}$. The   energy density $e$ then obeys
\begin{equation}
\partial_\tau e = -\frac{e+p}{\tau}
\end{equation}
and   $e \propto\tau^{-1-c_s^2}$ for $c_s^2=\text{const}$. 

\subsubsection{Navier-Stokes}
Despite the initial success of ideal hydrodynamics in heavy ion collisions, it 
was recognized that there are viscous corrections which contributes at first 
order in gradients~\cite{Danielewicz:1984ww,Romatschke:2007mq,Teaney:2003kp}. 
The only tensor 
structures built 
from energy and 
velocity 
gradients consistent with symmetries and \Eq{eq:fields} are
\begin{equation}
T^{\mu\nu}_\text{1st 
order}(\partial e,\partial u^\mu)= \Pi \Delta^{\mu\nu} + 
\pi^{\mu\nu}.
\end{equation}
where $\Pi(\partial e,\partial u^\mu)$ is a scalar and $\pi^{\mu\nu}(\partial 
e,\partial u^\mu)$ is a symmetric, traceless and 
$u^\mu$-orthogonal tensor~\cite{Kovtun:2012rj}.  The lower order equations of 
motion, 
\Eqs{eq:eomideal}, can be used to replace energy gradients $\partial e$ with 
gradients in velocity $\partial u^\mu$. Then the first order viscous 
corrections are determined uniquely 
up 
to the overall coefficient
\begin{equation}
\Pi = - \zeta \theta, \quad \pi^{\mu\nu}=-\eta \sigma 
^{\mu\nu},\label{eq:1storder}\footnote{For Bjorken flow $\theta=\tau^{-1}$, 
$\sigma^{xx}=\sigma^{yy}=-\frac{2}{3} \tau^{-1}$ and 
$\tau^2\sigma^{\eta\eta}=\frac{4}{3}\tau^{-1}$.}
\end{equation}
where $\sigma^{\mu\nu}$ is expressed in velocity gradients $D^\mu u^\nu$
\begin{equation}
\sigma^{\mu\nu}=2\left<D^\mu 
u^\nu\right>\equiv\Delta^{\mu\alpha}\Delta^{\nu\beta}(D_\alpha u_\beta+D_\beta 
u_\alpha)-\frac{2}{3}\Delta^{\mu\nu}\Delta^{\alpha\beta}D_\alpha u_\beta.
\end{equation}

Transport coefficients $\zeta$ and $\eta$ are called 
bulk and shear viscosity respectively and are of great interest in heavy ion 
collision physics.
The effective constant shear viscosity is found to be in the range  
$1\lesssim 4\pi \eta/s\lesssim   3\text{--}6$~\cite{Luzum:2012wu,Song:2010mg}, 
and quite close to 
the so called 
quantum bound of $\eta/s=1/(4\pi)\approx0.08$.
Constraining the temperature dependence of $\eta/s$ is an ongoing 
work~\cite{Bernhard:2016tnd}. Bulk viscosity $\zeta$ is expected to be small 
and perhaps become relevant close to the
QCD transition temperature 
$T_c\sim155\,\text{MeV}$~\cite{Karsch:2007jc,Arnold:2006fz, Meyer:2007dy}. 
Therefore $\zeta$ 
is often neglected in 
hydrodynamic simulations of heavy ion collisions (but see, for 
example,~\cite{Ryu:2015vwa}).

\subsubsection{Second order}

The dispersion relation $\omega(k)$ of relativistic Navier-Stokes equations  
predicts superluminal propagation of large wavenumber $k$  
modes~\cite{Hiscock:1983zz}. To render the equations of motion causal, 
the 
shear stress 
tensor $\pi^{\mu\nu}$ is promoted to a dynamical 
field, and the first order 
constitutive equations, \Eq{eq:1storder},  are imposed through a relaxation 
type 
equation~\cite{Israel:1976tn}
\begin{equation}
\partial_\tau \pi^{\mu\nu} \sim - \frac{1}{\tau_\pi}(\pi^{\mu\nu}+\eta 
\sigma^{\mu\nu}).
\end{equation}
This is equivalent to modifying constituent equations at second order in 
gradients.
In 
conformal field theories\footnote{At high energies  $T$ is the only relevant
 scale and the QCD is approximately conformal. The invariance 
under scale transformations in $d$ dimensions constrains the conformal equation 
of state 
$p=\frac{1}{d-1}e$, bulk viscosity $\zeta/s=0$, and $\eta/s=\text{const}$. } in 
flat space-time there 
are only four such independent gradient structures\footnote{For Bjorken flow 
$\left< 
D\sigma^{\mu\nu} 
\right>  + \frac{1}{3} \sigma^{\mu\nu} 
\theta=-\frac{2}{3}\tau^{-1}\sigma^{\mu\nu}$ and 
$\Omega^{\mu\nu}=0$.}~\cite{Baier:2007ix}
\begin{align}
T^{\mu\nu}_\text{2nd order} &=  \eta \tau_{\pi} \left[ \left< 
D\sigma^{\mu\nu} 
\right>  + \frac{1}{3} \sigma^{\mu\nu} \theta \right] 
+ \lambda_1 \left< \sigma_{\phantom{\mu}\lambda}^{\mu} \sigma^{\nu 
\lambda} \right> +\lambda_2 \left< 
\sigma_{\phantom{\mu}\lambda}^{\mu} \Omega^{\nu\lambda} \right>  +\lambda_3 
\left< \Omega_{\phantom{\mu}\lambda}^{\mu} \Omega^{\nu\lambda} 
\right>
\label{eq:2ndorder}
\end{align}
and four second order transport coefficients $\tau_\pi, \lambda_1,\lambda_2$ 
and $\lambda_3$. Here $\Omega^{\mu\nu}$ stands for vorticity tensor
\begin{equation}
\Omega^{\mu\nu} = \frac{1}{2} \Delta^{\mu\alpha}\Delta^{\nu\beta}(\Delta_\alpha 
u_\beta - \Delta_\beta u_\alpha).
\end{equation}

For  strongly coupled $\mathcal{N}=4$ 
supersymmetric Yang Mills theory  the 
second order transport coefficients can be calculated 
analytically~\cite{Baier:2007ix,Bhattacharyya:2008jc}
\begin{equation}
\frac{\tau_\pi}{\eta/(sT)}=4-\ln 4,\quad\frac{\lambda_1}{\tau_\pi 
\eta}=\frac{1}{2-\ln 2},\quad \frac{\lambda_2}{\tau_\pi \eta}=-\frac{\ln 
4}{2-\ln 2},\quad\frac{\lambda_3}{\tau_\pi \eta}=0,\label{secondorder}
\end{equation}
where the shear viscosity over entropy 
ratio saturates the quantum bound $\eta/s=1/(4\pi)$.

Using $\pi^{\mu\nu}=-\eta 
\sigma^{\mu\nu}$ in \Eq{eq:2ndorder} one obtains a  dynamical 
equation for the shear-stress tensor $\pi^{\mu\nu}$ 
\begin{align}
\pi^{\mu\nu} &=-\eta \sigma^{\mu\nu} - \tau_{\pi} \left[ \left< 
D\pi^{\mu\nu} 
\right>  + \frac{4}{3} \pi^{\mu\nu} \theta \right] 
- \frac{\lambda_1}{\eta} \left< \pi_{\phantom{\mu}\lambda}^{\mu} \sigma^{\nu 
\lambda} \right> -\frac{\lambda_2}{\eta} \left< 
\pi_{\phantom{\mu}\lambda}^{\mu} \Omega^{\nu\lambda} \right> +\lambda_3 
\left< \Omega_{\phantom{\mu}\lambda}^{\mu} \Omega^{\nu\lambda} 
\right>.\label{eq:dynamical}
\end{align}
Then $e$,  $u^\mu$ and five independent components of $\pi^{\mu\nu}$ are 
evolved according to  \Eqs{eq:cons} and \eq{eq:dynamical}. 
For studies concerning observables 
at the central rapidity region (as it is the case in this manuscript), one 
often assumes boost 
invariance along the $z$-direction. Then dependence on rapidity $\eta$ 
coordinate is dropped, reducing the problem to the so called 2+1D hydrodynamics.

In \Chap{chap:pca} we perform hydrodynamic simulations using second order 
conformal hydrodynamic equations of motion \Eqs{eq:cons} and \eq{eq:dynamical} 
with $\eta/s=1/(4\pi)$ and second 
order transport coefficients given in \Eq{secondorder}, but  we employ 
non-conformal lattice equation of state (see \Sect{sec:eos}).
In \Chap{chap:noise} for the analytic 
calculations we use Navier-Stokes 
equations with constant speed of sound $c_s^2=1/3$.

\subsection{Equation of state}
\label{sec:eos}
\begin{figure}
\centering
\includegraphics[width=0.48\linewidth]{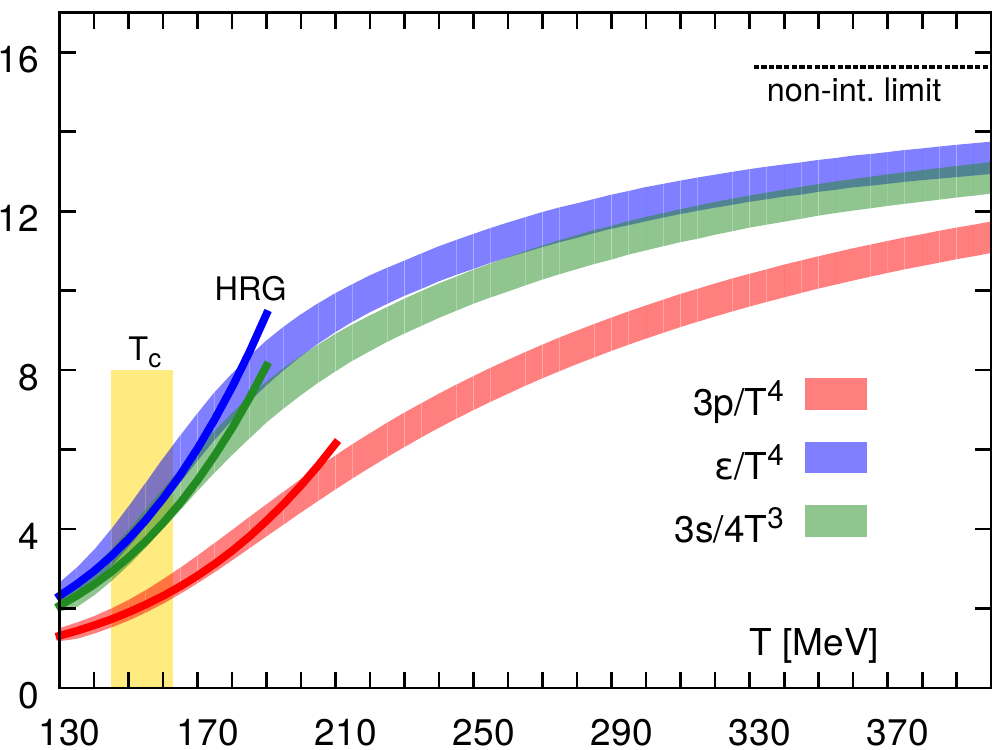}\quad
\includegraphics[width=0.48\linewidth]{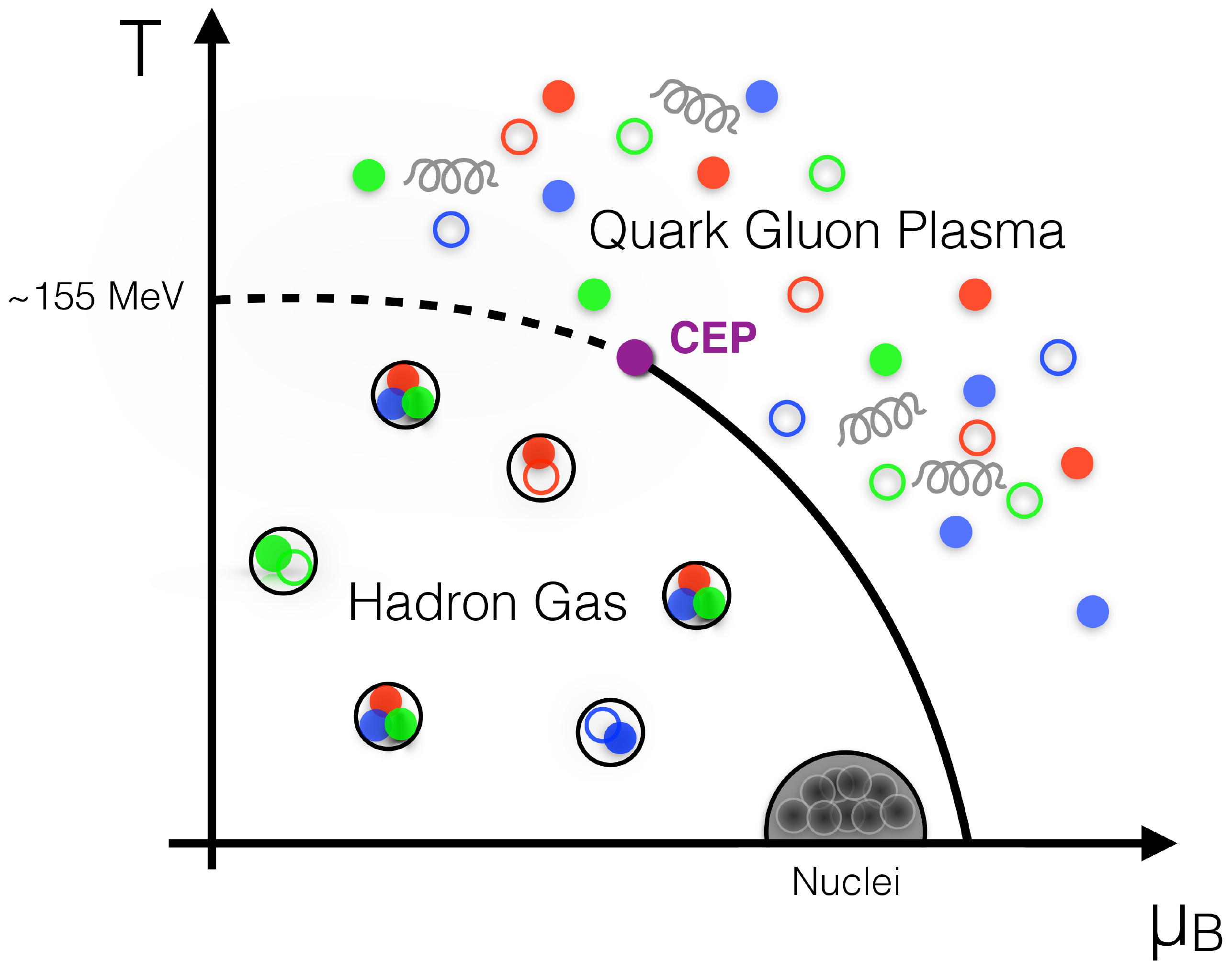}
\caption{
(a) State of the art lattice equation of state at zero chemical potential. 
Figure taken from 
\Ref{Bazavov:2014pvz}
(b) A sketch of QCD phase diagram as a function of temperature $T$ and 
baryon chemical potential $\mu_B$. The dashed line indicates a smooth 
crossover between hadronic and QGP phases, the solid line---a conjectured first 
order transition with second order critical end point (CEP). Figure taken from 
\Ref{Ding:2015ona}.}
\label{fig:eose-3pparam}
\end{figure}

One necessary ingredient for the hydrodynamic description of heavy ion 
collisions is 
the relation between energy density $e$,  baryon chemical potential $\mu_B$ and 
pressure $p$, i.e. the equation of state $p=p(e,\mu_B)$. Thanks to the enormous 
computational effort, the first principle 
calculations of the equation of state are available from 
Lattice QCD~\cite{Ding:2015ona}. Because of the 
famous sign problem for $\mu_B>0$, lattice 
computations are typically done at zero chemical potential. For high energy 
collisions the initial baryon number is negligible 
compared to the total produced number of particle-antiparticle pairs, therefore 
$\mu_B\approx 0$ is a reasonable approximation. 

At 
temperatures below 
the deconfinement temperature $T_c \sim 155\,\text{MeV}$, the main degrees of 
freedom of the QCD are 
hadrons and a hadron resonance gas  (HRG) model agrees well with the low 
temperature behavior of lattice equation of state (see 
\Fig{fig:eose-3pparam}(a)). At temperatures much higher than the deconfinement 
temperature $T_c$, 
the 
system is better 
described in terms of weakly interacting quarks and gluons and the equation 
of state is not too far from the massless gas limit.
 Lattice computations show 
that at zero chemical potential the transition between the two phases is a 
smooth 
crossover~\cite{Aoki:2006we}. However, for $\mu_B>0$ one can have a first order 
phase 
transition 
line, which terminates at the critical end point (CEP)~\cite{Stephanov:2004wx} 
(see 
\Fig{fig:eose-3pparam}(b)). This part of the phase diagram is accessible 
at medium energy nuclear collisions and is the target of the Beam Energy Scan 
program at RHIC~\cite{bes2star, bes2phenix}.

In \Chap{chap:pca}, we use hydrodynamic equations derived in the conformal 
limit, but with lattice equation of state  (s95p-v1
parametrization from \Ref{Huovinen:2009yb}). The effective kinetic theory of 
\Chap{chap:ekt} treats the QGP as a gas of weakly interacting massless 
particles, 
which 
automatically leads to a conformal equation of state $p=e/3$.
For the sake of 
simplicity, the conformal equation of state is also used in semi-analytical
computations of \Chap{chap:noise}.

\subsection{Freeze-out}

In heavy ion collisions the QGP phase exists only for $\tau\sim 10\,\text{fm}$ 
before 
evaporating into hadrons, and it is only the long lived particles (pions, kaons 
and 
protons) which are eventually detected~\cite{Abelev:2012wca}.  Transition 
from the deconfined quarks and gluons to the hadronic 
degrees of freedom  is described by a freeze-out procedure.

When the plasma temperate falls below the pseudo-critical 
temperature  $T_c\sim155\,\text{MeV}$, gluons and quarks recombines into  
hadrons (pions 
being the most abundant species). The 
change in the degrees of freedom is reflected in the rapid crossover of  the
equations of state (see \Fig{fig:eose-3pparam}(a)). As the system continues to 
expand and cool down, the hadronic rescatterings 
are 
not sufficiently rapid to keep the system in local equilibrium. The 
macroscopic 
hydrodynamic description has to be replaced with  microscopic kinetic   
theory evolution of hadrons until all interactions and decays terminate.

The transition to kinetic description is typically done at a constant 
temperature 
$T_\text{fo}\sim T_c$, where all particle species are assumed to be in 
chemical and thermal equilibrium. Then the particle distribution functions is 
given by the 
Cooper\nobreakdash-Frye formula~\cite{Cooper:1974mv}
\begin{equation}
E\frac{d^3 N_s}{d^3 p} = \frac{\nu_s}{(2\pi)^3}\int_\sigma f_s(x^\nu,p^\nu) 
p^\mu d\sigma_\mu,\label{eq:CF}
\end{equation}
where $\sigma$ is the freeze-out surface ($T=T_\text{fo}$) and 
$f_s(\mathbf{x},\p)$ is the 
distribution function 
for particle species $s$ produced by a fluid cell moving with velocity $u^\nu$. 
In ideal fluid dynamics, one simply takes boosted equilibrium distribution 
$f_s(\mathbf{x},\p) =f_\text{eq}(-u^\nu p_\nu/T)$, but in a dissipative fluid 
the 
distribution function $f_s$ acquires   viscous 
corrections~\cite{Dusling:2009df}.

The subsequent kinetic evolution of hadrons can be done in hadronic transport 
models, e.g. 
\texttt{UrQMD}~\cite{Petersen:2008dd}. In practice the kinetic interactions
does not modify the pion spectrum significantly therefore 
in qualitative  (and even quantitative) studies only the freeze-out spectrum, 
\Eq{eq:CF}, is used to estimate the observed flow 
harmonics~\cite{Dusling:2007gi, Molnar:2008xj,Qiu:2013wca}. 

In \Chap{chap:pca} we use  the pion freeze-out 
spectrum at $T_\text{fo}=140\,\text{MeV}$ to estimate the flow harmonics at the 
end 
of 
hydrodynamic evolution.
\subsection{Momentum anisotropies}
\label{momani}
In each heavy ion collision event thousands of particles are detected over 
several units of pseudo-rapidity $\eta$ and azimuthal angle $\varphi$. 
Particles 
can 
be further binned by species and the transverse momentum $p_T$ they 
carry. 
Particles with energy well above QGP temperature, 
i.e. $p_T\gtrsim 5\, \text{GeV}$, are produced at the 
initial instant of the heavy ion collision and are the source of the jet 
phenomena~\cite{CasalderreySolana:2007pr,dEnterria:2009xfs}. High energy parton 
jets interact with the QGP medium, but 
are 
generally not considered part of it. Meanwhile soft 
momentum particles $p_T\lesssim 5\, \text{GeV}$ are shown to exhibit collective 
behavior characteristic of an equilibrated medium evolving as a 
whole~\cite{Heinz:2013th,Teaney:2009qa,Luzum:2013yya,Gale:2013da,Romatschke:2009im}.
In this 
work we will concentrate on the properties of the medium and will not discuss 
the
high $p_T$-physics.

The key consequence of considering QGP as a locally
 equilibrated medium is that  particles  are produced according to some 
 underlying probability distribution in 
$p_T,\eta$ and $\varphi$~\cite{Voloshin:2008dg}
\begin{equation}
E\frac{d^3 N}{d^3 p} = \frac{d^3 N}{p_T dp_T d\eta d\phi}
\end{equation}
and each event is a finite realization of this 
distribution. However, each collision happens at a random impact parameter 
$\b$ (both the size and the orientation of $\b$ can 
fluctuate). 
Therefore events are classified in centrality bins ($\sim |\b|$), while the 
event 
plane angle ($\sim$ orientation of $\b$) is 
estimated from the measured particles (or observables are constructed in such a 
way that 
they do not depend on the event plane angle).

 In the QGP medium paradigm, all soft-physics information about 
the 
event is contained in the particle distribution function and accurate 
computation of the distribution function is the principal goal of heavy ion 
collision simulations~\cite{Luzum:2013yya}. 
Similarly, experimentalists try to determine the underlying distribution 
function from heavy ion collision measurements and separate it from the 
high\nobreakdash-$p_T$ physics~\cite{Voloshin:2008dg}. In the rest of this 
section we 
will define 
several  
observables on which theory to experiment comparisons are typically 
performed, and which play an important role in \Chap{chap:pca}.

\subsubsection{Theory definitions}

At rest a thermalized medium produces particles 
isotropically in the azimuthal angle $\varphi$. However, the non-isotropic 
expansion of the QGP fireball introduces azimuthal anisotropies in the particle 
spectrum known as flows, which are studied using the azimuthal Fourier 
series of particle distribution function~\cite{Voloshin:1994mz,Voloshin:2008dg}
\begin{equation}
E\frac{d^3N}{d^3p}=\frac{d^2N}{2\pi p_T dp_T d\eta}\left(1+2 
\sum_{n=1}^\infty 
v_n(p_T,\eta) \cos n(\varphi-\Psi_n(p_T,\eta))\right)\label{eq:distr}.
\end{equation}
Here particle distribution function is expressed in terms of azimuthally 
averaged particle 
yield ${d^2N}/{2\pi p_T dp_T d\eta}$ and  Fourier  coefficients 
$v_n(p_T,\eta)$---flow harmonics. The azimuthal angle $\Psi_n(p_T,\eta)$ 
defines the orientation 
of the particular anisotropy in the transverse plane. The index $n$ 
labels the 
order of Fourier harmonics and only the first few terms are significant and 
have associated terminology
\begin{align}
n=1&\quad\text{directed flow},\\
n=2&\quad\text{elliptic flow},\\
n=3&\quad\text{triangular flow},\\
n=4&\quad\text{quadrangular flow},\\
n=5&\quad\text{pentagonal flow}.
\end{align}
The $n=0$ Fourier coefficient is normalized to one in \Eq{eq:distr}, but it is 
by far the largest Fourier component of the underlying distribution function 
and is called the radial flow.

When boost invariant picture of heavy ion collisions is applicable (the case 
assumed throughout this manuscript), the rapidity dependence can be dropped or 
integrated over in \Eq{eq:distr}. Furthermore, it is also often assumed that 
the 
orientation of a particular flow component in the transverse plane, $\Psi_n$, 
is 
 $p_T$ independent~\cite{Luzum:2013yya}. Then
\begin{equation}
\frac{1}{\Delta \eta}\int_{\eta_\text{min}}^{\eta_\text{max}}\!\!d\eta 
\,E\frac{d^3N}{d^3p}=\frac{dN}{2\pi p_T dp_T}\left(1+2 
\sum_{n=1}^\infty 
v_n(p_T) \cos n(\varphi-\Psi_n )\right),
\end{equation}
and the momentum dependent $v_n(p_T)$ flow vector has momentum independent 
orientation $\Psi_n$ in the azimuthal plane. In \Chap{chap:pca} we show that 
this particular assumption 
does 
not hold exactly due to the existence of subleadings flows. Finally, a single 
number characterizing the strength of 
harmonic flow in a 
collision is the momentum integrated $v_n$
\begin{equation}
\frac{dN}{d\varphi} \propto 1 +2\sum_{n=1}^\infty v_n \cos 
n(\varphi-\Psi_n),\label{eq:intvn}
\end{equation}
where  $dN/d\varphi$ is particle yield over all rapidity $\eta$ and momentum 
$p_T$ bins.

\subsubsection{Experimental definitions}
Theoretically it is simpler to start with the most differential 
observable, \Eq{eq:distr}, and then narrow down to integrated quantities, 
\Eq{eq:intvn}. In 
the experiment, however, the finite number of measured particles in each event  
makes differential observables the most challenging ones to 
measure~\cite{Voloshin:2008dg}. On 
the 
other hand, the integrated $v_n$ in a 
single event can be estimated by averaging over all detected particles
\begin{equation}
v_n^\text{obs} e^{in\Psi_n^\text{obs}}=\left<e^{in\varphi}\right>.\label{eq:ep}
\end{equation}
This estimated value $v_n^\text{obs}$ fluctuates event-by-event around the true 
value $v_n$, but can be corrected through a number of 
methods~\cite{Voloshin:2008dg}. The 
number of particles in one event is 
not sufficient to have an accurate estimate of more differential flow 
observables, instead one uses the large number of similar events to do the 
event 
average.
For example, the event-plane method uses the orientation of the integrated 
$v_n$, \Eq{eq:ep}, to calculate the differential momentum anisotropy in the 
$v_n$ plane~\cite{Luzum:2013yya,Voloshin:2008dg}
\begin{equation}
v_n(p_T) = \left<\cos(n(\varphi(p_T)-\Psi_n)\right>.
\end{equation}
Such measurement does not depend on the random orientation of the 
impact parameter $\mathbf{b}$ in the lab frame (see \Fig{fig:rapidity}) and can 
be safely averaged over many different events.

An alternative method of extracting the underlying distribution function is to 
look at pair correlations in the collision~\cite{Luzum:2013yya,Voloshin:2008dg}
\begin{equation}
v_{n\Delta}(p_{T1},p_{T2}) =\left<\left< 
e^{in(\varphi_1-\varphi_2)}\right>\right>_\text{events},
\end{equation}
where $\varphi_1$ and $\varphi_2$ are azimuthal angles of two separate 
particles 
with momentum $p_{T1}$ and $p_{T2}$  respectively in the same event.
 The two-particle angular 
correlation is clearly independent of the collision orientation.

The equilibrated medium emits particles independently, thus for  one event 
the pair production yield factorizes into single particle distribution functions
\begin{equation}
\frac{d^6 N_\text{pairs}}{d^3p_1d^3p_2}=\frac{d^3 N}{d^3p_1}\frac{d^3 N}{d^3p_2}+\mathcal{O}(N).\label{eq:pairs}
\end{equation}
The factorizable term scales with the particle number as $\mathcal{O}(N^2)$ and 
is the source of so called collective or flow correlations. The second term 
denotes the 
non-flow 
correlations, which have typical $\mathcal{O}(N)$ scaling. These are the 
correlations  that are not collective for the entire medium, 
e.g. resonance decay chain, Bose-Einstein correlation, etc. The 
non-flow correlations are usually narrow in rapidity $\eta$, therefore they can 
be further suppressed by imposing  pseudo-rapidity gaps  between the correlated 
particles~\cite{Voloshin:2008dg}. We will 
neglect the non-flow contributions in further discussion.

\begin{figure}
\centering
\includegraphics[width=0.7\linewidth]{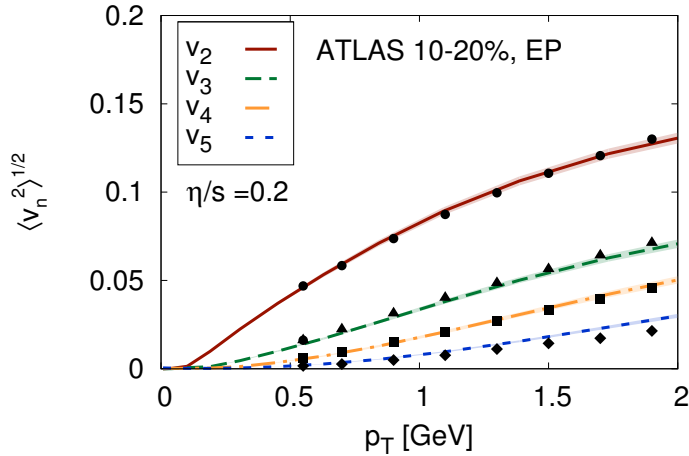}
\caption{Comparison of rms anisotropic flow coefficients 
$\left<v_n^2\right>^{1/2}$ as a function 
of transverse momentum in hydrodynamic simulation and experimental 
data~\cite{ATLAS:2012at}. 
Figure taken from \Ref{Gale:2012rq}.}
\label{fig:vn10-20}
\end{figure}

 Substituting single particle distribution function \Eq{eq:distr} in the pair 
 production yield formula \Eq{eq:pairs} 
and taking the average over  events results in the following relation between 
the two-particle and the single particle Fourier components 
\begin{equation}
v_{n\Delta}(p_{T1},p_{T2}) \equiv \left< \left< e^{ 
in(\varphi_{n1}-\varphi_{n2})}\right>\right>_\text{events}=\left< 
v_n(p_{T1})v_n(p_{T2}) \cos{ n 
(\Psi_{n1}-\Psi_{n2})}\right>_\text{events}\label{eq:vndelta}
\end{equation}
The correlation matrix on the left is experimentally 
measurable and encodes information about the single particle distribution 
function. For instance, the diagonal of the correlation matrix gives the 
root-mean-square of the harmonic flow $v_n(p_T)$
\begin{equation}
v_n\{2\}(p_T) \equiv v_{n\Delta}(p_{T},p_{T})= \sqrt{\left<v_n(p_T)^2\right>}.
\end{equation}
The rms harmonic flow vectors is well reproduced by hydrodynamic simulations of 
heavy ion collisions and agrees with experimental measurements as shown in 
\Fig{fig:vn10-20}.

\begin{figure}
\centering
\includegraphics[width=\linewidth]{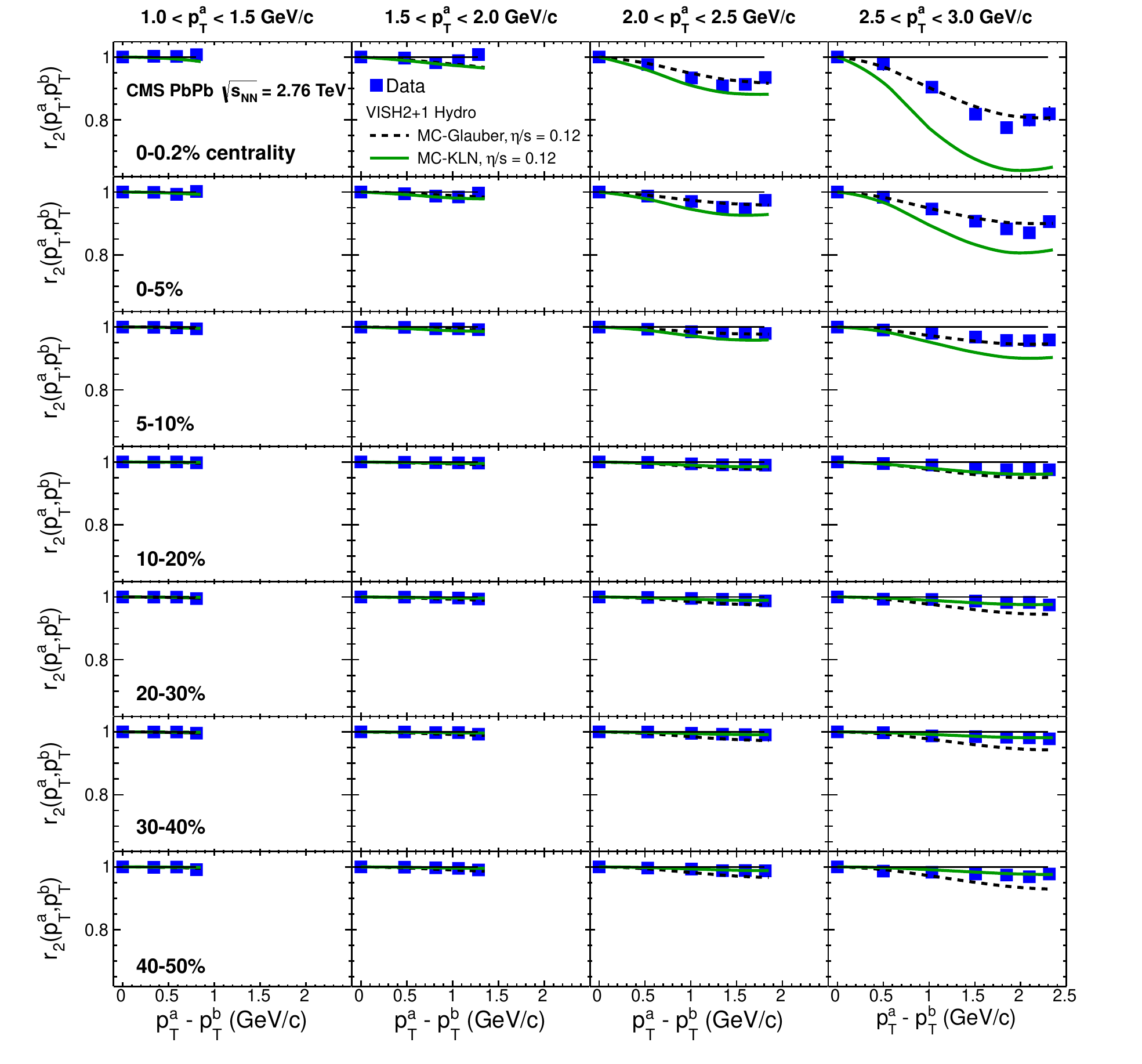}
\caption{\label{fig:cms-hin-14-012figure003} Factorization ratio, $r_2$, as a 
function of transverse momentum 
difference $p_T^a-p_T^b$ in bins of $p_T^a$ for  Pb+Pb collisions at 
$\snn=2.76\,\text{TeV}$. Figure taken from \Ref{Khachatryan:2015oea}.}

\end{figure}

If the flow angle $\Psi_n$ is independent of momentum $p_T$ and there are no 
event-by-event fluctuations in flow magnitude $v_n(p_T)$, then the right hand 
side of \Eq{eq:vndelta} factorizes
\begin{equation}
v_{n\Delta}(p_{T1},p_{T2}) = v_n(p_{T1})\times v_n(p_{T2}).
\end{equation}
Such flow factorization can be experimentally tested by 
measuring the factorization ratio
\begin{equation}
r_{n}(p_{T1},p_{T2})=\frac{v_{n\Delta}(p_{T1},p_{T2}) 
}{\sqrt{v_{n\Delta}(p_{T1},p_{T1}) v_{n\Delta}(p_{T2},p_{T2}) }} .
\label{eq:introrij}
\end{equation}
Experimental measurements show that at $\snn=2.76\,\text{TeV}$ Pb+Pb collisions 
factorization is broken~\cite{Khachatryan:2015oea,CMS:2013bza,Aamodt:2011by}. 
In particular, for 
the second harmonic flow $v_2$ factorization 
breaking reaches $\sim20\%$ for the most central Pb+Pb, but is small in 
peripheral 
collisions (see 
\Fig{fig:cms-hin-14-012figure003}).
Initially, factorization breaking was interpreted as a sign of non-flow 
effects, but later it was understood that even in purely hydrodynamic models 
initial fluctuations will cause 
factorization breaking~\cite{Gardim:2012im}. Studying the entire two-particle 
correlations  matrix, \Eq{eq:vndelta},  or factorization ratio, 
\Eq{eq:introrij},  provides an  access to initial fluctuations  not  
captured by the integrated flow observables.

 In \Chap{chap:pca} we use Principal Component Analysis to decompose the 
 two-particle correlations, \Eq{eq:vndelta}, into the most significant 
 components 
for boost invariant hydrodynamic simulations at $\snn=2.76\,\text{TeV}$. 
We show how principal components can be used to describe and interpret the 
transverse momentum factorization breaking.

\subsection{QCD medium at high temperature}
\label{highT}
One of remarkable 
properties of the QCD is the running of the coupling constant 
$\alpha_s=g^2/(4\pi)$. At large collision energies, the coupling 
appears small 
and the hard
cross-sections can be calculated using leading order QCD diagrams in 
vacuum~\cite{Burgess,Langacker,Peskin}. 
Indeed, many
calculations for proton-proton collisions based on the Standard Model compares 
excellently against the 
experimental measurements~\cite{Ellis}.
The increased collision energy of heavy ion 
collisions at LHC (from $\snn=200\,\text{GeV}$ at RHIC to  
$\snn=2.76\,\text{GeV}$ and $\snn=5.02\,\text{GeV}$ at LHC)  allows creation of 
matter with initial temperature 2-4 times 
above the QCD transition temperature 
$T_c\sim155\,\text{MeV}$~\cite{Adam:2015lda}. 
Therefore the bulk dynamics of the early stages of  the collision  might be 
described by the  perturbative processes of high temperature 
QCD~\cite{Blaizot:2003tw}. 

QCD processes in a medium are more complicated than those in vacuum, 
 because a 
naive diagrammatic expansion in coupling constant fails to correctly account for 
all contributing processes at the same order in 
$g$, as clarified by Braaten and Pisarski~\cite{Braaten:1989mz}. For 
a 
systematic  treatment of such contributions, 
one needs a separation of scales. At weak coupling $g\ll1 $ and  high 
temperature $T\gg T_c$ , the QGP is energetically dominated by high energy 
quasi-particles 
with typical (hard) momentum $p\sim T$. Loop corrections to the free 
propagator create an effective thermal mass $m_\text{th}\sim gT$ which is much 
smaller than the hard particle energy scale $p\sim T$, therefore hard partons 
can 
be treated as nearly massless excitations of the plasma traveling in  $\sim gT$ 
background. Over times much longer than 
duration of
any scattering  process, hard particles can be described by 
an effective kinetic theory; the phase space density $f(\x,\p)$ of hard 
particles then obeys
a Boltzmann equation,
\begin{equation}
(\partial_t + \mathbf{v}\cdot \mathbf{\nabla}_x ) f = -C[f],
\end{equation}
where $|\mathbf{v}|=1$ and $C[f]$ is the collision kernel. The leading order  
effective kinetic theory of high temperature gauge 
theories was laid out in the seminal paper by  Arnold, 
Moore and Yaffe~\cite{Arnold:2002zm}, and is known as the AMY formalism.

\begin{figure}
\centering
\includegraphics[height=0.18\linewidth]{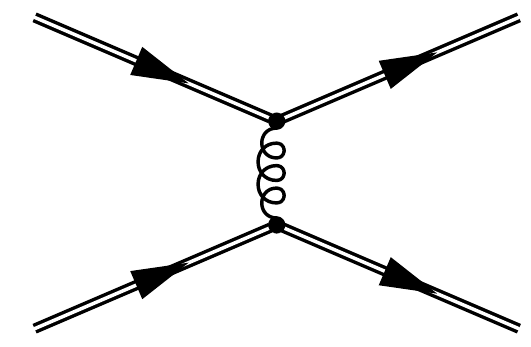}
\includegraphics[height=0.18\linewidth]{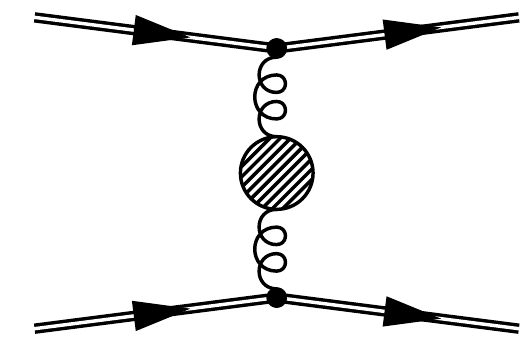}
\includegraphics[height=0.18\linewidth]{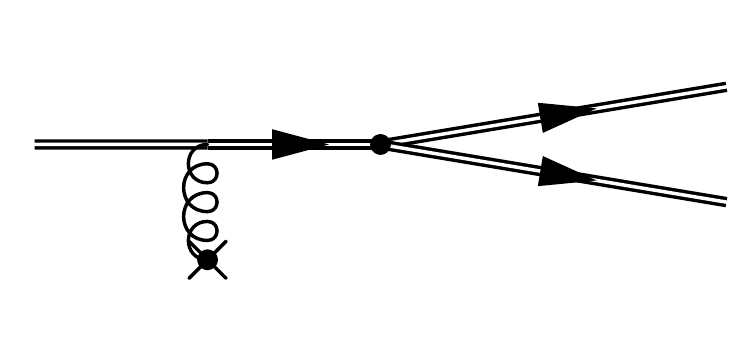}
\caption{(a) Hard elastic $2\rightarrow2$  scattering with $\sim T$ momentum 
exchange (b) Soft elastic $2\rightarrow2$  scattering with medium regulated  
$m_\text{th}\sim gT$  propagator (c)  
Collinear splitting of hard quasi-particle into two hard particles due to soft
$\sim gT$ momentum exchange with the medium. }
\label{fig:scatterings}
\end{figure}

At leading order the hard parton dynamics is governed by elastic 
$2\leftrightarrow2$  scatterings and inelastic $1\leftrightarrow2$ particle 
number changing QCD bremsstrahlung. When two hard 
partons exchange $\mathcal{O}(T)$ 
momentum, they undergo a large angle $\theta\sim 1$ 
scattering with a rate $\sim g^4 T$ (see \Fig{fig:scatterings}(a)). For a soft 
momentum exchange (small 
scattering angle), the cross-section is quadratically divergent  
in vacuum, but in the medium it is regulated by a thermal mass $m_\text{th}\sim 
gT$ (diagram \Fig{fig:scatterings}(b)). Therefore small angle scatterings 
$\theta \sim g$ occur at a rate $\sim 
g^2 T$, i.e. much more often than large angle scatterings, but the cumulative 
diffusive effect  is of the same order as a single hard scattering. In addition 
to 
elastic 
scatterings, hard quasi-particles can split into two 
near  collinear hard  particles thanks to soft momentum $\sim gT$ exchange with 
the medium (see \Fig{fig:scatterings}(c)). Such  radiative process is 
prohibited by energy-momentum conservation in vacuum and is crucial to the 
equilibration of the 
ultra-relativistic plasma~\cite{Baier:2000sb,Mueller:1999fp}. The perturbed 
hard parton has 
small 
virtuality and travels distance  $\sim (g^2 T)^{-1}$ before splitting. During 
this time multiple soft scatterings can occur and the destructive interference 
between different scattering  events
leads to suppression of collinear splitting known as the 
Landau-Pomeranchuk-Migdal 
(LPM) effect. Fully resummed  leading order effective $1\leftrightarrow2$ 
scattering rate 
can be obtained from an integral-differential equation~\cite{Arnold:2002ja}.

\begin{figure}
\centering
\includegraphics[width=0.6\linewidth]{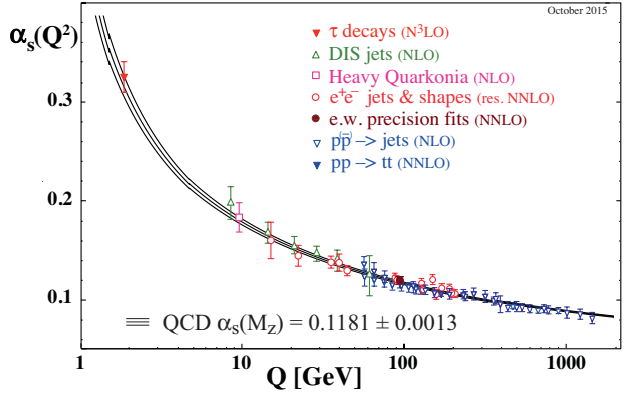}
\caption{Strong coupling constant $\alpha_s(Q)$ as a function of 
energy scale $Q$. Figure taken from 
\Ref{Olive:2016xmw}}
\label{fig:runinfcoupling2}
\end{figure}

With all leading order terms correctly accounted for, the kinetic theory 
provides an effective description of high temperature QCD medium and can be 
used to calculate, for example, the transport properties of the 
medium~\cite{Arnold:2003zc}. The AMY formulation of the effective kinetic 
theory has a 
number of requirements~\cite{Arnold:2002zm}:
\begin{itemize}
\item The temperature $T$ is sufficiently high that the coupling constant is 
small $g\ll1$ and all other QCD scales (quark masses and 
$\Lambda_{QCD}$) are negligible compared to the thermal masses $m_\text{th}\sim 
gT$.
\item The system is sufficiently close to local thermal equilibrium, which is  
defined by 
slowly varying local temperature $T(\x)$ and  velocity $u^\mu(\x)$ fields
\begin{equation}
f(\x,\p)=n(|\p|;T(\x), u^\mu(\x))+\delta f(\x,\p),\quad \delta 
f\ll f
\end{equation}
where $n(|\p|; T, u^\mu)$ is an equilibrium (Bose or Fermi) distribution.
\end{itemize}

As can be seen from 
\Fig{fig:runinfcoupling2}, the running of the 
coupling constant $\alpha_s$ is slow and $g=\sqrt{4\pi\alpha_s}$ is never 
parametrically small in the accessible energy range. 
Consequently,   ``hard" $\sim T$ and ``soft" $\sim gT$ 
momentum scales are not widely separated and the soft sector contributions have 
to be calculated non\nobreakdash-perturbatively~\cite{Braaten:1995ju}. Modern 
perturbative theory techniques were shown to reproduce various thermodynamic 
quantities for $T>2\text{--}3T_c$~\cite{Andersen:2011sf,Haque:2014rua} and 
there is a
considerable progress of extending dynamical calculations of effective kinetic 
theory beyond the leading order~\cite{Ghiglieri:2015ala}. Therefore applying 
the leading 
order AMY kinetic theory at realistic values of the coupling constant 
$\alpha_s\sim0.3$ in heavy ion collisions is an
extrapolation from the weak coupling regime, but it is built on  
first principle  systematically improvable calculations with  single 
parameter---the coupling constant.

At the initial moments after the collision, the QGP is believed to be in highly 
anisotropic state with longitudinal pressure $P_L$ much smaller than the 
transverse 
pressure $P_T$~\cite{Berges:2013eia,Berges:2013fga}. 
In AMY formulation of the kinetic theory the distribution function is 
restricted to 
parametrically small anisotropies, because of Weibel 
instability~\cite{Fukushima:2016xgg,Mrowczynski:2016etf}.
 The 
origin of Weibel instability is that charged particles respond  differently to 
magnetic 
field perturbations  depending on their direction of motion. In 
isotropic plasma the positive and negative feedbacks cancel out, but if the 
particle distribution function is anisotropic, certain field perturbations will 
be reinforced leading to an instability.
However, in the QCD the self interaction between 
the soft modes slows down the exponential growth of the instability and 
classical lattice simulations show that plasma instabilities are not important 
beyond the transient 
time~\cite{Arnold:2005vb,Berges:2013fga,Berges:2013eia,Berges:2013lsa,Berges:2014bba}.
This motivates applying isotropically screened kinetic theory even to systems 
with considerable anisotropy. However, the applicability of effective kinetic 
theory could be considerably widened if anisotropic screening would be taken 
into account 
non-perturbatively~\cite{Kajantie:2000iz,Braaten:1995jr,Kajantie:2002wa}.

In \Chap{chap:ekt} we use leading order effective kinetic theory with isotropic 
screening to model isotropization of gluon distribution function at realistic 
values of the coupling constant $\alpha_s\approx 0.26$.

\section{Outline}
\label{outline}
\subsection{\Chap{chap:pca}: Principal Component Analysis}
\label{prolog1}

\subsubsection{Problem}

The evolution of heavy ion collisions is a highly non-trivial map between 
initial 
conditions and  observables. Thanks to small QGP viscosity, many imprints 
of the initial state features can be seen in the measured particle spectrum.  
The observed flow factorization breaking (see \Fig{fig:cms-hin-14-012figure003} 
and \Sect{momani})
 can be explained by 
flow angle and magnitude decorrelations with the
transverse  momentum~\cite{Gardim:2012im,Heinz:2013bua,Kozlov:2014fqa}, 
indicating that that there are multiple independent sources of harmonic flow 
in heavy ion 
collisions.
These subleading flows are  a response  to 
different features of the initial geometry, and therefore their combined 
contribution to $v_n$ cannot be predicted by a single geometrical quantity 
like $\varepsilon_n$. 
 Only by correctly identifying the generating sources for 
all subleading flows, one can fully understand the map between initial 
conditions 
and 
final state observables.

\subsubsection{Idea}

To 
disentangle 
the separate causes of flow, one can use the full information contained in the
two-particle correlation matrix, \Eq{eq:vndelta}. A simple 
statistical method to achieve that is Principal Component 
Analysis (PCA)~\cite{citeulike:11071912,diva2:288565}. Given a $N\times N$ 
covariance 
matrix of 
$N$ 
observables, PCA identifies the largest statistically uncorrelated sources of 
variance in the system. This greatly 
reduces the dimensionality of the problem. Instead of the full $N 
\times N$ covariance matrix, one can deal with just few dominant 
principal components.  Schematically, event-by-event harmonic flow  $v_n(p_T)$ 
can be written as a sum of leading principal components
\begin{equation}
v_n(p_T)e^{in\Psi_n(p_T)} \propto \xi_1 v_n^{(1)}(p_T) +  \xi_2 v_n^{(2)}(p_T) 
+ \ldots
\end{equation}
where $\xi_1$ and $\xi_2$ are complex coefficients encoding
angle and magnitude  fluctuations of leading and subleading principal 
components.  PCA guarantees that different components are statistically 
uncorrelated, i.e.
\begin{equation}
\left<\xi_i \xi_j^* \right>= \delta_{ij}
\end{equation}
and that (real) principal component vectors $v_n^{(a)}(p_T)$ are mutually 
orthogonal with a certain measure $w(p_T)$, which has to be specified
\begin{equation}
\int dp_T w(p_T) v_n^{(a)}(p_T) v_n^{(b)}(p_T) \propto \delta^{ab}.
\end{equation}
The weight function $w(p_T)$ can be chosen to maximize the difference between 
physically independent sources of momentum dependent harmonic flow 
$v_n(p_T)$.  However, if the response has the same momentum dependence, one 
needs additional information to differentiate between flow sources. Since we 
are interested in collective flow, we weigh each momentum bin by corresponding 
multiplicity, thus giving the largest weight to correlations involving the 
largest number 
of particles.  
Therefore our choice for the weight function is 
\begin{equation}
w(p_T) = \left(\frac{dN}{2\pi p_T dp_T}\right)^2,
\end{equation}
but other choices are also possible. The weight function can be absorbed by 
defining 
multiplicity weighted harmonic flow\footnote{Note that in the literature 
$V_n(p_T)$ can  also
denote the standard (normalized) flow harmonic $v_n(p_T)$.}
\begin{equation}
V_n^{(a)}(p_T) \equiv v_n^{(a)}(p_T) \times \frac{dN}{2\pi p_T dp_T}.
\end{equation} 
With this definition $V_n$ flow is proportional to the observed particle count 
and is 
analogous to the experimentally measured $Q_n$ vector~\cite{Voloshin:2008dg}. 
Then the 
multiplicity weighted two-particle correlation function can be efficiently 
expanded in terms 
of principal components. For elliptic flow
\begin{figure}
\centering
\includegraphics[width=\linewidth]{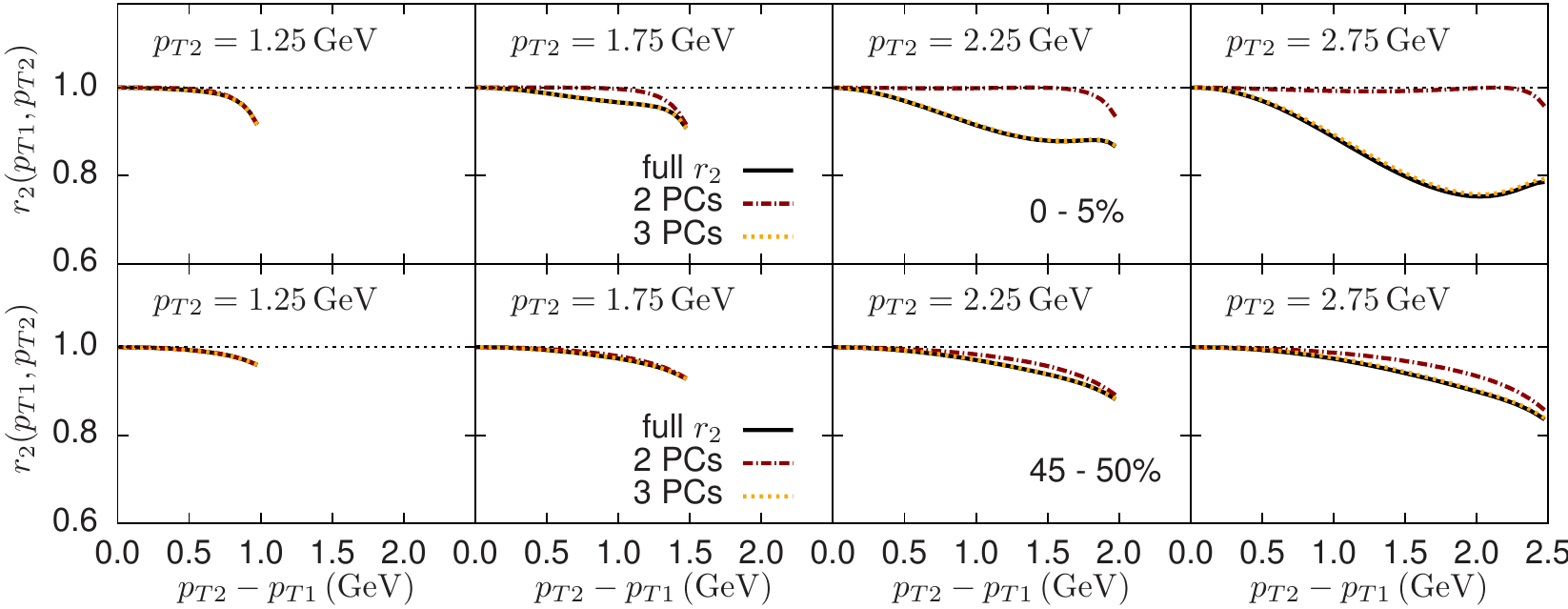}
\caption{Factorization ratio $r_2(p_{T1},p_{T2})$ [\Eq{introrij}] for elliptic 
flow and its 
approximations 
with principal components (PCs) in hydrodynamical simulations of central 
(0--5\%) and peripheral  (45--50\%) 
collisions.}
\label{introfig:rij}
\end{figure}
\begin{align}
\big<V_{2,p_{T1}} V^*_{2,p_{T2}}\big>  =
V_2^{(1)}(p_{T1}) 
   V^{(1)}_2(p_{T2})+
V_2^{(2)}(p_{T1}) 
   V^{(2)}_2(p_{T2})+\ldots.
\end{align}
Just two or three terms are needed to give an excellent approximation of the 
full 
factorization ratio $r_2(p_{T1}, p_{T2})$ defined now as
\begin{equation}
r_2(p_{T1}, p_{T2}) \equiv \frac{ \llangle V_2(p_{T1}) V_2^*(p_{T2}) 
\rrangle}{\sqrt{ \llangle |V_2(p_{T1})|^2 \rrangle
\llangle |V_2(p_{T2})|^2  \rrangle } }\,.\label{introrij}
\end{equation}
In  \Fig{introfig:rij} we see that indeed principal components efficiently 
describe factorization ratio in hydrodynamical simulations, which can be 
compared with experimentally measurements in \Fig{fig:cms-hin-14-012figure003}.

\subsubsection{Application}

\begin{figure}
\centering
\includegraphics[width=0.8\linewidth]{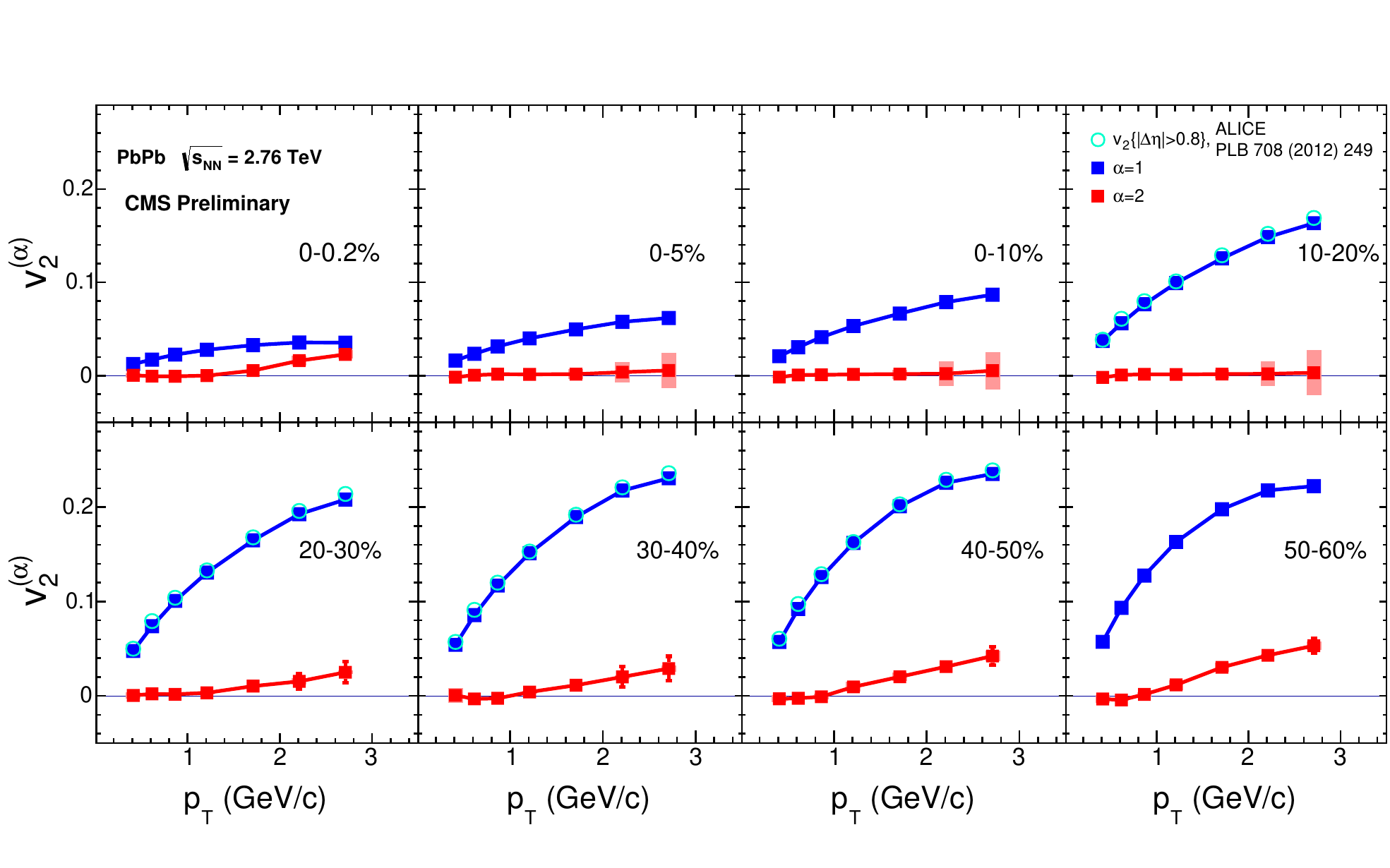}
\caption{First and second  principal component 
for elliptic flow as a 
function of $p_T$ in different centrality bins for Pb+Pb collisions at 
$\snn=2.76\,\text{TeV}$. Figure taken from \Ref{Milosevic:2016tiw}.}
\label{fig:figure002}
\end{figure}

Principal Component Analysis is a systematic way of expressing harmonic flow in 
statistically uncorrelated components. Because principal components are 
typically strongly ordered, only the leading two or three components are 
sufficient 
to describe  factorization breaking~\cite{Bhalerao:2014mua}. In 
\Chap{chap:pca}, we demonstrate that 
each component can be 
given an intuitive physical interpretation as a hydrodynamic response to 
particular initial state features (radial excitations of eccentricities) or a
nonlinear response to the lower order 
harmonic flow. 
The first phenomenological application of PCA in studying  factorization 
breaking in heavy 
ion collisions was introduced in \cite{Bhalerao:2014mua,Mazeliauskas:2015vea}, 
which was followed by the first
experimental 
measurements of principal components for elliptic and triangular flow in Pb+Pb 
and p+Pb collisions in 
\Ref{Milosevic:2016tiw}. As can be seen from the experimental results shown in  
\Fig{fig:figure002}, the subleading 
 elliptic flow is significant in peripheral and ultra-central collisions, 
which is consistent with findings in our work. However the subleading principal 
component for triangular flow was found to be negligible  for Pb+Pb collisions 
at 
$\snn=2.76\,\text{TeV}$~\cite{Milosevic:2016tiw}.

Another area of applications for PCA is the factorization 
breaking in pseudo-rapidity 
$\eta$~\cite{Khachatryan:2015oea,ATLAS:2012at,Aad:2014lta}. Here the flow 
decorrelation  at different rapidity bins was associated with forward-backward 
asymmetric in nuclear collisions~\cite{Bozek:2010vz,Bzdak:2012tp,Jia:2014ysa}. 
The $\eta$ dependent flow or 
multiplicity can be decomposed in terms of orthogonal 
polynomials~\cite{Bzdak:2012tp,Jia:2015jga} or principal 
components as an optimal basis~\cite{Bhalerao:2014mua}. Analogously to the case 
of  
transverse momentum correlations, principal component analysis presents itself 
as a data driven 
method of utilizing all information of the two particle correlations function.

In  \Chap{chap:pca}  we 
present a comprehensive Principal Component Analysis for flow harmonics 
$n=0\text{--}5$ in boost invariant hydrodynamic simulations of heavy ion 
collisions at 
$\snn=2.76\,\text{TeV}$.

\subsection{\Chap{chap:ekt}: Weak coupling equilibration}
\label{prolog2}

\subsubsection{Problem}

The process of equilibration in heavy ion collisions is currently a hotly 
debated 
topic~\cite{Romatschke:2016hle,Kurkela:2016vts,Chesler:2016vft,Gelis:2016rnt}.
Conventionally, for  hydrodynamics to be applicable, one 
needs an approximate local thermal equilibrium or at least pressure 
isotropization~\cite{Arnold:2004ti}, but phenomenological 
simulations use relativistic second order hydrodynamics at times
as early as $\tau=0.2\,\text{fm}$, when pressure anisotropies are still 
large~\cite{Gale:2012rq,Niemi:2015qia}.
 This 
naturally leads to a question of how QGP approaches hydrodynamic behavior in 
such a short time. In 
addition, the initial out-of-equilibrium evolution determines the starting 
conditions for hydrodynamics, which is one of the main sources of uncertainty 
in the extraction of QGP transport coefficients~\cite{Luzum:2012wu,Song:2010mg}.
 A 
clear understanding of pre-equilibrium dynamics  is necessary to have a 
consistent picture of heavy ion collisions and can be crucial in interpreting 
signals of collectivity in small collisions systems like p+Pb and p+p.

\subsubsection{Idea}

At asymptotically high temperatures, but moderate densities, the QGP can be 
viewed as a gas of weakly coupled particles with effective kinetic 
description, which is also applicable out-of-equilibrium. By 
following the evolution of these hard partons in kinetic theory, one can 
understand the equilibration from the first principle QCD calculations.
 For heavy ion collisions such ``bottom-up" path to thermalization in weakly 
coupled QCD was laid out in the seminal paper by Baier, Mueller, Schiff and 
Son~\cite{Baier:2000sb}. The paper assumes the Color Glass Condensate (CGC) 
picture 
of the initial state of  heavy ion collision, which is characterized by large
saturation scale $Q_s\gg\Lambda_{QCD}$,  parametrically high density $f\sim 
\alpha_s^{-1}$ and strongly interactions (despite the coupling constant 
$\alpha_s\ll1$ being small). 
By the time $\tau Q_s> 1$ the  longitudinal expansion dilutes 
the system sufficiently that  
it 
can be described as a collection of quasi-particles with well defined, but 
highly anisotropic distribution function $f(\x,\p)$. Most of energy is 
carried by hard $p\sim Q_s$ gluons, which undergo elastic 
$2\leftrightarrow 2$ scatterings, but  thanks to longitudinal
 expansion the anisotropy continues to grow. 
Eventually the number density of hard gluons drops below one and they start 
cascading to the soft sector through collinear splitting. The soft gluons 
(relatively to the  saturation scale $Q_s$) collide often and quickly 
thermalize. The 
remaining hard partons loose their energy to this thermal bath and the 
system reaches equilibrium.

\def\ubr#1#2{\underbrace{#1}_{\text{#2}}}
\definecolor{uibred}{RGB}{167, 38, 47}
\definecolor{MyDarkGreen}{rgb}{0.0, 0.30, 0.00}
\definecolor{MyDarkBlue}{rgb}{0.0, 0.00, 0.7}
\newcommand{\punch}[1]{{\color{uibred}\emph{#1}}}
\newcommand{\auth}[1]{{\color{MyDarkGreen}\scriptsize #1}}
\newcommand{\bblock}[1]{{\color{MyDarkBlue}#1}}
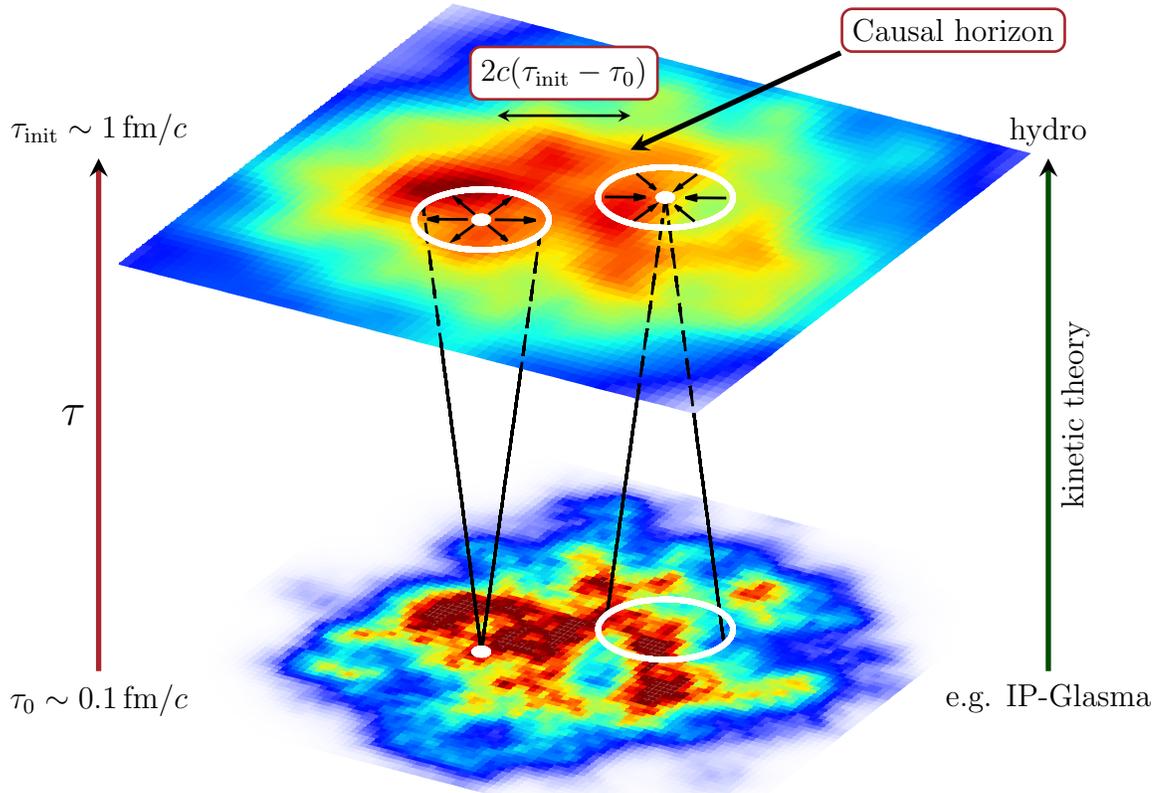
\begin{figure}
\centering
\begin{tikzpicture}[>=stealth]
{\node[anchor=south west] (image) at (0,0) 
{\includegraphics[width=0.9\linewidth]{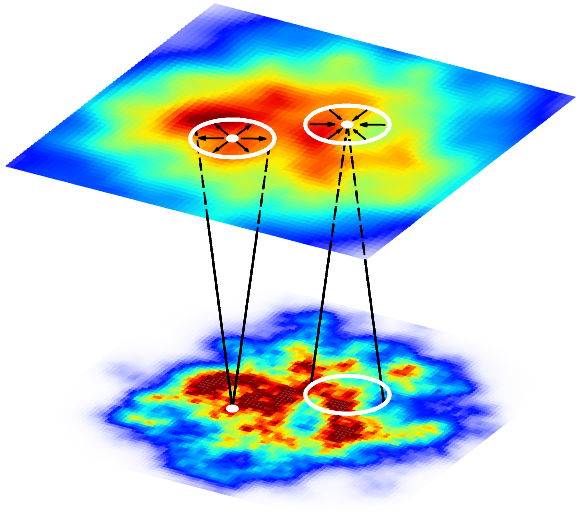}};}
\begin{scope}[x={(image.south east)},y={(image.north west)}]
\draw[->,line width = 2pt, draw=uibred, overlay]  (0,0.2) node[below] {$\tau_0 
\sim 0.1\, \text{fm}/c$} to node[left] {\Large $\tau$} (0,0.8) node[above] 
{$\tau_\text{init}\sim 1\,\text{fm}/c$};
{\draw[->,line width = 2pt, draw=MyDarkGreen,overlay]  (1,0.2) 
node[below] {e.g. IP-Glasma} to node[below, rotate=90] { 
kinetic theory} (1,0.8) node[above] { 
hydro};}
{
\draw[<->,line width = 1pt,draw=black] (0.42,0.85) to  node[above, fill=white, 
draw=uibred, rectangle, rounded corners, yshift=0.2cm] 
{ 
$2c(\tau_\text{init}-\tau_0)$} (0.56,0.85);
\node[overlay, draw=uibred, rectangle, rounded corners, line width=1pt] (c) at 
(0.9,0.95) {Causal horizon}; 
\draw[->, line width = 2pt] (c.south west) to (0.56,0.81);
}
\end{scope}
\end{tikzpicture}
\caption{Kinetic theory describes the evolution from the microscopic formation 
time  $\tau_0$ to the equilibration time $\tau_\text{init}$, when 
hydrodynamics 
becomes applicable~\cite{Kurkela:2016vts}. By causality, for a given point in 
the transverse plane it is sufficient to analyze the pre-equilibrium evolution 
within the causal neighborhood of that point.\label{causal}}
\end{figure}

Another important aspect of the pre-thermal evolution for  heavy ion collision 
is the propagation of initial  fluctuations in 
the transverse plane (see \Fig{causal}).
In weak coupling framework the different stages of 
thermalization are 
parametrically 
separated by the inverse powers of the coupling constant $\alpha_s\ll 
1$ and 
the  kinetic theory describes the  evolution of the 
system 
from the microscopic formation time $\tau_0\sim Q_s^{-1}$ to the (much later) 
times when 
hydrodynamics becomes applicable  $\tau_\text{init}\sim 
\tau_\text{equilibrium}$~\cite{Kurkela:2016vts}.
In practice (and for realistic values of the
coupling constant $\alpha_s\approx 0.26$ in kinetic 
theory~\cite{Kurkela:2015qoa}), 
the hydrodynamization time 
is short, $\tau_\text{init}\sim 1\,\text{fm}$, and initial perturbations 
interact locally. In other words, for a given point $\x_0$ in the transverse 
plane the causally connected region $|\x-\x_0|\leq 
c(\tau_\text{init}-\tau_0)\sim 
1\,\text{fm}$  is 
much smaller than 
the total transverse nuclear geometry 
$R_\text{Pb}\sim 5\,\text{fm}$, but comparable to a single nucleon scale 
$R_p\sim 1\,\text{fm}$.
Therefore heavy ion collisions evolution from $\tau_0$ to $\tau_\text{init}$ 
can be decomposed into 
equilibration of
translationally invariant background and small perturbations around it in the 
causally connected region $|\x-\x_0|\leq c(\tau_\text{init}-\tau_0)$ (see 
\Fig{causal}).
 For linear perturbations the 
pre-equilibrium evolution can be conveniently expressed in terms of Green 
functions, which map initial perturbations to energy momentum tensor at the 
 times when hydrodynamics is applicable. By convolving kinetic theory response 
 functions with microscopic initial state models like IP-Glasma, we can obtain 
 initial conditions for hydrodynamics with the complete pre-thermal evolution 
 history.

\subsubsection{Application}

The early time dynamics in heavy ion collisions is under active investigations 
in both strong coupling holographic 
approach in model theories~\cite{Chesler:2016vft} and weak coupling 
framework in QCD~\cite{Gelis:2016rnt}.
Realistic collisions happen in neither of these idealized  limits, but both can 
be used as physically motivated bounds for the equilibration in heavy ion 
collisions~\cite{Keegan:2015avk}. Recent work on classical Yang-Mills 
evolution~\cite{Berges:2013fga,Berges:2013eia}, showed that the   
``bottom-up"  scenario~\cite{Baier:2000sb} is the preferred thermalization 
picture in a weak coupling framework. It was put to practice for uniform boost 
invariant background in \Ref{Kurkela:2015qoa}, showing that kinetic theory 
(extrapolated to 
realistic values of the coupling constant $\alpha_s\approx0.26$) describes 
equilibration and approach to hydrodynamics in 
phenomenologically feasible, i.e. short enough,  time. In \Chap{chap:ekt} we 
extend this work to 
include the equilibration of transverse perturbations and demonstrate that they 
too hydrodynamize.

One of the goals of early time dynamics is to complement the successful 
hydrodynamic simulations of heavy ion collisions with the complete 
pre-equilibrium evolution. Various approaches taken include continuing 
classical 
Yang-Mills evolution~\cite{Schenke:2012wb}, using free 
streaming~\cite{Broniowski:2008qk,Liu:2015nwa}, taking holography inspired 
initial 
conditions~\cite{vanderSchee:2013pia,Romatschke:2015gxa,Kurkela:2016vts} or 
different formulations of hydrodynamics~\cite{Martinez:2012tu}. The main 
advantages of kinetic theory 
equilibration are that it is based on the first principle QCD and automatically 
leads isotropization and approach to hydrodynamics. Kinetic theory is also a 
natural continuation of the successful microscopic initial state 
models based on CGC framework.

 In \Chap{chap:ekt} we use effective 
kinetic theory to construct Green functions for the linearized perturbations 
around boost invariant background and demonstrate equilibration and 
hydrodynamization of CGC inspired out-of-equilibrium initial conditions.

\subsection{\Chap{chap:noise}: Non-equilibrium thermal 
fluctuations}
\label{prolog3}

\subsubsection{Problem}

Since fluctuations in heavy ion collisions is one of the main tools of 
inferring 
medium properties,  understanding the 
various sources of fluctuations is central to heavy ion physics. A considerable 
effort went into modeling the creation and propagation of initial state 
fluctuations, e.g. second order hydrodynamics or IP-Glasma initial conditions. 
However, the QGP (as any 
dissipative medium) must also have thermal fluctuations~\cite{LandauStatPart1, 
LandauStatPart2}. Ordinarily, such 
fluctuations are suppressed by the large number of constituents, but a 
 QGP droplet is on the edge of what can be called a 
macroscopic system with of order $\sim 10^4$ of produced 
particles~\cite{ATLAS:2011ag,Chatrchyan:2011pb}.
Secondly, thermal fluctuations have been long known to create long time tails 
$\propto t^{-\frac{3}{2}}$ 
in  the two point correlation 
functions~\cite{velocity_auto,wainwright2,bixon_zwanzig}. Formally such 
fractional power terms break the gradient expansion of relativistic 
hydrodynamics at the second order~\cite{Kovtun:2011np} and therefore a 
consistent higher order description of heavy ion collisions needs to assess  
the 
nonlinear noise contributions.
 Finally, the discovery of the QGP opened up a completely new domain in the 
phase diagram of 
hot and dense nuclear matter. Naturally, chartering out this new territory 
requires determining the transition line between the QGP and ordinary hadronic 
matter. At high collision energies, i.e. small baryon chemical potential, 
lattice 
computations indicate a smooth transition between the two phases, however at 
$\mu_B>0$ a first order transition is 
possible~\cite{Stephanov:2004wx}. The 
ongoing 
Beam Energy Scan (BES) program at RHIC is aimed to  find the signs 
of the first order transition and the critical end point---the 
point where smooth crossover turns into the first order transition. Critical 
points 
in the phase diagrams are characterized by  and dominated by stochastic 
fluctuations~\cite{LandauStatPart1}. Therefore thermal fluctuations must be an 
important 
ingredient 
in the correct interpretation of the experimental signatures of the critical 
point.

\subsubsection{Idea}

Any system in thermal equilibrium at temperature $T$ must satisfy 
fluctuation-dissipation 
theorem, which in frequency space $\omega$ is
\begin{equation}
G_S(\omega) =  \frac{2T}{\omega}\text{Im}\,G_R(\omega),
\end{equation}
where $G_S(\omega)$ is symmetrized correlator capturing the magnitude of 
fluctuations 
and $G_R(\omega)$ is the retarded Green function describing system response to 
perturbations~\cite{LandauStatPart1}. Microscopically the meaning of 
fluctuation-dissipation theorem is
that a dissipative system can loose energy only to the  
microscopic degrees of freedom, however in equilibrium, the opposite process 
must occur and microscopic degrees of freedom induces macroscopic thermal 
fluctuations in the system.

Hydrodynamics is an effective theory describing long wavelength physics 
$\lambda \gg l_\text{mfp}$, where $l_\text{mfp}$ is microscopic medium scale, 
e.g. mean free path. The medium properties like rest frame energy density $e$ 
and pressure 
$p$ is the manifestation of kinetic and potential energy contained in the 
microscopic 
degrees of freedom. Similarly, transport coefficients like viscosity $\eta$ 
characterize the energy transfer from modes with $\lambda>l_\text{mfp}$ to 
those below the hydrodynamic cut-off, $\lambda < l_\text{mfp}$.
The hydrodynamics discussed in \Sect{hydro} does not include thermal 
fluctuations, 
however it is well known how to incorporate stochastic terms in 
hydrodynamic equations of motion~\cite{LandauStatPart1, 
LandauStatPart2} (for a recent review see~\cite{Kovtun:2012rj}) 
\begin{align}
D_\mu T^{\mu\nu}=0\label{eq:eomzero},\quad 
T^{\mu\nu}=T^{\mu\nu}_\text{ideal}+T^{\mu\nu}_\text{visc.}+S^{\mu\nu},
\end{align}
where $S^{\mu\nu}$ is a noise term with zero mean, and variance $\llangle 
S^{\mu\nu} S^{\rho\sigma}\rrangle{\sim}2 T\eta \delta(t- 
t')$.
Solving hydrodynamics with noise gives the complete energy density and velocity 
evolution for all wavelengths $\lambda>l_\text{mfp}$, for which \Eq{eq:eomzero} 
is valid, but the solution is not deterministic and depends on the  stochastic 
noise.  Short wavelength fluctuations make numerical simulations very 
challenging and one is typically interested in the 
average noise effect to the long wavelength physics. Averaging \Eq{eq:eomzero} 
over noise 
gives an effective equation of motion for physical perturbations over 
length scales much larger than the dominant distances of noise
\begin{equation}
D_\mu \left<T^{\mu\nu}(e, u^\sigma)\right>=0.
\end{equation}
\begin{figure}
\centering
\includegraphics[width=0.8\linewidth]{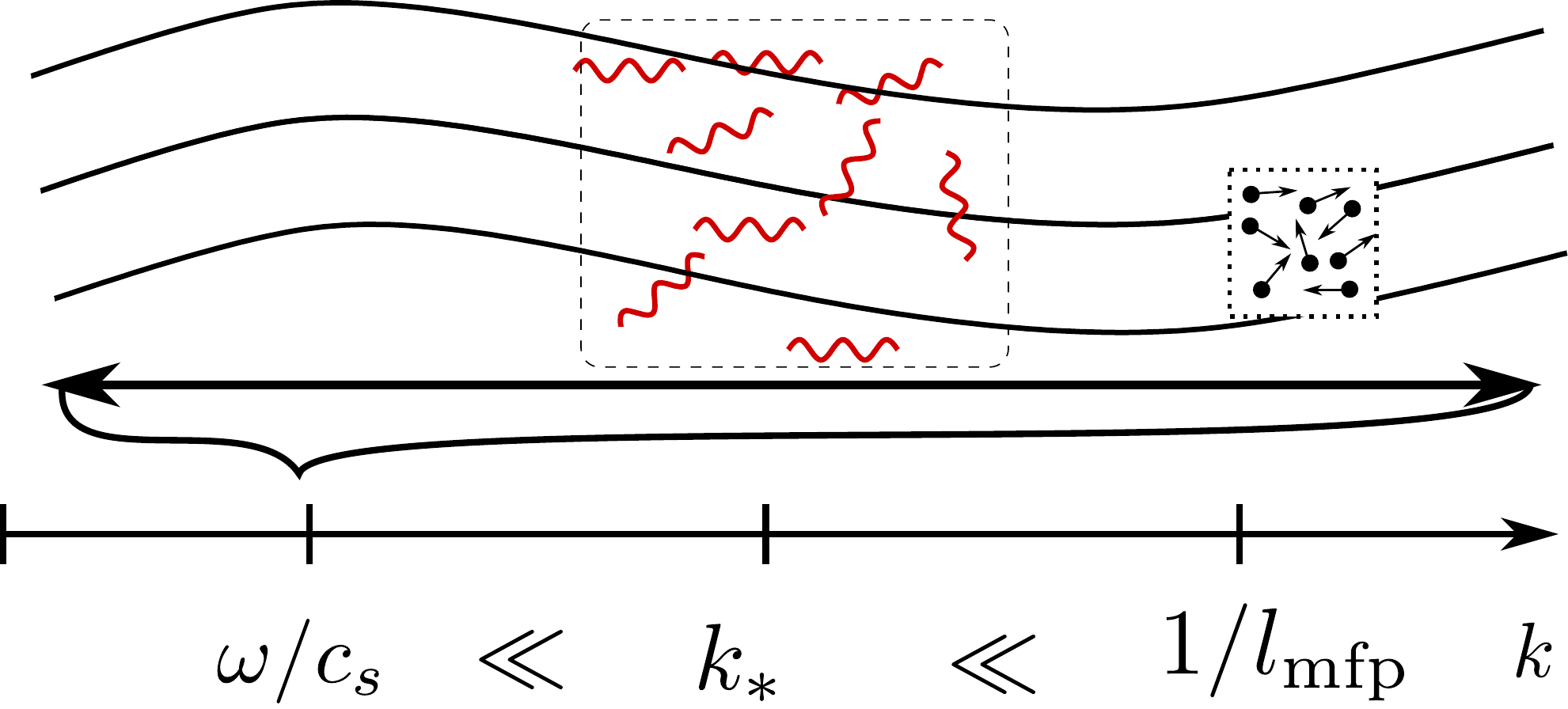}
\caption{\label{intro:scales} The hydro-kinetic description of noise is 
based on 
the  separation of scales between the long wavelength hydrodynamic background 
(with $k \sim \omega/c_s $), and shorter wavelength hydrodynamic
fluctuations (with $k \sim k_{*}\equiv \sqrt{\omega/\gamma_\eta}$). 
The wavelengths of the hydrodynamic
fluctuations  are still much longer than microscopic mean free path. 
The hydrodynamic fluctuations are driven out of
equilibrium by the expanding background, and this 
deviation is the origin of the long-time tail correction to the stress tensor. }
\end{figure}
Energy momentum tensor is a non-linear 
function in background fields $e$ and $u^\sigma$, therefore expanding 
$T^{\mu\nu}$ around the averaged fields gives 
\begin{equation}
\left<T^{\mu\nu}(e, u^\sigma)\right>=T^{\mu\nu}(\left<e\right>, 
\left<u^\sigma\right>)+\mathcal{O}(\left<v^i 
v^j\right>)\ldots\label{eq:averagedeom}
\end{equation}
As we see fluctuations modify the background constitutive equations  
by the quadratic contributions from two point correlation functions $N^{ij}\sim 
\left<v^i v^j\right>$. In equilibrium such 
terms can be calculated using thermal correlation functions and it
amounts to a constant renormalization of background energy and transport 
coefficients.
Rest frame energy density renormalization can be simply understood as including 
kinetic 
energy of noise fluctuations in the average energy density. After the 
appropriate redefinition of background fields and transport coefficients 
ordinary hydrodynamic equations can be solved.

The situation is different in systems where  the evolution, characterized by 
frequency 
$\omega$, is fast enough to drive 
hydrodynamic fluctuations out of equilibrium. 
The dissipative 
relaxation  timescale of modes with wavenumber $k$ is given by $\gamma_\eta 
k^2$, where $\gamma_\eta=\eta/(e+p)$ is the momentum diffusion coefficient. 
Balance between  the driving frequency and relaxation rate
\begin{equation}
\omega \sim \gamma_\eta k_*^2 
\end{equation}
defines a dissipative scale $k_*$ separating the long wavelength physics at 
$\omega/c_s$ and short wavenumber hydrodynamic fluctuations, which remain in 
equilibrium (see \Fig{intro:scales})
\begin{equation}
\frac{\omega}{c_s}\ll k_* \ll \frac{1}{l_\text{mfp}}\label{sepscales}.
\end{equation}

Importantly, out-of-equilibrium
hydrodynamic fluctuations at $k\sim k_*$ obey the 
known hydrodynamic equations of motion, \Eq{eq:eomzero}, and we show that by 
using the separation of 
scales given by \Eq{sepscales} it is possible to derive hydro-kinetic equations 
for 
expectation  of two point 
correlation functions $N^{ij}\sim\left<v^i v^j\right>$ for $k\sim k_*$
\begin{equation}
\partial_\tau N = -\ubr{ Dk^2 (N - N_0)}{relaxation to equilibrium} + 
\ubr{FN}{forcing}.\label{eq:relax}
\end{equation}
Deviations of a stationary solution of \Eq{eq:relax} from the equilibrium value 
$N_0$ around the dissipative scale $k_*$ determines modifications to the 
nonlinear noise contributions in 
\Eq{eq:averagedeom}, which is the principal outcome of hydrodynamics with noise.

\subsubsection{Application}

Thermal fluctuations occur naturally in dissipative systems, but in the 
context of heavy ion collisions  hydrodynamic fluctuations  have been 
considered only recently~\cite{Kovtun:2003vj,Kovtun:2011np}. The direct 
dependence of thermal noise on the medium parameters makes hydrodynamic 
fluctuations an attractive alternative probe of viscous coefficients in QGP.  
Previous calculations considered one dimensional averaged noise effects for the 
expanding 
Bjorken flow~\cite{Gavin:2006xd,Kapusta:2011gt} and numerical 
simulations of 
hydrodynamics with 
noise on smooth~\cite{Yan:2015lfa,Young:2014pka,Nagai:2016wyx} and 
event-by-event fluctuating backgrounds~\cite{Murase:2016rhl}.
The short range of noise correlations makes it problematic to simulate noise 
correctly on numerical grids. Instead in \Chap{chap:noise} we propose effective 
hydro\nobreakdash-kinetic equations, which can be in principle coupled to 
existing 
hydrodynamic simulations to determine the nonlinear noise contributions to the 
relevant long wavelength physics. In the framework of hydro-kinetics, the  
3D momentum phase space of hydrodynamic fluctuations can be straightforwardly 
organized to give a physically intuitive interpretation of different  noise 
effects. Finally, the hydro-kinetics is not restricted to close-to-equilibrium 
systems and can be applied to general hydrodynamic backgrounds.

 In \Chap{chap:noise} we derive effective hydro-kinetic 
equations 
for noise 
correlations in the  expanding Bjorken background and calculate the nonlinear 
noise modifications of background constitutive  equations.

\noappendicestocpagenum

\chapter{Subleading flows and factorization breaking with principal components}
\label{chap:pca}

 The 
material in the following sections  previously 
appeared in:
\begin{itemize}
\item A. Mazeliauskas and D. Teaney, \emph{Subleading harmonic flows in 
hydrodynamic 
simulations of heavy ion collisions}, Phys. Rev. C91, 044902
(2015)~\cite{Mazeliauskas:2015vea}.  Copyright (2015) by the American Physical 
Society
\item A. Mazeliauskas and D. Teaney, \emph{Fluctuations of harmonic and radial
flow in heavy ion collisions with principal components}, Phys. Rev. C93,
024913 (2016)~\cite{Mazeliauskas:2015efa},  Copyright (2016) by the American 
Physical Society
\end{itemize}

\section{Introduction}

Two-particle correlation measurements are of paramount importance in studying
ultra-relativistic heavy ion collisions, and provide an extraordinarily
stringent test for theoretical models. Indeed, the measured two-particle
correlations exhibit elliptic, triangular, and higher harmonics flows,  which 
can be used to constrain the transport properties of the quark gluon plasma 
(QGP)~\cite{Heinz:2013th,Luzum:2013yya}.  The remarkable
precision of the experimental data 
as a function of transverse momentum and pseudorapidity
has led to  new analyses of factorization breaking, nonlinear mixing, event 
shape selection, and
forward-backward
fluctuations~\cite{Khachatryan:2015oea,Aad:2014fla,Aad:2015lwa,Adam:2015eta,Adamczyk:2015xjc,Adare:2015ctn}.
In this paper we analyze the detailed structure of two-particle transverse
momentum correlations by using  event-by-event (boost-invariant) hydrodynamics
and principal component analysis (PCA) 
\cite{Bhalerao:2014mua}.  Specifically, we
decompose the event-by-event harmonic flow $V_n(p_T)$
into  principal components and investigate the physical
origin of each of these fluctuations. Here we present an extensive theoretical 
study of  
$n=0\text{--}5$ flow harmonics at the LHC (Pb+Pb at $\snn=2.76\,{\rm TeV}$) . 
In particular, 
we 
demonstrate the importance of radial flow fluctuations for subleading flows of 
higher harmonics.

Taking the second harmonic for definiteness,  the two-particle correlation
matrix of momentum dependent elliptic flows, $C_2(p_{T1},p_{T2}) \equiv 
\llangle V_2(p_{T1}) V_2^*(p_{T2}) \rrangle$ is traditionally parametrized by  
factorization ratio
$r_2(p_{T1}, p_{T2})$~\cite{Gardim:2012im},
\begin{equation}
r_2(p_{T1}, p_{T2}) \equiv \frac{ \llangle V_2(p_{T1}) V_2^*(p_{T2}) 
\rrangle}{\sqrt{ \llangle |V_2(p_{T1})|^2 \rrangle
\llangle |V_2(p_{T2})|^2  \rrangle } }\,.
\end{equation}  

If there is only one source of elliptic flow in the event [for example if
in each event
$V_2(p_T) = f(p_T) \varepsilon_2$ 
with $\varepsilon_2$ a complex eccentricity and
$f(p_T)$ a fixed real function of $p_T$]  then the correlation matrix of 
elliptic
flows $C_2(p_{T1},p_{T2})$ factorizes into a product of functions, and the
$r_2$ parameter is unity.
However, if there are multiple independent sources of elliptic flow in the 
event, then the correlation matrix does not factorize, and the 
$r_2$ parameter is less than unity~\cite{Gardim:2012im}.  
The $r_2$ parameter
 has been extensively studied 
both 
experimentally~\cite{Khachatryan:2015oea,Aamodt:2011by,ATLAS:2012at,Aad:2014lta}
 and 
theoretically~\cite{Gardim:2012im,Heinz:2013bua,Kozlov:2014fqa}.
PCA is a statistical
technique that decomposes the flow correlation matrix into eigenvectors and
eigenvalues. The procedure naturally identifies the
most important contributions to flow fluctuations.   Typically  only two or 
three modes
are needed to give an excellent  description of the full covariance matrix (see 
\Fig{fig:rij}).  When there are
only two significant eigenvectors, the $r_n$ matrix
can be expressed as~\cite{Bhalerao:2014mua}
\begin{equation}
\label{pcar}
r_n(p_{T1},p_{T2})           \simeq 1 - \frac{1}{2} \left( 
\frac{V_n^{(2)}(p_{T1})}{V_n^{(1)}(p_{T1}) } -  
\frac{V_n^{(2)}(p_{T2})}{V_n^{(1)}(p_{T2})}\right)^2 \, ,
\end{equation}
where  $V_{n}^{(1)}(p_T)$  and $V_{n}^{(2)}(p_T)$ are the 
first and second eigenvectors. In particular, in the case of triangular flow, 
factorization breaking in 
event-by-event
hydrodynamics arises because the simulated triangular flow is predominantly the 
result of two
statistically uncorrelated contributions---the linear response to 
$\varepsilon_3$~\cite{Alver:2008aq} and the
linear response to the first radial excitation of 
$\varepsilon_3$. The extension of this understanding of factorization breaking 
to the other harmonics was surprisingly subtle due to the quadratic
mixing between the leading and subleading harmonic flows. 

Experimentally, it is observed that factorization breaking is largest for 
elliptic flow in
central collisions (see in particular Fig.~28 of Ref.~\cite{ATLAS:2012at} and 
Fig.~1 of Ref.~\cite{Khachatryan:2015oea}). Indeed, the $r_2$ parameter 
decreases
rather dramatically from mid-central  to central collisions. This indicates
that the relative importance of the  various initial state fluctuations which
drive elliptic flow are changing rapidly as a function of centrality.  The
current manuscript explains the rapid centrality dependence of factorization
breaking in $v_2$ as an interplay between the linear response to the
fluctuating elliptic geometry, and the nonlinear mixing of the radial
flow and average elliptic flow. 
This quadratic mixing is similar to the mixing between $v_5$ and $v_2,v_3$ 
~\cite{Borghini:2005kd,Qiu:2011iv,Gardim:2011xv,Teaney:2012ke}, and 
this picture can be confirmed experimentally by measuring specific three point 
correlations analogous  to the three plane correlations
measured in the $v_5,v_2,v_3$ case~\cite{Aad:2014fla,Aad:2015lwa}.

To understand the linear and nonlinear contributions quantitatively, we will 
break
up the fluctuations in hydrodynamics into their principal components, and
analyze the linear and nonlinear contributions of each principal component to
the simulated harmonic spectrum. In Sects.~\ref{principals} and 
\ref{simulations} we review the 
analysis definitions,  and the key features of simulations.
To gain insight into the results of PCA analysis, we will first look at the 
leading and subleading harmonics of triangular flow (which is sourced entirely 
by 
fluctuations). 
Section \ref{subleading} 
studies
the basic properties of the subleading triangular flow, such as its dependence
on centrality and viscosity. 
In \Sect{avg_geometry} we show that the subleading triangular flow arises
(predominantly) from the radial excitation of the triangular geometry.  
 In 
\Sect{predictors} we discuss the strategy for constructing the best linear 
predictor for leading and subleading flows. We define a generalized 
eccentricity $\varepsilon_n\{\rho(r)\}$ in \Eq{epsgen} and use radial Fourier 
modes 
to optimize the radial weight $\rho(r)$. The 
geometric predictors described 
above are ultimately based on the
assumption of linear response. At least for the third harmonic, these 
assumptions are checked in \Sect{linear_responsepca1}.

After gaining experience with the triangular flow, we summarize the key 
results of a comprehensive principal component analysis  of flow 
harmonics
$n=0\text{--}5$ in the second part of 
our paper. In
\Sect{pca:results} we give individual 
discussions 
for each harmonic flow and construct optimal predictors for leading and 
subleading flows based on initial geometry 
and nonlinear mixing. First, we discuss radial flow 
fluctuations in 
\Sect{radial} and then demonstrate their importance in generating subleading 
elliptic flow in \Sect{elliptic}. In 
\Sect{v1v3} we describe our PCA results for direct and triangular 
flows. 
Finally, in \Sect{v4v5} we discuss the quadrangular and pentagonal flows and 
how the nonlinear mixing of lower order principal components adds to 
these flows. 
We put forward some experimental observables in the discussion in 
\Sect{discussion}. For convenience, we present a catalog of 
figures showing PCA results for each harmonic in the 
\hyperref[lof]{Appendix}.

\section{Principal components}
\label{principals}

PCA is a statistical technique for extracting the dominant components in 
fluctuating data. In the context of flow in heavy ion collisions it was first 
introduced in Ref.~\cite{Bhalerao:2014mua}  to quantify the 
dominant momentum space fluctuations of harmonic
flows in transverse momentum and rapidity in a precise way.  This section
provides a brief review of this statistical technique.

Paraphrasing Ref.~\cite{Bhalerao:2014mua}, in the flow picture of heavy ion
collisions the particles in each event are drawn independently from a single
particle distribution which  fluctuates from event to event.  The
event-by-event single particle distribution is expanded in a Fourier series
\begin{equation}
\frac{\dd N}{\dd \p}   = V_0(p_T) + \sum_{n=1}^{\infty} V_n(p_T) e^{-in 
\varphi}  + {\rm H.c.}\, ,\label{fseries}
\end{equation}
where $\dd\p =  \dd y \,\dd p_{T} \, \dd\varphi$ notates the phase space,
$\varphi$  is the azimuthal angle of the distribution, and H.c. denotes
Hermitian conjugate. $V_n(p_T)$ is a complex Fourier coefficient recording the
magnitude and orientation of the $n$th harmonic flow.  This definition
deviates from the common practice of normalizing the complex Fourier
coefficient by the multiplicity, $v_n(p_T) = V_n(p_T)/V_0(p_T)$. 

Up to non-flow corrections of order the multiplicity $N$, the long-range part
of the two-particle correlation function is determined by the statistics of the
event-by-event fluctuations  of the single distribution
\begin{equation}
\llangle \frac{\dd N_\text{pairs}}{\dd\p_1 \dd \p_2 }  \rrangle  = \llangle 
\frac{\dd N}{\dd \p_1} \frac{\dd N}{\dd \p_2} \rrangle + \mathcal 
O\left(N\right) \, .
\end{equation}
If the two-particle correlation function  is also expanded in a Fourier series
\begin{equation}
\llangle \frac{\dd N_\text{pairs}}{\dd\p_1 \dd \p_2 }  \rrangle  = \sum_{n} 
V_{n\Delta}(p_{T1}, p_{T2}) e^{-in(\varphi_1 - \varphi_2)} \, , 
\end{equation}
then this series determines the statistics of $V_n(p_T)$
\begin{equation}
 C_n(p_{T1},p_{T2})\equiv  V_{n\Delta}(p_{T1}, p_{T2}) -\left< 
   V_{n,p_{T1}}\right>\left< 
   V^*_{n,p_{T2}}\right> = \big< \big(V_{n,p_{T1}}-\left< 
   V_{n,p_{T1}}\right>\big) \big( V^*_{n,p_{T2}}-\left< 
   V^*_{n,p_{T2}}\right>\big) \big> \, 
   .
\end{equation}
The covariance matrix  $C_{n}(p_{T1},p_{T2})$, which is real,  symmetric,
and positive-semidefinite, can be decomposed into real orthogonal eigenvectors 
$V_n^{(a)}(p_T)$,
\begin{align}
  C_n(p_{T1},p_{T2}) =&  \sum_{a} \lambda^{a} \psi^{(a)}_n(p_{T1}) 
   \psi^{(a)}_n (p_{T2})   = \sum_{a} V_n^{(a)}(p_{T1}) 
   V^{(a)}_n(p_{T2})\label{dec2},
\end{align}
where $V_n^{(a)}(p_T)\equiv\sqrt{\lambda^{a} } \,\psi^{(a)}_n(p_T)$ is the 
square root of the eigenvalue times a normalized eigenvector 
$\int_0^\infty\!dp_T\,\psi^{(a)}\psi^{(b)}=\delta_{ab}$. The eigenvalue records 
the squared variance of a given fluctuation.

The principal components $V_n^{(1)}(p_{T}), V_n^{(2)}(p_{T}),\ldots$ of a given 
event 
ensemble 
can be used as optimal basis 
for event-by-event expansion of harmonic flow
\begin{equation}
V_n(p_{T}) -\left<V_n(p_{T})\right> = \xi_n^{(1)} V^{(1)}_n(p_{T}) + 
\xi_n^{(2)} 
V^{(2)}_n(p_{T}) +\ldots \label{def_xi} \, .
\end{equation}
The complex coefficients $\xi_n^{(a)}$ are the projections of 
harmonic flow onto principal component basis and record  the orientation and 
event-by-event amplitude of their respective flows.
Principal components are mutually uncorrelated
\begin{equation}
\left<\xi_n^{(a)}\xi_n^{*(b)}\right> = \delta^{ab}.\label{orth}
\end{equation}

Typically the eigenvalues of $C_n(p_{T1},p_{T2})$ are strongly ordered and only 
the 
first 
few terms in the expansion are significant. Often the large components
have a definite physical interpretation. We define the 
scaled magnitude of the flow vector $V_n^{(a)}(p_T)$ as
\begin{equation}
\label{pcamagnitudes}
\| v^{(a)}_n\|^2  \equiv \frac{\int \big(V_n^{(a)}(p_T)\big)^2dp_T}{\int 
\left<dN/dp_T\right>^2dp_T},
\end{equation}
which is a measure of the size of the fluctuation without
trivial dependencies on the mean multiplicity in a given centrality class.
\begin{figure}
\centering
\includegraphics[width=\linewidth]{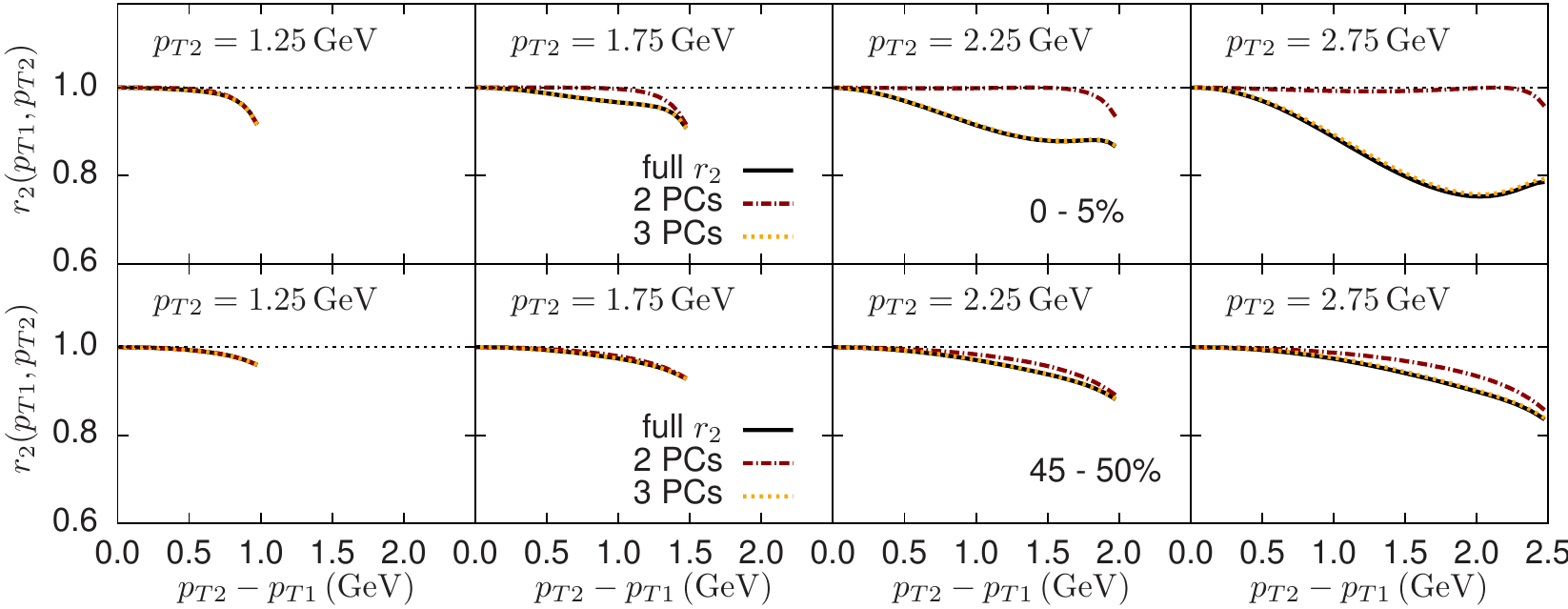}
\caption{Factorization ratio $r_2(p_{T1},p_{T2})$ [\Eq{rij}] for elliptic 
flow and its 
approximations 
with principal components (PCs) in central (0--5\%) and peripheral  (45--50\%) 
collisions.}
\label{fig:rij}
\end{figure}

The leading flow vector $V^{(1)}_n(p_T)$ corresponds to fluctuations 
with the largest root-mean-square amplitude, while subsequent components 
maximize the 
variance  in the remaining orthogonal directions. This yields a very efficient 
description of the full covariance matrix $C_n(p_{T1},p_{T2})$ and 
factorization ratio
\begin{equation}
r_n(p_{T1}, p_{T2}) \equiv \frac{ C_n(p_{T1}, p_{T2})}{\sqrt{ C_n(p_{T1}, 
p_{T1}) 
C_n(p_{T2}, p_{T2})}}\leq 1\, \label{rij}.
\end{equation}
$r_n(p_{T1},p_{T2})$ is bounded by unity  within 
hydrodynamics~\cite{Gardim:2012im}. By truncating series expansion 
of the covariance matrix
[\Eq{dec2}] at two or three principal components we can 
approximate $C_n(p_{T1},p_{T2})$ and $r_n(p_{T1},p_{T2})$. Truncating  
at the leading term would constitute complete flow factorization, i.e., 
$r_n=1$. 
The factorization matrix for elliptic flow is
shown in \Fig{fig:rij}. We see that in peripheral collisions at low momentum 
$p_T<2.5\,\rm{GeV}$ just two 
principal components are sufficient to describe momentum dependence of 
factorization ratio $r_2$. In central collisions (as explained in 
\Sect{elliptic}) the subleading and subsub-leading elliptic flows are 
comparable 
and both are needed to given an excellent description of $r_2$.

 Analogous decompositions of two-particle correlations
into principal components exist for all harmonics and 
all centralities. 
Interpreting these large flow components physically is the goal of this work.

\section{Simulations}
\label{simulations}
We used boost-invariant event-by-event viscous hydrodynamics to simulate 5000 
Pb-Pb  collisions at the CERN Large Hadron Collider (LHC) ($\snn=2.76\,{\rm 
TeV}$) in fourteen 5\% centrality classes 
selected by impact parameter.  Our simulations are
boost invariant and implement second order viscous
hydrodynamics~\cite{Baier:2007ix}, using a code base which has been developed
previously~\cite{Dusling:2007gi,Teaney:2012ke}.  For the initial conditions we
use the Phobos Glauber Monte Carlo \cite{Alver:2008aq}, and we distribute the
entropy density in the transverse plane according to a two-component model.
Specifically, for the $i$th  participant we assign a weight
\begin{equation}
A_i \equiv \kappa \left[ \frac{ (1- \alpha) }{2}   + \frac{\alpha}{2} (n_{\rm 
coll})_i \right] \, ,
\end{equation}
with $\alpha=0.11$,  $\kappa=35.1$ for $\eta/s=0.08$, and $\kappa = 32.8$
for $\eta/s=0.16$. $(n_{\rm coll})_i$ is the number of binary collisions
experienced by the $i$th participant; so the total number of binary
collisions is $ N_{\rm coll} = \half \sum_{i} (n_{\rm coll})_i$.  The entropy
density in the transverse plane at initial time $\tau_o$ and transverse
position $\x=(x,y)$ is taken  to be
\begin{equation}
s(\tau_o,\x)  = \sum_{i\element {\parts}} s_{i}(\tau_o, \x - \x_i) \, ,
\end{equation}
where $\x_i = (x,y)$ labels the transverse coordinates of the $i_{\rm th}$ 
participant, and
\begin{equation}
s_{i}(\tau_o, \x) = A_i \, \frac{1}{\tau_o (2\pi \sigma^2) } e^{- 
\frac{x^2}{2\sigma^2} -\frac{y^2}{2\sigma^2} } \, , 
\end{equation}
with $\sqrt{2}\sigma = 0.7\,{\rm fm}$.
The parameters $\kappa$ and $\alpha$ are marginally different from  Qiu's
thesis \cite{Qiu:2013wca}, and we have independently verified that this choice
of parameters reproduces the average multiplicity in the event.%
\footnote{ More precisely we have verified that for these parameters
    hydrodynamics with averaged initial conditions reproduces $\left. \dd
    N_{\rm ch}/\dd \eta\right|_{\eta=0}$ as a function of centrality after all
    resonance decays are included.  Assuming that the ratio of the charged
    particle yield to the direct pion yield is the same as in the averaged
    simulations, the current event-by-event simulations reproduces $\dd N_{\rm
ch}/\dd \eta$.}

The equation of state is motivated by lattice QCD
calculations~\cite{Laine:2006cp} and has been used previously by Romatschke
and  Luzum~\cite{Luzum:2008cw}.  In this paper we  compute ``direct" pions
(i.e. pions calculated directly from the freeze-out surface) and we do not
include resonance decays. We use a freeze-out temperature of $T_{\rm fo} =
140\,{\rm MeV}$.

Simulation results were generated for fourteen 5\% centrality classes with
impact parameter up to $b={12.4}\rm{fm}$ and at two viscosities, $\eta/s=0.08$
and $\eta/s=0.16$. Unless specified, the results are for $\eta/s=0.08$.  We
generated 5000 events per centrality class.%
\footnote{
We thank Soumya Mohapatra for collaboration during the initial stages of this
project.}
We then performed PCA for the $n$th harmonic $V_n(p_T)$ by discretizing
$V_n(p_T)$ results from hydrodynamics into 100 equally spaced bins between
$p_{T} = 0\ldots 5\,{\rm GeV}$, and finding the eigenvalues and eigenvectors of
the resulting Hermitian matrix.

Table~\ref{glaubertable} records the Glauber data which is used in this
analysis.  Event-by-event averages with the initial entropy density are notated
with square brackets, e.g.
\begin{equation}
[r^2]  \equiv \frac{1}{\overline S_{\rm tot} } \int d^2\x \, \tau_o 
s(\tau_o,\x) r^2,   
\end{equation}
where $\overline S_{\rm tot}$ is the average total entropy in a given
centrality class, $\left<\int d^2\x \, \tau_o s(\tau_o,\x)\right>$.  Averages
over events are notated with $\llangle\,\rrangle$, so that the root mean square
radius is
\begin{equation}
  R_{\rm rms}\equiv \sqrt{\left<[r^2]\right>}\, .
\end{equation}
As a technical note, here and below the radius is measured from the center of
entropy, so $[\x]=0$.  $\varepsilon_{n,m}$  is defined
in a somewhat unorthodox fashion in \Eq{eps3all}, with 
$\varepsilon_{3,3}^{\rm
rms} \equiv \sqrt{\llangle |\varepsilon_{3,3}|^2  \rrangle}$.
$\overline{r}_{\rm max}$ is the averaged maximum participant radius, ${\rm max}
\, |\x_i| $.
\begin{table}
\begin{center}
\begin{tabular}{|c |c |c | c | c | c |} \hline
   $\vphantom{\overline{\overline{H}}}$ Centrality& ($b_{\rm min}, b_{\rm 
   max}$) & $\overline N_{\rm part}$ & $R_{\rm rms}$ & $\overline{r}_{\rm max}$ 
   & $\varepsilon_{3,3}^{\rm rms}$ \\ \hline \hline
0--5  \%  & (0.0, \,  3.3) & 384 &  4.1 &  8.1 & 0.11\\
5--10 \%  & (3.3, \,  4.7) & 335 &  3.9 &  7.8 & 0.14\\
10--15\%  & (4.7, \,  5.7) & 290 &  3.7 &  7.5 & 0.17\\
15--20\%  & (5.7, \,  6.6) & 250 &  3.6 &  7.3 & 0.20\\
20--25\%  & (6.6, \,  7.4) & 215 &  3.4 &  7.0 & 0.22\\
25--30\%  & (7.4, \,  8.1) & 184 &  3.3 &  6.7 & 0.25\\
30--35\%  & (8.1, \,  8.8) & 156 &  3.2 &  6.4 & 0.28\\
35--40\%  & (8.8, \,  9.4) & 132 &  3.1 &  6.2 & 0.32\\
40--45\%  & (9.4, \,  9.9) & 110 &  3.0 &  5.9 & 0.35\\
45--50\%  & (9.9, \, 10.5) & 91  &  2.9 &  5.7 & 0.39\\
50--55\%  & (10.5,\, 11.0) & 74  &  2.7 &  5.4 & 0.44\\
55--60\%  & (11.0,\, 11.5) & 60  &  2.7 &  5.1 & 0.48\\
60--65\%  & (11.5,\, 11.9) & 47  &  2.6 &  4.8 & 0.52\\
65--70\%  & (11.9,\, 12.4) & 37  &  2.4 &  4.4 & 0.55\\\hline
\end{tabular}
\end{center}
\caption{
   Table of parameters from the Glauber model (all distances are measured in 
   fm).
   \label{glaubertable}}
\end{table}

\section{Subleading triangular flow}
\label{subleading}

To gain insight into principal components of harmonic flow, we first analyze  
triangular flow, since it is a strong signal and is driven entirely by 
fluctuations~\cite{Alver:2010gr}. As a first step, we list the (scaled) 
magnitudes of flows $\|v_{3}^{(a)}\|$ 
[\Eq{pcamagnitudes}] in central collisions for the simulations described 
above.  
	\begin{center}
		\begin{tabular}{|c ||c |c | c | c |} \hline
			$a$ & 1 & 2 & 3 & 4\\\hline\hline
			$\|v_3^{(a)}\|$ & $1.5\times 10^{-2}$ & $2.6\times 10^{-3}$ & 
			$4.8\times 
			10^{-4}$ & 
			$1.1\times 10^{-4}$ \\ \hline
		\end{tabular}
	\end{center}

   Note that the scaled magnitudes are proportional to the square-root of
   the eigenvalues, $\|v_3^{(a)}\| \propto \sqrt{\lambda_a}$.
From the decreasing magnitudes  of the listed (scaled) magnitudes, we see that
the first two eigenmodes account for 99.9\% of the squared variance, which can
be represented as a  sum of the eigenvalues
\begin{align}
   \int_0^{\infty} \dd p_{T} \llangle V_{3}(p_T)  V_{3}^*(p_T) \rrangle =&  
   \sum_{a}\lambda_a  \propto \sum_{a} \|v_3^{(a)}\|^2 \, .
\end{align}

\begin{figure}
	\includegraphics[width=0.48\linewidth]{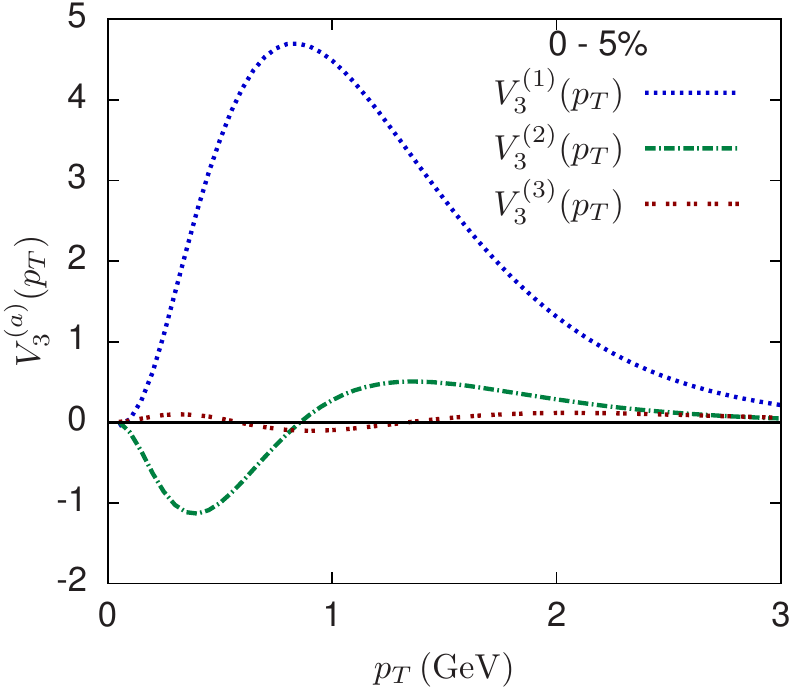}\quad
	\includegraphics[width=0.48\linewidth]{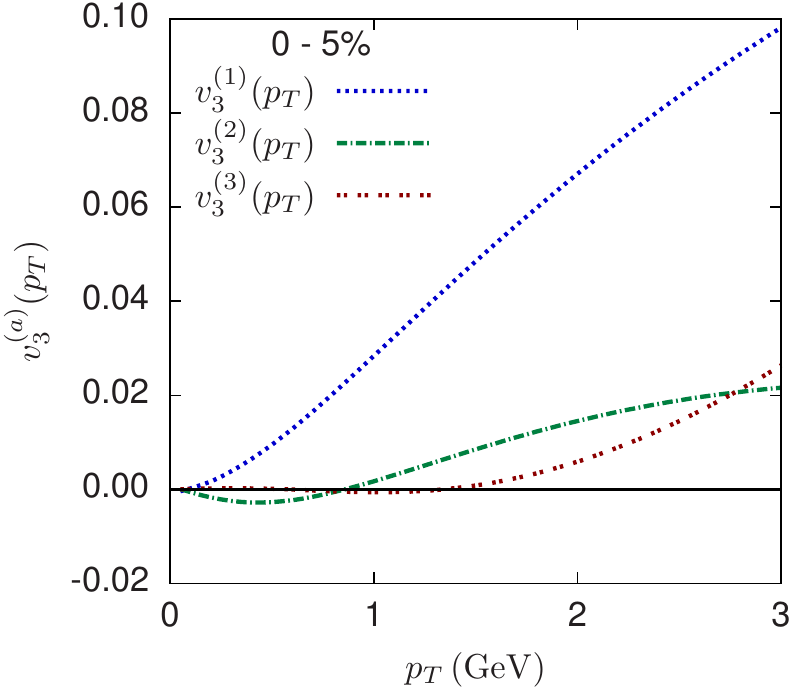}%
	\caption{Momentum dependence of triangular flow components in central 
	collisions. a) 
	Principal flow vectors, $V^{(a)}_3(p_T)$. b) Principal flow vectors divided 
	by 
		the average multiplicity, 
    $v_3^{(a)}(p_T) \equiv V_3^{(a)}(p_T) /\llangle \dd N/\dd p_T 
    \rrangle$.\label{PCA_Vn_1_nonorm}}
\end{figure}

Figure \ref{PCA_Vn_1_nonorm}(a) displays the eigenvectors, $V_{3}^{(a)}(p_T)$, 
for the
leading and first two subleading modes.  We see that for triangular flow  only 
the first two flow
modes are significant. To make contact with the more traditional definitions of 
$v_3(p_T)$,
we divide by $\llangle \dd N/\dd p_T \rrangle$ and present the same eigenmodes
in \Fig{PCA_Vn_1_nonorm}(b).

\begin{figure}
\centering
	\includegraphics[width=0.5\linewidth]{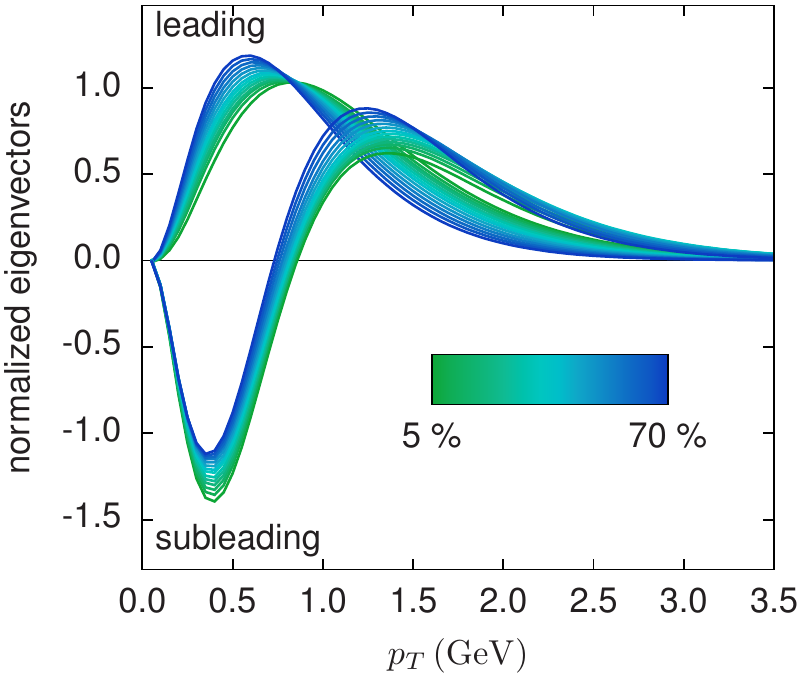}
	\caption{Centrality dependence of triangular flow eigenvectors 
	$\psi^a(p_T)$.\label{PCA_Vn_2}}
\end{figure}

We also investigated the  centrality and viscosity dependence of the principal
components. The normalized principal flow eigenvectors $\psi^{(a)}_3(p_T)$ are
approximately independent of viscosity (not shown). In \Fig{PCA_Vn_2}, we show
the centrality dependence of these normalized eigenvectors. In more central
collisions the eigenvectors shift to larger transverse momentum, which can be
understood with the system size scaling introduced in Ref.~\cite{Basar:2013hea}.

\begin{figure}
\centering
	\includegraphics[width=0.5\linewidth]{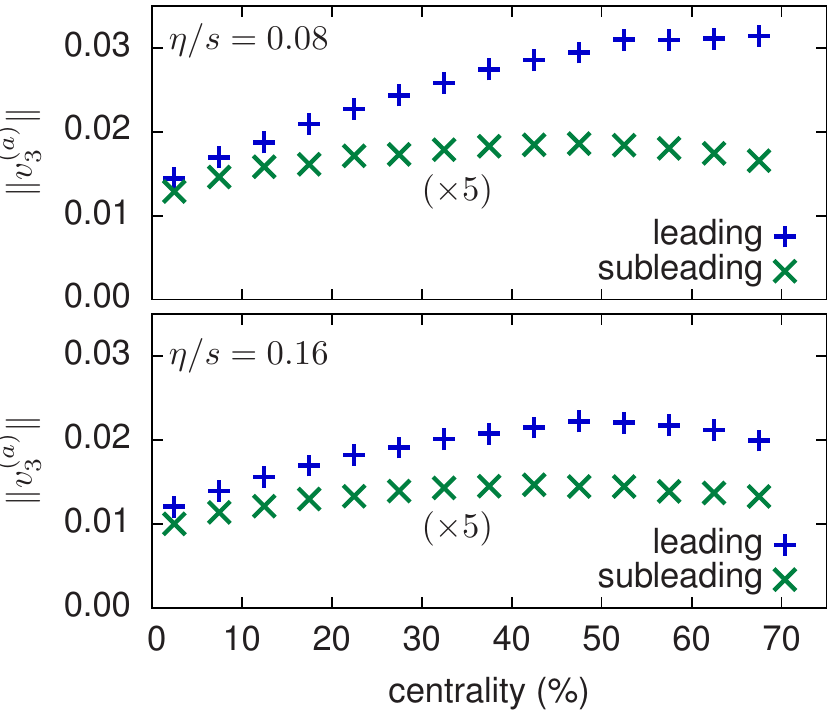}
	\caption{Centrality and viscosity dependence of scaled eigenvalues 
	$\|v^{(a)}_3\|$. 
    (The subleading flow has been magnified 5 times to bring to scale with 
    leading flow.)
\label{PCA_eval_2}}
\end{figure}

The magnitude of the flow, i.e. the squared integral $\int\,
(V_{3}^{(a)}(p_T))^2dp_T$, depends on both centrality and viscosity. To
factor out the trivial multiplicity dependence of $V_3(p_T)$,  we plot the
scaled flow eigenvalues $\|v_n^{(a)}\|$ [see \Eq{pcamagnitudes}] in 
\Fig{PCA_eval_2}.
Going from $\eta/s=0.08$ to $\eta/s=0.16$ we see significant suppression of the
leading mode.  In general the subleading scaled flow $\|v_3^{(a)}\|$ depends
weakly on centrality.

We present PCA results on all harmonic $n=0\text{--}5$ in the \app{lof}. 
Namely, we show 
the  flow magnitude dependence on centrality for the largest three principal 
components, and principal eigenvectors in central 0-5\% bin. Individual 
discussions for each harmonic are given in 
\Sect{pca:results}.

\section{Average geometry in the subleading plane}
\label{avg_geometry}

In this section, we clarify the physical origin of the subleading flow by
correlating the subleading hydrodynamic response with the geometry.

\begin{figure}
\centering
\includegraphics[width=0.9\linewidth]{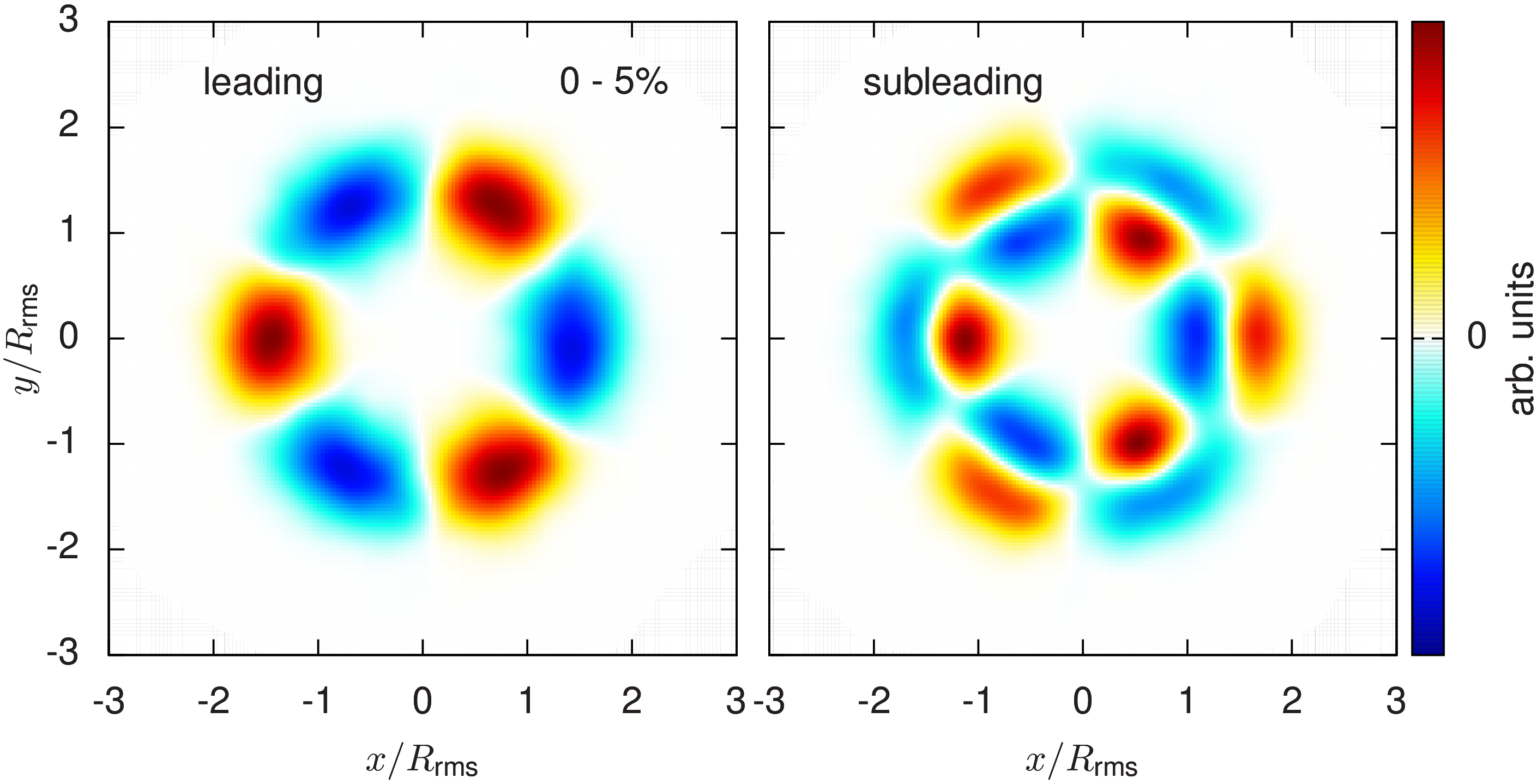}
\caption{Average geometry$\times r^3$ in the leading and subleading
 principal component planes in central collisions 
minus an averaged radially symmetric background, $r^3 
(\overline{S}(\x;\,\xi^{(a)}_3) 
- \left<S(\x)|\xi^{(a)}_3|\right>)$.
Peak fluctuations are $\pm$10--20\% above the background.\label{PCA_S_1}}
\end{figure}

As a first step, we determined the average initial geometry in the principal
component plane.  Specifically, for each event the phase of the principal
component $\xi^{(a)}_n$ [see \Eq{def_xi}] defines orientation of the flow. We 
then 
rotate each event
into $\xi^{(a)}_n$ plane and average the initial entropy density, $S(\x)\equiv 
\tau_o
s(\tau_o, \x)$. More precisely, the event-by-event geometry in the principal
component plane  is defined to be 
\begin{equation}
	S(r,\phi;\,\xi^{(a)}_n) \equiv 
	\frac{1}{n}\sum_{\ell=0}^{n-1}S\left(r,\phi+(\arg 
	\xi^{(a)}_n +2\pi \ell)/n\right) \, , 
\end{equation}
where we have averaged over the phases of $\sqrt[n]{\xi^{(a)}_n}$. Next, we 
average
$S(r,\phi;\,\xi^{(a)}_n)$ over all events weighted by the magnitude of the 
flow\footnote{Radial flow $V_0(p_T)$ does not have a particular orientation in 
the 
transverse plane. Instead we average event-by-event geometry with respect to the
sign and magnitude of  the radial flow  $\overline{S}(r,\phi;\,\xi^{(a)}_0) 
\equiv\left<S(r,\phi) 
	\xi^{(a)}_0\right>$ } 
\begin{equation}
	\overline{S}(r,\phi;\,\xi^{(a)}_n) \equiv\left<S(r,\phi;\,\xi^{(a)}_n) 
	|\xi^{(a)}_n|\right>\label{Srphibar}.
\end{equation} 
Figure \ref{PCA_S_1} shows the in-plane averaged geometry
$\overline{S}(r,\phi;\,\xi^{(a)}_3)$ for the leading and subleading principal
components of the triangular flow in central collisions.  
Clearly, the leading triangular principal component $V_{3}^{(1)}$ is strongly 
correlated
with the triangular components of the initial geometry, while the subleading
component $V_{3}^{(2)}$ is correlated with the radial excitations of this
geometry. In central collisions, subleading flows of other harmonics 
$n=0\text{--}5$ are also
predominantly correlated with radial excitations of the corresponding order 
eccentricity.  However, in peripheral collisions, flow harmonics 
with significant nonlinear mixing, e.g. subleading 
$v_2$, $v_4$ and $v_5$, are not entirely due to radially excited geometry (as 
explained in \Sect{results}) and 
thus the averaged geometry  in the subleading 
principal component plane does not 
show sharp features of radial excitations.

\begin{figure}
\centering
\includegraphics[width=0.5\linewidth]{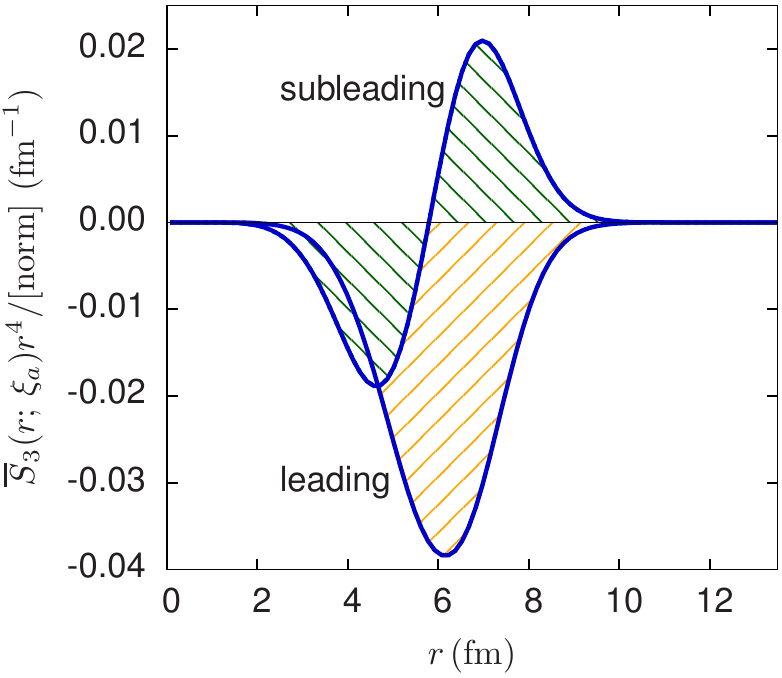}
\caption{\label{PCA_S_2} 
Correlation between the principal components and the 
triangular geometry, $\llangle S_3(r) \xi^{*(a)}_3 \rrangle$, for
the leading and subleading flows in central collisions.
The result has been multiplied by $r^4$ and normalized by $\overline{S}_{\rm 
tot}R_{\rm rms}^3$, so that the area under the leading curve is approximately 
$\varepsilon_{3,3}^{\rm rms}$.}
\end{figure}

To give a one-dimensional projection of \Fig{PCA_S_1}, we integrate
\Eq{Srphibar} over the azimuthal angle to define
\begin{equation}
	\overline{S}_n(r;\,\xi^{(a)}_n)\equiv \int_0^{2\pi}\!\!\!\text{d}\phi\, 
	\overline{S}(r,\phi;\,\xi^{(a)}_n)\,e^{in\phi}\label{Srbar}.
\end{equation}
This is equivalent to defining $S_n(r)$,
\begin{equation}
S_{n}(r) \equiv \int_0^{2\pi} \!\!\! \dd \phi \, S(r,\phi) \,  e^{in\phi}  \, , 
\end{equation}
and correlating this with the flow fluctuation $\xi^{(a)}_n$
\begin{equation}
\overline{S}_n(r;\,\xi^{(a)}_n) = \llangle S_n(r) \xi^{*(a)}_n \rrangle \, .
\end{equation}
Results for the triangular geometry $\overline{S}_3(r;\,\xi^{(a)}_3)r^4$ are 
shown by the blue (gray) 
curves 
in
\Fig{PCA_S_2}. 
Again we see that the leading  triangular flow originates from a geometric 
fluctuation with
a large integrated eccentricity, while the subleading flow is sensitive to the
radial excitation of the triangularity. We conclude that the  relatively small
subleading flow can correspond to a fairly significant fluctuation of the 
initial geometry.

\section{Geometrical predictors}
\label{predictors}

It is clear from \Fig{PCA_S_2} that usual geometric predictors based on a 
single coarse grained quantity like $\varepsilon_{n,n}$ cannot capture all the 
features of the radially excited geometry.
In this section we will construct several geometric predictors for the leading
and subleading flows following strategy outlined in Ref.~\cite{Gardim:2011xv}. 
Keeping the 
discussion general, let $\xi^{(a)}_{n\,\text{pred}}$ be a geometric quantity 
which predicts the event-by-event amplitude and phase of 
the corresponding flow $\xi^{(a)}_n$ .
For example, for the leading $n=3$ component
the triangularity $\varepsilon_{3,3}$ (defined below) is an excellent choice 
for $\xi_{3\,\text{pred}}^{(1)}$.

The geometric predictors are 
designed to maximize the correlation
between a particular flow signal and the geometry. Specifically,
the predictors maximize the Pearson correlation coefficient between the 
event-by-event magnitude and orientation of $a$th principal 
component, $\xi^{(a)}_n$, and the geometrical predictor 
$\xi^{(a)}_{n\,\text{pred}}$
\begin{align}
	\text{max}\quad	Q^{(a)}_n=\frac{\big<\xi^{(a)}_n 
	\xi^{*(a)}_{n\,\text{pred}}\big>}{\sqrt{\big<\xi^{(a)}_n 
	{\xi^*_n}^{(a)}\big>\big<\xi^{(a)}_{n\,\text{pred}}\xi^{*(a)}_{n\,\text{pred}}\big>}}.\label{pears}
\end{align}

We constructed several predictors 
for the flow $\xi^{(a)}_n$ by 
assuming a linear relation between the flow and the geometry.
The simplest predictor consists of linear combinations of the
first two eccentricities of the initial geometry. These are defined as 
\begin{subequations}
	\label{eps3all}
	\begin{align}
		\varepsilon_{n,n} &\equiv   -\frac{[r^n e^{in \phi }]}{ R_{\rm rms}^n } 
		\, , \label{eps33} \\
		\varepsilon_{n,n+2} &\equiv   -\frac{ [r^{n+2} e^{in \phi }]}{ R_{\rm 
		rms}^{n+2}}  
		\, ,\label{eps35} 
	\end{align} 
\end{subequations} 
where the square brackets $[\,]$ denote an integral over the initial entropy 
density for a specific event, normalized by the average total entropy 
$\bar{S}_\text{tot}$.  $R_{\rm rms}\equiv\sqrt{\llangle[r^2]\rrangle}$ is 
the event averaged root-mean-square radius. Note that our definitions of
$\varepsilon_{n,n}$ and $\varepsilon_{n,n+2}$ are chosen to make 
the event-by-event quantities $\varepsilon_{n,n}$ and $\varepsilon_{n,n+2}$ 
linear in the fluctuations. In this notation, the 
geometric predictor based on 
these eccentricities is
\begin{equation}
\xi_{n\,\text{pred}}^{(a)} = \varepsilon_{n,n} + c_1 \, \varepsilon_{n,n+2},
\end{equation}
where $c_1$ is adjusted to maximize the correlation coefficient in 
\Eq{pears}, 
and 
the overall normalization is irrelevant.
While the first two eccentricities provide an excellent predictor for
the leading flow, they do not predict  the subleading flow very 
well. This is  in part because the radial weight $r^{n+2}$ is too 
strong at large $r$.

More generally, one can define eccentricity as a functional of radial weight 
function $\rho(r)$:
\begin{equation}
\varepsilon_n\{\rho(r)\} \equiv - \frac{[\rho(r) 
e^{in\phi}]}{R_\text{rms}^n}.\label{epsgen}
\end{equation}
It is the goal of this paper to find the optimal radial weight
function $\rho(r)$ for predicting both leading and subleading flows. 

It is evident from  \Fig{PCA_S_2} that the leading and subleading geometries
have different characteristic wave numbers. 
To find the optimal radial weight we expand $\rho(r)$ in radial Fourier modes
\begin{equation}
\rho(r)= \sum_{b=1}^{n_k} w_b \frac{2^n n!}{k_b^n} J_n(k_b r),\label{exp}
\end{equation}
where $J_n(x)$ is a Bessel function of order $n$, $w_b$ are expansion
coefficients, and $k_b$ are definite wavenumbers  specified below.
The prefactor is chosen 
so that for a single $k$ mode ($w_1=1,w_{b>1}=0$) at small $k$ ($k R_{\rm rms} 
\ll 1$) the 
generalized
eccentricity approaches $\varepsilon_{n,n}$ 
\begin{equation}
\lim_{k \rightarrow 0} \varepsilon_n\{\rho(r)\} = \varepsilon_{n,n}.
\end{equation}
At small $k$, we expand the $J_{n}(k r)$ and find
\begin{equation}
    \varepsilon_n\{\rho(r)\} \simeq \varepsilon_{n,n}  + c_1 \, 
    \varepsilon_{n,n+2}, 
    \end{equation}
    where $c_1 = -(k R_{\rm rms}/2)^2/(1+n)$.
Thus the functional form of $\rho(r)$  adopted here
 yields a tunable linear combination of the eccentricities in \Eq{eps3all},
 but the wave number 
parameter regulates the behavior at large $r$.

\begin{figure}
\centering
\includegraphics[width=0.5\linewidth]{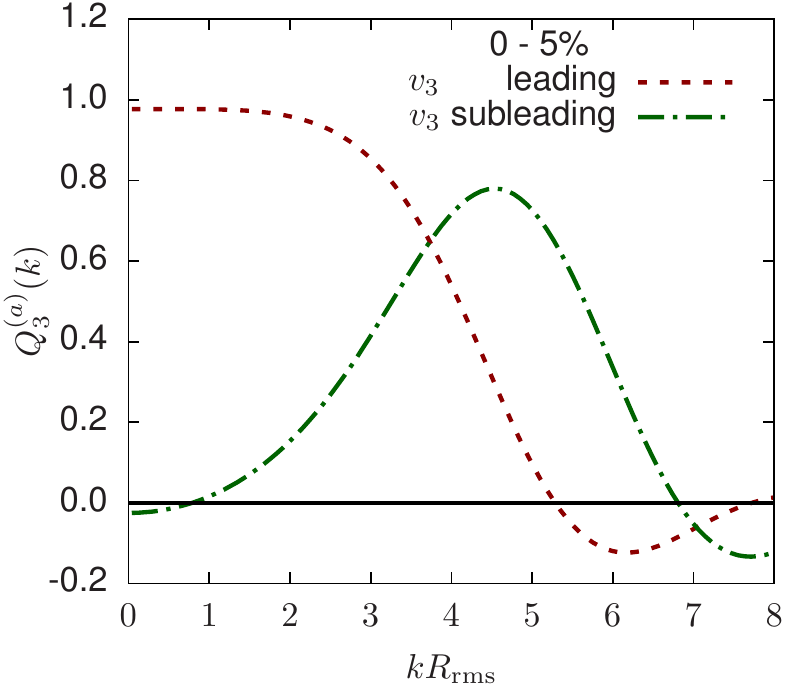}
\caption{Quality plot (or Pearson correlation coefficient)  for 
as a single $k$ mode predictor for principal triangular flows (central 
collisions). 
\label{oneterm}}
\end{figure}
\begin{figure}
\centering
   \includegraphics[width=0.7\textwidth]{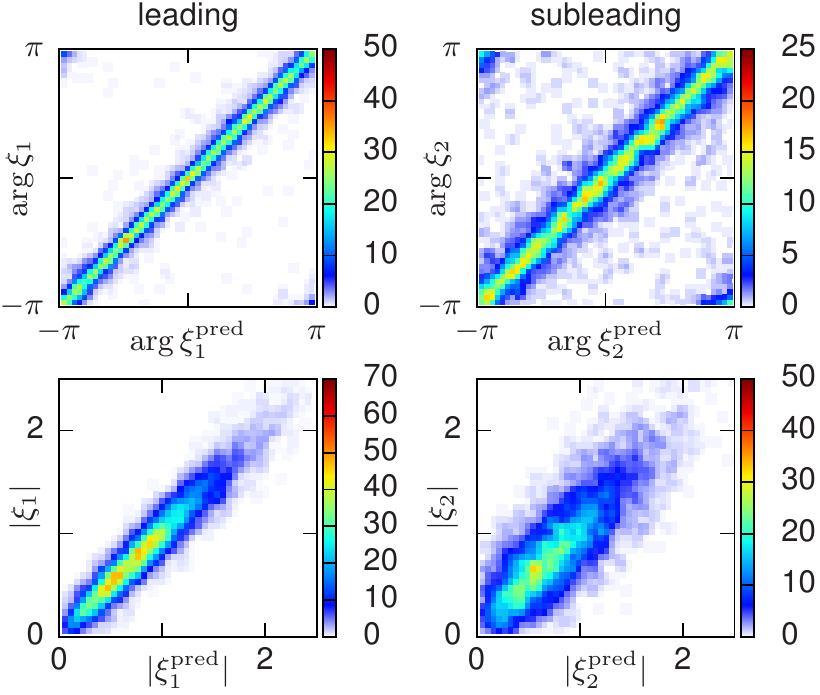}
		\caption{Angle and magnitude correlations between the leading and 
		subleading triangular flow and the  optimal linear predictor based on 
		two $k$ modes, \Eq{exp}.\label{phi}}
\end{figure}

In  \Fig{oneterm} we show Pearson correlation coefficient $Q^{(a)}_3(k)$ 
between the triangular flow and a 
single $k$ mode predictor. We see that the leading triangular 
component is 
produced
by low-$k$ fluctuations, while subleading flow originates from fluctuations at
larger $k$.
By using only two well chosen 
$k_b$ values for the Fourier expansion in \Eq{exp} an approximately optimal 
radial weight
can be found (see \Fig{phi}). 
Including additional $k$ modes in the functional
form of $\rho(r)$ does not significantly improve the predictive power
of the generalized geometric eccentricity.
For the two $k$ modes we required (somewhat
arbitrarily) that the ratio of $k$ values would 
be fixed to the ratio of the first two Bessel zeros:
\begin{equation}
\frac{k_1}{k_2}=\frac{j_{n,1}}{j_{n,2}}\label{j0}  \, .
\end{equation}
With this choice our basis functions were orthogonal in the interval 
$[0,R_o]$,  where $k_1 = j_{n,1}/R_o$.   We then adjusted $R_o$ to maximize the 
correlation coefficient 
between $\varepsilon_{n}\{\rho(r)\}$ and the flow $\xi^{(a)}_n$. To account for 
changing system size with centrality, we used a fixed
$R_o/R_\text{rms}$ ratio. In most cases we used $R_o/R_\text{rms}\approx 3.0$, 
but for all directed flow components ($\xi_1^{(1)}$ and $\xi_1^{(2)}$) and the 
second elliptic flow component ($\xi_2^{(2)}$), we found that 
$R_o/R_\text{rms}\approx 2.0$ optimized the correlation between the flow and 
the geometry.

Ultimately, the assumption that the amplitude and phase of the flow is 
determined at least approximately by initial eccentricity,
$\varepsilon_{n}\{\rho(r)\}$,  
is based on linear response.  
If nonlinear physics becomes important
(as in the case of $v_4$ and $v_5$) then the predictors should be 
modified to incorporate this physics (see below and Ref.~\cite{Gardim:2011xv}). 
Thus, below
we will refer to the $\varepsilon_{n}\{\rho(r)\}$ (with an optimized
radial weight) as the \emph{best linear predictor} and incorporate
quadratic nonlinear corrections to the predictor as needed.

\subsection{Testing linear response}
\label{linear_responsepca1}

The success of the linear flow predictors discussed in this work depends
on the applicability of linear response.  A straightforward way to check this
assumption is to compare the averaged response of event-by-event hydrodynamics
to the  hydrodynamic response to suitably averaged initial conditions.

\begin{figure}
\centering
\includegraphics[width=0.5\linewidth]{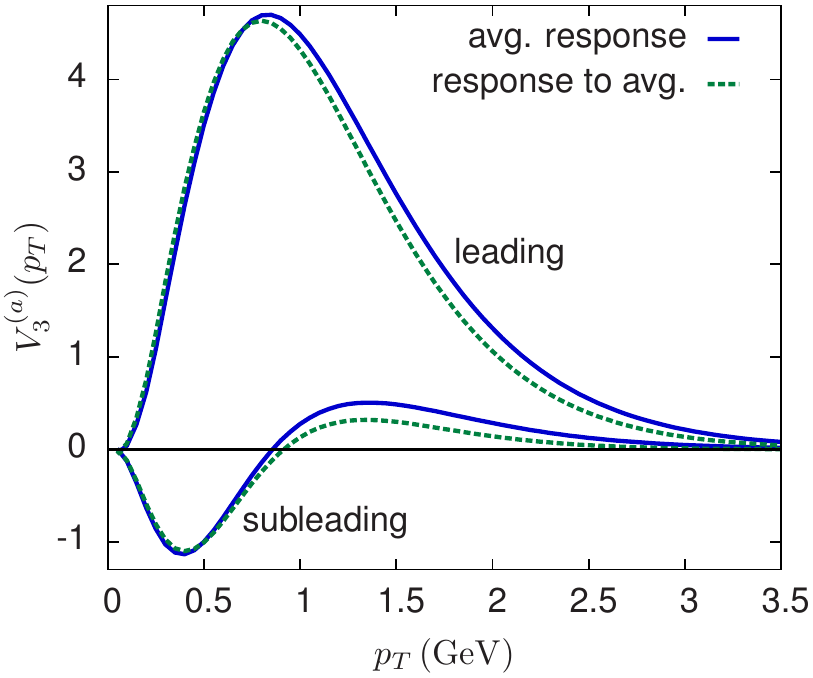}
\caption{Comparison of event-by-event hydro (averaged response) and single-shot 
hydrodynamics
(response to average geometry) in central collisions. The singe-shot 
hydrodynamic results are generated
from the initial conditions in \Fig{PCA_S_1}\label{PCA_Vn_res_1}.}
\end{figure}

In \Sect{avg_geometry} we computed the  average geometry in the event planes of
the leading and subleading triangular flows (see \Fig{PCA_S_1}).  It is 
straightforward to
simulate this smooth initial condition and to compute the associated
$V_{3}(p_T)$. This is known as ``single-shot" hydrodynamics in the
literature~\cite{Qiu:2011iv}.  In \Fig{PCA_Vn_res_1}  we compare  $V_{3}(p_T)$
from the leading and subleading average geometries to the principal components
$V_{3}^{(1)}(p_T)$ and $V_{3}^{(2)}(p_T)$ of event-by-event hydro.
The qualitative features of both principal components are reproduced well by
single-shot hydrodynamics, especially for the leading flow.  It is particularly
notable how the single-shot evolution reproduces the  change of sign in
$V_3^{(2)}(p_T)$.  However,  in an important $p_T$ range, $p_T\sim 1.2 \,{\rm
GeV}$, the single-shot evolution misses the event-by-event curve for subleading
flow by $\sim30\%$.

\begin{figure}
\includegraphics[width=\linewidth]{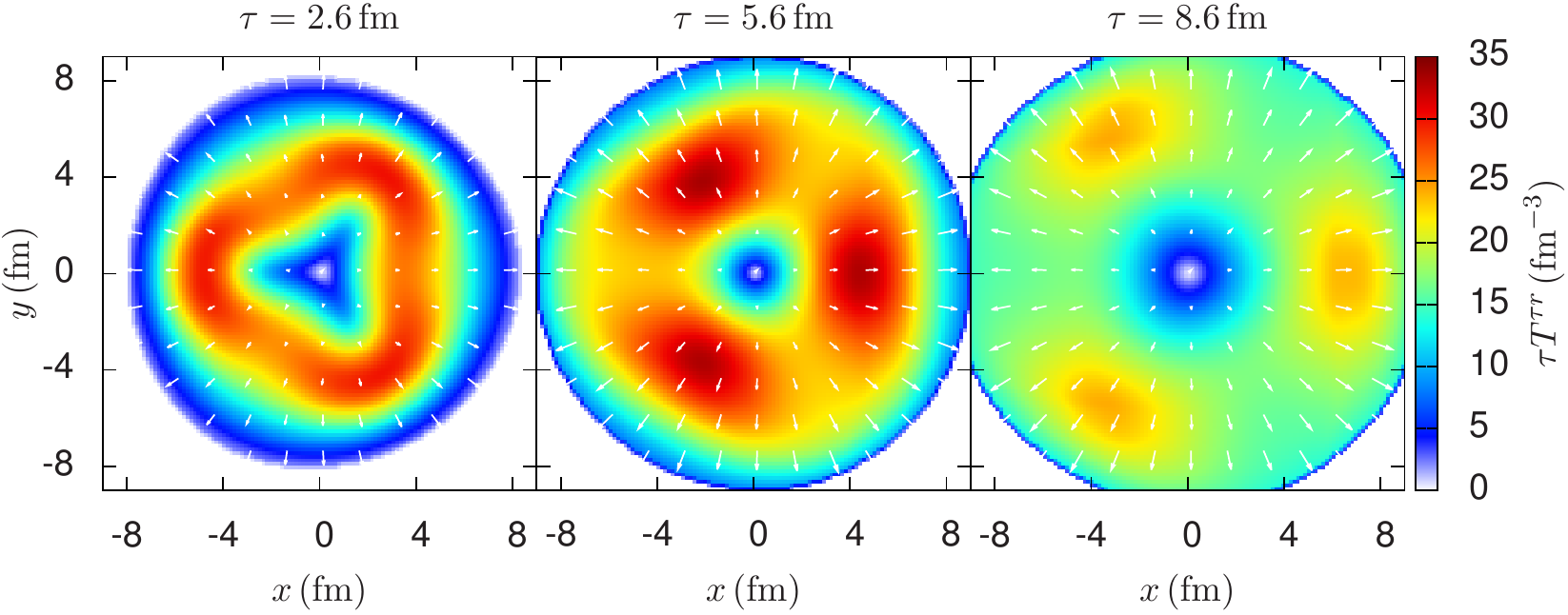}
\caption{ Hydrodynamic evolution of the subleading triangular flow for the
averaged initial conditions shown in \Fig{PCA_S_1}(b).  The color contours
indicate the radial momentum density per rapidity, $\tau T^{\tau r}$, while the
arrows indicate the radial flow velocity.\label{evolv}}
\end{figure}

It is useful to examine the time development of the subleading triangular flow 
in the
single-shot hydrodynamics.  In \Fig{evolv}, we present three snapshots of the
subleading flow evolution. The color contours show the radial momentum density
per rapidity,
\begin{equation}
	\tau T^{\tau r}=\tau (e+p) u^\tau u^r \,, \label{tauT0r}
\end{equation} 
as a function of proper time $\tau$.

Shortly after the formation of the fireball, at $\tau = 2.6\, {\rm fm}$ we
observe negative triangular flow in \Fig{evolv}(a). This flow is produced by
the excess of material at small radii flowing into the ``valleys" at larger
radii [see \Fig{PCA_S_1}(b)]. However, the radial flow has not developed yet,
and therefore this phase of  the evolution creates negative flow at small
transverse momentum.  After this stage, we see typical flow evolution of a
triangular perturbation, i.e. the negative geometric eccentricity at small
radii is transformed into positive triangular flow at large transverse momentum
[see Figs.~\ref{evolv}(b) and (c)]. The inner eccentricity dominates over the 
outer
eccentricity at high $p_T$ because the radial flow has more time to develop
before freeze-out, and because there is more material at small radii.

\section{Results}
\label{pca:results}

In this section we report the principal component analysis summary and selected 
plots for flow harmonics $n=0\text{--}5$. The comprehensive list of figures 
for each harmonic is given in \app{lof}. Following the discussion of triangular 
flow in previous sections, we first identify the dominant principal components 
and then clarify their physical origin by finding an optimized geometric  
predictor as explained in \Sect{predictors}. A new feature absent for the 
triangular flow is the nonlinear flow coupling to lower order principal 
components. We find that the 
subleading elliptic flow in peripheral collisions has a strong nonlinear 
coupling to the radial flow fluctuations $\xi_0^{(2)}$. We also find the 
nonlinear mixing in subleading $n=4$ and $n=5$ flow components.

\subsection{Radial flow}
\label{radial}

Radial flow (or $V_0(p_{T})$) is the first term in the Fourier series, 
\Eq{fseries},  
and 
is 
by far the largest harmonic. Traditionally, the experimental and theoretical study of the fluctuations of 
$V_0(p_{T})$ (i.e., multiplicity and $p_{T}$ fluctuations) has been distinct 
from elliptic and triangular flow. There is no reason for this distinction.

\begin{figure}
\centering
{\includegraphics[width=0.48\linewidth]{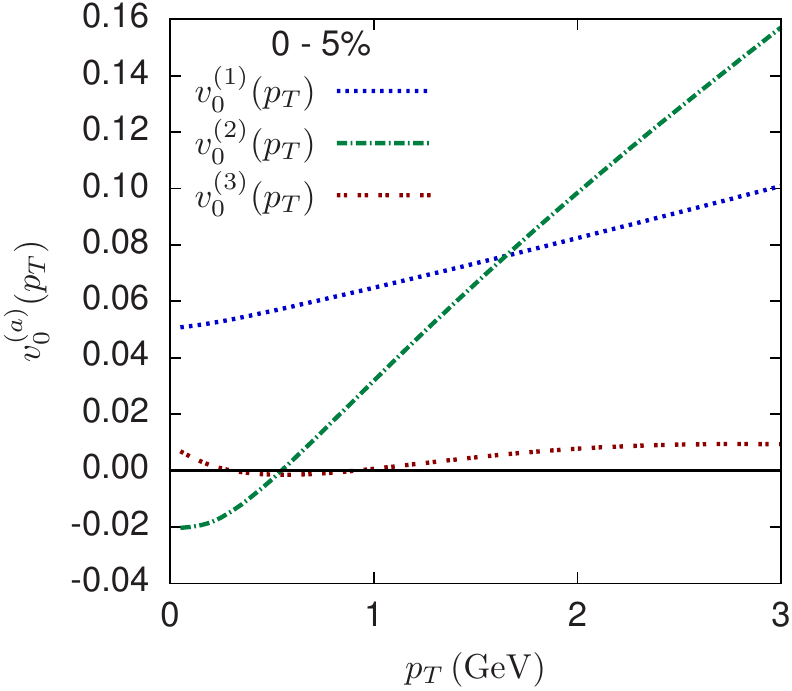}}\quad
{\includegraphics[width=0.48\linewidth]{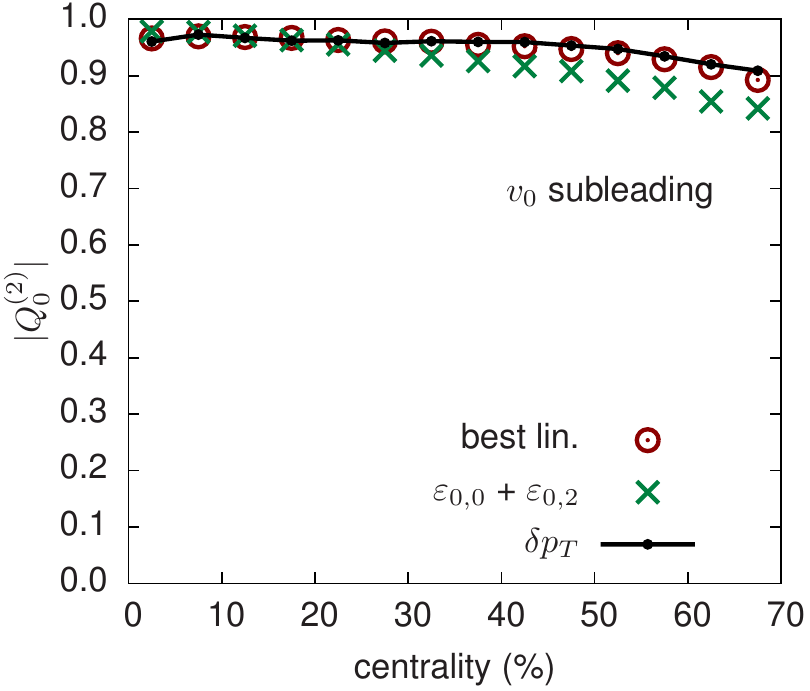}}
\caption{(a) The $p_T$ dependence of the principal components of radial flow normalized by the average multiplicity, 
$v_0^{(a)}(p_T)\equiv 
V_0^{(a)}(p_T)/\left<dN/dp_T\right>$. (b) The Pearson correlation coefficient [\Eq{pears}]  between
the subleading radial flow and various predictors versus centrality. The best linear predictor is described in \Sect{predictors}.\label{v0-fig}}
\end{figure}

Examining the scaled $V_0(p_T)$ eigenvalues shown in \Fig{v0-fig}(a), we see 
that there are
two large  principal components.
The first principal component is sourced by multiplicity fluctuations,
i.e., the magnitude of $V_0(p_{T})$ fluctuates (but not its shape)
due to the impact parameter variance in a given centrality bin.  
Corroborating this inference,
\Fig{v0-fig}(a) shows the momentum dependence of the leading principal 
component, which is approximately flat.\footnote{There is a small upward 
tending slope in our
simulations of this component, because multiplicity and mean $p_T$ fluctuations 
only approximately factorize into leading and subleading principal components. 
Using different definitions of centrality bins could perhaps make this separation cleaner.
} Clearly this principal component is not
particularly interesting, and the PCA procedure gives a practical method for isolating these trivial geometric fluctuations in the data set.
The second principal component is of much greater interest, and shows a linear  
rise with $p_T$ that  is  indicative of the fluctuations in the radial
    flow velocity of the fluid~\cite{Bhalerao:2014mua}.

In  early insightful papers~\cite{Broniowski:2009fm,Bozek:2012fw}, the 
fluctuations in the flow velocity (or mean $p_T$) 
were associated with the fluctuations in the initial fireball radius. 
These radial fluctuations are well described by both the
eccentricities $(\varepsilon_{0,0}$, $\varepsilon_{0,2})$, \Eq{eps3all}, and the 
optimized 
eccentricity $\varepsilon_{0}\{\rho(r)\}$, \Eq{epsgen}.   Therefore,
as seen in \Fig{v0-fig}(b),
the subleading flow signal is strongly correlated with these linear
geometric predictors.

Also shown in  \Fig{v0-fig}(b) is the correlation between subleading 
radial flow $\xi_0^{(2)}$ and mean transverse momentum fluctuations around the average
\begin{equation}
\delta p_T\label{dp}\equiv [p_T] - \left<[p_T]\right>.
\end{equation}
Indeed, the subleading radial flow correlates very well with mean momentum 
fluctuations in all centrality bins.

In the next sections we will study the nonlinear mixing between the radial flow
$\xi_0^{(2)}$ and all other harmonics.

\subsection{Elliptic flow}

\subsubsection{Nonlinear mixing and elliptic flow}
\label{elliptic}

\begin{figure}
\centering
\includegraphics[width=0.5\linewidth]{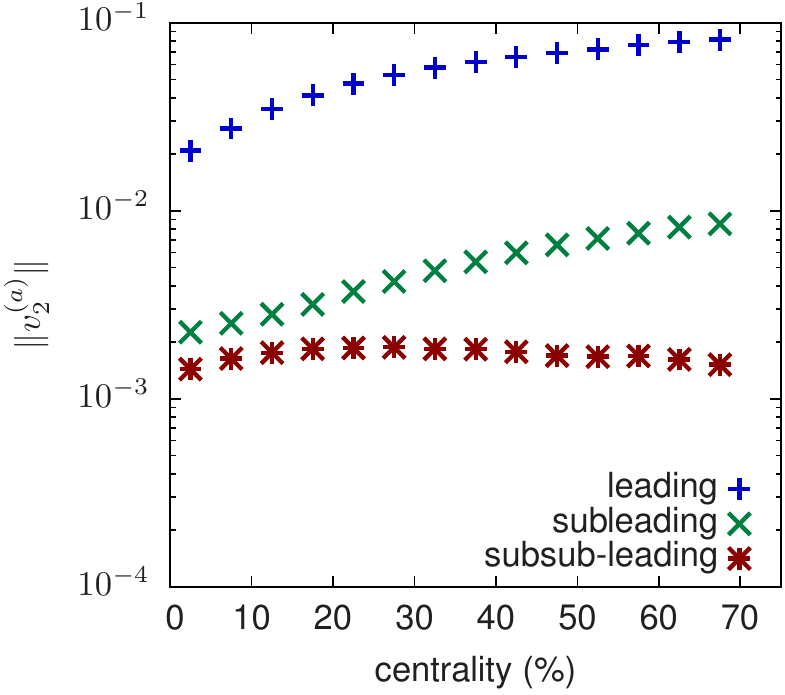}
\caption{The magnitudes of the principal
components of elliptic flow, $\|v^{(a)}_2\|$, versus centrality [see 
\Eq{pcamagnitudes}].
\label{pv2-eval}}
\end{figure}
\begin{figure}
\centering
{\includegraphics[width=0.48\linewidth]{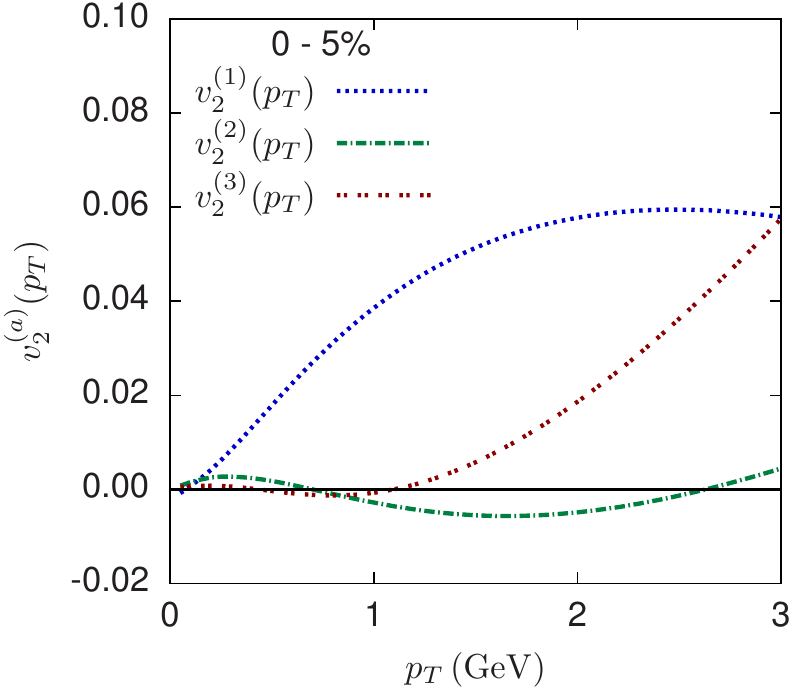}}\quad
{\includegraphics[width=0.48\linewidth]{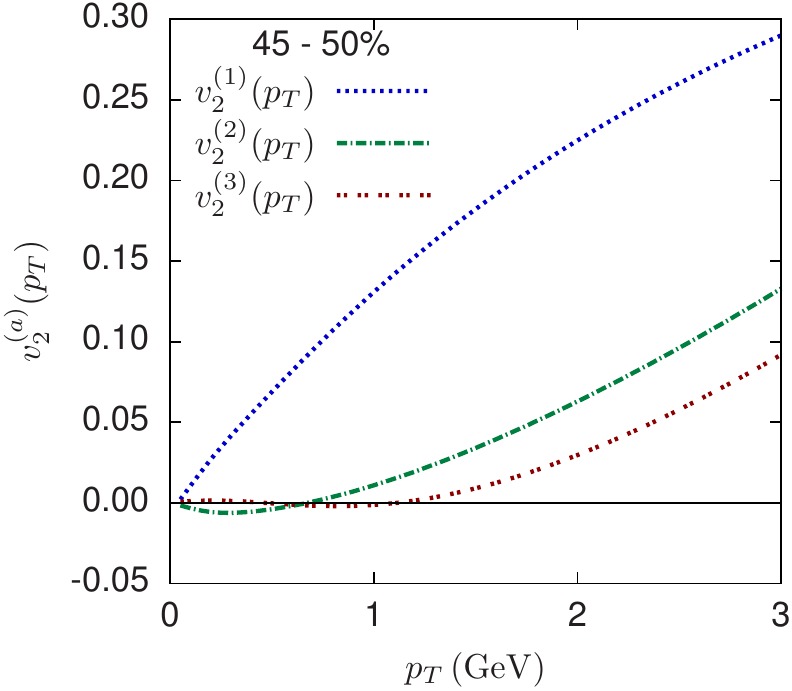}}
\caption{The $p_T$ dependence of the principal components of elliptic flow normalized by the average multiplicity, 
$v_2^{(a)}(p_T)\equiv 
V_2^{(a)}(p_T)/\left<dN/dp_T\right>$, for central (0--5\%) and peripheral
collisions (45--50\%). 
\label{v2-vn_norm-p}
}
\end{figure}

We next study the fluctuations of $V_2(p_T)$ as function of centrality.
As seen in \Fig{pv2-eval}, the principal component spectrum of elliptic flow 
in 
central collisions 
consists of two nearly degenerate subleading 
components in addition to the dominant leading component. This degeneracy is lifted in more peripheral 
bins. Comparing the $p_T$ dependence of the principal flows
shown in \Figs{v2-vn_norm-p}(a) and \ref{v2-vn_norm-p}(b), we see that 
going 
from 
central (0\nobreakdash--5\%) to peripheral (45\nobreakdash--50\%) collisions, 
the magnitude of the second principal component  increases 
in size and its momentum dependence changes
dramatically. By contrast, the growth of the third principal component is
much more mild. This strongly suggests that the average elliptic geometry
is more important for the subleading  than the subsub-leading mode.

\begin{figure}
\centering
{\includegraphics[width=0.48\linewidth]{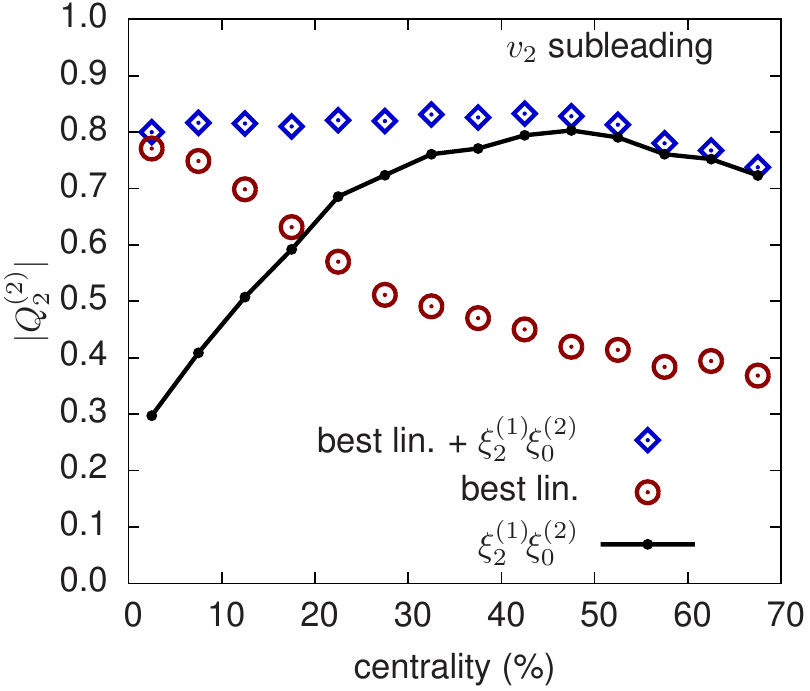}}\quad
{\includegraphics[width=0.48\linewidth]{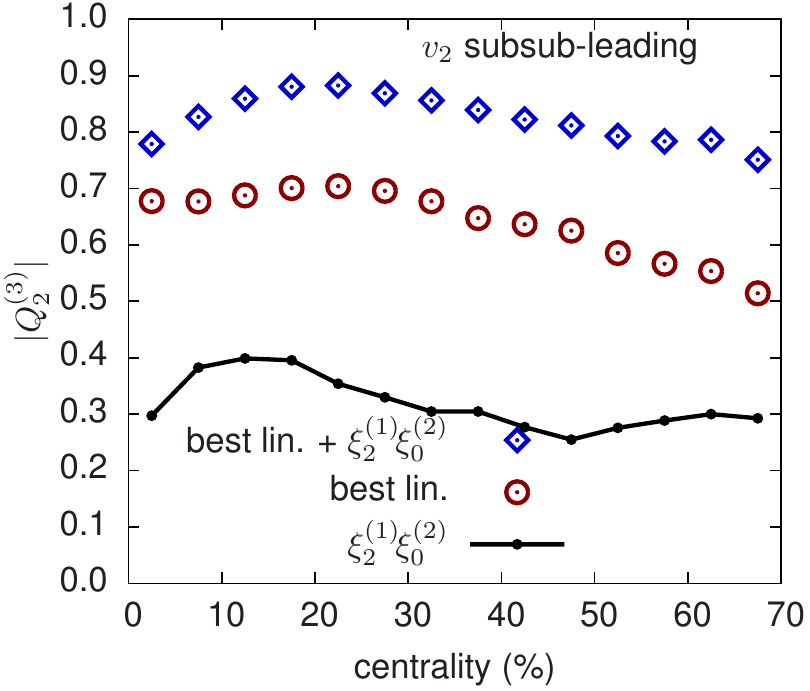}}
\caption{
    Pearson correlation coefficient 
    between the subleading elliptic flows and the best linear predictor [\Eq{epsgen}] 
    with and without the nonlinear
    mixing between the radial and leading elliptic flows, 
    $\xi_2^{(1)}\xi_0^{(2)}$. (a) and (b) show the correlation 
    coefficient for $v_2$ subleading and $v_2$ subsub-leading flows 
    respectively.
\label{pv2-corr}
}
\end{figure}

To find a geometrical predictor for the sub- and sub-sub-leading modes
we first tried the best linear predictor $\varepsilon_2\{\rho(r)\}$. 
In \Fig{pv2-corr}(a) (the red circles), we see that the correlation coefficient between this optimal linear predictor and the subleading flow signal drops precipitously as a function
of centrality. As we will explain now, this is because nonlinear mixing
becomes important for the subleading mode.

The ellipticity of the almond shaped geometry in peripheral collisions is 
traditionally parametrized by eccentricity $\varepsilon_{2,2}$
and it serves as an excellent predictor for the leading elliptic flow.
However, $\varepsilon_{2,2}$ does not completely fix the 
initial geometry, and the radial size of the fireball can fluctuate at fixed 
eccentricity. As 
explained in \Sect{radial}, the radial size fluctuations 
modulate the momentum spectrum of the produced particles, and for a background 
geometry with large {\em constant} eccentricity 
this generates
fluctuations in the $p_T$ dependence of the elliptic flow, i.e., subleading 
elliptic flow. This subleading flow lies in the reaction plane following the
average elliptic flow, but its sign (which is determined by $\delta p_T$) is 
uncorrelated with $\varepsilon_{2,2}$.

The orientation of the 
reaction plane in peripheral bins is strongly correlated with the 
integrated 
$v_2$ or the 
leading elliptic principal component $\xi_2^{(1)}$, while the mean 
$p_T$ fluctuations are tracked by the subleading radial flow component $\xi_0^{(2)}$. Therefore we 
correlated the sub- and sub-sub-leading  elliptic flows
with the 
product of the  leading elliptic and radial flows, i.e. 
we computed the correlation coefficient in \Eq{pears} with
$\xi_{2,\rm pred}^{(2)}=\xi_2^{(1)}\xi_0^{(2)}$. 
Examining \Fig{pv2-corr}(a) (the black line), we see 
see that 
the correlation between the subleading elliptic flow and
the nonlinear mixing rises with
centrality,  as the correlation with best linear predictor drops. Examining \Fig{pv2-corr}(b) on the other hand, we see that the 
subsub-leading elliptic flow has 
stronger correlation with the initial geometry than the nonlinear mixing.  Combining 
best linear geometric predictor and quadratic mixing terms  in the 
predictor, i.e.
\begin{equation}
\xi_{2\,\rm pred}^{(2)}  = \varepsilon_{2}\{\rho(r)\} + c_1 
\xi_2^{(1)}\label{pred2} 
\xi_0^{(2)}, 
\end{equation}
we achieve consistently 
high correlations for all centralities [the blue diamonds in \Fig{pv2-corr}(a) 
and (b)].

\subsubsection{Dependence on viscosity}
\label{viscosity}

\begin{figure}
\centering
\includegraphics[width=0.5\linewidth]{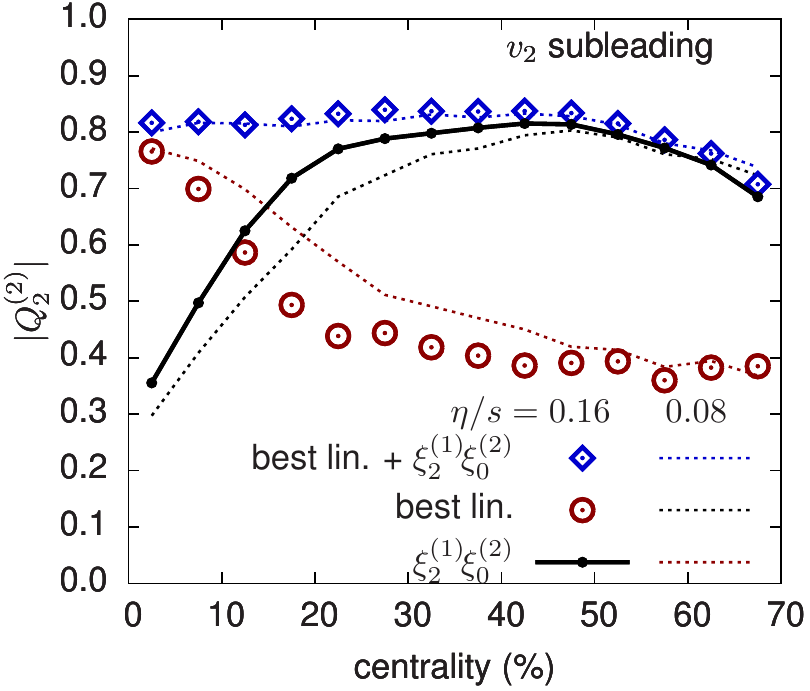}
\caption{
     Pearson correlation coefficients  for the subleading elliptic flow at 
     viscosity over entropy ratio $\eta/s=0.16$. Dashed lines repeat 
     $\eta/s=0.08$ results from \Fig{pv2-corr}(a) for the ease of comparison.
\label{pv2s16-corr}
}
\end{figure}

Before leaving this section we will briefly comment  on the viscosity
dependence of these results. Figure \ref{pv2s16-corr} shows a typical 
result  for
a slightly larger shear viscosity, $\eta/s=0.16$.
As discussed above, the subleading elliptic 
flow [i.e., the event-by-event fluctuations in $V_2(p_T)$] is a result of the
linear response to the first radial excitation of the elliptic eccentricity,
and a nonlinear mixing of radial flow fluctuations and the leading elliptic
flow. In \Fig{pv2s16-corr} we see that a slightly larger shear viscosity 
tends to preferentially damp the linear response leaving a stronger nonlinear
signal. This is because
the initial geometry driving the linear response has a significantly larger 
gradients due to the combined azimuthal and radial variations. Thus in 
\Fig{pv2s16-corr} the linear response  dominates the 
subleading flow only in 
very central collisions.
These trends with centrality are qualitatively familiar from previous analyses of
the effect of shear viscosity on the nonlinear mixing of
harmonics~\cite{Teaney:2012ke,Qiu:2012uy}.

\subsection{Triangular and directed flows}
\label{v1v3}

\begin{figure}
\centering
{\includegraphics[width=0.48\linewidth]{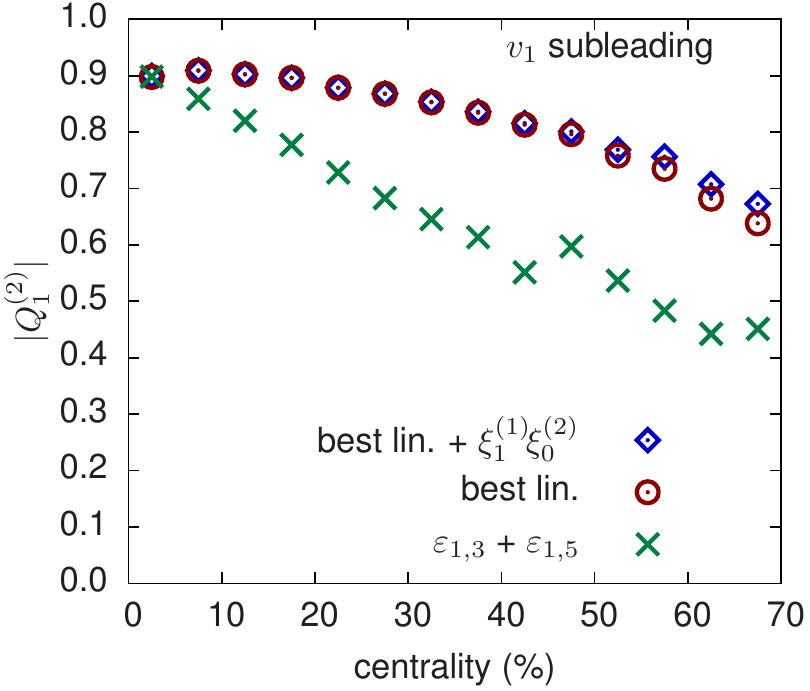}}\quad
{\includegraphics[width=0.48\linewidth]{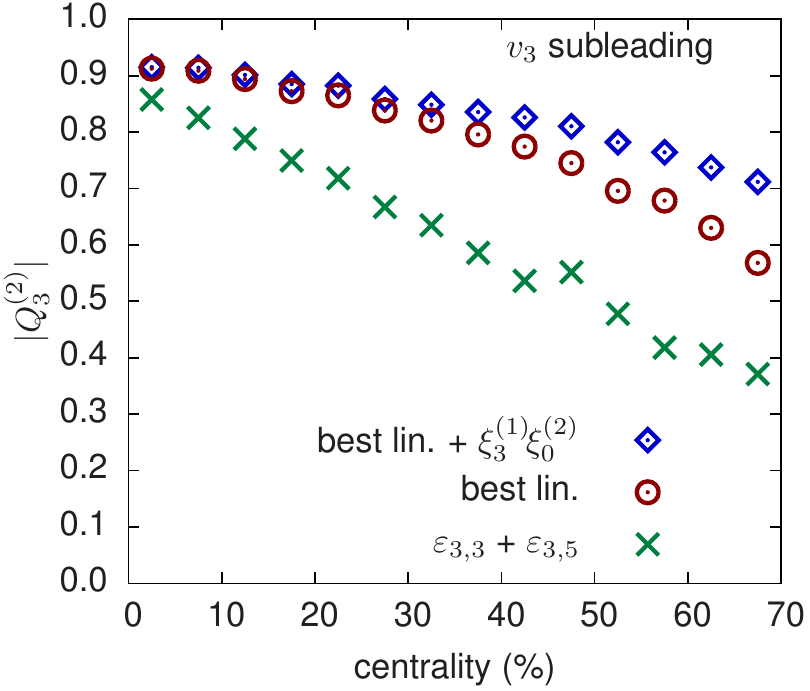}}
\caption{
    Pearson correlation coefficient 
    between the subleading (a) directed and (b) triangular flows and the best 
    linear predictor
    with and without radial flow mixing. 
\label{pv1v3-corr}}
\end{figure}

In \Fig{v3-corr1} in the appendix we show the correlation coefficient between 
the
leading flow amplitudes $\xi_3^{(1)}$ and the predicted amplitudes
$\xi_3^{(1)\rm pred}$ using the optimal linear predictor and the
$\varepsilon_{3,3}$, $\varepsilon_{3,5}$ fit. 
As is well known, the leading triangular
mode is very well predicted  by $\varepsilon_{3,3}$ and $\varepsilon_{3,5}$,
though the quality degrades towards peripheral 
collisions~\cite{Gardim:2014tya}.    
As shown in \Fig{pv1v3-corr}(b), for the
subleading flow the linear correlation coefficient is reduced relative to the
leading flow, and a high degree of correlation is only achieved for the 0--40\%
centrality range.  The simple geometric predictor based on $\varepsilon_{3,3}$
and $\varepsilon_{3,5}$  is reasonably correlated with the subleading flow in
central collisions,  but this correlation rapidly deteriorates in more
peripheral collisions.  The optimal linear predictor 
$\varepsilon_{3}\{\rho(r)\}$ based on two judiciously chosen $k$ values 
generally out performs all other predictors we studied for both leading and 
subleading flows.
Adding the nonlinear mixing term $\xi_{0}^{(2)}\xi_3^{(1)}$ to the best linear 
predictor 
marginally improves the already good correlation with the subleading flow  in 
peripheral collisions. 

Directed flow exhibits many similarities to triangular flow.  Specifically,
the subleading directed flow is reasonably well correlated with the optimal
linear predictor characterizing the radially excited dipolar geometry.  
Nonlinear mixing
between the  leading directed flow and the radial flow is unimportant [see 
\Fig{pv1v3-corr}(a)].

\subsection{The $n=4$ and $n=5$ harmonic flows}
\label{v4v5}
\begin{figure}
\centering
{\includegraphics[width=0.48\linewidth]{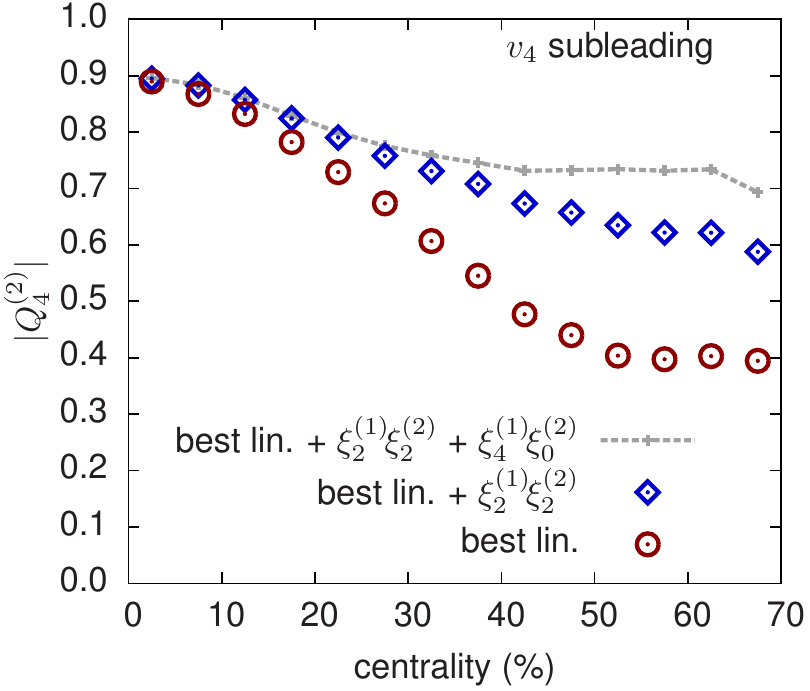}}\quad
{\includegraphics[width=0.48\linewidth]{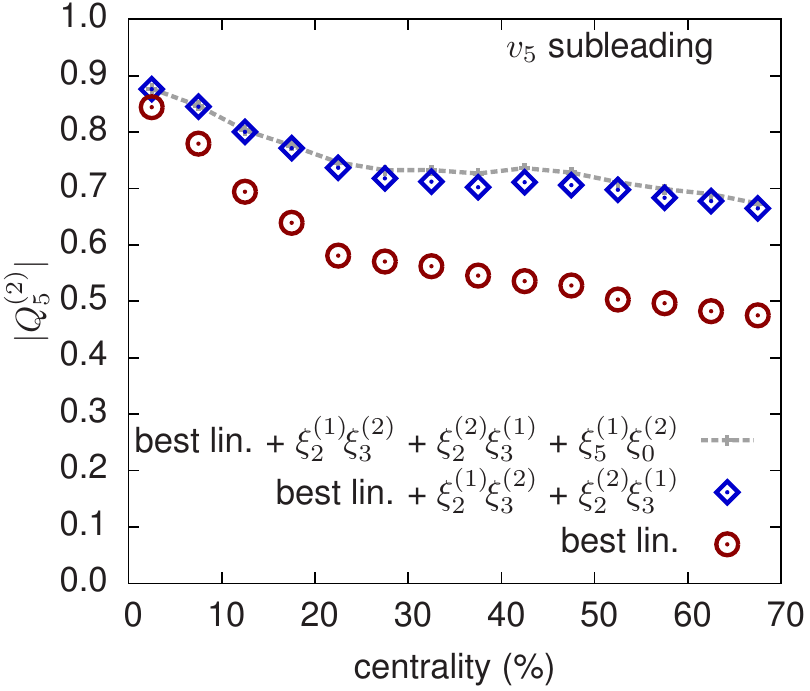}}
\caption{
    Pearson correlation coefficient 
    between  the subleading $v_4$ and $v_5$ flows and the best linear 
    predictor  with and without several nonlinear terms [see 
    \Eqs{subv4v5_predictor1} and \eq{subv4v5_predictor2}].
\label{pv4v5-corr}}
\end{figure}

It is well known
that the leading components of the $n=4$ and $n=5$ harmonics
are determined by the nonlinear mixing of lower order harmonics in peripheral collisions~\cite{Borghini:2005kd,Qiu:2011iv,Gardim:2011xv,Teaney:2012ke,Aad:2015lwa}.

For comparison with other works~\cite{Gardim:2011xv,Gardim:2014tya}, in the 
Appendix in \Figs{v4-corr1} and \ref{v5-corr1} we 
construct a predictor based on
a linear combination of the eccentricities
\begin{align}
\varepsilon_{4,4} + c_1 \varepsilon_{2,2}\varepsilon_{2,2}\quad\text{for }n=4,\\
\varepsilon_{5,5}+ c_1 \varepsilon_{2,2}\varepsilon_{3,3}\quad\text{for }n=5,
\end{align}
where here and below the coefficient $c_1$ is adjusted to 
maximize  the correlation with the flow.
This predictor is compared
to a  linear combination of the 
optimal eccentricity $\varepsilon_n\{\rho(r)\}$ and 
the corresponding nonlinear mixings of the leading principal components
\begin{subequations}
    \label{leadingv4v5}
\begin{align}
    \varepsilon_{4}\left\{\rho(r)\right\} + c_1 \xi_2^{(1)}\xi_2^{(1)} \quad\text{for }n=4,\\
    \varepsilon_{5}\left\{\rho(r)\right\} + c_1 \xi_2^{(1)}\xi_3^{(1)}\quad\text{for }n=5.
\end{align}
\end{subequations}
Both sets of predictors perform reasonably well, though the second set has
a somewhat stronger correlation with the flow.

Returning to the subleading components, 
we first correlated  the best linear
predictors, $\varepsilon_{4}\{\rho(r)\}$ and $\varepsilon_5\{\rho(r)\}$, 
with the corresponding subleading flow signals. As seen in \Fig{pv4v5-corr} (the red circles), the correlation decreases rapidly with centrality, especially for $v_4$.  
Motivated by \Eq{leadingv4v5} which predicts the event-by-event leading $v_4$ and $v_5$ in terms $v_2$ and $v_3$, we construct a predictor for the 
subleading $v_4$ and $v_5$ in terms of the fluctuations of $v_2$ and $v_3$ (see 
Secs.~\ref{elliptic} and \ref{v1v3}, respectively). The full predictor reads
\begin{subequations}
    \label{subv4v5_predictor1}
\begin{align}
    &\varepsilon_{4}\left\{\rho(r)\right\} + c_1 \xi_2^{(1)}\xi_2^{(2)}, & \text{for }n=4,\\
    &\varepsilon_{5}\left\{\rho(r)\right\} + c_1 \xi_2^{(1)}\xi_3^{(2)} + c_2 \xi_3^{(1)}\xi_2^{(2)}, &\text{for }n=5.\label{v5nl}
\end{align}
\end{subequations}
Including the mixings between the  subleading $v_2$ and $v_3$ and the corresponding leading components greatly improves the correlation in mid-central bins (the blue diamonds).
Finally, in an effort to improve the  $v_4$ predictor in the most
peripheral bins we have added additional nonlinear mixings between
the radial flow and the leading principal components
\begin{subequations}
\label{subv4v5_predictor2}
\begin{align}
    &\varepsilon_{4}\left\{\rho(r)\right\} + c_1 \xi_2^{(1)}\xi_2^{(2)} + c_2 \xi_{4}^{(1)} \xi_0^{(2)}, & \text{for }n=4,\\
    &\varepsilon_{5}\left\{\rho(r)\right\} + c_1 \xi_2^{(1)}\xi_3^{(2)} + c_2 \xi_3^{(1)}\xi_2^{(2)} + c_3 \xi_5^{(1)} \xi_0^{(2)}, &\text{for }n=5.
\end{align}
\end{subequations}
As seen in \Fig{pv4v5-corr}(a) (the grey line) the coupling
to the radial flow improves the  correlation between the subleading $v_4$  and the
predictor in peripheral collisions. 
On the other hand, for $v_5$, \Fig{pv4v5-corr}(b), 
all of the information about the coupling to the radial flow is already included in \Eq{v5nl} and adding $v_0$ does not improve the correlation.

\section{Discussion}
\label{discussion}
 In this paper  we classified the event-by-event fluctuations of the momentum 
 dependent Fourier harmonics $V_n(p_T)$ for $n=0\text{--}5$ by performing a 
 principal component analysis of the two-particle correlation matrix in 
 hydrodynamic simulations of heavy ion collisions.
The \emph{leading} principal component for each harmonic is very strongly 
correlated 
with the  integrated flow,
and therefore this component
 is essentially the familiar $v_n(p_T)$  measured in the event plane. 
The \emph{subleading} components describe 
additional $p_T$ dependent fluctuations of the magnitude and phase of 
$v_n(p_T)$. The subleading flow is uncorrelated with the integrated $v_n$ 
and thus it is 
projected out in analyses of harmonic
flow based on the scalar product or event plane methods.
This paper focuses on 
the physical origins of the subleading flows, which are the largest source of
factorization breaking in hydrodynamics.

First,  we first studied the basic 
properties of principal components for the specific case of triangular flow 
such as its
dependence on transverse momentum (\Fig{PCA_Vn_1_nonorm}), and centrality and
shear viscosity (\Fig{PCA_eval_2}).  
 The leading
principle component of the triangular flow was found to be the hydrodynamic 
response to the participant
triangularity, while the subleading flow (which is uncorrelated with the
leading flow) is the hydrodynamic response to the first radial excitation of
the triangularity.  This conclusion was reached by averaging the event-by-event
geometry in the event plane of the subleading flow (\Fig{PCA_S_1}).  The
magnitude of this radial excitation is on par with the  magnitude of the
triangularity (\Fig{PCA_S_2}), although the hydro response is smaller in
magnitude. The  triangular flow response is approximately linear
to the  geometrical deformation.  This was checked by simulating the response
to the average in-plane geometry with ``single-shot" hydrodynamics
(\Fig{evolv}), and comparing this result to event-by-event hydrodynamics;
i.e., we compared the response to the average with the averaged response
(\Fig{PCA_Vn_res_1}).

Having gained experience with triangular flow principal components, we 
conducted an extensive survey of principal flows for $n=0\text{--}5$ harmonics. 
We summarized the main results in \Sect{pca:results}, and for convenience 
reproduced PCA plots for all harmonics  in the \app{lof}. 
Our systematic study started by placing radial flow (the $n=0$ 
harmonic) in 
the same framework as the other harmonic flows in \Sect{radial}. We identified 
the subleading 
$n=0$ principal component  with mean $p_T$  fluctuations and confirmed (as is well known~\cite{Broniowski:2009fm,Bozek:2012fw})
that these fluctuations are predicted by the variance of the radial size of
the fireball.

Next, we investigated the nature of the subleading elliptic flows in 
\Sect{elliptic}. 
The 
principal component analysis reveals  that in central collisions there are 
two comparable sources of subleading elliptic flow,  
but they have strikingly 
different centrality dependence (see \Figs{pv2-eval} and \ref{v2-vn_norm-p}).
In mid-peripheral collisions 
 the first subleading component 
mainly reflects
a nonlinear mixing between 
elliptic and radial flows,
and  this component is only weakly correlated with 
the radially excitations of the elliptic geometry.
The second subleading component in this centrality range is substantially smaller and more closely reflects the radial excitations.
In more central collisions, however, the nonlinear mixing with the average elliptic 
flow becomes small, and the sub and subsub-leading principal components
become comparable in size.
Thus, the rapid centrality dependence of factorization
breaking in $v_2$ is the result of an interplay between the linear response to the
fluctuating elliptic geometry, and the nonlinear mixing of the radial
and average elliptic flows. Larger shear viscosity only strengthens the trend 
as shown in \Fig{pv2s16-corr}.

This nonlinear mixing can be confirmed experimentally by measuring 
the correlations between the principal components 
$\llangle \xi_{2}^{(2)} (\xi_2^{(1)} \xi_0^{(2)})^* \rrangle$ 
which is predicted in \Fig{pv2-corr}.  
The prediction is that
three point correlation between
the subleading elliptic event plane, the mean $p_T$ fluctuations, and 
the leading elliptic event plane defined by the $Q_2$ vector, i.e.,
\begin{equation}
       \frac{\llangle \xi_2^{(2)} \delta p_T Q_2^* \rrangle }{ \sqrt{\llangle (\delta p_T)^2 \rrangle \llangle |Q_2|^2 \rrangle } },
\end{equation}
changes rapidly from central to midperipheral collisions.
This correlation is analogous to 
the three plane correlations such as 
$\llangle V_5 (V_2 V_3)^* \rrangle$
measured previously~\cite{Aad:2015lwa}. 

In \Sect{v1v3} the subleading directed and triangular flows 
were shown to be a linear response to the radial 
excitations of the corresponding eccentricity of the initial geometry. 
In these cases a generalized eccentricity $\varepsilon_{n}\{\rho(r)\}$ 
with an optimized  radial weight (describing the radial excitation) 
provides a good predictor for the subleading flows (\Fig{pv1v3-corr} and  
\ref{phi}).

Finally, in \Sect{v4v5} we studied factorization breaking in $v_4$ and $v_5$. 
With the comprehensive understanding of the fluctuations of $v_2$ and $v_3$
described above, the corresponding fluctuations in $v_4$ and $v_5$ were
naturally explained as the nonlinear mixing of subleading  $v_2$ and $v_3$ 
with their leading counterparts, together with linear response to the 
quadrangular and pentagonal geometries (see \Fig{pv4v5-corr}). 

The study of the fluctuations in the harmonics spectrum presented here
shows the power of the  principal component method in elucidating the physics which drive the event-by-event flow. We hope that this motivates
a comprehensive experimental program measuring the principal components and their correlations for $n=0-5$. Such an analysis would clarify the initial state in typical and ultra-central events with unprecedented precision,  and would strongly constrain the dynamical response of the quark gluon plasma.

\begin{subappendices}

\section{List of figures \label{lof}}
Here we present a comprehensive catalog of PCA plots for each harmonic 
$n=0\text{--}5$.
Centrality dependence of flow magnitudes for $n=0$, appearing as \Fig{pv2-eval} 
in the text above,  is repeated here as \Fig{v0-eval}, and analogous plots for 
other harmonics are given in
Figs.~\ref{v1-eval}-\ref{v5-eval}.
The $p_T$ dependence of normalized principal components for radial and elliptic 
flows in central (0-5\%) collisions shown in \Figs{v0-fig}(a) and 
\ref{v2-vn_norm-p}(a) are reproduced as 
\Figs{v0-vn_norm} and \ref{v2-vn_norm} and complemented with 
\Figs{v1-vn_norm} and \ref{v3-vn_norm}-\ref{v5-vn_norm}. Additionally, 
Figs.~\ref{v0-Vn_nonorm}-\ref{v5-Vn_nonorm}
depict the same principal components, but without 
normalization by average multiplicity $\left<dN/dp_T\right>$. Finally, in the 
paper we 
showed the Pearson correlation coefficients for the subleading flows for each 
harmonic $n=0\text{-}5$ in
\Figs{v0-fig}(b),\ref{pv2-corr}(a),\ref{pv1v3-corr}(a), \ref{pv1v3-corr}(b), 
\ref{pv4v5-corr}(a) and \ref{pv4v5-corr}(b), while in this appendix we show 
results for both leading and subleading flows in 
the series of figures  Figs.~\ref{v0-corr1}-\ref{v5-corr1} and Figs.~\ref{v0-corr2}-\ref{v5-corr2}.

\newcommand{\figwidth}{0.40\linewidth}
\newcommand{\FLOWN}{0}
\begin{figure}
\centering
\subfigure[
{\,Centrality dependence of the (scaled) magnitudes of flows  
$\|v^{(a)}_\FLOWN\|$.}\label{v\FLOWN-eval}]{
\includegraphics[width=\figwidth]{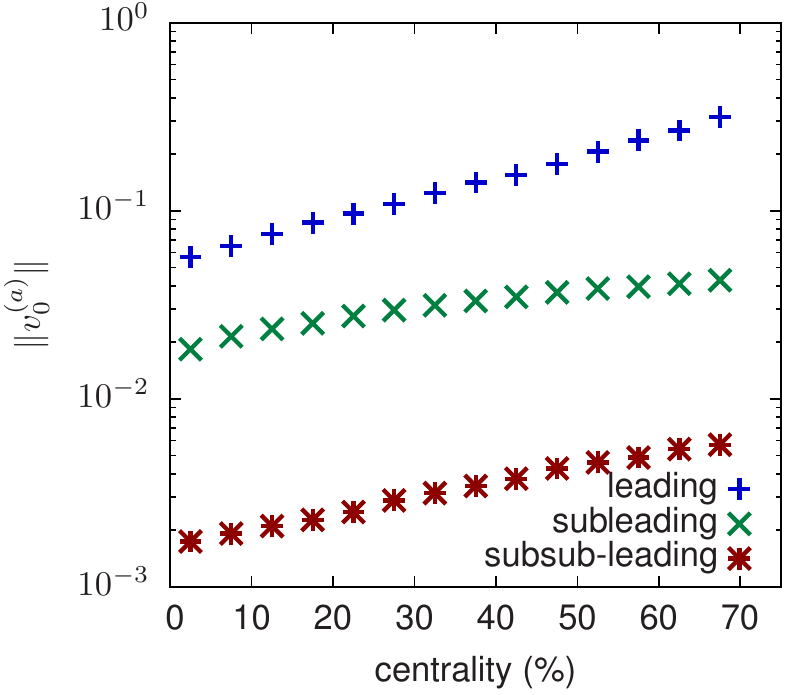}}\\
\subfigure[{\,Momentum dependence of principal flow vectors 
$V_\FLOWN^{(a)}(p_T)$ 
in 
central collisions.}\label{v\FLOWN-Vn_nonorm}]{
\includegraphics[width=\figwidth]{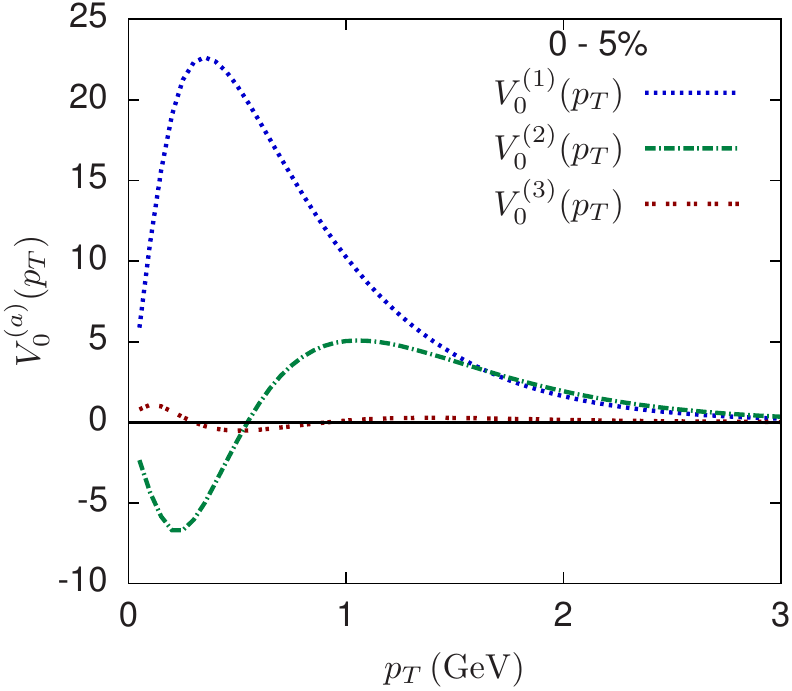}}\quad
\subfigure[{\,Principal flow vectors divided by the average multiplicity, 
$v_\FLOWN^{(a)}(p_T)\equiv 
V_\FLOWN^{(a)}(p_T)/\left<dN/dp_T\right>$.}\label{v\FLOWN-vn_norm}]{
\includegraphics[width=\figwidth]{std_A0_a0_B0_b0_CENT00_vn_norm}}
\subfigure[{\,Pearson correlation coefficient between the leading flow (zero 
suppressed for clarity) and several predictors.}\label{v\FLOWN-corr1}]{
\includegraphics[width=\figwidth]{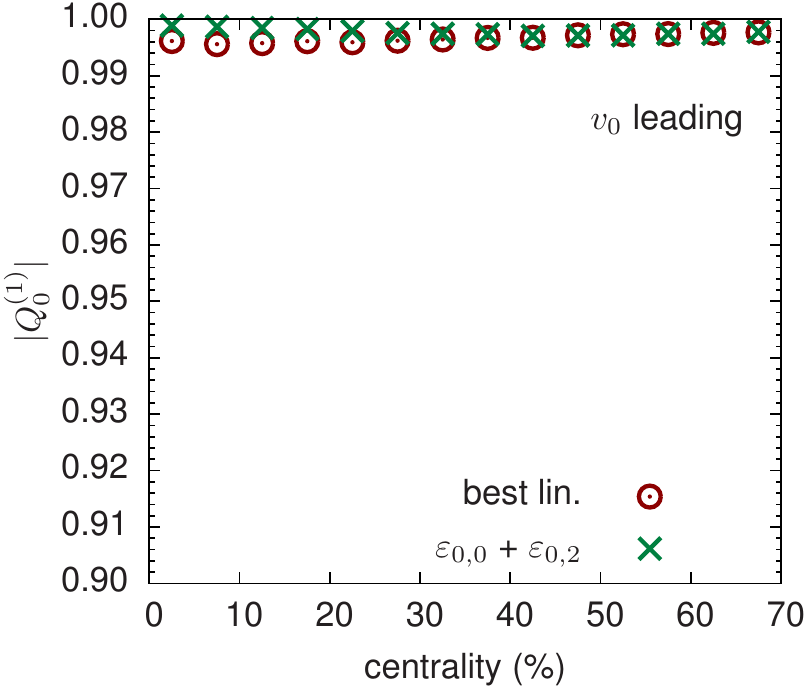}}\quad
\subfigure[{\,Pearson correlation coefficient between the subleading 
flow and several predictors.}\label{v\FLOWN-corr2}]{
\includegraphics[width=\figwidth]{std_A0_a0_B0_b0_corr2_c}}
\caption{Principal component analysis for $n=\FLOWN$ harmonic flow.}
\label{v\FLOWN}
\end{figure}

\renewcommand{\FLOWN}{1}
\begin{figure}
\centering
\subfigure[
{\,Centrality dependence of the (scaled) magnitudes of flows  
$\|v^{(a)}_\FLOWN\|$.}\label{v\FLOWN-eval}]{
\includegraphics[width=\figwidth]{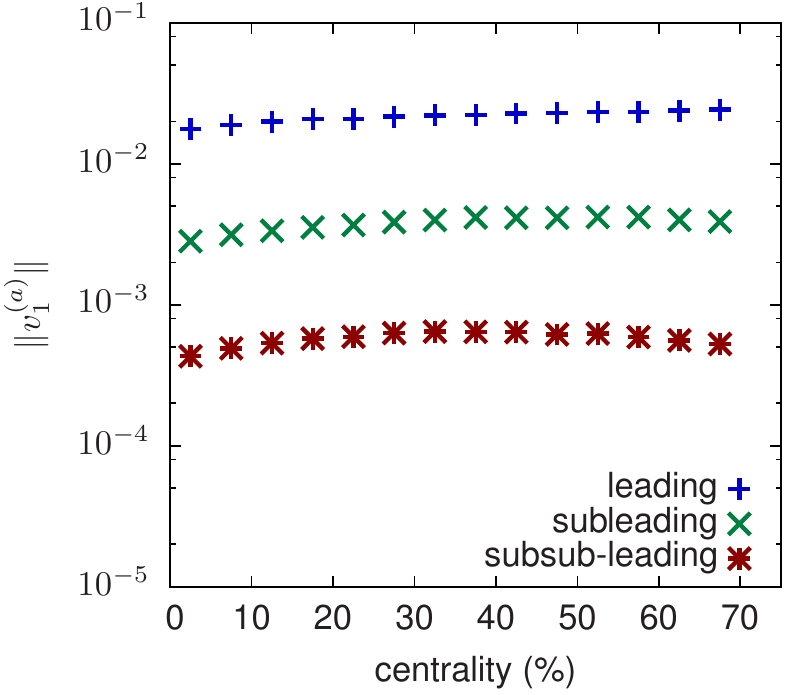}}\\
\subfigure[{\,Momentum dependence of principal flow vectors 
$V_\FLOWN^{(a)}(p_T)$ 
in 
central collisions.}\label{v\FLOWN-Vn_nonorm}]{
\includegraphics[width=\figwidth]{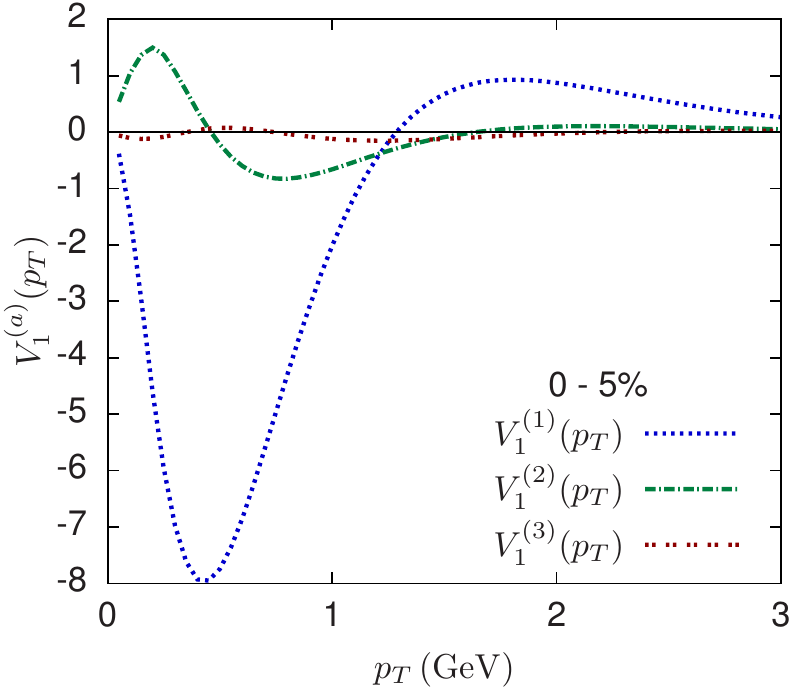}}\quad
\subfigure[{\,Principal flow vectors divided by the average multiplicity, 
$v_\FLOWN^{(a)}(p_T)\equiv 
V_\FLOWN^{(a)}(p_T)/\left<dN/dp_T\right>$.}\label{v\FLOWN-vn_norm}]{
\includegraphics[width=\figwidth]{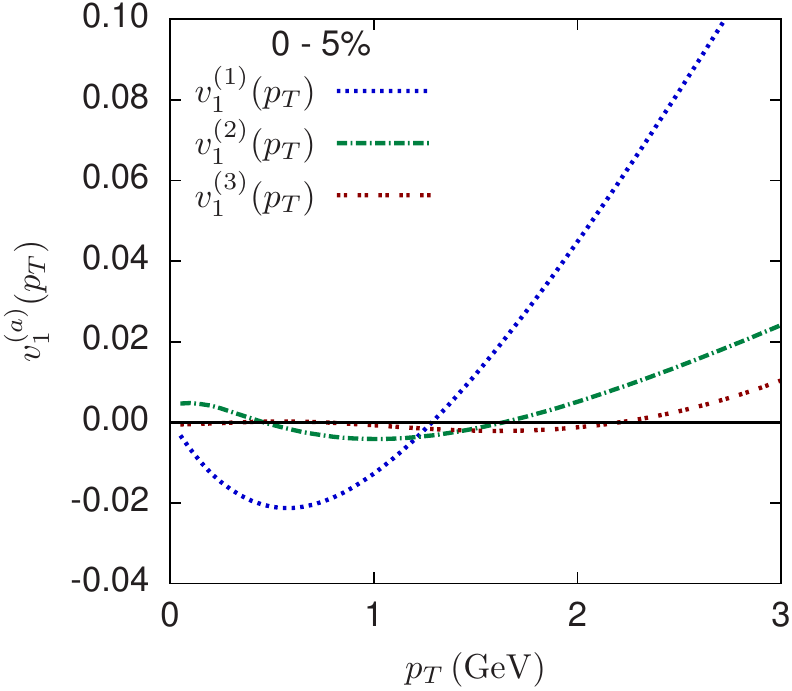}}
\subfigure[{\,Pearson correlation coefficient between the leading flow (zero 
suppressed for clarity) and several predictors.}\label{v\FLOWN-corr1}]{
\includegraphics[width=\figwidth]{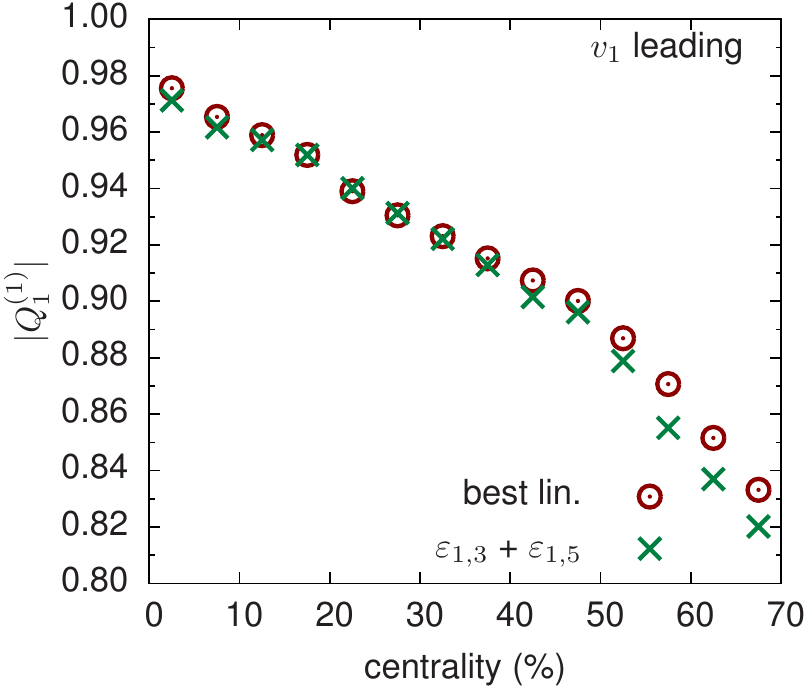}}\quad
\subfigure[{\,Pearson correlation coefficient between the subleading 
flow and several predictors.}\label{v\FLOWN-corr2}]{
\includegraphics[width=\figwidth]{std_A1_a0_B1_b0_corr2_c}}
\caption{Principal component analysis for $n=\FLOWN$ harmonic flow.}
\label{v\FLOWN}
\end{figure}

\renewcommand{\FLOWN}{2}
\begin{figure}
\centering
\subfigure[
{\,Centrality dependence of the (scaled) magnitudes of flows  
$\|v^{(a)}_\FLOWN\|$.}\label{v\FLOWN-eval}]{
\includegraphics[width=\figwidth]{std_A2_a0_B2_b0_eval_s}}\\
\subfigure[{\,Momentum dependence of principal flow vectors 
$V_\FLOWN^{(a)}(p_T)$ 
in 
central collisions.}\label{v\FLOWN-Vn_nonorm}]{
\includegraphics[width=\figwidth]{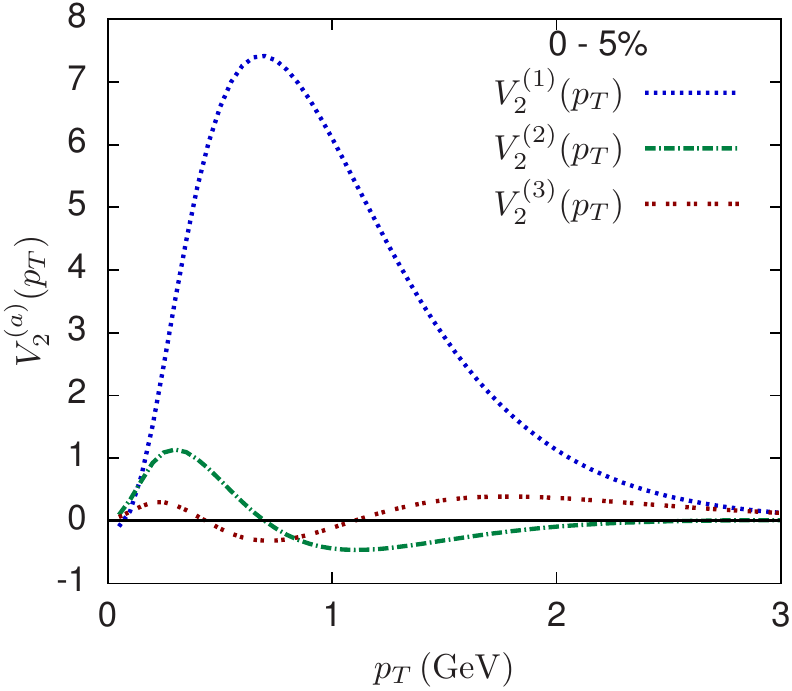}}\quad
\subfigure[{\,Principal flow vectors divided by the average multiplicity, 
$v_\FLOWN^{(a)}(p_T)\equiv 
V_\FLOWN^{(a)}(p_T)/\left<dN/dp_T\right>$.}\label{v\FLOWN-vn_norm}]{
\includegraphics[width=\figwidth]{std_A2_a0_B2_b0_CENT00_vn_norm}}
\subfigure[{\,Pearson correlation coefficient between the leading flow (zero 
suppressed for clarity) and several predictors.}\label{v\FLOWN-corr1}]{
\includegraphics[width=\figwidth]{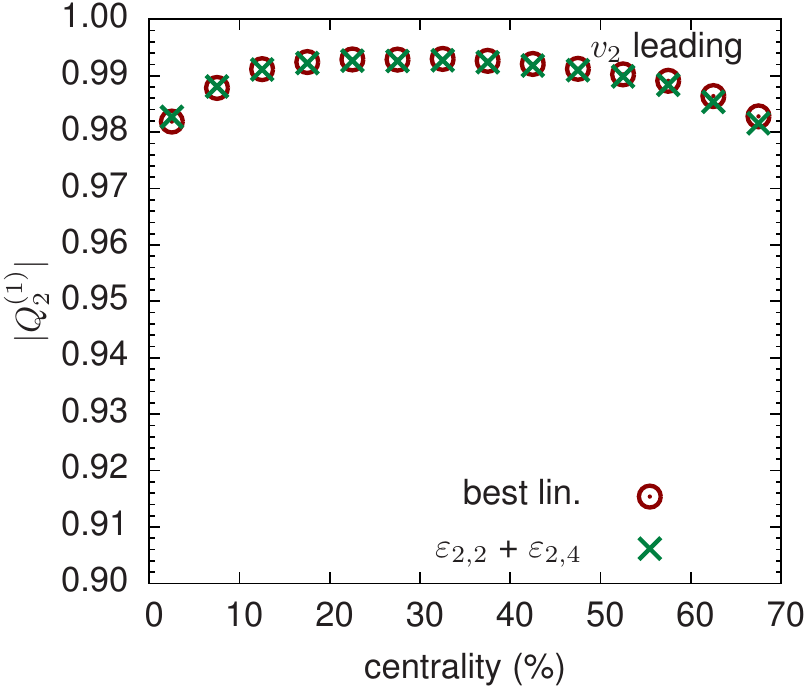}}\quad
\subfigure[{\,Pearson correlation coefficient between the subleading 
flow and several predictors.}\label{v\FLOWN-corr2}]{
\includegraphics[width=\figwidth]{std_A2_a0_B2_b0_corr2_c}}
\caption{Principal component analysis for $n=\FLOWN$ harmonic flow.}
\label{v\FLOWN}
\end{figure}

\renewcommand{\FLOWN}{3}
\begin{figure}
\centering
\subfigure[
{\,Centrality dependence of the (scaled) magnitudes of flows  
$\|v^{(a)}_\FLOWN\|$.}\label{v\FLOWN-eval}]{
\includegraphics[width=\figwidth]{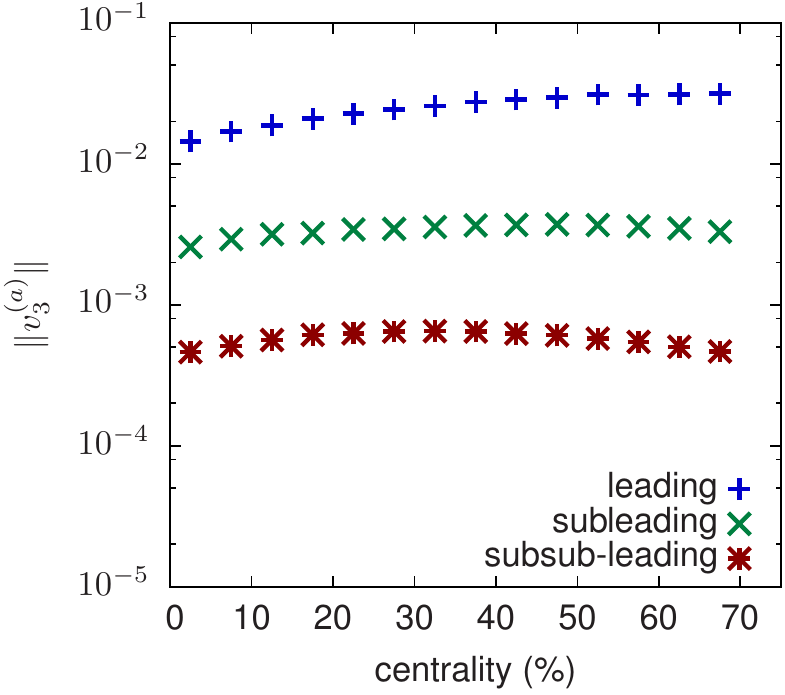}}\\
\subfigure[{\,Momentum dependence of principal flow vectors 
$V_\FLOWN^{(a)}(p_T)$ 
in 
central collisions.}\label{v\FLOWN-Vn_nonorm}]{
\includegraphics[width=\figwidth]{std_A3_a0_B3_b0_CENT00_Vn_nonorm}}\quad
\subfigure[{\,Principal flow vectors divided by the average multiplicity, 
$v_\FLOWN^{(a)}(p_T)\equiv 
V_\FLOWN^{(a)}(p_T)/\left<dN/dp_T\right>$.}\label{v\FLOWN-vn_norm}]{
\includegraphics[width=\figwidth]{std_A3_a0_B3_b0_CENT00_vn_norm}}
\subfigure[{\,Pearson correlation coefficient between the leading flow (zero 
suppressed for clarity) and several predictors.}\label{v\FLOWN-corr1}]{
\includegraphics[width=\figwidth]{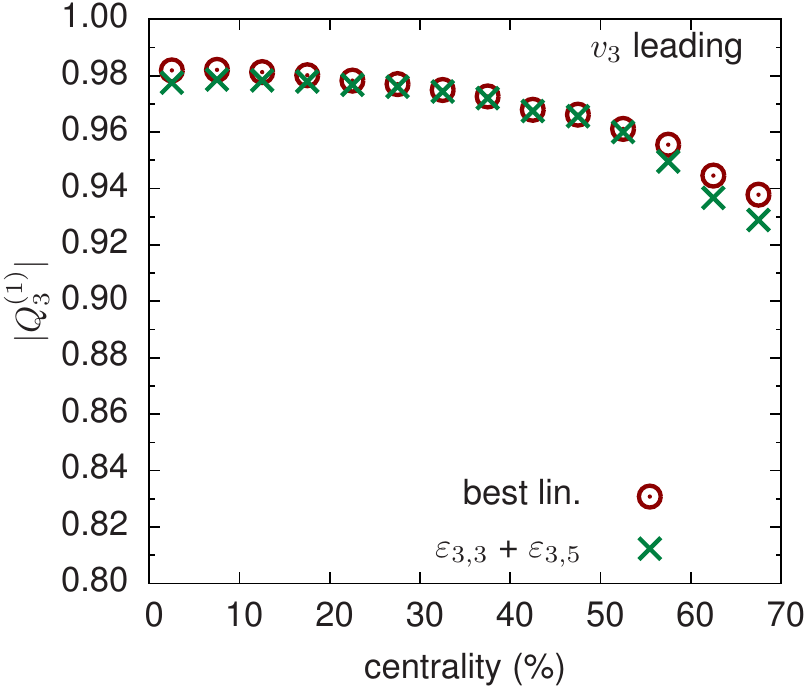}}\quad
\subfigure[{\,Pearson correlation coefficient between the subleading 
flow and several predictors.}\label{v\FLOWN-corr2}]{
\includegraphics[width=\figwidth]{std_A3_a0_B3_b0_corr2_c}}
\caption{Principal component analysis for $n=\FLOWN$ harmonic flow.}
\label{v\FLOWN}
\end{figure}

\renewcommand{\FLOWN}{4}
\begin{figure}
\centering
\subfigure[
{\,Centrality dependence of the (scaled) magnitudes of flows  
$\|v^{(a)}_\FLOWN\|$.}\label{v\FLOWN-eval}]{
\includegraphics[width=\figwidth]{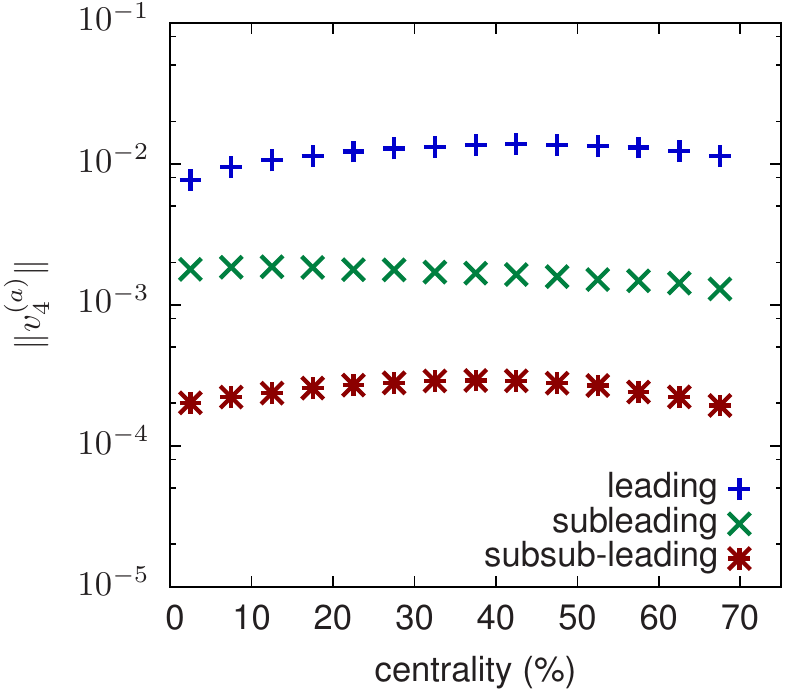}}\\
\subfigure[{\,Momentum dependence of principal flow vectors 
$V_\FLOWN^{(a)}(p_T)$ 
in 
central collisions.}\label{v\FLOWN-Vn_nonorm}]{
\includegraphics[width=\figwidth]{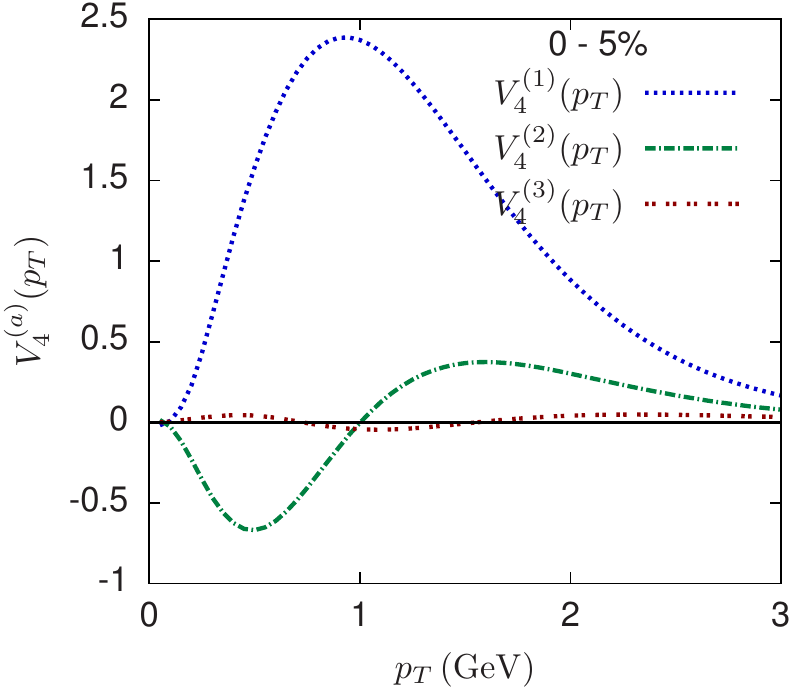}}\quad
\subfigure[{\,Principal flow vectors divided by the average multiplicity, 
$v_\FLOWN^{(a)}(p_T)\equiv 
V_\FLOWN^{(a)}(p_T)/\left<dN/dp_T\right>$.}\label{v\FLOWN-vn_norm}]{
\includegraphics[width=\figwidth]{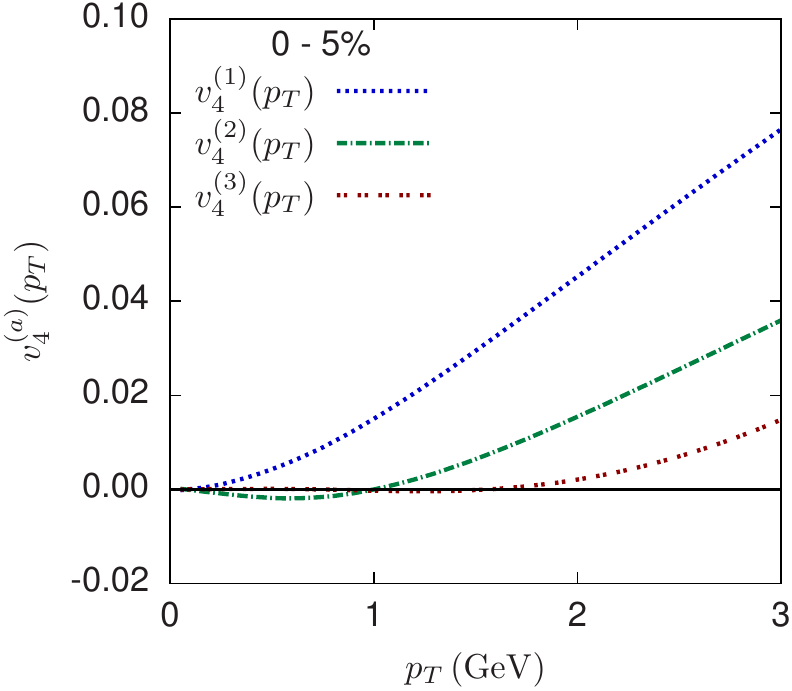}}
\subfigure[{\,Pearson correlation coefficient between the leading flow (zero 
suppressed for clarity) and several predictors.}\label{v\FLOWN-corr1}]{
\includegraphics[width=\figwidth]{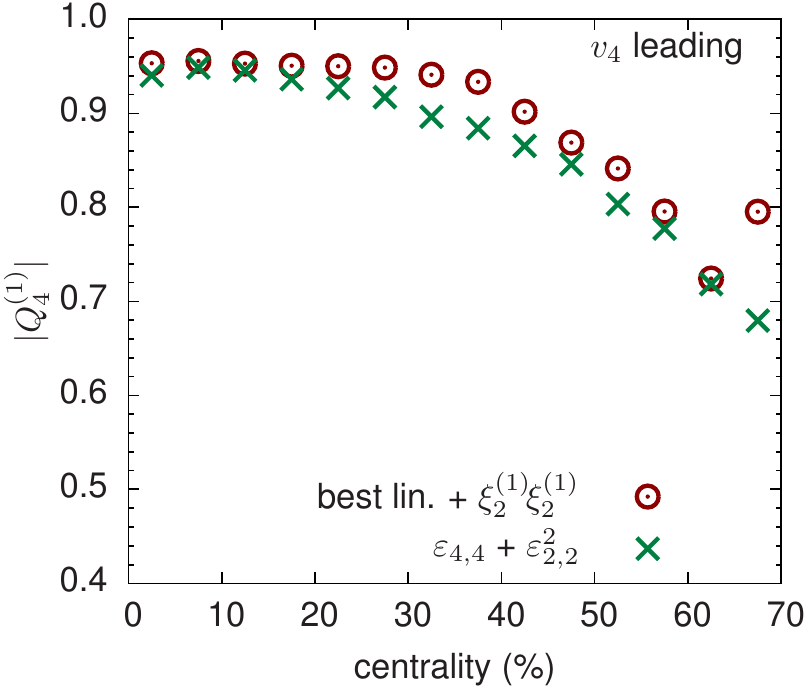}}\quad
\subfigure[{\,Pearson correlation coefficient between the subleading 
flow and several predictors.}\label{v\FLOWN-corr2}]{
\includegraphics[width=\figwidth]{std_A4_a0_B4_b0_corr2_c}}
\caption{Principal component analysis for $n=\FLOWN$ harmonic flow.}
\label{v\FLOWN}
\end{figure}

\renewcommand{\FLOWN}{5}
\begin{figure}
\centering
\subfigure[
{\,Centrality dependence of the (scaled) magnitudes of flows  
$\|v^{(a)}_\FLOWN\|$.}\label{v\FLOWN-eval}]{
\includegraphics[width=\figwidth]{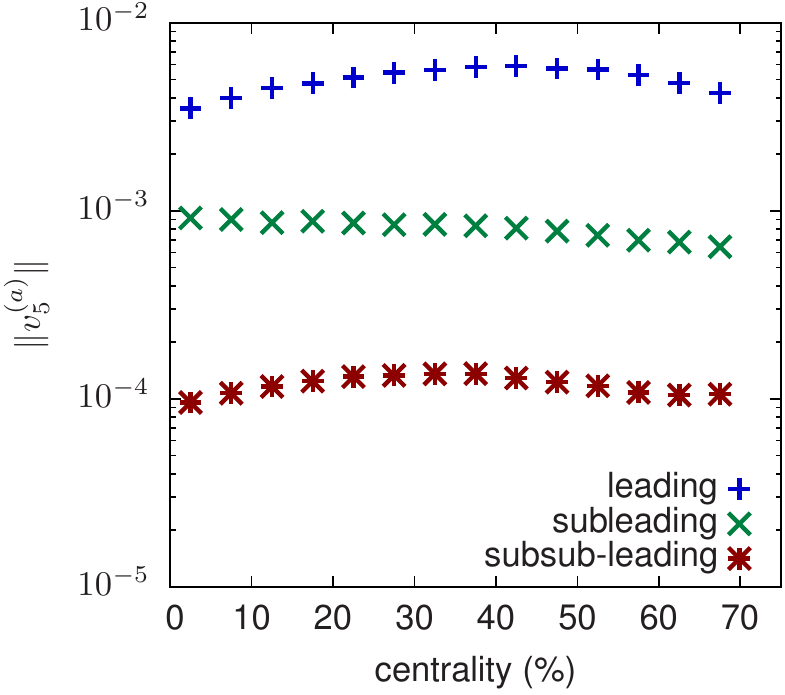}}\\
\subfigure[{\,Momentum dependence of principal flow vectors 
$V_\FLOWN^{(a)}(p_T)$ 
in 
central collisions.}\label{v\FLOWN-Vn_nonorm}]{
\includegraphics[width=\figwidth]{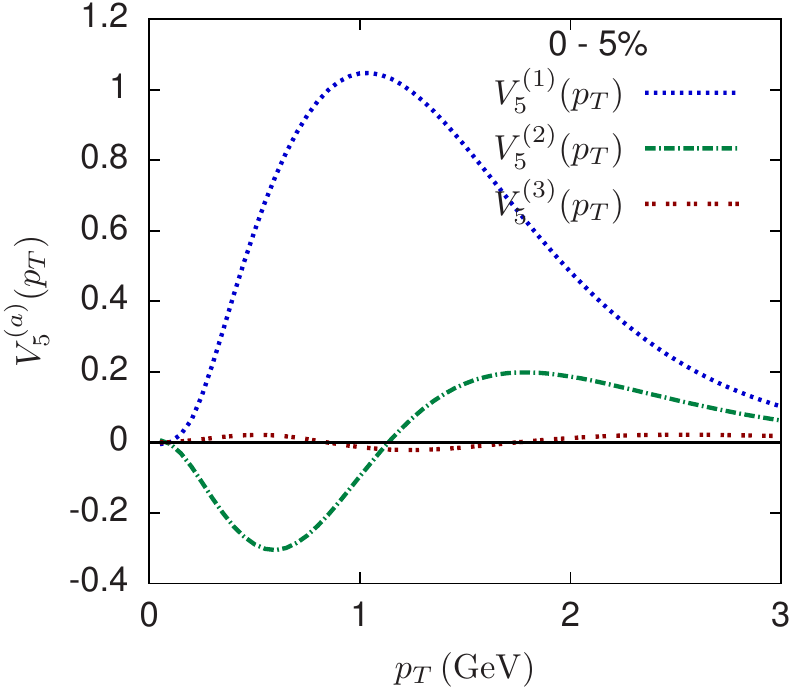}}\quad
\subfigure[{\,Principal flow vectors divided by the average multiplicity, 
$v_\FLOWN^{(a)}(p_T)\equiv 
V_\FLOWN^{(a)}(p_T)/\left<dN/dp_T\right>$.}\label{v\FLOWN-vn_norm}]{
\includegraphics[width=\figwidth]{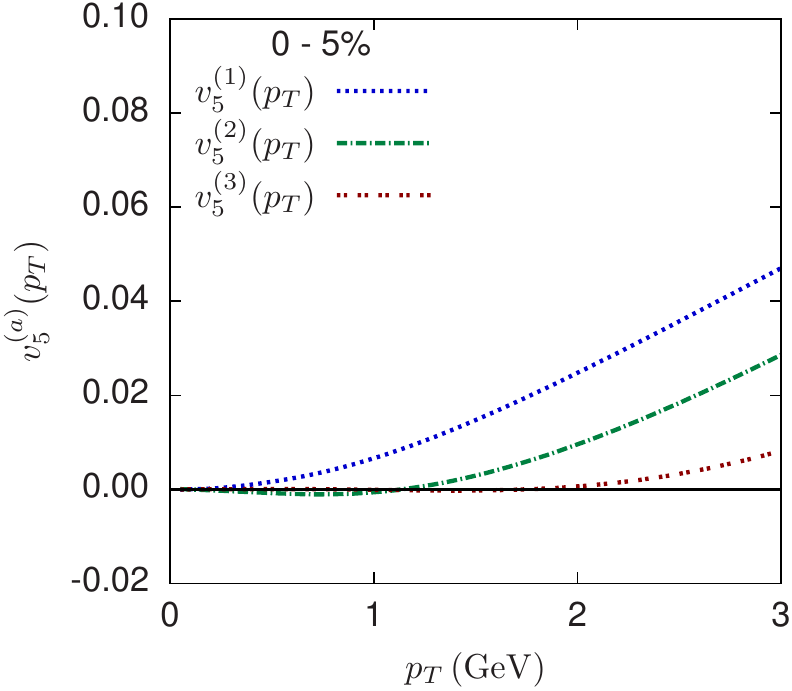}}
\subfigure[{\,Pearson correlation coefficient between the leading flow (zero 
suppressed for clarity) and several predictors.}\label{v\FLOWN-corr1}]{
\includegraphics[width=\figwidth]{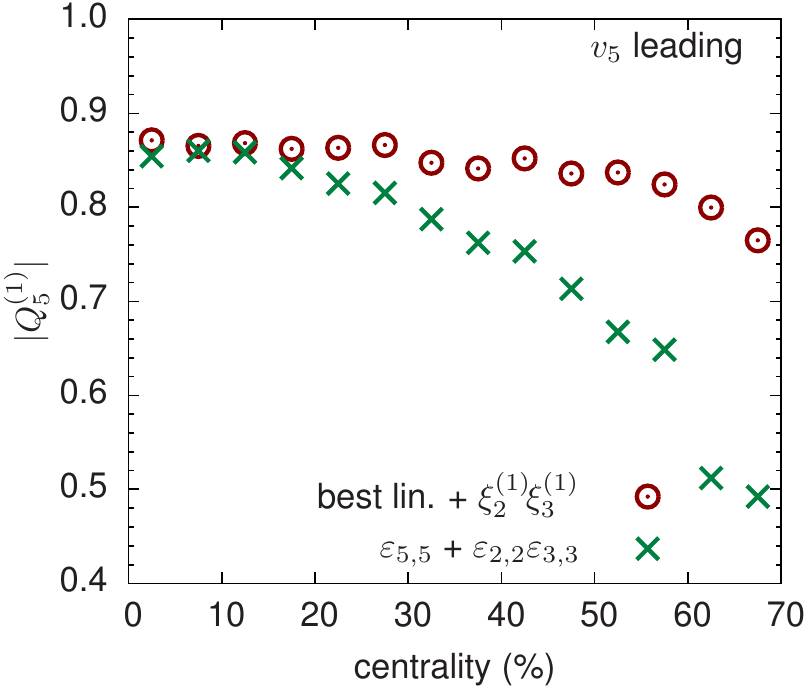}}\quad
\subfigure[{\,Pearson correlation coefficient between the subleading 
flow and several predictors.}\label{v\FLOWN-corr2}]{
\includegraphics[width=\figwidth]{std_A5_a0_B5_b0_corr2_c}}
\caption{Principal component analysis for $n=\FLOWN$ harmonic flow.}
\label{v\FLOWN}
\end{figure}

\end{subappendices} 

\chapter{Equilibration in weakly coupled effective kinetic theory}
\label{chap:ekt}

This chapter contains material published in 
\begin{itemize}
\item L. Keegan, A. Kurkela, A. Mazeliauskas and D. Teaney, \emph{Initial 
conditions for hydrodynamics from weakly coupled pre-equilibrium evolution}
J. High Energ. Phys. 08, 171 (2016)~\cite{Keegan:2016cpi}. Copyright (2016) by 
authors.
\end{itemize}
The discussion of the Green functions for the initial momentum 
perturbations, i.e. the Appendix C of the original publication, was omitted 
here.

\section{Introduction}
Viscous relativistic hydrodynamics provides a remarkably detailed and 
phenomenologically successful description of the expansion of the Quark Gluon Plasma (QGP)
in the ultra-relativistic heavy-ion 
collisions realized at the BNL Relativistic Heavy Ion Collider (RHIC) and at 
the 
CERN 
Large 
Hadron Collider (LHC)~\cite{Heinz:2013th,Luzum:2013yya,Teaney:2009qa}.
Hydrodynamics is an effective theory based on an assumption that the medium is
sufficiently close to  local thermal equilibrium
that the full stress tensor can be  
expanded in gradients of the energy and momentum densities~\cite{Baier:2007ix}. 
However, due to
the singular geometry of heavy ion collisions, the gradients diverge at early times, and the hydrodynamic approach does not apply during the initial stages of
the evolution.
Indeed, hydrodynamic simulations start at some sufficiently late
\emph{initialization time} $\ti\sim 1\,\text{fm}/c$, when the gradient 
expansion
becomes a useful approximation scheme. 
The initial conditions for hydrodynamics at $\ti$ are generally unknown, and
must be parametrized and fitted to data~\cite{Bernhard:2015hxa}. This procedure
often neglects any prethermal
evolution, and 
limits the empirical determination of the transport coefficients of the
QGP~\cite{Liu:2015nwa}.

A useful prethermal model  should smoothly and
automatically approach hydrodynamics. If this is the case, 
the combined pre-thermal and hydrodynamic evolutions will be
independent of the initialization
time~\cite{vanderSchee:2013pia,Romatschke:2015gxa,Kurkela:2016vts}. 
In most simulations the prethermal evolution is
either completely neglected \cite{Niemi:2015qia}, or modelled in a way that 
does not contain the
correct physics to produce  hydrodynamic 
flow~\cite{Broniowski:2008qk,Schenke:2012wb,Liu:2015nwa}.
 In addition, in
some models (such as the successful IP-glasma model \cite{Schenke:2012wb} motivated by parton saturation) the initial conditions  
contain strong gradients 
which limit the effectiveness of the hydrodynamic derivative expansion~\cite{Niemi:2014wta,Noronha-Hostler:2015coa}. 
Different hydrodynamic codes regulate these  extreme initial conditions in
different ad hoc ways, e.g. by arbitrarily setting the shear stress
tensor to zero when the hydrodynamics is initialized.  Again, these ambiguities 
limit the ability of hydrodynamic simulations to determine the
transport properties of the QGP.

In the limit of weak coupling $\alpha_s \ll 1$ the approach to hydrodynamics, or \emph{hydrodynamization}, is 
described by an Effective Kinetic Theory (EKT) \cite{Arnold:2002zm}, which 
takes into account the non-trivial 
in-medium dynamics of screening and the Landau-Pomeranchuk-Migdal suppression of
collinear radiation.
In ref.~\cite{Kurkela:2015qoa} (which includes one of the authors), it 
was shown that the EKT, starting with 
initial 
conditions motivated by the Color-Glass Condensate (CGC) saturation framework~\cite{Iancu:2002xk,Iancu:2003xm,Gelis:2010nm, Gelis:2007kn, Lappi:2011ju}, reaches hydrodynamics in a phenomenologically reasonable time
scale of $\sim 10/Q_s$, where $Q_s$ is the (adjoint representation) saturation scale, which is estimated to be of order of few GeV
for central heavy-ion collisions at the LHC.  This first calculation used the EKT
to monitor the equilibration of a uniform plasma of infinite transverse extent  during a Bjorken expansion.

Transverse gradients in the profile 
will initiate flow during the equilibration process. This \emph{preflow}
and the accompanying modifications
of  the initial energy density profile will influence the subsequent hydrodynamic evolution. The goal of the current paper is to use the EKT 
to precisely determine the preflow and the components of the 
energy momentum tensor $T^{{\mu}\nu}(\ti, {\bf x}_\perp)$ 
that should be used to initiate the hydrodynamic evolution for
a given initial energy density profile.
Although the kinetic theory calculation can be used to match different models for the initial energy profile to hydrodynamics,  the weak coupling
approximations made in the IP-glasma model lead naturally to
effective kinetic theory.

\begin{figure}
    \begin{center}
        \includegraphics[width=0.7\textwidth]{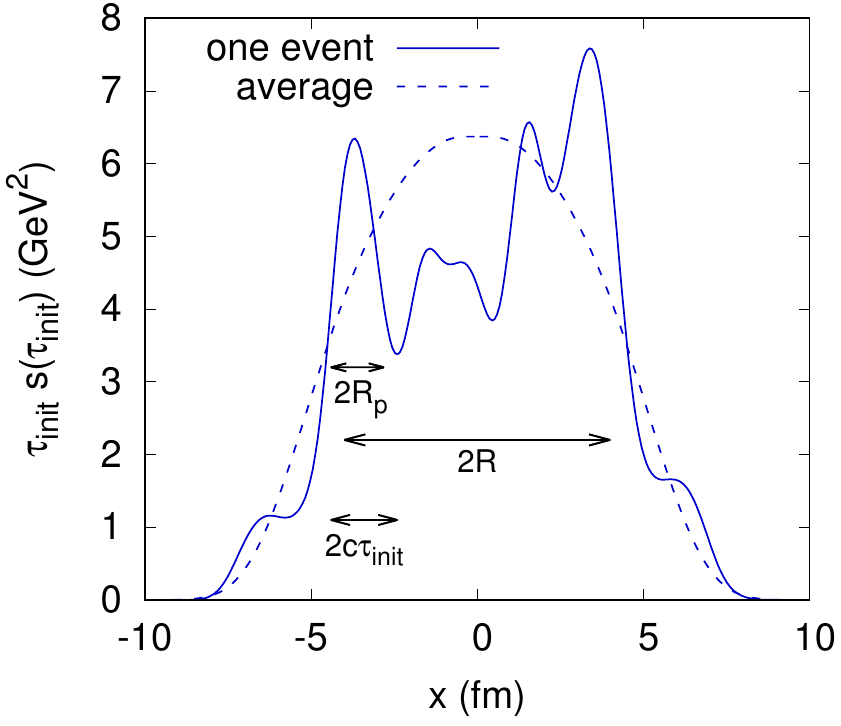}
        \caption{A typical entropy density profile (times ${\ti}$) for 
            a single event used 
            as an initial condition in 
            current hydrodynamic simulations at the LHC for a 0-5\% centrality class~\cite{Mazeliauskas:2015vea}.  
            An
        event averaged initial condition is shown by the dashed line. 
            Often the initial flow velocity is set to zero. The different 
            scales are discussed in the text.
    \label{ichydro}}
    \end{center}
\end{figure}

\Fig{ichydro} shows a typical transverse (entropy) profile that is used  in
current hydrodynamic simulations~\cite{Mazeliauskas:2015vea}. Clearly
during the equilibration process the profile will change and generate initial 
flow.
The equilibration time, $c\ti$, is short compared to the nuclear radius,  $R$.
For this reason the prethermal evolution is
insensitive to the global collision geometry.
Indeed, we may decompose the transverse plane into
causally disconnected patches of size $c\ti \ll R $  whose prethermal evolution 
can be separately determined. 
In these patches, the global nuclear geometry determines
a small gradient that can be considered as a linear perturbation over a translationally invariant background. 
Thus, corrections to initial conditions for hydrodynamics
from the global geometry are of order $c\ti/R$~\cite{Vredevoogd:2008id}.
In addition to the global geometry, the initial energy density profile includes event-by-event fluctuations at smaller 
scales set by the nucleon size $R_p$,  which is comparable to the 
causal horizon $R_p \sim c \ti$.
Event-by-event fluctuations at these length scales are  suppressed by 
$1/\sqrt{N_\text{part}}$ where $N_{\text{part}}$ is the number of participating 
nucleons in the event,  $N_{\text{part}}\sim 100-300$.
Therefore,  such fluctuations can also be treated
in a linearized way as fluctuations over a translationally invariant
background.
The
structure of the initial profile at even smaller
scales is less well known, but in models based on CGC, one expects fluctuations 
to subnuclear scales of order the saturation momentum, $Q_s^{-1} \sim 
0.1\,\text{fm}$. 

Finally, 
an important scale is set by the mean free path, which in a weakly 
coupled theory is of order $1/\lambda^2 T_{\rm eff}$ for states
not too far from equilibrium.  In practice, this length scale is comparable,
though slightly shorter than the causal horizon and the nucleon scales. Without the scale separation, the medium prethermal response  to initial perturbations in the transverse plane can only be 
computed by a calculation within the EKT. Fortunately, as discussed
above linearized kinetic theory is sufficient to determine this response. 

To summarize, our strategy is to use linearized kinetic theory to follow 
the hydrodynamization  of energy perturbations on top of a 
far-from-equilibrium  Bjorken background with translational symmetry in the 
transverse directions.  This determines
the stress tensor for hydrodynamics  at the initialization time.
The length scales of relevance are 
the nuclear-geometry, the nucleonic scale, the causal horizon $c\ti$, and the mean free path
\begin{equation}
    R \gg R_{p} \sim c\ti \sim \frac{1}{\lambda^2 T_{\rm eff}} \, .
\end{equation}
By linearizing the problem and solving for the response,
we will determine a Green function describing how an energy fluctuation  
at the earliest moments, $\tau\sim 1/Q_s$,
evolves  during the equilibration process to the  hydrodynamic
fields, i.e. the energy and momentum 
densities, $\delta T^{00}(\ti, \x_\perp)$ 
and $\delta T^{0i}(\ti,\x_\perp)$ respectively. We will verify 
that  the subsequent evolution is described by second order hydrodynamics to
certifiable precision.

In \Sect{linearized} we outline the linearized EKT, and study the linear 
response of the EKT in equilibrium. In \Sect{Bjorken} we systematically study
the approach to equilibrium of Fourier modes of specified $k$, starting
with a far from equilibrium initial state.  
In \Sect{Green} we Fourier 
transform these results and determine a coordinate space Green function which
produces the appropriate
initial conditions for hydrodynamics at $\ti$
when convolved with a specified initial state. 
We also analyze
the long wavelength limit of these Green functions, making contact and providing additional insight into previous work on preflow~\cite{Vredevoogd:2008id}. Finally,
we discuss our conclusions in \Sect{discuss}.

\section{Linearized kinetic theory}
\label{linearized}

\subsection{Setup}
At weak coupling the non-equilibrium evolution of the boost invariant color and 
spin averaged
gluon distribution function
is described in 
terms of an effective kinetic equation \cite{Arnold:2002zm}
\begin{align}
\partial_\tau f_{\x_\perp,\p}+ \frac{\bf p}{|p|}\cdot \nabla_{\x_\perp} 
f_{\x_\perp,\p} - \frac{p_z}{\tau}\partial_{p_z} f_{\x_\perp,\p}= 
-\mathcal{C}[f_{\x_\perp,\p}],
\end{align}
where the effective collision kernel $\mathcal{C}[f]$ incorporates the elastic $2\leftrightarrow 2$ and inelastic $1\leftrightarrow 2$ processes as required for a leading order description in the coupling constant $\lambda =4\pi \alpha_s N_c$, which is the only parameter of the EKT. 
The kinetic theory is valid when the occupancies are perturbative $\lambda f \ll 1$ and when the relevant
distance scales are larger than the typical Compton wavelength of the particles 
$\Delta x \gtrsim \langle p \rangle^{-1}$. 
The details 
of the scattering kernel have been discussed in Refs.~\cite{Arnold:2002zm, 
Kurkela:2015qoa,Keegan:2015avk} and are briefly repeated here in the 
\app{colker}.
We use the isotropic screening approximation from \cite{Kurkela:2015qoa} which is leading order 
accurate  for parametrically isotropic systems $\mathcal P_L/\mathcal P_T \approx 1$. (Here and below $\mathcal P_{L}$ and $\mathcal P_T$ denote the longitudinal and transverse pressures.)
In the current paper we will consider only gluonic degrees of freedom and assume that 
the contribution of quarks is suppressed during the pre-equilibrium evolution.%
\footnote{The initial far-from-equilibrium state is parametrically dominated by gluons. Once the plasma
has thermalized it should contain also fermionic degrees of freedom. However, the production of fermions is suppressed
by larger color factors $C_F/C_A$,  and 
by Pauli blocking factors (while scattering of gluons is Bose enchanced). 
It is therefore plausible that the system hydrodynamizes before
it is chemically equilibrated.}

We split the 
distribution function into a translationally symmetric background and a 
linearized perturbation with a 
wavenumber $\k_\perp$ in the transverse plane
\begin{equation}
f_{\x_\perp,\p}=\bar f_\p + \int \frac{d^2 \k_\perp}{(2\pi)^2} A({\bf k}_\perp) 
e^{i \k_\perp \cdot \x_\perp}\delta f_{\k_\perp,\p},
\end{equation}
where $A({\bf k}_\perp)$ characterizes the initial density profile. The kinetic 
equations for the background
and the (complex) fluctuation then read
\begin{subequations}
\label{boltz}
\begin{eqnarray}
(\partial_\tau - \frac{p_z}{\tau} \partial_{p_z})  \bar f_{\p} &=& - \mathcal{C}[\bar f], \\
(\partial_\tau - \frac{p_z}{\tau} \partial_{p_z} + \frac{i \p_\perp\cdot \k_\perp}{p} )  \delta f_{\k_\perp,\p} &=& - \mathcal{C}[\bar f, \delta f],
\end{eqnarray}
\end{subequations}
where $\mathcal{C}[\bar f, \delta f]$ is the collision kernel linearized in 
$\delta f$ (see \app{colker}  for details).

\begin{figure}
\centering
\includegraphics[width=0.7\textwidth]{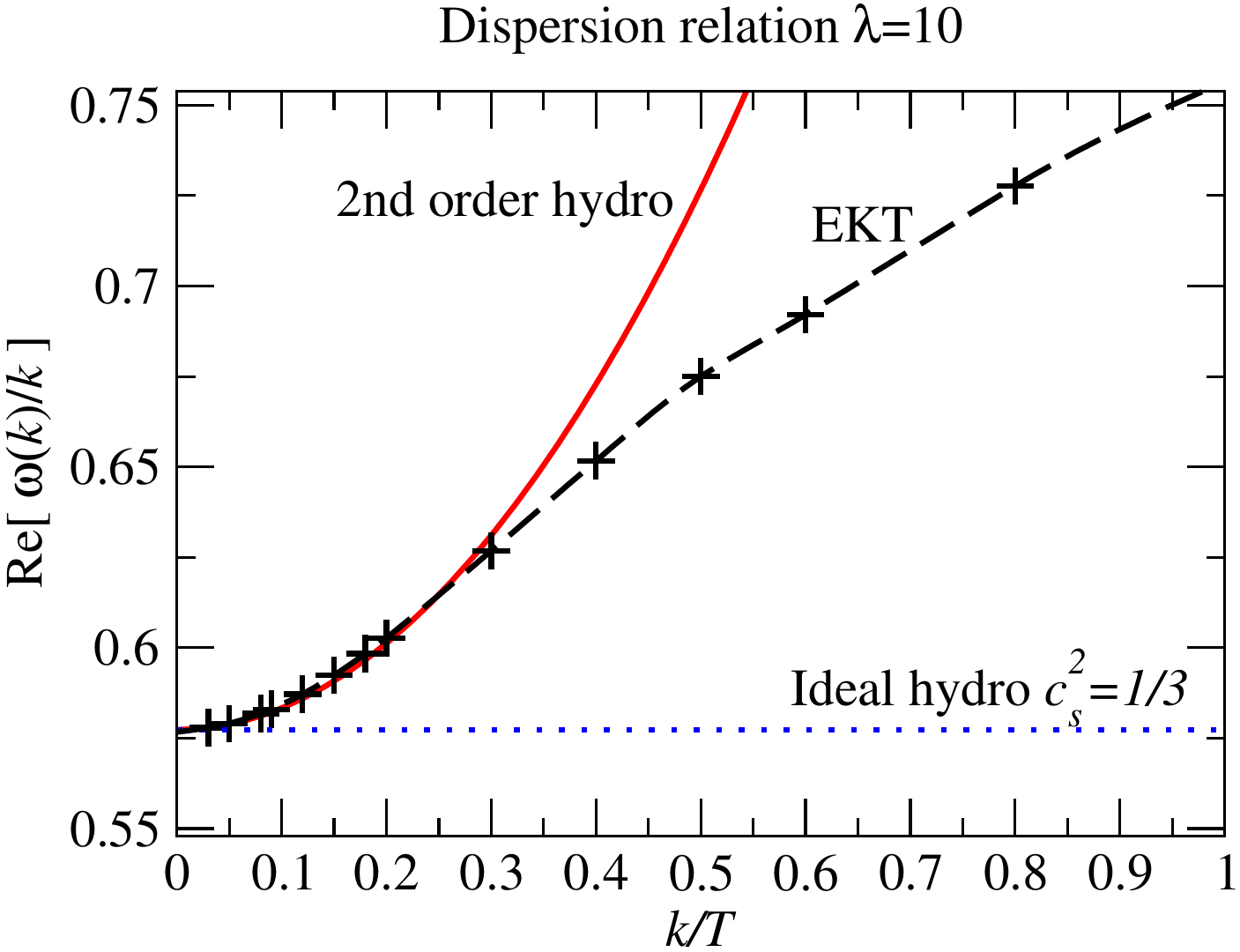}
\caption{\label{disprel} The dispersion relation of sound modes with thermal background from the EKT.
The long wavelength modes are described by ideal hydrodynamics with $\omega = c_s k$ and $c_s^2 = 1/3$, and the approach to ideal 
hydrodynamics is well described by 2nd order hydrodynamics. For modes with wave numbers larger than $k \gtrsim 0.4 T$,
the dispersion relation differs significantly from the hydrodynamic expectation.
}
\end{figure}

\subsection{Hydrodynamization close to equilibrium}
\label{kspace}
Before studying the equilibration  process,
we will analyze the linear response of the EKT close
to equilibrium, corresponding to the $\tau\rightarrow \infty$ limit of \eq{boltz}.
Our goal in this section is to determine at what wavenumbers (characterized by $k/T$) linearized energy-momentum perturbations  are described by hydrodynamics for an equilibrated background.

The 
dispersion relation for the sound mode to second order in the hydrodynamic expansion 
reads~\cite{Baier:2007ix}
\begin{equation}
\omega = c_s k - i\frac{4}{3}\frac{\eta }{e+p}k^2 + \frac{4}{3}\frac{\eta}{e+p}\left( c_s \tau_\pi - \frac{2}{3 c_s}\frac{\eta}{e+p}\right)k^3,
\label{eq:disp_rel}
\end{equation}
where $c_s^2 = 1/3$ for conformal equation of state and $\eta, \tau_\pi$, are known transport coefficients at weak coupling \cite{Arnold:2003zc, York:2008rr}. For $\lambda = 10$ (corresponding to $\alpha_s \approx 0.26$) the hydrodynamic coefficients read $\eta/s =  0.62$, 
$\tau_\pi = 5.1 \eta/sT$, and $\lambda_1 = 0.8 \eta \tau_\pi$.
We will quantify at what numerical 
values of  $k/T$ the corrections to \eq{eq:disp_rel} become sizeable. 

To this end, the kinetic theory is initiated  in local 
thermal equilibrium with a spatially varying temperature, 
$T(\x) = T + \delta T e^{i \k_\perp \cdot \x_\perp}$, and  corresponding  phase space distribution
\begin{equation}
\delta f^{(1)}_{\k_\perp,\p} =  -\frac{\delta T}{T} p \, \partial_p \bar 
f_{\p}, \quad \textrm{ and}\quad
\bar f_\p = \frac{1}{e^{p/T}-1}.
\end{equation}
Gradients in the energy density drive
momentum perturbations in due course, and the
frequency of the subsequent (damped) oscillations determines the real part of the
dispersion relation. Numerically, we obtained the oscillation frequency by measuring the time
interval between the successive nodes. 

The results
for various $k/T$ are depicted in \Fig{disprel} for $\lambda =10$.  We see that 
at small $k$,  $k/T \lesssim 0.1$, the 
dispersion relation is well described by the $c_s^2 = 1/3$ result of ideal hydrodynamics, and the approach to ideal hydrodynamics
is described by the 2nd order corrections of \eq{eq:disp_rel}
(note that 
the real part of the frequency does not get a first order correction). 
Indeed, 
for $k/T \lesssim 0.4$, the second order hydrodynamic theory matches well 
with the EKT. 

At higher values of $k/T$, the EKT finally saturates at $\omega = 
k $ in contrast to the strict (unresummed) second order hydrodynamics. For $\lambda = 10$ this 
happens 
only 
at rather large values of $k$, $k > T$. For these wavenumbers, the wavelength of the perturbation is  comparable  to the typical gluon 
Compton wavelength, and the linear response of the system
cannot be reliably computed with kinetic theory in this regime.

We conclude that for $\lambda = 10$, the smallest scales that hydrodynamize have $k \sim 0.4 \, T$. Varying the 
value of $\lambda < 10$ (not shown), we find that the scale where hydrodynamics breaks downs tracks the shear viscosity, $k \sim 0.4 \, [\eta(\lambda = 10)/\eta(\lambda)]\,T $ with varying $\lambda$. Note that for smaller $\lambda$, the saturation to $\omega \sim k$ can take place within the
regime of validity of the effective theory.

\section{Hydrodynamization of fluctuations far from equilibrium}
\label{Bjorken}
We now move on to study the hydrodynamization of spatially dependent fluctuations 
on top of a far-from-equilibrium boost invariant background.
As discussed in \cite{Kovner:1995ja, Lappi:2006fp, Gelis:2013rba, Kurkela:2015qoa}, at very 
early times $\tau \lesssim Q_s^{-1}\,$ the 
far-from-equilibrium gluonic system in the midrapidity region is parametrically over-occupied $\lambda f \sim 1$,
and the dynamics is described with coherent classical gauge fields rather than with particles. This part of the evolution is 
characterized by negative values of the longitudinal pressure $\mathcal P_L$, which is a result of the coherence of
the approximately boost invariant fields. However, classical numerical
simulations~\cite{Lappi:2011ju, Gelis:2013rba} (as well as analytical series
solutions to the classical equations of motion~\cite{Chen:2015wia,Li:2016eqr})
show that  in a timescale $Q_s \tau \sim 1$, 
the coherence is lost, the longitudinal pressure approaches zero $\mathcal P_L \sim 0$, and the occupancies become perturbative \cite{Kurkela:2011ti, Kurkela:2011ub,Berges:2013fga, Baier:2000sb}. At this point the system may be passed
to the EKT \cite{Mueller:2002gd,Jeon:2004dh, York:2014wja, Kurkela:2015qoa}. 

Following \cite{Kurkela:2015qoa}, we take as our initial condition at $\tau_0 = 1/Q_s$ a parametrization
\begin{align}
    f(p_z,p_\perp) &= \frac{2}{\lambda} A f_0(p_z \xi/p_0,p_\perp/p_0), \label{eq:init_cond}\\
f_0(\hat p_z,\hat p_\perp) & = \frac{1}{\sqrt{\hat p_\perp^2+\hat p_z^2}} e^{-2 (\hat  p_\perp^2 + \hat  p_z^2)/3}, \label{f0}
\end{align}
where $p_0 = 1.8\,Q_s$, $\xi=10$, and $\lambda=10$.  
The parameters $p_0$ and $\xi$ are motivated 
by classical simulations where $\sqrt{\llangle p_T^2 \rrangle}\approx 1.8\, Q_s$ 
and $\llangle p_z^2 \rrangle \ll \llangle p_T^2 \rrangle$.
The amplitude, $A$, is adjusted so 
that energy per rapidity
\begin{align}
    \tau_0 e(\tau_0) =  \tau_0 \nu_g \int \frac{d^3p}{(2\pi)^3} \, |p|\, f(p_z, p_\perp) \, ,
\end{align}
matches the results
of classical simulations \cite{Lappi:2011ju}, where
\begin{align}
    \label{Qsdef}
    \tau_0 e(\tau_0)  \simeq  0.358\,\frac{\tau_0 \nu_g Q_s^4}{\lambda} \, .
\end{align}
Here $\nu_g = 2 d_A= 16$ is the number of gluonic degrees of freedom.
With these parameters, the number of gluons and the mean $p_T$ in the EKT at $\tau_0$  are
\begin{align}
    \frac{dN}{d^2\x_\perp dy} =&  0.232\,\frac{\nu_g Q_s^2}{\lambda} ,  \qquad \sqrt{\llangle p_{T}^2\rrangle}  = 1.8\,Q_s,
 \end{align}
 which roughly matches the classical Yang-Mills simulations.

We will follow response to the specific initial perturbations of these initial 
conditions describing   energy density fluctuations in the transverse plane
\begin{align}
\delta f^{(1)}_{\k_{\perp},\p}(\tau_0) &= - \frac{\delta Q_s}{Q_s} p \, 
\partial_p \bar 
f_{\p} \label{initialdf},
\end{align}
which results from varying 
the saturation scale in \eq{eq:init_cond}, $Q_s(\x_\perp) \sim Q_s + \delta Q_s e^{i 
\k_\perp \cdot \x_\perp}$.

\begin{figure*}
\centering
    \includegraphics[width=1\textwidth]{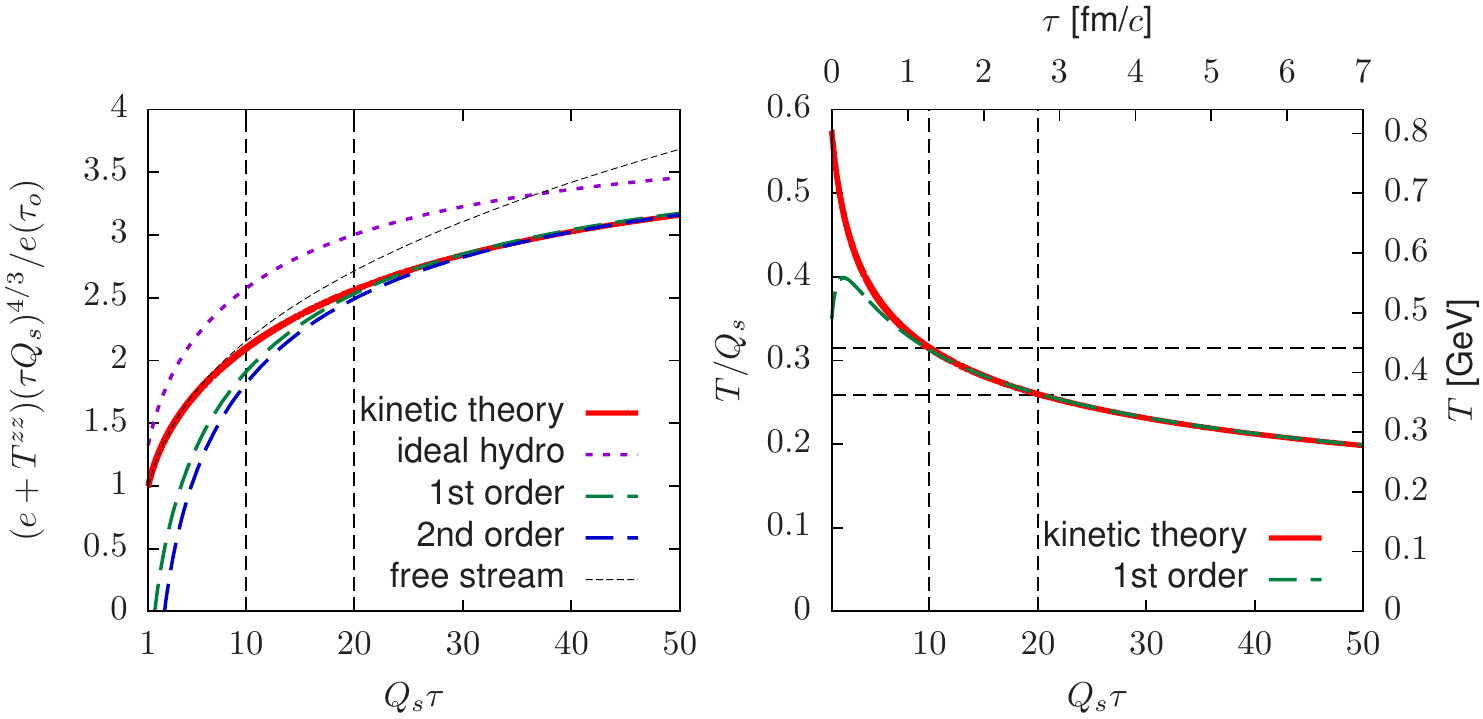}
\caption{(a) A comparison of the relevant combination $e(\tau) + T^{zz}(\tau)$ for the  
kinetic theory background with the hydrodynamic constitutive equations of \eq{eqTzz}.
(b) The background effective temperature as obtained from the Landau matching condition
$e = \nu_g \frac{\pi^2}{30} T^4$.  Extrapolation of first order hydro (fitted 
at asymptotic late 
times) is shown for comparison, \Eq{asympT}. The scales in physical units 
correspond to 
$Q_s = 1.4\,{\rm GeV}$ which yields the entropy required by hydrodynamic 
simulations (see text).
    \label{background}
}
\end{figure*}

Without loss of generality, we can choose the wave vector $\hat{k}_\perp=(k,0)$ 
to 
point in $x$-direction. Then at any time, the energy and momentum 
perturbations are defined 
as
\begin{align}
\delta e(\tau,k)  & \equiv \delta T^{00}=\nu_g \int \frac{d^3\p}{(2\pi)^3 
}p^0 \delta f,\\
g^x(\tau, k) &\equiv \delta T^{0x} =  \nu_g \int \frac{d^3\p}{(2\pi) 
^3}p^x \delta f,
\end{align}
and their evolution is governed by the linearized conservation equations
\begin{subequations}
    \label{eq:conservation}
\begin{align}
&\partial_\tau e(\tau)=-\frac{e(\tau)+
T^{zz}(\tau)}{\tau}, \label{econs}\\
&\partial_\tau \delta e(\tau,k)+ik  g^x (\tau,k)=-\frac{\delta 
e(\tau,k) +  \delta T^{zz}(\tau,k)}{\tau},\label{decons}\\
&\partial_\tau g^x(\tau,k)+ik \delta T^{xx}(\tau,k)= -\frac{g^x 
(\tau,k)}{\tau},\label{dgcons}
\end{align}
\end{subequations}
where $\delta T^{\mu\nu}$ is the energy-momentum tensor perturbation caused by 
$\delta f$.  If the system is described by  
hydrodynamics then $T^{zz}$, $\delta T^{zz}$ and $\delta T^{xx}$ 
are determined through the constitutive equations by the first moments of the particle distribution function $e$, 
$\delta e$ 
and $g^x$. For conformal second order 
viscous hydrodynamics these relations are
\begin{subequations}
    \label{eq:constit}
\begin{align}
T^{zz}(\tau) &= 
\frac{1}{3}e-\frac{4}{3}\frac{\eta}{\tau}-\frac{8}{9}\frac{\tau_\pi 
\eta - 
\lambda_1}{\tau^2},\label{eqTzz}\\
\delta T^{xx}(\tau,k) = &\frac{\delta e(\tau, k)}{e} 
\left[\frac{1}{3}e + 
\frac{1}{3}\eta \tau_\pi 
k^2 + \frac{1}{2\tau}\eta -\frac{2\left( 
\lambda_1-\eta\tau_\pi\right)}{9\tau^2} \right] \nonumber \\
&- i \frac{kg^x(\tau,k)}{e}\left[
\eta-\frac{1}{\tau}\left( \frac{\eta^2}{2e} + \frac{\eta 
\tau_\pi}{2} -\frac{2}{3}\lambda_1 \right)\right]\label{eqdTxx}\\
\delta T^{zz}(\tau,k) = &\frac{\delta e(\tau, k)}{e} 
\left[\frac{1}{3}e - 
\frac{1}{6}\eta \tau_\pi 
k^2 - \frac{1}{\tau}\eta +\frac{4\left( 
\lambda_1-\eta\tau_\pi\right)}{9\tau^2} \right] \nonumber \\
&+ i \frac{kg^x(\tau,k)}{e}\left[
\frac{1}{2}\eta-\frac{1}{\tau}\left( \frac{\eta^2}{4e} +\frac{2}{3}\lambda_1 
\right)\right]\label{eqdTzz},
\end{align}
\end{subequations}
where  the constitutive equations for ideal (or first order viscous) hydrodynamics 
can 
be recovered by 
setting $\eta=\tau_\pi=\lambda_1=0$  (or $\tau_\pi=\lambda_1=0$).

\subsection{Evolution of the background energy density}

Before studying the perturbations, we will study the equilibration of
the background energy density, elaborating on  the original 
study~\cite{Kurkela:2015qoa}. In \Fig{background}(a) we compare the energy 
momentum tensor combination $e+T^{zz}$ 
in  the kinetic theory simulation
to the constitutive equation, \eq{eqTzz}. The $e+T^{zz}$ combination is motivated by the  conservation law in \eq{econs}. 

At early times the system evolves approximately according to  free streaming%
\footnote{During this part of the evolution, the system evolves according to the nonthermal attractors
discussed in e.g. \cite{Baier:2000sb,Kurkela:2011ub,Berges:2013fga,Kurkela:2015qoa}. The nonthermal attractors are characterized by  $T^{zz}\ll e$, and for the current discussion the fine details of the attractor are irrelevant, and the evolution resembles that of free streaming.},
with $T^{zz}\sim 0$ and $e \propto \tau^{-1}$.
As already noticed in \cite{Kurkela:2015qoa}, the constitutive equations give 
an increasingly accurate description of the EKT
stress tensor as a function of time. While the ideal constitutive equations are 
rather far from the EKT at all relevant times, the viscous and 2nd order 
equations quickly converge to the EKT. Note that an accidental (approximate)
cancellation of $\lambda_1 - \eta \tau_\pi$ makes the 
second order correction anomalously small~\cite{Luzum:2008cw}, and only 
at rather late times  after non-hydrodynamic modes have
almost completely decayed does second order hydrodynamics finally improve the first order result (not shown).
By times $Q_s
\tau =\{ 10, 20 \}$, the viscous constitutive equations agree
with the EKT within $\{10\%,2\%\}$.

It is noteworthy that the EKT interpolates smoothly between the free streaming
and viscous hydrodynamic evolutions without an extended period during which the
evolution is not approximately described by one or the other approximation
scheme. At $Q_s\tau{=}10$ the evolution is somewhat closer to free streaming,
and using hydrodynamics at this point is a rough, though perhaps acceptable,
approximation. At $Q_s\tau{=}20$ hydrodynamics is a better approximation, but at this time the causal horizon is becoming comparable to the nuclear radius.

Given the agreement with the constitutive equations, one  can use
the hydrodynamic equations to propagate the system forward in time. 
At late times ideal hydrodynamics is valid, and the entropy per area 
per rapidity approaches a constant
\begin{equation}
\label{finalentropy}
\lim_{\tau \rightarrow \infty} \tau s(\tau) \equiv \frac{\nu_g \Lambda_s^2}{\lambda}.
\end{equation}
(Here the $\Lambda_s$ parametrization is motivated by the scaling of the 
initial multiplicity with the saturation scale in \eq{Qsdef}.)
Dimensional reasoning indicates that $\Lambda_s^2$ is proportional to 
$Q_s^2$. Taking the data presented in \Fig{background}(a) we may extrapolate
  $\tau \rightarrow \infty$ to determine the proportionality coefficient 
\begin{equation}
\label{asymptotic}
\Lambda_{s}^2 = 1.95 \, Q_s^2.
\end{equation}
Here we have used the ideal equation of state to convert energy density 
to entropy density.  Since the entropy per gluon of an ideal gluon 
gas is 
$
3.6$, \eq{asymptotic} implies the asymptotic number of gluons per area  per rapidity
is more than a factor of two larger than the input number of gluons at $\tau_0$
\begin{equation}
\left. \frac{dN}{d^2\x_\perp d\eta}  \right|_{\tau\rightarrow\infty}  
=2.33\left. \frac{dN}{d^2\x_\perp d\eta}  \right|_{\tau=\tau_0} .
\end{equation}
At $Q_s \tau = \{10, 20\}$ the entropy and gluon multiplicity have reached only \{72, 82\}\% of their asymptotic values, corresponding to  gluon multiplication factors of $\{1.6,1.9\}$ respectively.  

Finally, let us make phenomenological contact with more complete hydrodynamic simulations of
heavy ion collisions, and estimate the saturation momentum required by 
phenomenology.
The initial entropy in hydrodynamics is normally adjusted to 
reproduce the mean multiplicity. Using the computer code from one such 
hydrodynamic simulation at the LHC~\cite{Mazeliauskas:2015vea},  we computed 
the average  entropy per area at the hydrodynamic initialization 
time\footnote{Specifically, for the event-by-event 
hydro code described in ref.~\cite{Mazeliauskas:2015vea} we first 
    created a smooth entropy density profile, $\overline {s(\x_\perp)} $, by
    averaging over events for a 0-5\% centrality class, $b=[0,3.3]\,{\rm fm}$. 
    This average is shown in \Fig{ichydro}. We then computed a single averaged 
    entropy
density by averaging $\overline{ s(\x_\perp) }$ with 
$\overline{s(\x_\perp)}$ as a radial weight. } 
\begin{equation}
\llangle \ti s (\ti)  \rrangle  = 4.13\,{\rm GeV}^2.
\end{equation}
Entropy production during the subsequent hydrodynamic evolution
is small, approximately 15\%, and therefore this constant is approximately 
independent of the  initialization time. With \Eqs{finalentropy} and 
(\ref{asymptotic}),
setting $Q_s\simeq 1.4\,{\rm GeV}$ in the EKT will roughly reproduce the entropy
in hydrodynamic simulations provided the system is passed to hydrodynamics at 
$Q_s \tau=10$  where the entropy in the EKT has reached 70\% of its asymptotic 
value in \eq{finalentropy}. 

In \Fig{background}(b) we present the time evolution of the effective 
temperature (determined from the energy density and the equation of state)
in physical units for $Q_s \simeq 1.4\,{\rm GeV}$. Its time dependence at late 
times is well described by first order viscous hydrodynamics with the 
asymptotic value
\begin{equation}
\label{asympT}
\lim_{\tau \rightarrow 
\infty}\left(T+\frac{2}{3}\frac{\eta}{s\tau}\right)\tau^{1/3}= 0.763\, 
Q_s^{2/3}.
\end{equation}
The temperature at $Q_s\tau{=}10$ is $T\simeq 430\,{\rm MeV}$,
or close to three times the pseudo-critical temperature. At these
temperatures, modern weakly coupled techniques can be expected to
work reasonably, justifying our approximation scheme. Similarly,
 with this value of $Q_s$
an initialization time of $Q_s\tau\simeq 10$  corresponds to $\ti \simeq 
1.4\,{\rm fm}$.  
The elliptic and 
triangular flows in central events develop on a later time scale, of order 
$R/c_s \sim 8\,{\rm fm}$, and therefore
an initialization time of this order may be acceptable.
 A more complete study 
including fermions will be needed to definitively answer this question. 

\subsection{Evolution of the perturbations}
\begin{figure*}
    \begin{center}
\includegraphics[width=0.47\textwidth]{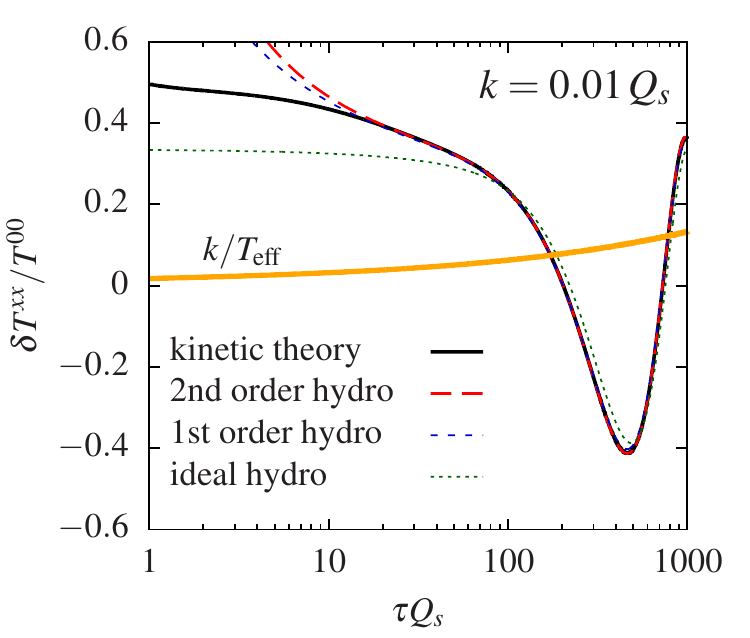}
\includegraphics[width=0.47\textwidth]{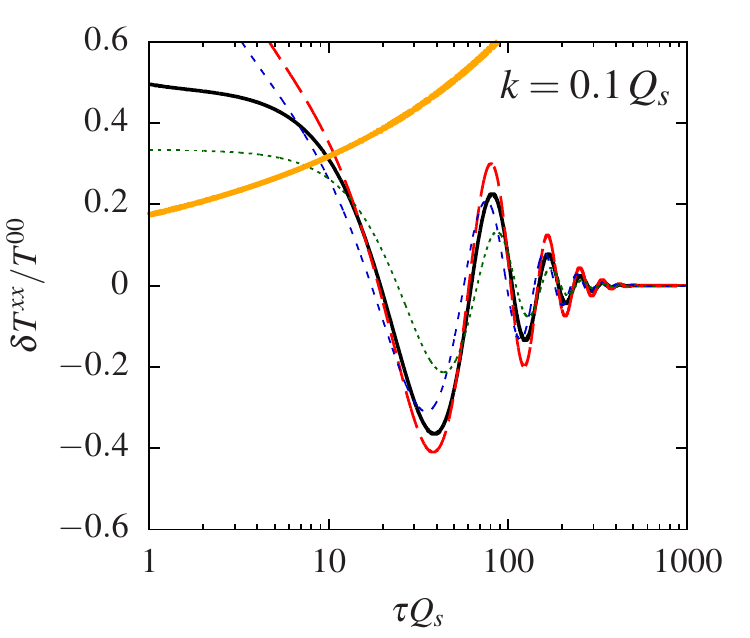}\\
\includegraphics[width=0.47\textwidth]{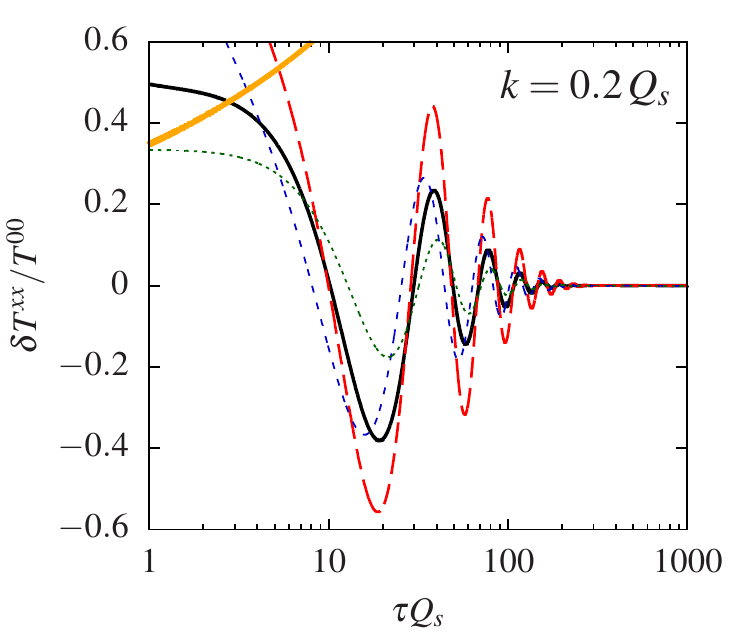}
\includegraphics[width=0.47\textwidth]{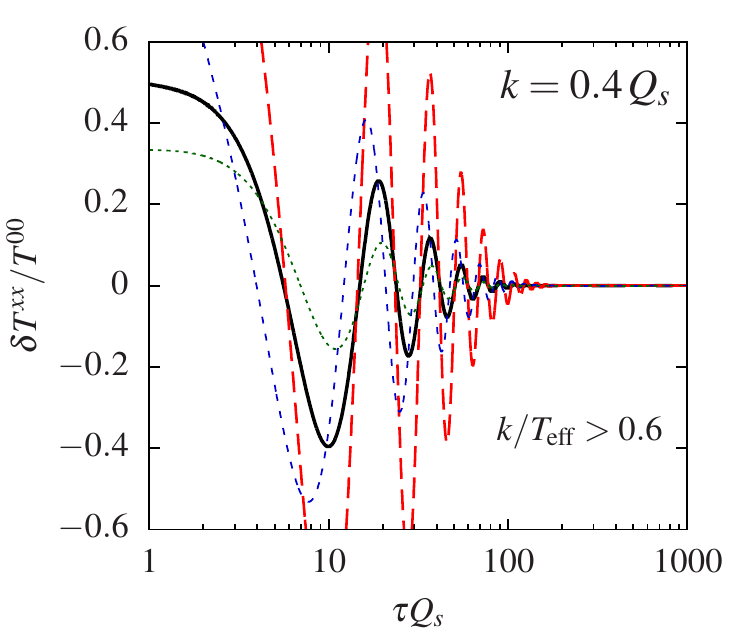}
\end{center}
\caption{$\delta T^{xx}/ T^{00}$ compared with hydrodynamic constitutive
equations (the curves have been normalized by the magnitude of
the initial perturbation $\delta T^{00}(\tau_0)/T^{00}(\tau_0)$). Long wavelengths with  $k\lesssim 0.1\,T$ are
described by the hydrodynamics at approximately the same time as the background
$Q_s \tau \sim 10$. Shorter wavelengths with $k\sim 0.4\,Q_s$ are never well
described by hydrodynamics.  }\label{figdTxx}
\end{figure*}

We now move on to describe the evolution of the linearized energy density 
perturbations on top
of the thermalizing non-equilibrium background. We start with the initial 
condition of \eq{initialdf} with different values of $k/Q_s$.
In figures~\ref{figdTxx}(a)-(d) we show $xx$\nobreakdash-component of 
perturbation energy 
momentum tensor $\delta T^{xx}$ compared with ideal, viscous, and second order 
hydrodynamic constitutive equations of \Eq{eqdTxx}.
The lines have been normalized by the background energy density $T^{00}(\tau)$, so that any observed damping 
is due to nontrivial dynamics associated to the spatial inhomogeneity. 

We consider fixed values of $k/Q_s$, which do not correspond to
fixed values of $k/T$, as the effective temperature is changing due to 
the expansion (see \Fig{figdTxx}). 
For  very large wavelengths with $k/Q_s = 0.01$, we  
observe  that while ideal constitutive equations have rather 
large corrections, these are well accounted for by the viscous and 2nd order 
equations, roughly at the same time as the background constitutive equation is 
satisfied, i.e. at times after $\tau \sim 10/Q_s$. Determining the temperature 
of 
the background from Landau matching condition $T^4 = \nu_g\frac{\pi^2 }{30} e$ 
(see \Fig{background} (right)), we find that at times $\tau = \{10, 20\}/Q_s$, 
the 
wavelength $k=0.01Q_s$ in units of temperature $T=\{  0.31,0.26 \}Q_s$ is 
$k/T(\tau) =  \{0.032, 0.039 \}$. 
As discussed in \Sect{kspace} and in \Fig{disprel} these values 
of $k/T$ are accurately described by second order hydrodynamics.

We see that even for larger $k/Q_s = \{0.1,0.2,0.4\}$ the hydrodynamic constitutive relations are approximately 
fulfilled when the background has hydrodynamized around $\tau \sim 10/Q_s$. However, larger values of $k/Q$ 
correspond to larger values of $k/T$, and even at late times there are corrections to the constitutive equations. While these 
corrections are moderate for $k/Q_s=0.2$ for which $k/T(\tau = 10/Q_s) \approx 0.6$, they remain $\mathcal{O}(1)$ for $k/Q_s = 0.4$.
It is therefore questionable whether it is justifiable to pass these short scales to hydrodynamic description at any time. 

\section{A Green function for hydrodynamics}
\label{Green}

We now move on to describe how the response to the linearized perturbations in 
EKT can 
be used in a hydrodynamic simulations to encapsulate the far-from-equilibrium 
dynamics of transverse perturbations
during the time scales between $\tau \sim 1/Q_s$ and $\{10,20\}/Q_s$.

\begin{figure*}
\includegraphics[width=0.48\linewidth]{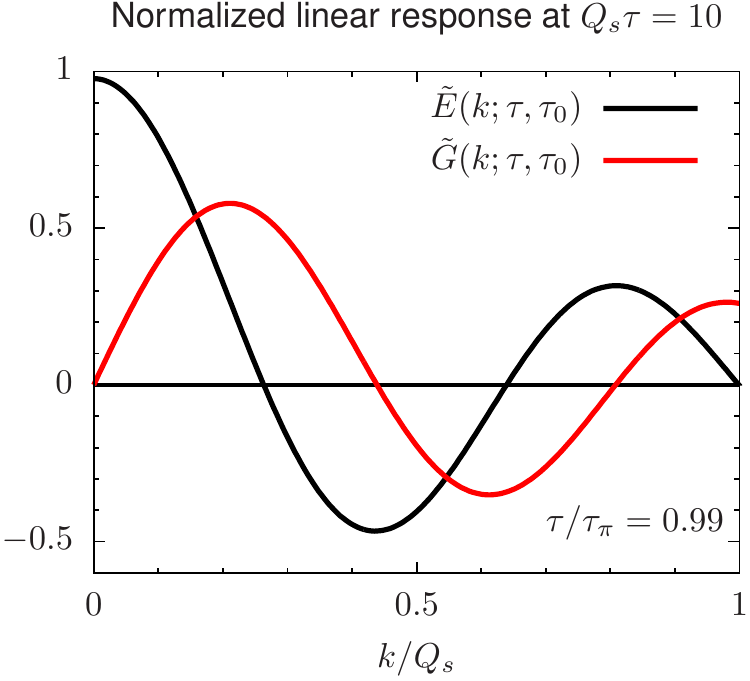}\quad
\includegraphics[width=0.48\linewidth]{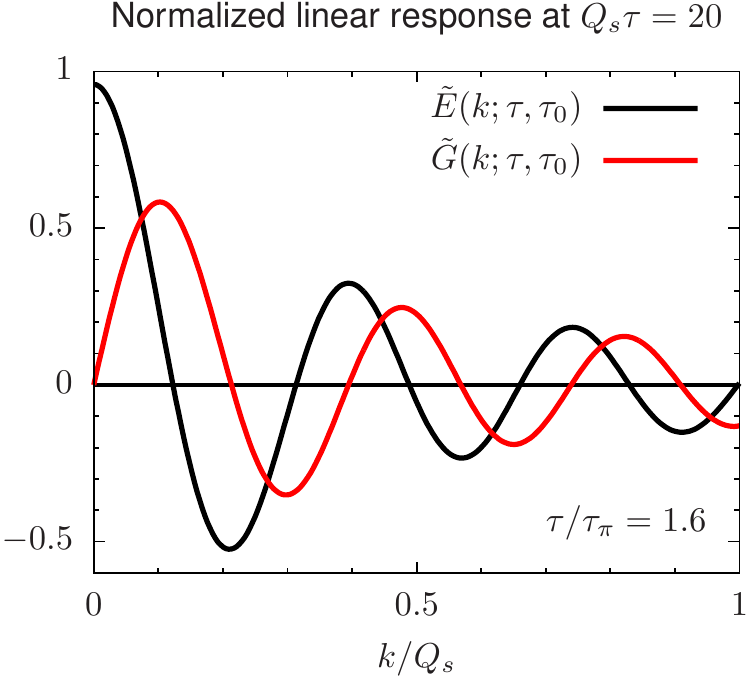}
\caption{Normalized linear response functions in $k$-domain [\Eqs{E} and (\ref{G})] for the initial energy perturbation (a) at $Q_s \tau=10$ and (b) at 
$Q_s \tau=20$ .\label{linear_response}}
\end{figure*}

In order to construct the initial state for hydrodynamics at $\ti$ from a given geometry at $\tau_0$,
the linear response 
of the components of $T^{\mu\nu}$ to the initial perturbation are needed. 
The constitutive relations reduce the number of independent components
of the energy momentum tensor, so it suffices to specify only $\delta T^{00}$ and $\delta T^{0x}$.
\Fig{linear_response}(a) displays the energy  
and momentum  response functions 
($\tilde{E}(k; 
\tau,\tau_0)$ and $\tilde{G}(k; \tau,\tau_0)$  respectively)
to 
an initial energy perturbation $\delta e(\tau_0,k)$ in $\k$-space
\begin{align}
\frac{\delta e(\tau,k)}{ e(\tau)} &\equiv \tilde{E}(k; \tau,\tau_0)\frac{\delta e(\tau_0,k)}{ 
e(\tau_0)},\label{E}\\
\frac{ g^x(\tau,k)}{ e(\tau) }&\equiv-i\tilde{G}(k; 
\tau,\tau_0)\frac{\delta 
e(\tau_0,k)}{ e(\tau_0) }.\label{G}
\end{align}
The results are presented at two suggested initial times $\ti = \{10,20\}/Q_s$.
We will analyze the response functions at asymptotically small $k$ and in coordinate space
in the next two subsections.

\subsection{The kinetic theory response at asymptotically small $k$}
\label{smallk}

The most important contribution to the flow arises from the 
average nuclear geometry, which is smooth on  spatial scales of order $c\ti$.
For this reason the flow due to the average geometry is determined by the $k\rightarrow 0$ limit of the response functions. This section will provide
an analytic understanding of this limit, i.e. the intercept of $\tilde{E}(k; 
\tau,\tau_0)$ and the slope of $\tilde{G}(k; \tau,\tau_0)$ in 
\Fig{linear_response}.

First, we will determine how the long wavelength energy perturbations in the transverse plane
change as a function of time.
Returning to the conservation 
equations, \eq{econs} and (\ref{decons}), and setting $k=0$, we have
\begin{align}
&\partial_\tau e(\tau)=-\frac{e(\tau)+
T^{zz}(\tau)}{\tau}, \label{econs2}\\
&\partial_\tau \delta e(\tau)=-\frac{\delta 
e(\tau) +  \delta T^{zz}(\tau)}{\tau}.\label{decons2} 
\end{align}
From these equations the fractional perturbations in the
transverse plane $\delta e/e$ remain constant in time 
in the free streaming limit (where $T^{zz}$ and $\delta T^{zz}$ are
zero), and in the hydrodynamic limit (where $T^{zz}$ and $\delta T^{zz}$ 
are one third  $e$ and $\delta e$). 
Outside
of these limits  $\delta e/e$ is not constant in time.

However, a constant of the motion at $k=0$ can be constructed 
whenever the hydrodynamic gradient expansion is applicable. Indeed, by dimensional analysis, an all order constitutive 
equation at $k=0$  must take the following form
\begin{equation}
T^{zz} = e f(e^{1/4}\tau),
\end{equation}
where $f(x)$ is an order one function and $\delta T^{zz}=\partial_e T^{zz}\delta e$.
Then straightforward steps show
that to all orders in the gradient expansion
\begin{equation}
    \label{detzz}
\lim_{k\rightarrow 0}\frac{\delta e(\tau,k)}{\,3 e- T^{zz}}=\text{const.}
\end{equation} 
For conformally invariant theories with $T^{xx}{=}T^{yy}$ this can be written as
\begin{equation}
    \label{detxx}
    \lim_{k\rightarrow 0}\frac{\delta e(\tau,k)}{e+T^{xx}}=\text{const.}
\end{equation} 

\begin{figure*}
\centering
\includegraphics[width=0.45\linewidth]{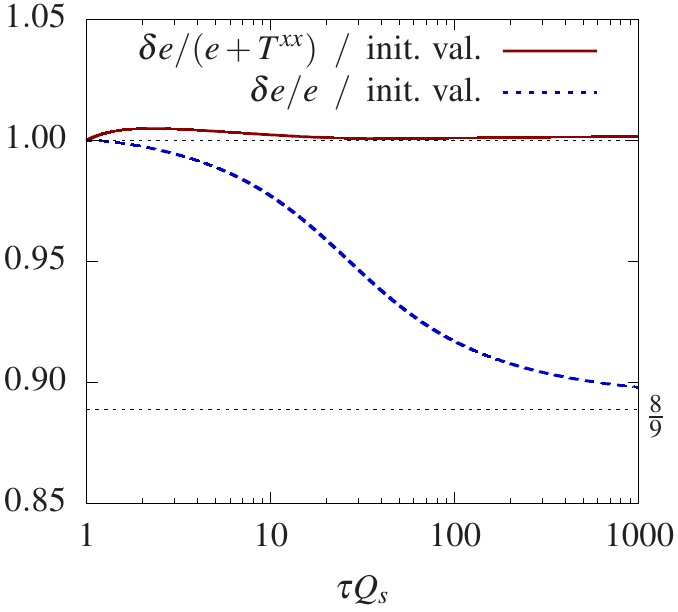}
\hspace{0.03\linewidth}
\includegraphics[width=0.46\linewidth]{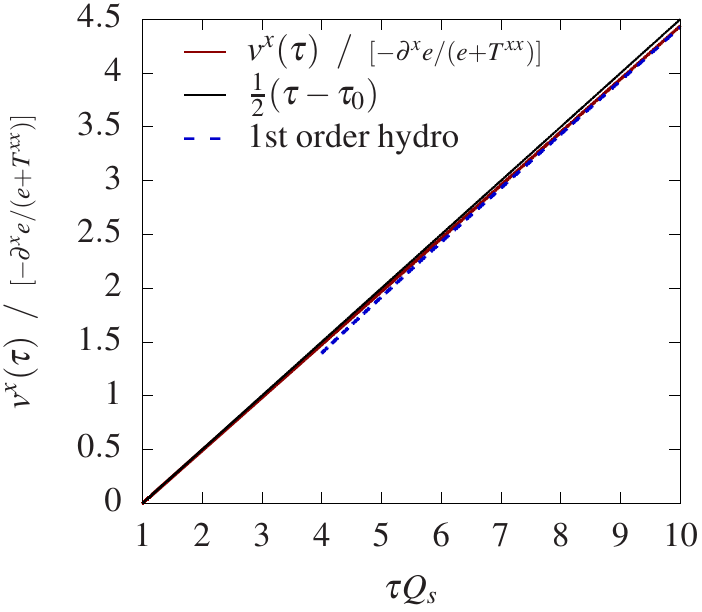}
\caption{ (a) Normalized energy perturbation versus time in the (asymptotically) small $k$ limit. $8/9$ is the 
change in $(e+T^{xx})/e$ between free streaming and ideal hydrodynamic limits (see \eq{eratio}).
(b) The velocity perturbation versus time in the (asymptotically) small $k$ limit scaled
by $-\partial^{x}e /(e+T^{xx})$ (see \eq{vresult2}). The result is compared to 
$\tfrac{1}{2} (\tau-\tau_0)$ (see also ref.~\cite{Vredevoogd:2008id}) and first 
order 
hydrodynamics.}
\label{fig:intde}
\end{figure*}

In \Fig{fig:intde}(a) we present the time evolution of $\delta e/e$ and $\delta 
e/(e + T^{xx})$ relative to their initial values. For our initial conditions 
$\delta e /(e + T^{xx})$ remains very nearly constant throughout the entire evolution.
Using this result, the
 change in $\delta e/e$ can be determined by the ratio of 
 $(e+T^{xx})/e$ at the initial and final times, when $T^{xx}$ is approximately $e/2$ and $e/3$ respectively. This reasoning leads to 
 an asymptotic relation between the initial and final energy perturbations
 \begin{equation}
 \label{eratio}
 \lim_{\tau \rightarrow \infty} \frac{\delta e(\tau)}{e(\tau)} = \frac{8}{9} \frac{\delta e(\tau_0) }{e(\tau_0)},
 \end{equation}
 which is shown in \Fig{fig:intde}(a).

Next, we will determine the velocity at (asymptotically) small $k$ as
a function of time from the pre-thermal evolution.
From the conservation equations for perturbations, \Eqs{decons} and 
(\ref{dgcons}),
the  momentum perturbations at small $k$ satisfy
\begin{equation}
    \label{grelation}
\partial_\tau \left(\frac{\tau g^x}{ik} + \frac{1}{2}\delta e 
\tau^2\right)=  
-\frac{\tau}{2} \left(2\delta T^{xx} + \delta T^{zz} -\delta T^{00}\right).
\end{equation}
For conformal theories with $\delta T^{xx}{=}\delta T^{yy}$ the right 
hand side of \eq{grelation} is zero and 
\begin{equation}
\label{momentumeq}
\frac{\tau g^x}{ik} + \frac{1}{2}\delta e 
\tau^2= \mbox{const}.
\end{equation}
At late times and in coordinate
space this condition reads 
\begin{equation}
\frac{T^{0x}(\tau)}{T^{00}(\tau)}=-\frac{1}{2}\tau\frac{\partial_x 
T^{00}(\tau)}{T^{00}(\tau)} \, ,
\end{equation} 
which was first noted in~\cite{Vredevoogd:2008id}. Here 
we have shown that this relation is a consequence of conformal symmetry (see also \cite{Vredevoogd:2008id}) and the small $k$ limit.

Using \eq{momentumeq} and the definition $g^{x}=(e+T^{xx})v^{x}$, the velocity as a function of time is given by
\begin{equation}
    \label{vresult1}
    \frac{v^x}{ik} = -\frac{\tau}{2} \frac{\delta e}{e + T^{xx} }
\, \left(1-\frac{\delta e(\tau_0)\tau_0^2}{\delta e(\tau)\tau^2}\right).
\end{equation}
Thus, after a brief transient period of order $\tau_{0}$, the velocity 
is directly proportional to time
\begin{equation}
\label{vresult2}
v^{x} =\frac{\tau}{2} \, \left(\frac{-\partial^x e(\tau, \x)}{e(\tau) + 
T^{xx}(\tau)}\right)\,,  \qquad \frac{-\partial^x e(\tau,\x)}{e(\tau) + 
T^{xx}(\tau)} = {\rm const}.
\end{equation}
In \Fig{fig:intde}(b) we compare the growth of the velocity with time
given by \eq{vresult1} with a simple estimate based on \eq{vresult2}.
The simple estimate does a remarkably good job for all times.

\subsection{Response in coordinate space}
\label{response}
To construct the initial conditions for hydrodynamics with 
the correct prethermal evolution, we determine the 
Green functions $E(|\x|;\tau,\tau_0)$ and $G(|\x|;\tau,\tau_0)$ which convert 
the initial  profile of energy perturbations $\delta e(\tau_0,\x)$ to the required 
energy and momentum fluctuations at thermalization time
\begin{subequations}
    \label{kerneleqs}
\begin{align}
\frac{\delta e(\tau,\x)}{e(\tau)} &= 
\int 
d^2\x'
\frac{\delta 
e(\tau_0,\x')}{ e(\tau_0) }
E( 
|\x-\x'|; 
\tau,\tau_0),\\
\frac{ g^i(\tau,\x)}{e(\tau)} &= \int 
d^2\x'
 \frac{\delta 
e(\tau_0,\x')}{ e(\tau_0)}
\frac{(\x-\x')^i}{|\x-\x'|}G( |\x-\x'|; 
\tau,\tau_0).
\end{align}
\end{subequations}
Currently hydrodynamic simulations often smooth the initial conditions
before
starting the hydrodynamic evolution
by convolving the energy density with a Gaussian%
\footnote{See ref.~\cite{Noronha-Hostler:2015coa} for a current
discussion of the observables that are influenced by this arbitrary 
regulator.}.  In contrast, \Eq{kerneleqs}  
smooths the initial conditions and generates pre-flow in a physical way, and 
provides
an attractive alternative to this ad hoc procedure.
\begin{figure*}
\centering
\includegraphics[width=0.47\linewidth]{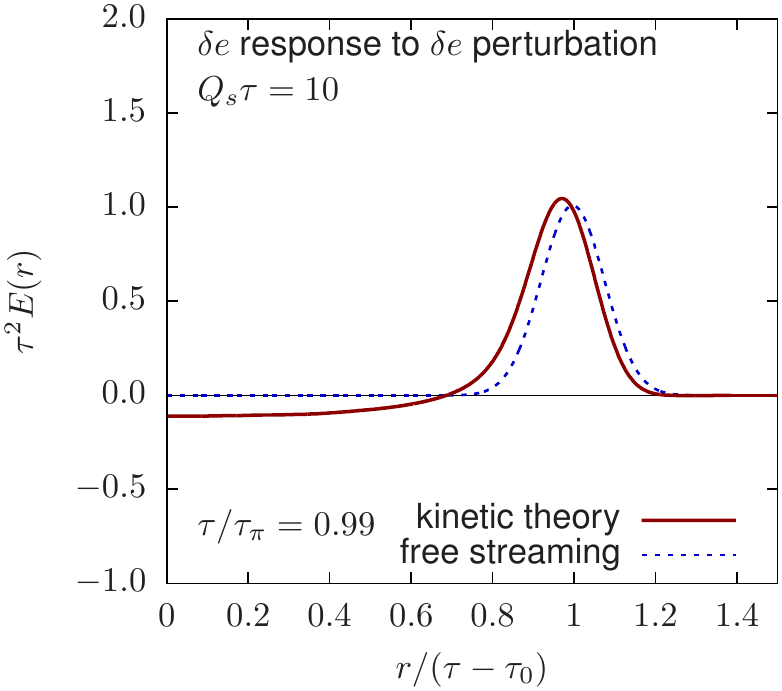}
\includegraphics[width=0.47\linewidth]{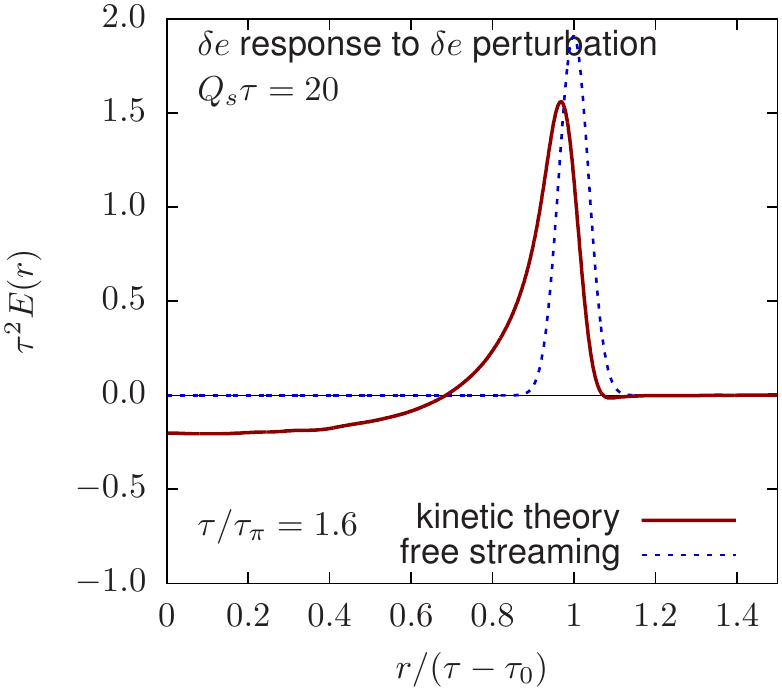}
\includegraphics[width=0.47\linewidth]{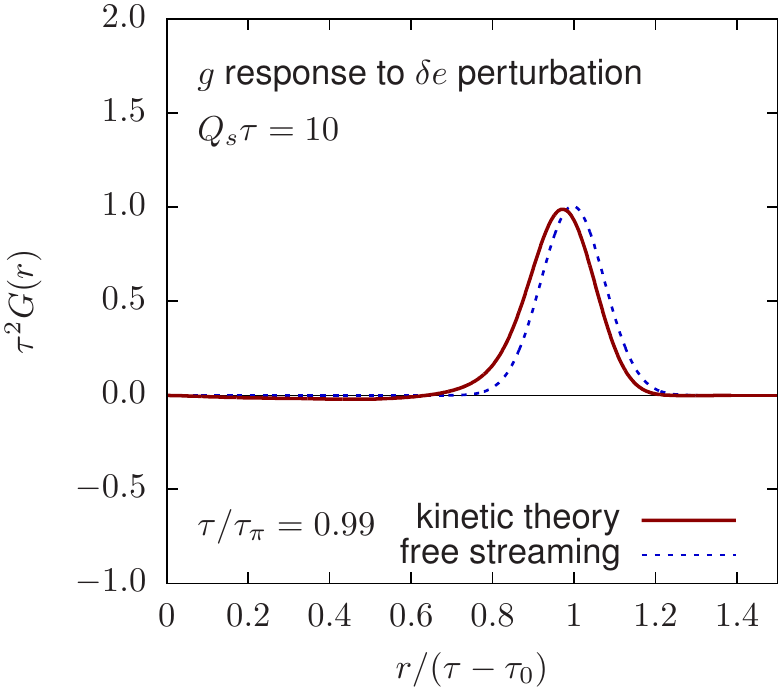}
\includegraphics[width=0.47\linewidth]{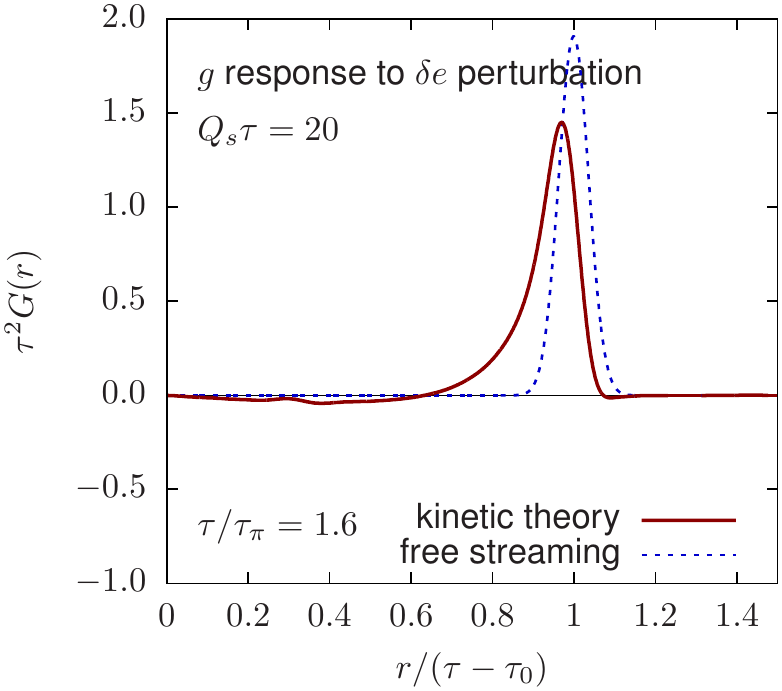}
\caption{(top) Energy and  (bottom) momentum Green functions,     
\Eq{kerneleqs}, for initial \emph{energy perturbation} in coordinate space at 
(left) $Q_s\tau=10$  and (right) $Q_s \tau=20$. 
}
\label{fig:kernelv0}
\end{figure*}

\begin{figure*}
\centering
\includegraphics[width=0.48\linewidth]{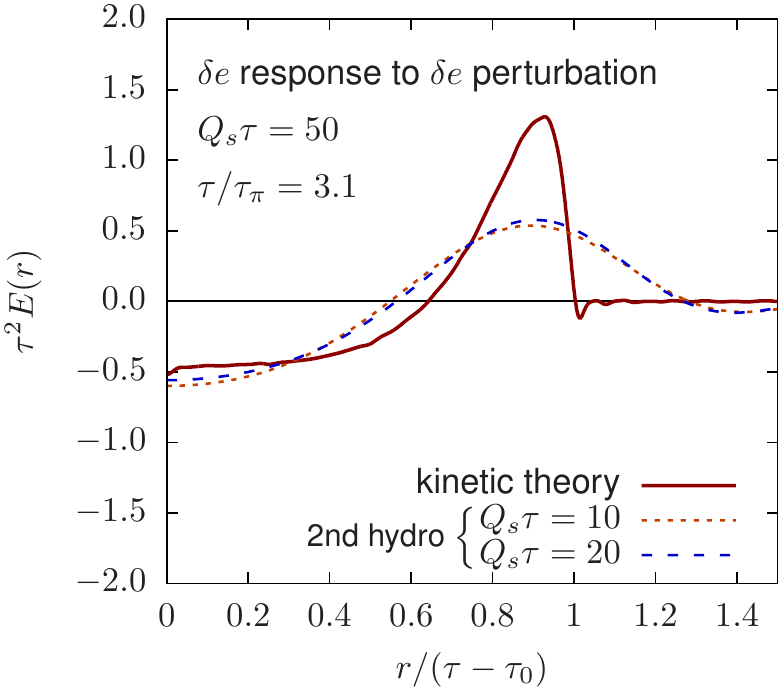}\quad
\includegraphics[width=0.47\linewidth]{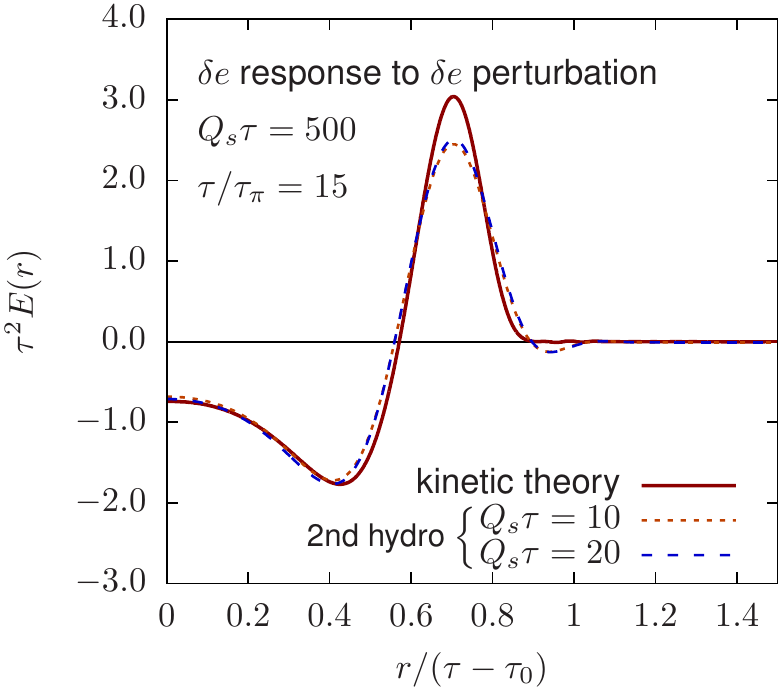}
\caption{Energy Green functions for initial energy perturbations in coordinate space at late times (a) $Q_s \tau=50$ and (b) $Q_s \tau=500$. 
    The results are compared to linearized second order hydrodynamics (\Eqs{eq:conservation} and (\ref{eq:constit})), with the initial
conditions obtained from kinetic theory at $Q_s\tau{=}10$ and $Q_s \tau{=}20$ 
(see \Fig{linear_response}). }
\label{fig:kernelv0_late}
\end{figure*}

The EKT is applicable for distance scales  that are larger than the 
 Compton wavelength of the particles $\sim 1/Q_s$. This limits the accuracy of the 
 Green function in spatial domain that can be reached in a computation based on kinetic theory. 
 In order to fold this uncertainty into our result, we regulate our Green function  by convoluting with a Gaussian weight,  $e^{-r^2/2\sigma^2}/(2\pi\sigma^2)$, with $r=|\x|$, and with a 
 width of the order of the initial Compton wavelength $\sigma Q_s = 
 0.7$.
In momentum space this corresponds to suppressing the large $k$ contributions 
by an exponential
envelope $\exp(-  \sigma^2 k^2 / 2)$
\begin{align}
E(|\x|; \tau,\tau_0) &= \int \frac{d^2 \k}{(2\pi)^2} e^{i \k\cdot \x} 
e^{-\sigma^2 k^2/2} E(|\k| ; \tau,\tau_0), \label{ft1} \\
G(|\x|; \tau,\tau_0) &= \int \frac{d^2 \k}{(2\pi)^2}(-i 
\hat{k}\cdot\hat{x}) e^{i 
\k\cdot \x} 
e^{-\sigma^2k^2/2} G(|\k| ; \tau,\tau_0) \label{ft2}.
\end{align}
The regulated Green functions  are shown in \Fig{fig:kernelv0} at the  
initialization times $\ti Q_s=\{10,20\}$
(for details of the Fourier transform see \app{kernelFT}.) At $\tau Q_s=10$ the 
system has spent a significant proportion of the total evolution time with small
longitudinal pressure
$T^{zz}\approx 0$, and therefore the resulting response is similar to the free 
streaming prediction (see \app{kernelFT}). However, the Green function  in 
\Fig{fig:kernelv0}(a) 
is peaked for $r{<}c|\tau-\tau_0|$,  suggesting  a slight deflection 
from the free streaming trajectory.
Additionally,  
the energy perturbation is negative at small $r$,  which is indicative 
of a nascent approach to hydrodynamics. At later times, such as $Q_s\tau=20$ in 
\Fig{fig:kernelv0}(b), 
these differences become more pronounced. 
Similar features are visible in the momentum response to an initial energy 
perturbation shown \Fig{fig:kernelv0}(c) and (d).
Finally, in \Fig{fig:kernelv0_late} we 
show 
the Green functions at later times $Q_s\tau=50$ and $Q_s\tau=500$, 
and compare to linearized  second order hydrodynamics (\Eqs{eq:conservation} and (\ref{eq:constit})) with initial conditions
taken from the $Q_s\tau=\{10,20\}$ results.  Between $Q_s\tau{=}20$ and $Q_s\tau{=}50$, the hydrodynamics overdamps the high $k$ modes  (see also \cite{Romatschke:2015gic}), and the response
is broader than the predictions of kinetic theory. However,
these Green functions will be convolved with the initial conditions,  
and thus the resulting hydrodynamic initial state is mostly sensitive to the first moments of
these kernels. The moments of the EKT and hydro kernels are
determined by the small $k$ behaviour of the response functions, which
agree to a few percent (not shown). At later times $Q_s\tau=500$, the
response is largely determined by Fourier modes in the hydrodynamic 
regime $k \lesssim 0.1\,Q_s$, and the EKT and hydro kernels are visually similar.

To summarize, the hydrodynamic evolution sets in early at rather large
anisotropies, and the hydrodynamic constitutive equations are approximately satisfied as soon as the
$T^{zz}$ starts to significantly deviate from the free streaming expectation,
$T^{zz}\approx 0$. For this reason 
the time interval when the evolution is not described by 
free streaming or hydrodynamics is comparatively brief (see 
\Fig{background}(a)),
and as hydrodynamics becomes marginally applicable at $Q_s\ti \sim 10$, the Green function closely resembles the free streaming result.
Therefore, an approach where the evolution is described by free streaming until $\ti$ 
seems well motivated \cite{Broniowski:2008qk,Liu:2015nwa}, provided that the correct value of $\ti$ is used. However, such  
an ad hoc approach does not account for some of the qualitative details of the Green function, such as the depletion of the energy density in the interior region.

\section{Discussion}
\label{discuss}
In this paper, we have provided a bridge between the far-from-equilibrium
initial conditions of heavy-ion collisions and hydrodynamized plasma. Our main
result is the coordinate space Green functions (see \Eq{kerneleqs} and
\Fig{fig:kernelv0}), which can be used to filter the pre-equilibrium
energy density to find the full energy-momentum tensor for hydrodynamics at the
initialization time.  The procedure can be implemented in complete hydrodynamic
simulations, removing one source of uncertainty.  Perhaps more importantly, the
approximations in the EKT are compatible with the IP-Glasma setup, and thus the
whole evolution from saturated nuclei to hydrodynamics can be comprehensively
modelled within a perturbatively controlled framework.

We provide the coordinate space Green functions at two different suggested initialization times, $Q_s \ti=\{10,20\}$. 
At the earlier initialization time, $Q_s\ti=10$,
there are  significant (though bearable) corrections 
to the constitutive relations due to non-hydrodynamic modes (see 
\Fig{background}(a) and \Fig{figdTxx}). By $Q_s\ti=20$
the constitutive relations at small $k$ are well satisfied, and the subsequent
evolution is reasonably
captured by  second order hydrodynamics%
\footnote{When examining \Fig{fig:kernelv0_late}, one must remember that the 
full Green functions
will be convolved with the initial conditions, and thus the response
of the system is mostly sensitive to the first moments of the  
kernels in \Fig{fig:kernelv0_late}.
The first EKT moments (i.e. the small $k$ behavior of the response) agree with the hydrodynamics to the percent level. }
(see \Fig{fig:kernelv0_late}).
The approximate overlap of the two 2nd order viscous lines in 
\Fig{fig:kernelv0_late}, which correspond to initializing the hydro at $\ti 
Q_s= \{10,20\}$,
demonstrates that the subsequent hydrodynamical evolution is indeed rather insensitive to the initialization time. 
In \Sect{Green} we examined the (asymptotically) small $k$ 
limit of the Green functions, and confirmed (and clarified)
a preflow estimate by Vredevoogd and Pratt~\cite{Vredevoogd:2008id} (see 
\Fig{fig:intde}).

The hydrodynamics that the EKT follows is characterized by the weak coupling
value of $\eta/s \approx 0.62$~\cite{Arnold:2003zc},  which is
significantly higher than the AdS/CFT result $\eta/s\simeq 0.08$~\cite{Policastro:2001yc},  and
current phenomenological estimates, which assume that
$\eta/s$ is independent of the temperature.
Recent analyses have relaxed the temperature independence of $\eta/s$, and
shown that 
the value of $\eta/s$ at higher temperatures $T \sim 3 T_c$ is poorly
constrained by data~\cite{Niemi:2015qia}. Since it is the high temperature regime that is most relevant for the transition to hydrodynamics,  we believe
that the current kinetic theory results for the initial stages can be
consistent with hydro phenomenology, provided $\eta/s$ decreases towards the
strong coupling result as the system cools towards $T_c$.

Nevertheless, to apply our results to a hydro simulation with  lower viscosity than
the perturbative expectation, we note that the two  initialization
times $\ti Q_s = \{10,20\}$ correspond in units of the hydro parameters  to 
$\ti = \{.99,1.6\} \tau_\pi$ or 
$T_\text{init}\tau_\text{init}/(4\pi\eta/s)=\{0.4,0.7\}$.
The scaled times $\tau/\tau_\pi$ or $\tau T/(4\pi\eta/s)$ can be used 
to initialize simulations when the transport coefficients differ.
Such an approach is supported by the reasonably good
scaling properties of the hydrodynamization times and prethermal evolution as
function of the coupling constant when expressed in terms of the hydrodynamic
variables~\cite{Kurkela:2015qoa,Keegan:2015avk}. 

Although our EKT description can be further improved by inclusion of fermionic
degrees of freedom and by improving the connection to the early classical
evolution, we believe that it already provides a physically sound picture of
the approach to hydrodynamics and can be used to initialize all components of
the energy-momentum tensor for subsequent hydrodynamic evolution.
This eliminates a source of uncertainty in current simulations, and provides a 
satisfyingly complete
description of the early time evolution in heavy ion collisions.

 Microscopic initial state models 
like IP-Glasma also produce a 
non zero initial transverse flow at early times $\tau_o\sim1/Q_s$, which should 
be  propagated by EKT response functions analogous to \Eq{kerneleqs}. Then in 
linear response theory the energy and momentum flow at hydrodynamization time 
$\tau_\text{init}$ is a linear combination of contributions from initial energy 
and momentum perturbations. The demonstration of a smooth connection between 
initial conditions and hydrodynamics with kinetic theory pre-equilibrium in 
a realistic heavy 
ion simulation will appear in the follow up publication.

\begin{subappendices}

\section{Collision kernel}
\label{colker}
In this appendix we provide additional details on the collision
kernels used in  \eq{boltz}. The collision kernel for the uniform 
background 
contains terms arising  from 
elastic $2 \leftrightarrow 2$ scatterings and inelastic $1\leftrightarrow 2$ 
collinear splittings
\begin{equation}
\mathcal{C}[f] = \mathcal{C}_{2 \leftrightarrow 2}[f] + \mathcal{C}_{1 \leftrightarrow 2}[f].
\end{equation}
The two collision terms read\,~\cite{Arnold:2002zm, 
Kurkela:2015qoa,Keegan:2015avk}
\begin{align}
\label{2to2}
 \mathcal{C}_{2\leftrightarrow 2}[f](\p)&= \frac{1}{4|\p| \nu_g}\int \frac{d^3 k}{2k (2\pi)^3}\frac{d^3 p'}{2p' (2\pi)^3}\frac{d^3 k'}{2k' (2\pi)^3}|\mathcal{M}(\p,\k;\p',\k')|^2 (2\pi)^4 \delta^{(4)}(P+K-P'-K') \nonumber \\
 &\times \big\{ f_\p f_\k[1+ f_{\p'}][1+ f_{\k'}]-f_{\p'}f_{\k'}[1+ f_{\p}][1+ 
 f_{\k}] \big\}
\end{align}
and
\begin{align}
\label{1to2}
\mathcal{C}^{1\leftrightarrow 2}[f](\p) &= \frac{(2\pi)^3}{2|\p|^2 
\nu_g}\int_0^{\infty} dp' dk' \, \delta(|\p|-p' 
-k')\gamma(\p;p'\hat{\p},k'\hat{\p})\nonumber 
\times
\big\{f_\p[1+ f_{p' \hat{\p}}][1+ f_{k' \hat{\p}}] - f_{p' \hat{\p}}f_{k' \hat{\p}}[1+ f_{\p}]\Big\}\nonumber\\
&+ \frac{(2\pi)^3}{|\p|^2 \nu_g}\int_0^{\infty} dp' dk \, \delta(|\p|+k 
-p')\gamma(p' \hat{\p};\p,k \hat{\p})
\times
\big\{ f_{\p}f_{k \hat{\p}}[1+ f_{p' \hat{\p}}] -f_{p' \hat{\p}}[1+ f_\p][1+ f_{k\hat{\p}}]\Big\},
\end{align}
where $\hat{\p}$ is the unit vector parallel to $\p$, and capital letters 
denote null 4-vectors, i.e. $P^0 \equiv |\p|$. The effective elastic 
$|\mathcal{M}|^2$ and inelastic $\gamma$ scattering matrix elements contain 
non-trivial structures arising from the soft and collinear divergences, which  
are dynamically regulated by the in-medium physics. 

For the most 
of kinematics the effective elastic scattering element is given 
by\footnote{\label{fnote}  Equations \Eq{c2to2} and 
\Eq{spliting} have some minor typos corrected compared to 
refs.~\cite{Kurkela:2015qoa,Keegan:2015avk}.}
\begin{align}
|\mathcal{M}|^2 = 2 \lambda^2 \nu_g \left( 9 + \frac{(s-t)^2}{u^2}+ 
\frac{(u-s)^2}{t^2}+ \frac{(t-u)^2}{s^2}\right).\label{c2to2}
\end{align}
For a soft gluon exchange  with the momentum 
transfer $q = |\p' - \p|$  in $t$-channel (or $q = |\p' - \k|$ in $u$-channel)
the collision 
matrix is proportional to 
$\propto 1/(q^2)^2$, and thus suffers from a soft Coulomb divergence.  It is 
regulated by replacing 
\begin{equation}
q^2 t \rightarrow (q^2 + 2 \xi_0^2 m^2)t,
\end{equation}
in the denominators of divergent terms (similarly for the $u$-channel). Here 
$m^2$ is the thermal asymptotic 
mass of the gluon defined as
\begin{equation}
m^2 = 2 \lambda \int \frac{d^3 p}{(2\pi)^3}\frac{f_\p}{|\p|}.
\end{equation}
The coefficient $\xi_0= e^{5/6}/\sqrt{8}$ is fixed so that the matrix element 
reproduces the 
drag and momentum diffusion properties of soft scattering at leading order for isotropic distributions $f_{\p}$ \cite{York:2014wja}.  

The effective splitting kernel reads 
\begin{align}
\gamma(p {\bf \hat p}; p' {\bf \hat p},k' {\bf \hat p}) = \frac{p^4 + 
p'^4+k'^4}{p^3 p'^3 k'^3 }\frac{\nu_g \lambda}{8 (2\pi)^4}\int \frac{d^2 
h}{(2\pi)^2} 2 {\bf h} \cdot {\rm Re} {\bf F},\label{c1to2}
\end{align} 
where the equation for ${\bf F}$ accounts for splitting due to multiple 
scatterings with transverse momentum exchange ${\bf} q$, and momentum 
non-collinearity
\begin{align}
2 {\bf h} = &  i \delta E({\bf h} ){\bf F}({\bf h})   +\frac{\lambda 
T_*}{2}\int \frac{d^2  q_\perp}{(2\pi)^2} \mathcal{A}({\bf 
q}_\perp)\label{spliting}  \\ & 
\times \Big[ 3{\bf F({\bf h})}-{\bf F}({\bf h}-p' {\bf q}_\perp)-{\bf F}({\bf 
h}-k' {\bf q}_\perp)-{\bf F}({{\bf h}+p {\bf q}_\perp})\Big]\nonumber.
\end{align}
with
$
T_* = \frac{\lambda}{m^2}\int \frac{d^3 p}{(2\pi)^3}f_\p(1+f_\p) ,
$
and $\delta E = m^2/2p'+m^2/2k'-m^2/2p + {\bf h}^2/2p k' p'$. In 
the isotropic screening approximation
\begin{equation}
\mathcal{A}({\bf q}_\perp) = \left( \frac{1}{{\bf q}_\perp^2} - \frac{1}{{\bf 
q}_\perp^2 + 2m^2}\right).
\end{equation}
Both $m^2$ and $T_*$ are self-consistently evaluated at 
each time step.

The linearized collision kernels are obtained trivially  by replacing $f 
\rightarrow \bar f + \delta f$ in the integrands of \ref{2to2} and  
\ref{1to2}  and 
linearizing in $\delta f$. In addition one has to take into account the 
linear variation of the thermal mass $\delta m^2 $ and the effective 
temperature 
$\delta T_*$ in the scattering matrix elements \Eqs{c2to2} and (\ref{c1to2})
\begin{align}
\delta m^2 &= 2 \lambda \int \frac{d^3 p}{(2\pi)^3} \frac{\delta f_\p}{|\p|},\\
\delta T_*  &= \frac{\lambda}{m^2} \int \frac{d^3 p}{(2\pi)^3}  \delta f_\p 
(1+ 2 f_\p) -\frac{\delta m^2}{m^2} T_*.
\end{align}
where $m^2$ and $T_* $ are evaluated from the unperturbed  background 
distribution.

\section{Fourier transform of Green functions}
\label{kernelFT}
Here we provide details of performing Fourier transforms in \Eqs{ft1} and (\ref{ft2}) to obtain spatial Green  functions shown in 
\Figs{fig:kernelv0} and \ref{fig:kernelv0_late}. The two dimensional Fourier 
transforms can be straightforwardly reduced to 
one dimensional
Hankel transforms
\begin{align}
E( |\x|; \tau,\tau_0) &= \int_0^\infty \frac{dk}{2\pi} k \tilde{E}(k; 
\tau,\tau_0) 
e^{-\sigma^2 k^2/2}
J_0(k 
|\x|),\label{H1}\\
G(|\x|; \tau,\tau_0) &= \int_0^\infty \frac{dk}{2\pi} k  \tilde{G}(k; 
\tau,\tau_0) 
e^{-\sigma^2 k^2/2}\label{H2}
J_1(k |\x|).
\end{align}
Integrals in \Eqs{H1} and (\ref{H2})  were done numerically by using cubic 
interpolation for $\tilde{E}$ 
and $\tilde{G}$ within the available range of wavenumbers $k\in [0,4]Q_s$ 
To avoid the oscillatory behaviour due to a sharp $k$ cut-off at $k=4Q_s$, we 
extrapolated the Green functions until the Gaussian envelope
$e^{-\sigma^2 k^2/2}$ 
smoothly cuts off the integral. 
 For extrapolation at large $k$ we used functional forms motivated by free streaming results: $C_0 J_{0}(v_0|k|(\tau-\tau_0))$ and $C_1 J_{1}(v_1|k|(\tau-\tau_0))$, where coefficients 
$C_i$ and $v_i$ were fitted to match the oscillatory  behaviour of Green functions at the largest available $k$.
For $Q_s\tau=\{10,20,50\}$ we used $Q_s \sigma = 0.7$ for the envelope 
corresponding roughly to the smallest scales the EKT can resolve. 
 For $Q_s \tau=500$, perturbations with large wavenumbers were sufficiently 
 suppressed by EKT evolution that no extrapolation was 
 necessary.

For early times and large values of $k$ the collision terms in the Boltzmann 
equation \eq{boltz} can be neglected and the system is freely streaming. For 
particle distributions that are highly anisotropic in $z$ direction 
($\mathcal{P}_L\ll \mathcal{P}_T$), but isotropic in $xy$-plane, energy 
perturbations are propagating in circular wavefronts at the velocity $v$ of constituent particles  (for massless gluons $v=c$). In such free streaming evolution energy perturbations at time $\tau$ and position $\x$ are equal to the average of energy perturbations at $\tau_0$ on a circle
$|\x-\x'|=c|\tau-\tau_0|$\cite{Liu:2015nwa}. Thus, free streaming Green functions in coordinate space are
\begin{equation}
E( |\x|; \tau,\tau_0)=G( |\x|; \tau,\tau_0) = \frac{1}{2\pi 
|\x|}\delta(|\tau-\tau_0|-|\x|).
\end{equation}
Free streaming Green functions shown in \Fig{fig:kernelv0} 
were also folded in with a Gaussian regulator as 
discussed  above.

\end{subappendices} 

\newcommand{\dlangle}{\langle\!\langle}
\newcommand{\drangle}{\rangle\!\rangle}
\newcommand{\ihat}{{\hat{\imath}}}
\newcommand{\jhat}{{\hat{\jmath}}}
\def\half{\tfrac{1}{2}}
\def\third{\tfrac{1}{3}}
\def\xp{{\vec{x}_\perp}}
\def\yp{{\vec{y}_\perp}}

\def\A{{\mathcal A}}
\def\kp{{\vec{k}_\perp}}
\def\chit{{\chi_{\tau_0}^{gg}}}
\def\xh{{\hat{x}}}
\def\yh{{\hat{y}}}
\def\zh{{\hat{z}}}
\def\ih{{\hat{\imath}}}
\def\jh{{\hat{\jmath}}}
\def\kh{{\hat{k}}}
\def\bkap{{\bm \kappa}}
\def\Refs#1{Refs.~\cite{#1}}
\chapter{Non-linear noise corrections in Bjorken expansion}
\label{chap:noise}

The following sections of this chapter are near verbatim reproduction of 
\begin{itemize}
\item Y. Akamatsu, A. Mazeliauskas and D. Teaney, \emph{A kinetic regime of 
hydrodynamic fluctuations and long time tails for a Bjorken expansion}
Phys. Rev. C95, 014909 (2017)~\cite{Akamatsu:2016llw}. 
Copyright (2017) 
by the American Physical Society
\end{itemize}

\section{Introduction}

\subsection{Overview}
The purpose of the current paper is to develop a set of kinetic equations for
hydrodynamic fluctuations, and to use these kinetic equations to study
corrections to Bjorken flow arising from thermal fluctuations.  The specific
test case of Bjorken flow (which is a hydrodynamic model for the longitudinal
expansion of a nucleus-nucleus collision~\cite{Bjorken:1982qr}) is motivated by
the 
experimental 
program of ultra-relativistic heavy-ion collisions at RHIC and the LHC. 
Detailed  measurements of two particle correlation
functions  have provided overwhelming evidence that the evolution of the excited
nuclear material is remarkably well described by the hydrodynamics of the Quark 
Gluon Plasma (QGP) with a small shear viscosity to entropy ratio of order 
$\eta/s \sim 2/4\pi$~\cite{Heinz:2013th,Luzum:2013yya}.
The typical relaxation times of the plasma, while short enough to support
hydrodynamics, are not vastly smaller than the inverse expansion rates of the 
collision.  For
this reason
the gradient expansion underlying the hydrodynamic formalism has been
extended to include first and second order viscous 
corrections~\cite{Baier:2007ix}, and 
these corrections systematically improve the agreement
between hydrodynamic simulations and measured two particle 
correlations~\cite{Heinz:2013th}.
Additional corrections, which have not been systematically included, arise 
from thermal fluctuations of the local energy and momentum densities and could  
be 
significant in  nucleus-nucleus collision where only $\sim 20000$ particles are 
produced. 
This has prompted a keen practical interest in the heavy ion community in
simulating relativistic hydrodynamics with stochastic
noise~\cite{Gavin:2006xd,Kapusta:2011gt,Yan:2015lfa,Young:2014pka,Kapusta:2012zb,Murase:2016rhl,Nagai:2016wyx}.
In a non-relativistic context such simulations have reached a fairly mature
state~\cite{bell2007numerical,donev2011diffusive,balboa2012staggered}. 
For a static fluid, thermal fluctuations 
give rise through the nonlinearities of the equations of 
motion to
fractional powers in the fluid response function
at small frequency, $G_{R}(\omega) \propto
\omega^{3/2}$.  Indeed,
the ``long-time tails" 
first observed in molecular-dynamics 
simulations~\cite{velocity_auto,wainwright2,bixon_zwanzig}  are a 
consequence of this non-analytic $\omega^{3/2}$ behavior.
For Bjorken flow, the same nonlinear stochastic physics leads to
fractional powers in the gradient expansion for the longitudinal pressure of
the fluid. One of the goals of this manuscript is to compute the
coefficient of the first fractional power in this expansion.

The measured two particle
correlations in heavy ion collisions reflect both the fluctuations in the 
initial conditions and
thermal fluctuations. 
Thermal fluctuations are believed to be a small (but conceptually important)
correction to non-fluctuating
hydrodynamics~\cite{Kapusta:2011gt,Young:2014pka,Yan:2015lfa}.  In addition,
thermal fluctuations can become significant close to the QCD critical 
point~\cite{Stephanov:1998dy,Kapusta:2012zb} and  in smaller colliding systems 
such as
proton-nucleus and proton-proton collisions~\cite{Yan:2015lfa}, which show
remarkable signs of collectivity~\cite{Loizides:2016tew}.

In the current manuscript, rather than simulating nonlinear
fluctuating hydrodynamics directly, we will reformulate fluctuating
hydrodynamics as non-fluctuating hydrodynamics (describing a
long wavelength background) coupled to a set of kinetic equations 
describing the phase space distribution of short wavelength hydrodynamic
fluctuations. For Bjorken flow this set of equations can be solved to determine 
the first fractional powers in the gradient expansion. 

\subsection{Hydrodynamics with noise and fractional powers in the gradient 
expansion}
\label{sec:kstar}

At finite temperature, real-time dynamics in each regime of scales has an 
efficient and systematic description by an effective theory 
\cite{Arnold:1997gh}.
Hydrodynamics is a long wavelength effective theory which describes 
the evolution of conserved quantities by organizing corrections 
in powers of gradients.
For the hydrodynamic expansion to apply we require  
frequencies under consideration to be small compared to the microscopic 
relaxation rates
\begin{equation}
\epsilon \equiv  \frac{\omega \eta}{(e+p) c_s^2} \ll 1 
     \label{epsilon}  \, ,
\end{equation}
where we have estimated the microscopic relaxation time with the hydrodynamic 
parameters, $\tau_R\equiv \eta/(e + p) c_s^2$~\cite{Teaney:2009qa} and 
for 
later convenience defined $\epsilon\equiv\omega\tau_R$.

For definiteness, we follow 
precedent~\cite{Son:2007vk,Baier:2007ix,Kovtun:2011np} and consider a conformal 
neutral
fluid driven from equilibrium by a small metric perturbation $h_{xy}(\omega)$ 
of 
frequency $\omega$. 
Within the framework of linear response (see \Sect{gravitysec} and 
\Ref{Hong:2010at} 
for further details), 
the stress tensor at low frequency 
takes the form 
\begin{equation}
\label{naiveresponse}
\delta T^{xy} = -h_{xy}(\omega) \Big( p - i\omega \eta 
  + 
  \left(\eta \tau_{\pi} - \frac{\kappa}{2} \right) \omega^2  \Big).
\end{equation}
The first term is the prediction of ideal hydrodynamics $\delta T^{xy} = -p 
h_{xy}$; 
the middle term is the prediction of first order viscous
hydrodynamics~\cite{Son:2007vk}, where $\eta$ is the shear viscosity; 
finally, the last term is the prediction of second order hydrodynamics, where 
$\tau_{\pi}$ and $\kappa$  are 
the associated second order parameters~\cite{Baier:2007ix}.  

In writing 
\Eq{naiveresponse} we have neglected additional contributions stemming from 
fluctuations which will be described below. 
Thermal fluctuations can be incorporated into the hydrodynamic description by 
including
stochastic terms into the equations of 
motion~\cite{LandauStatPart1, LandauStatPart2}\cite[for a recent review:][]{Kovtun:2012rj}
\begin{align}
d_\mu T^{\mu\nu}=0\label{eomzero},\quad 
T^{\mu\nu}=T^{\mu\nu}_\text{ideal}+T^{\mu\nu}_\text{visc.}+S^{\mu\nu},
\end{align}
where variance of the
noise, $\llangle S^{\mu\nu} S^{\rho\sigma}\rrangle{\sim}2 T\eta \delta(t- 
t')$,  is determined by the fluctuation dissipation theorem at temperature $T$ 
and introduces no
new parameters into the effective theory\footnote{
    We follow a standard notation for hydrodynamics summarized in 
    Ref.~\cite{Teaney:2009qa}.  $d_{\mu}$ 
    notates a covariant derivative using the ``mostly-plus" metric convention. 
    $T^{\mu\nu}_{\rm ideal} = (e + p) u^{\mu} u^{\nu} + pg^{\mu\nu}$
    and $T^{\mu\nu}_{\rm
    visc} = - \eta \sigma^{\mu\nu}$ where $\sigma^{\mu\nu}  = \Delta^{\mu\rho}
    \Delta^{\nu\sigma} (d_{\rho} u_{\sigma} + d_{\sigma} u_{\rho} - \frac{2}{3}
    g_{\rho\sigma} d_{\gamma} u^{\gamma})$, with $\Delta^{\mu\nu} = g^{\mu\nu} 
    + u^{\mu} u^{\nu}$.
The noise correlator is fully specified in \Eq{noisecorrelator} of 
\Sect{gravitysec}.}. After including these stochastic
terms, the correlators of momentum and energy evolve 
to their equilibrium values  in the absence of the external force, 
$h_{xy}(\omega)$. 
Specifically, the
equilibrium two point functions of the energy and momentum densities, 
$\delta e(t,\x)\equiv T^{00}(t,\x) - \llangle T^{00} \rrangle$ and $g^i(t,\x) 
\equiv T^{0i}$ respectively, approach the textbook result~\cite{LandauStatPart1}
\begin{subequations}
\label{etcorrelators}
\begin{align}
 \llangle \delta e(t,\k) \delta e(t,-\k') \rrangle &= \frac{(e + p) T}{c_s^2} 
 \; (2\pi)^3 \delta^3(\k - \k'), \\
 \llangle g^i(t,\k) g^j(t,-\k')) \rrangle &=  (e + p) T \; \delta^{ij}(2\pi)^3 
 \delta^3(\k - \k'),
\end{align}
\end{subequations}
where $c_s$ is the speed of sound, and $\delta e(t,\k)$ notates the spatial 
Fourier transform of $\delta e(t,\x)$.
In the presence of an external force or a non-trivial expansion these 
correlations are driven 
away from equilibrium. The purpose of hydrodynamics with 
noise is to describe in detail these deviations from equilibrium.

Due to the nonlinear character of hydrodynamics the thermal
fluctuations change the evolution of the system. Indeed, a diagrammatic
analysis of the hydrodynamic response at one-loop order
shows that the stress in the presence of a weak external field (or the retarded 
Green function) is 

\begin{equation}
{\llangle T^{xy}(\omega) \rrangle} = -h_{xy}(\omega) \left(p - i \omega 
\eta + (i+1)
\frac{\left(7 + \left(\tfrac{3}{2}\right)^{3/2}\right) }{240\pi}
T \left(\frac{\omega}{\gamma_{\eta}} \right)^{3/2}  + \mathcal
O(\omega^2) \right)  , \label{Txy}
\end{equation}

where $p$, $e$, and $\eta$ are renormalized physical quantities 
(see \Sect{grav1} and \Sect{grav2} for further 
discussion of the renormalization),
and
\begin{equation}
  \gamma_{\eta} \equiv \frac{\eta}{e + p} \, ,
\end{equation} 
is the momentum diffusion coefficient
~\cite{Kovtun:2003vj,Kovtun:2011np}.
As emphasized and estimated previously,
the fractional order $\omega^{3/2}$ is parametrically  larger 
than second order hydrodynamics~\cite{Kovtun:2011np}. However, the coefficient 
of 
the $\omega^{3/2}$ terms is vanishingly small in weakly coupled theories
and in strongly coupled theories at large $N_c$, and 
therefore second order hydrodynamics may be an effective approximation scheme  
except at very small frequencies.
In the context of holography,
the $\omega^{3/2}$ term can only be determined by performing a one 
loop calculation in the bulk~\cite{CaronHuot:2009iq}. 

In the current paper we will rederive \Eq{Txy}  using a kinetic description of 
short wavelength hydrodynamic fluctuations.
For an external driving frequency of order
$\omega$,
we identify an important length scale set by equating  the damping rate and  
the external frequency
\begin{equation}
\label{kstarw}
\gamma_{\eta}  k_*^2 \sim \omega \, ,\quad   k_{*} \sim  
      \left(\frac{ 
      \omega  }{\gamma_\eta} \right)^{1/2} \, .
\end{equation}
We will refer to the $k_*$ as the \emph{dissipative scale} below (see also
\Ref{CaronHuot:2009iq}).
Modes with wavenumbers significantly larger than the dissipative scale, $k\gg 
k_{*}$,  are  damped
and reexcited by the noise on a time scale which is short compared to period 
$2\pi/\omega$, and
this rapid competition leads to the equilibration of these shorter wavelengths, 
i.e. 
their equal  time correlation functions are given by \Eq{etcorrelators}. 
By contrast, modes with wavenumbers of order $k\sim k_{*}$ have equal time 
correlation functions which deviate from the equilibrium expectation values.

\begin{figure}
\centering
\includegraphics[width=0.8\linewidth]{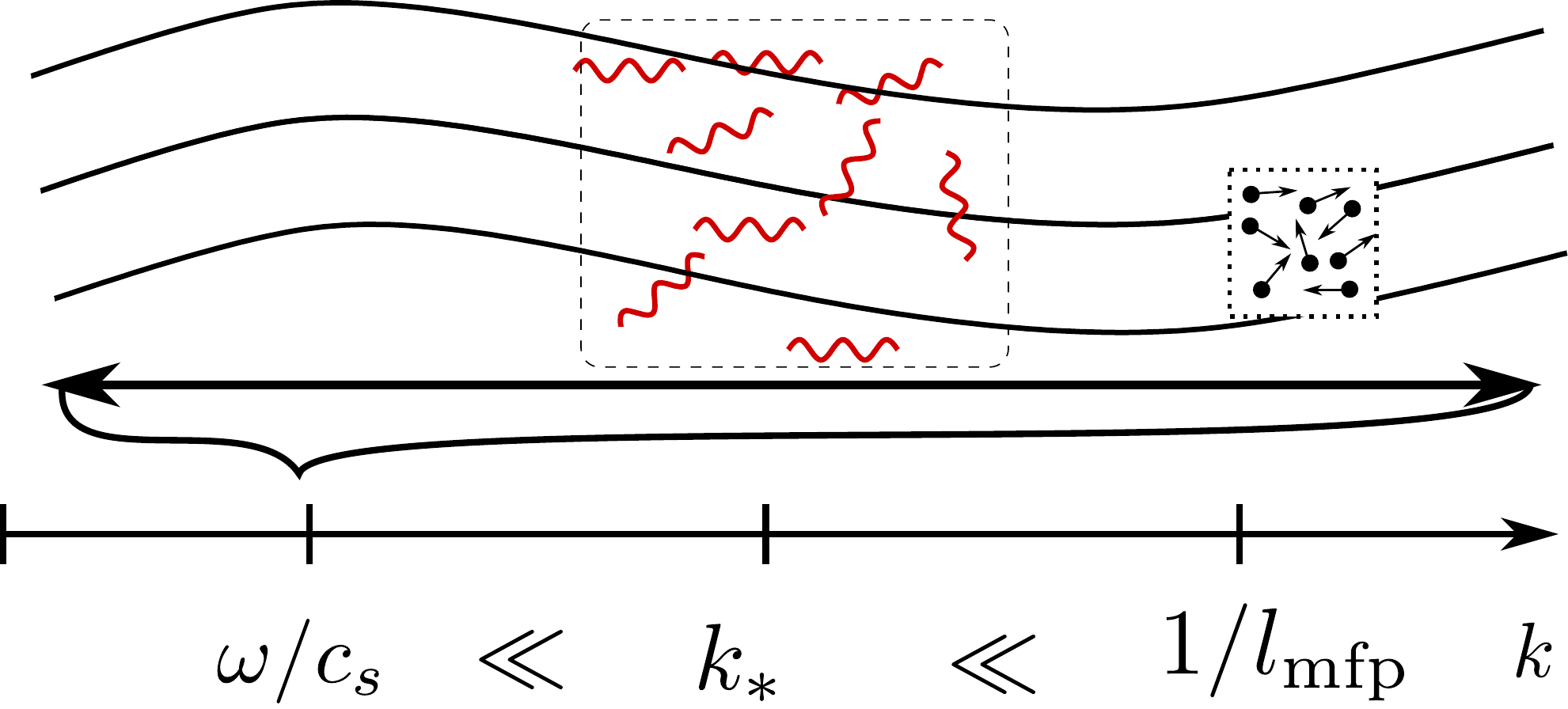}
\caption{\label{scales} The hydro-kinetic description of noise is 
based on 
the  separation of scales between the long wavelength hydrodynamic background 
(with $k \sim \omega/c_s $), and shorter wavelength hydrodynamic
fluctuations (with $k \sim k_{*}\equiv \sqrt{\omega/\gamma_\eta}$). 
The wavelengths of the hydrodynamic
fluctuations  are still much longer than microscopic mean free path. 
The hydrodynamic fluctuations are driven out of
equilibrium by the expanding background, and this 
deviation is the origin of the long-time tail correction to the stress tensor. }
\end{figure}
It is notable that the wavenumbers of interest $k_*$ are large 
compared to $\omega/c_s$, 
but still small compared to microscopic
wavenumbers of order the inverse mean free path
\footnote{
The effect of second-order hydrodynamics is suppressed compared to the 
first-order hydrodynamics as long as the derivative expansion works, i.e. $k\ll 
1/\ell_{\rm mfp}$.
The causal property of the second-order hydrodynamics is gained by modifying 
the dispersions at $k\sim 1/\ell_{\rm mfp}$.
}.
Estimating the mean free path as $\ell_{\rm mfp} = c_s\tau_R$, we see that
the strong inequalities
\begin{equation}
\label{scalesep0}
\frac{\omega}{c_s} \ll k_{*}  \ll  \frac{1}{\ell_{\rm mfp} },
\end{equation}
can be written as 
\begin{equation}
\label{scalesep}
\frac{\omega}{c_s} \ll  \frac{\omega}{c_s} \, \frac{1}{\sqrt{\epsilon}} \ll 
\frac{\omega}{c_s} \, \frac{1}{\epsilon} ,
\end{equation}
and thus holds whenever hydrodynamics is applicable, $\epsilon \ll 1$.
The scale separation 
illustrated in \Fig{scales}
can be used to set up an 
approximation scheme where modes of order $k_*$ on a soft ($k\sim 
\omega/c_s$) background are treated with
a kinetic or Wentzel-Kramers-Brillouin (WKB) type approximation scheme. We 
will develop the appropriate 
kinetic equations in \Sect{gravitysec}.  These kinetic equations
can be solved and used to determine how the two point functions of 
energy and momentum with wavenumbers of order $k_*$ deviate from equilibrium
when driven by an external perturbation. The $\omega^{3/2}$ term
in \Eq{Txy} roughly  represents  the contribution of $\int k^2 dk  {\sim}k_*^3$ 
slightly out of equilibrium hydro-kinetic modes per volume, with each mode 
contributing  $\half T$ of energy to the stress tensor. Note that the 
contribution to the 
stress tensor of 
modes outside of the kinetic regime $k\ll k_*$  is suppressed by phase 
space.

Similar kinetic equations can be derived for much more general flows. 
We will establish the appropriate kinetic equations for a Bjorken
expansion~\cite{Bjorken:1982qr}, which is a useful model for the 
early stages of a heavy ion collision. 
The ideal, first, and second order terms in the gradient expansions
have been given in Refs.~\cite{Bjorken:1982qr},\; \cite{Danielewicz:1984ww},\; 
and \cite{Baier:2007ix,Bhattacharyya:2008jc} respectively. 
For a conformal (non-fluctuating) fluid the longitudinal pressure during a 
Bjorken expansion takes the form
\begin{equation}
\tau^2T^{\eta\eta} = p - \frac{4}{3} \frac{\eta}{\tau} +  \frac{8}{9\tau^2} 
(\lambda_1 - \eta\tau_\pi) + \ldots  \, .
\end{equation}
The expansion rate is $\partial_{\mu} u^{\mu} = 1/\tau$, and each
higher term in the gradient expansion is suppressed by an 
integer power of $1/\tau T$.
For  Bjorken flow the expansion rate plays the role of frequency, and
the distribution of sound modes are characterized  by a dissipative scale 
analogous
to \Eq{kstarw} of order\footnote{
    The quantities $k_{*}(\tau)$, $\gamma_{\eta}(\tau)$, $s(\tau)$, $\ldots$   
    are
    all functions of time for a Bjorken expansion, e.g.  for a conformal 
    equation of state and an ideal expansion, $k_{*}(\tau)
    \propto \tau^{-2/3}$, $\gamma_\eta \propto \tau^{1/3} $, $s(\tau) \propto
    \tau^{-1}$,  etc.  Throughout the paper $k_{*}$,$\gamma_{\eta}$, $s$,
    $\ldots$ (without a time argument) will denote the physical quantity
    at the final time of consideration. The explicit time argument will 
    be used when needed, e.g. $k_{*}(\tau') = k_{*} (\tau/\tau')^{-2/3} $.
}
\begin{equation}
k_{*}   \sim \frac{1}{(\gamma_\eta \tau)^{1/2}} 
.\label{kstar}
\end{equation}
At this scale the viscous damping rate balances the expansion rate. These 
hydrodynamic modes satisfy the inequality
\begin{equation}
\label{scalesep2}
\frac{1}{c_s \tau} \ll  k_{*}  \ll \frac{1}{\ell_{\rm mfp} }, 
\end{equation}
and this strong set of inequalities can be used to determine 
a kinetic equation for hydrodynamic modes of order $k_*$. The equal
time correlation functions for wavenumbers of this order deviate 
from their equilibrium form in \Eq{etcorrelators}, and the kinetic equations 
precisely determine
the functional dependence of this deviation. Finally, these
modes contribute to the longitudinal pressure and determine
first fractional power  in the longitudinal pressure of a conformal fluid 
(analogous to \Eq{Txy}). In \Sect{bjorken} we will establish that 
this nonlinear correction to the longitudinal component of the 
stress tensor is
\begin{multline}
\frac{\llangle \tau^2T^{\eta\eta}\rrangle}{e + p} =   \Big[ \frac{p}{e+p} - 
\frac{4}{3} 
 \frac{\gamma_\eta}{\tau}
+   \frac{1.08318}{s\, (4\pi \gamma_\eta \tau)^{3/2} } +  \mathcal O\left( 
\frac{1}{(\tau T)^2} \right) \Big].
\end{multline}
Noise also contributes to transverse momentum 
fluctuations, and this contributes at quadratic order to $\llangle T^{\tau\tau} 
\rrangle$ as we discuss
in \Sect{bjorken}. Thus,
a complete description of a Bjorken expansion with noise must also reexamine 
the 
relationship between the background energy density $e$, and the one point 
function $\llangle T^{\tau\tau} \rrangle$.

An outline of the paper is as follows. In \Sect{gravitysec} we consider a 
static fluid perturbed by an 
external gravitational perturbation. The purpose of this section is to introduce
the kinetic equations, and to reproduce
the results of the diagrammatic analysis of 
Refs.~\cite{Kovtun:2003vj,Kovtun:2011np} using the hydro-kinetic
theory adopted here.  In \Sect{bjkinetic} we linearize the hydrodynamic 
equations
of motion to determine the appropriate kinetic equations for a Bjorken 
expansion. In \Sect{bjstress} and \ref{sketch} we determine the solutions to 
the kinetic theory and use these solutions to evaluate the contribution of 
hydrodynamic modes to the stress tensor. We give an intuitive physical 
interpretation of the main results of the paper in 
\Sect{qualitative}. 
Finally we 
conclude with results and discussion in \Sect{results}.

\section{Hydrodynamic fluctuations in a static fluid}
\label{gravitysec}

We will first derive the kinetic equations for 
hydrodynamic fluctuations in homogeneous flat space in \Sect{grav1}.
The purpose here is to introduce notation, and to discuss the 
kinetic approximations in the simplest context. Then in \Sect{grav2}
we will perturb the system with a gravitational field and derive 
the appropriate kinetic theory in this case. 
We then use this hydro-kinetic theory to reproduce the results of 
loop calculations~\cite{Kovtun:2003vj,Kovtun:2011np}
for the renormalization of the shear viscosity 
and the long-time tails which characterize the hydrodynamic response due to 
nonlinear noise effects.

\subsection{Relaxation equations for hydrodynamic fluctuations}
\label{grav1}

To illustrate the approximations that follow and to introduce
notation, 
we first will derive 
kinetic equations for the two point functions for energy 
and
momentum density perturbations around a static homogeneous background.
The basics of the techniques adopted in our analysis is reviewed in 
\cite{Pomeau:1974hg,Fox1978}.
The (bare) background quantities of the hydrodynamic effective theory, such as 
the energy density, pressure, and shear
viscosity ($e_0(\Lambda)$, $p_0(\Lambda)$, and $\eta_0(\Lambda)$ respectively) 
are calculated
by integrating out fluctuations above a scale $\Lambda$, i.e. by excluding the 
contributions of hydrodynamic fluctuations with wavenumber $k < \Lambda$ to the 
stress tensor.  This is important because modes with $k < \Lambda$ will not 
be in equilibrium when the system is perturbed by a driving force.
The relation between the bare parameters and the physical quantities (which may 
be computed in infinite volume with lattice QCD for instance) is discussed in 
\Sect{grav2} and in \Ref{Kovtun:2011np}, where $\eta_0(\Lambda)$ is referred to 
as $\eta_{\rm cl}(p_{\rm max})$.

To derive 
a relaxation equation for the two point functions we linearize the equations of 
stochastic hydrodynamics and
study the eigenmodes of the system. The correlations between eigenmodes
with vastly different frequencies are neglected in a kinetic
(or coarse graining) approximation.
For the constant background $e_0=\text{const}$, and to linear order in 
field perturbations and stochastic
fluctuations, the equations of motion (\Eq{eomzero}) become
\begin{subequations}
\label{eomflat}
\begin{align}
\partial_t \delta e +ik_i g^i &=0,\label{eom21}\\
 \partial_t g_{i} +i k_i \delta p+\gamma_\eta k^2 
g_i+\frac{1}{3}\gamma_\eta k_ik_j g^j
&=-\xi_i\label{eom22},
\end{align}
\end{subequations}
where $\gamma_\eta \equiv \eta_0/(e_0+p_0)$ is
computed with bare quantities, 
and ${-}\xi_i$ is the stochastic force, ${-}ik_j 
S^{j}_{\phantom{k}i}(t,\k)$. Here $S^{j}_{\phantom{k}i}(t,\k)$ are spatial 
components of the noise tensor 
with equilibrium correlation given by~\cite{LandauStatPart2}

\begin{align}
    \label{noisecorrelator}
\langle S^{\mu\nu}(t_1,\k)S^{\alpha\beta}(t_2,-\k')\rangle =
2T \eta_0\left[
\left(\Delta^{\mu\alpha}\Delta^{\nu\beta} + 
\Delta^{\mu\beta}\Delta^{\nu\alpha}\right)
- \frac{2}{3}\Delta^{\mu\nu}\Delta^{\alpha\beta}
\right]\nonumber \\\times(2\pi)^3\delta^3(\k-\k') \delta(t_1 - t_2).
\end{align}

 It is convenient to combine 
\Eq{eomflat}  into a single matrix 
equation for an amalgamated field $\phi_a=(c_s \delta e, g_j)$
\begin{equation}
\partial_t \phi_{a}(t, \k)=-i \mathcal{L}_{ab} \phi_b-\mathcal{D}_{ab} \phi_b 
- \xi_a 
\label{eom3},
\end{equation}
where ideal and dissipative terms are 
\begin{equation}
 \mathcal{L}_{ab}=\begin{pmatrix}0 &c_s k_j \\c_sk_i & 0 \end{pmatrix},\quad  
 \mathcal{D}_{ab}=\gamma_\eta\begin{pmatrix}0 &0\\0 & k^2 
 \delta_{i  j}+\frac{1}{3}k_i k_j
 \end{pmatrix},
\end{equation}
and the stochastic noise $\xi_a$ satisfies correlation equation
\begin{multline}
    \left<\xi_a(t_1,\k) \xi_b(t_2,-\k')\right>= 2T 
    (e_0+p_0)\mathcal{D}_{ab}  (2\pi)^3   \delta^3(\k-\k')   
    \delta(t_1-t_2).\label{xixi}
\end{multline}
At the dissipative scale
the acoustic matrix $\mathcal{L}\sim c_sk_*$ 
originating from ideal equations of motion dominates over the competing 
dissipation  $\mathcal{D}$ and 
fluctuation $\xi_a$ terms. $\mathcal 
L_{ab}$   has 
four 
eigenmodes: two 
longitudinal 
sound modes 
with $\lambda_{\pm} = \pm c_s |\k|$  and two transverse 
zero modes ($\lambda_{T_1}= \lambda_{T_2} = 0$). 
Since $\mathcal{L}$ drives evolution of $\phi_a$, it will be convenient to 
analyze the dynamics in terms of eigenmodes 
of $\mathcal L_{ab}$:
\footnote{
Another reason why analysis in terms of eigenmodes of $\mathcal L_{ab}$ is 
convenient is that
they form a real and orthonormal basis and the projection onto each mode is 
easily handled.
}
\begin{equation}
   (e_{\pm})_a = \frac{1}{\sqrt{2}}
\begin{pmatrix}
    1 \\
  \pm\hat{k}
\end{pmatrix},
\quad
(e_{T_1})_a =
\begin{pmatrix}
    0 \\
    \vec{T}_1 \\
\end{pmatrix},
\quad
(e_{T_2})_a =
\begin{pmatrix}
    0 \\
    \vec{T}_2 \\
\end{pmatrix},\label{vectors}
\end{equation}
where $\hat{\k}=\k/|\k|$, and $\vec{T}_1$  and  $\vec{T}_2$ are
two orthonormal spatial vectors perpendicular to $\hat\k$
\begin{subequations}
\begin{align}
\hat{k}&=(\sin \theta \cos \varphi, \sin \theta \sin \varphi, \cos \theta),\\
\vec{T}_1&=(-\sin \varphi, \cos \varphi, 0),\\
\vec{T}_2&=(\cos \theta \cos \varphi, \cos \theta \sin \varphi, -\sin \theta).
\end{align}
\end{subequations}
Now we will derive a relaxation equation for the two point correlation 
function of hydrodynamic fluctuations by defining a density matrix 
$N_{ab}(t,\k)$
\begin{equation}
\left<\phi_a(t,\k) 
\phi_b(t,-\k')\right>\equiv 
N_{ab}(t,\k)(2\pi)^3\delta^3(\k-\k'),\label{defN}
\end{equation}
and analyzing the time evolution of $N_{ab}(t,\k)$.

The analysis is most transparent in the eigenbasis, $\phi_{A} \equiv \phi_a 
\left(e_{A}\right)_a$ with $A=+,-,T_1, T_2$,
and below we will determine the equation of motion for  $N_{AB}\equiv\llangle 
\phi_A \phi_B\rrangle $ where $A,B=+,-,T_1,T_2$.
We note that the positive and negative sound modes $\phi_{+}$ and $\phi_{-}$
are related since the hydrodynamic
fields are real, $\phi_{-}^*(\k,t) = \phi_{+}(-\k,t)$.

Using the equations of motion for $\phi_A$ we calculate the infinitesimal 
change of 
$N_{AB}(t+dt)-N_{AB}(t)$,  and use the equal time correlator for the 
noise  (\Eq{xixi}) to find a differential equation for $N_{AB}$ 
\begin{equation}
\partial_t N=-i[\mathcal{L},N] 
-\{\mathcal{D}, N\}+2T (e_0+p_0)\mathcal{D},\label{eomAB}
\end{equation}
where $[X,Y]\equiv XY-YX$, $\{X,Y\}\equiv XY+YX$, and $[\mathcal{L},N]_{AB} 
=(\lambda_A-\lambda_B) 
N_{AB}$.
We are interested in the evolution of two point 
correlation functions over time scales much larger than acoustic oscillations,
$\Delta t \gg 1/(c_s k_*)$. 
On these timescales the off-diagonal matrix elements
of the density matrix, $N_{+T_1}$ for example,  
rapidly oscillate reflecting the large difference in eigenvalues, $\lambda_{+} 
- \lambda_{T_1} \sim c_s k_*$.
In a coarse graining approximation the contributions of 
these off-diagonal matrix elements to physical quantities can be neglected when 
averaged over times
long compared to $1/(c_s k_*)$.  
This reasoning does 
not apply to the diffusive modes 
$A,B=T_1,T_2$ where both eigenvalues are zero,  but rotational symmetry in the 
transverse $xy$-plane requires $N_{T_1T_2}$ to vanish\footnote{Rotational 
symmetry in the 
transverse $xy$-plane requires that $\left<g_i g_j\right>\sim A \delta_{ij} + B 
\hat{k}_i \hat{k}_j$, where $i,j=x,y$. Such a tensor structure has vanishing 
$T_1T_2$ projection.}.

With these approximations, the non-trivial relaxation equations of two point 
correlation functions in \Eq{eomAB} are
\begin{subequations}
\label{Nflat}
\begin{align}
\partial_t N_{\pm\pm}(t,\k)  &= -\frac{4}{3}\gamma_\eta k^2 
    (N_{\pm\pm}-N_0)\label{N++},\\
\partial_t N_{T_1T_1}(t,\k)  &= -2\gamma_\eta k^2 
(N_{T_1T_1}-N_0),\\
\partial_t N_{T_2T_2}(t,\k)  &= -2\gamma_\eta k^2 
(N_{T_2T_2}-N_0)\label{NT2T2},
\end{align}
\end{subequations}
where 
\begin{equation}
N_0=T (e_0+p_0)
\end{equation} 
is the equilibrium value for $N_{AA}$ 
(c.f. \Eq{etcorrelators}). In the absence of external perturbations, two point 
correlation functions relaxes to their equilibrium values.
The next step towards general kinetic equations is to study how equal time 
correlations are driven out of equilibrium by the presence of external fields.

\subsection{Linear response to gravitational perturbations}
\label{grav2}
In this section we will study the evolution of two point energy and momentum 
correlators in the presence of time varying gravitational field. We determine
the kinetic equations in the time dependent background, and use these
equations to reproduce the modifications of
the retarded Green function (\Eq{Txy}) due to thermal 
fluctuations, which were previously found by a one-loop 
calculation~\cite{Kovtun:2003vj,Kovtun:2011np}.

A straightforward way of introducing an external source to equations of motion 
is to study fluctuating hydrodynamics in the presence of a small metric 
perturbation, $g_{\mu\nu} = \eta_{\mu\nu} + h_{\mu\nu}$.
The Green function records the response of $T^{\mu\nu}$ to the metric 
perturbation
\begin{equation}
\delta \llangle T^{\mu\nu}(\omega) \rrangle = 
-\frac{1}{2}G^{\mu\nu,\alpha\beta}_R(\omega)h_{\alpha\beta}(\omega).
\end{equation}

For a constant homogeneous background with time dependent metric perturbation 
$h_{ij}(t)$, symmetry 
constrains the form of the retarded Green function
\begin{multline}
    G_R^{ij, kl}(\omega)= 
    \mathring{G}_R(\omega)\, ( \delta^{ik}\delta^{jl}+ 
    \delta^{il}\delta^{jk}-\frac{2}{3}\delta^{ij}\delta^{kl}) +
\overline{G}_{R}(\omega)\, \delta^{ij}\delta^{kl}\label{Gsym},
\end{multline}
and therefore we can obtain the Green function in \Eq{Txy}, i.e. 
$\mathring{G}_R(\omega)$, 
by considering a 
diagonal traceless metric perturbation,
$h_{ij}(t)=h(t)\, \text{diag}\,(1,1,-2)$.

In the presence of metric perturbations and thermal fluctuations, the energy 
momentum tensor is 
\begin{equation}
    \label{tijeq}
    \delta \llangle T^{ij}(t)\rrangle = -p_0h^{ij}-\eta_0  \partial_th^{ij} 
+ \frac{\left<g^i(t,\x) g^j(t,\x)\right>}{e_0+p_0},
\end{equation}
where the nonlinear term stems from the constitutive relation of 
ideal hydrodynamics, $T^{ij} = p_0 \delta^{ij} + (e_0 + p_0) u^{i} u^{j}$.
The averaged squared momentum, $\llangle g^i(t,\x) g^j(t,\x)\rrangle$, is 
related to the two-point 
functions of $g^{i}$ in $\k$ space as
\begin{equation}
\left<g^i(t,\x) g^j(t,\x)\right> =\int 
\frac{d^3k}{(2\pi)^3}N^{ij}(t,\k).\label{gg}
\end{equation}
In this integral, the equilibrium value of $N^{ij}$ and its first viscous 
correction will 
renormalize $p_0$ 
and $\eta_0$ (see below), while the finite remainder 
will determine the first fractional power in the stress tensor correlator 
$\propto \omega^{3/2}$.

Studying the hydrodynamic equations in \Eq{eomzero}, and neglecting  metric 
perturbations of the dissipative terms, 
we find  that the
linearized equations of motion  are identical to flat background \Eq{eomflat}, 
but now 
there is a difference between covariant and contravariant indices
\begin{subequations}
\begin{align}
\partial_t \delta e +ik_i g^i &=0,\\
 \partial_t g_{i} +i k_i \delta p+\gamma_\eta k^2 
g_i+\frac{1}{3}\gamma_\eta k_ik_j g^j
&=-\xi_i.
\end{align}
\end{subequations}
To avoid this complication,
we use a  vielbein formalism and scale  the spatial components of momentum and 
wavenumber by 
$\sqrt{g_{ij}}$, i.e.  
$g^i$ and $k_j$ are replaced by
\begin{subequations}
\begin{align}
G_{\hat{\imath}}&=(1+\frac{1}{2}h_{ij})g^j,\\
K_{\hat{\imath}}&=(1-\frac{1}{2}h^{ij})k_j,
\end{align}
\end{subequations}
where now the position of hatted indices is unimportant.
Analogously 
to \Eq{eom3}, we obtain a matrix equation for $\phi_a=(c_s \delta e, G_{\hat 
\imath})$
\begin{equation}
\partial_t \phi_{a}(t, \k)=-i \mathcal{L}_{ab} \phi_b-\mathcal{D}_{ab} \phi_b - 
\xi_a -
\mathcal{P}_{ab}\phi_b,
\label{eom4}
\end{equation}
with an additional metric dependent source term   
\begin{equation}
 \mathcal{P}_{ab}=\begin{pmatrix}0 &0 \\0 & \frac{1}{2}\partial_0 h_{\hat{i} 
 \hat{j}} 
 \end{pmatrix},
\end{equation}
which drives the hydrodynamic fluctuations away from equilibrium.
The eigenbasis of $\mathcal{L}$ (see \Eq{vectors}) is now defined with respect 
to the time dependent vector
$\vec{K}(t)$, but remains orthonormal at all times. Furthermore, the metric 
perturbation preserves rotational symmetry in the transverse $xy$-plane, and 
this 
guarantees that the $T_1$ and $T_2$ modes are not mixed by the time-dependent 
perturbation. Thus, the only non-trivial 
diagonal components of the symmetrized energy and momentum two point functions  
are 
\begin{subequations}
    \begin{align}
\partial_t N_{\pm\pm}  &= -\frac{4}{3}\gamma_\eta K^2 
(N_{\pm\pm}-N_0)  
  -\frac{1}{2}\partial_t h \, (\sin^2\theta_K-2\cos^2 \theta_K)N_{\pm\pm},\\ 
\partial_t N_{T_1T_1}  &= -2\gamma_\eta K^2 (N_{T_1T_1}-N_0)-\partial_t h \,
N_{T_1T_1},\\
\partial_t N_{T_2T_2} & = -2\gamma_\eta K^2 (N_{T_2T_2}-N_0) 
   -\partial_t h \,
(\cos^2 \theta_K-2\sin^2\theta_K) N_{T_2T_2}.
  \end{align}
\end{subequations}
We can find a perturbative solution to these equations for a small periodic 
metric perturbation, e.g.
\begin{equation}
    N_{T_2T_2}(\omega,\k) \simeq N_0\left(2\pi\delta(\omega){+}\frac{i\omega 
    h(\omega) (\cos^2\theta_K {-} 2 \sin^2\theta_K)}{-i\omega + 2\gamma_\eta 
    K^2}\right).
\end{equation}
To find the correction to the energy momentum tensor due to the nonlinear 
momentum 
fluctuations in \Eq{tijeq}, we need to perform the $k$ 
space integral in \Eq{gg}
\begin{align}
\langle\phi_a(x)\phi_b(x)\rangle
=&{\int} \frac{d^3 K}{(2\pi)^3} N_{ab}(\tau,\k)= {\int} \frac{K^2 dK d\cos 
\theta_K d\varphi_K}{(2\pi)^3} (e_A)_a 
N_{AB}(\tau,\k) (e_B)_{b} \, .
\end{align}
Note, care should be taken when transforming the zeroth 
order value $N_{AA}=N_0$ to original unhatted basis as it produces terms linear 
in metric perturbation.
The modification of the response function $\mathring{G}_R(\omega)$ due to the 
momentum fluctuations (i.e. the last term in \Eq{tijeq}) is
\begin{align}
    \mathring{G}_R(\omega) &= -\frac{1}{6}(\delta T^{xx}+\delta T^{yy}-2\delta 
T^{zz})/h(\omega),\nonumber\\
&\supset -\frac{T}{6}\int\frac{d^3 K}{(2\pi)^3}
\left(-6+  i\omega  
 \frac{(\sin^2\theta_K-2\cos^2\theta_K)^2}{-i\omega +\frac{4}{3}\gamma_\eta 
 K^2}\right. \nonumber \\
& \quad\qquad \left.  
 + \; \;  
 i\omega\frac{1+(\cos^2\theta_K-2\sin^2\theta_K)^2}{-i\omega+2\gamma_\eta 
 K^2}\right).
\end{align}
Performing $K$-space integral with UV cutoff, $K_\text{max}=\Lambda$, and
adding the remaining terms in \Eq{tijeq},
we find
\begin{multline}
\label{flatans}
    \mathring{G}_R(\omega)  =
    \left( p_0 + \frac{\Lambda^3}{6\pi^2} T \right)
        {-}i \left(\eta_0  + \frac{\Lambda}{\gamma_\eta} \frac{17}{120\pi^2}T 
        \right) \omega  \\
 +(1+i)\frac{1}{\gamma_\eta^{3/2}}  
\frac{ (\frac{3}{2})^{3/2}
+ 7 }{240\pi} T\omega^{3/2},
\end{multline}
in agreement with previous work~\cite{Kovtun:2003vj,Kovtun:2011np}.
The first two terms in \Eq{flatans} are the 
renormalized pressure ($p \equiv p_0(\Lambda) + O(T\Lambda^3)$) and shear
viscosity ($\eta \equiv \eta_0(\Lambda) + O(\Lambda T^2)$) as discussed 
previously~\cite{Kovtun:2011np}. 
In general, $\Lambda\ll 1/\ell_{\rm mfp} \leq T$ holds and the renormalization 
only slightly shifts the quantities in the thermodynamic limit ($\Lambda\to 0$).
Further
discussion of the renormalization of these quantities is given in the next 
section when the expanding
case is presented.

The last term is the finite nonlinear modification of the medium response,
and agrees with loop calculations in equilibrium. The
kinetic approach outlined in this section has the advantage that it can be 
readily applied to more general backgrounds, and we will exploit this advantage 
to calculate the analogous correction for a Bjorken 
expansion in the next section.  In contrast to the linear response described 
here,
the deviation from equilibrium in the expanding case is of order unity. 
Consequently, computing the first fractional power in an expanding system
with the diagrammatic formalism would require
an extensive resummation, which would invariably reproduce kinetic calculation 
described 
in the next section~\cite{Jeon:1995zm}.

\section{Hydrodynamic fluctuations for a Bjorken expansion}
\label{bjorken}

In this section we will derive the kinetic evolution equations for hydrodynamic 
fluctuations during 
a 
Bjorken expansion.
We consider a neutral conformal fluid, for which $c_s^2=1/3$, $\zeta=0$, and 
$\mu_{\rm B}=0$.
In Bjorken coordinates
the energy and momentum conservation laws are
\begin{eqnarray}
\partial_{\mu}T^{\mu\nu}+\frac{1}{\tau}T^{\tau\nu} + 
\Gamma^{\nu}_{\mu\beta}T^{\mu\beta}=0,
\end{eqnarray}
with $\Gamma^{\tau}_{\eta\eta}=\tau$ and 
$\Gamma^{\eta}_{\tau\eta}=\Gamma^{\eta}_{\eta\tau}=1/\tau$~\cite[see for 
example:][]{Teaney:2009qa}.
For hydrodynamics without noise  the background
flow fields  are independent of transverse coordinates
and rapidity and satisfy 
\begin{align}
    \frac{d (\tau T^{\tau\tau})}{d\tau} &= - \tau^2T^{\eta\eta}, \\
    \frac{d (\tau T^{\tau i})}{d\tau} &= 0,
\end{align}
where roman indices, $i,j \ldots $, run over transverse coordinates $x,y$.
The transverse momentum $T^{\tau i}$ is constant, and can be
chosen to be zero. In hydrodynamics
$T^{\tau\tau}$ and $\tau^2 T^{\eta\eta}$  are related by 
constitutive equations 
\begin{align}
    T^{\tau\tau} &= e ,\\
    \tau^2T^{\eta\eta} &= c_s^2 e - \frac{4\eta}{3\tau}.
\end{align}
Note  that in $\tau^2 T^{\eta\eta}$ the viscous correction is of order 
$\epsilon=\eta/(e+p)\tau\ll1$ smaller than the ideal part, and the solution is 
approximately $e(\tau) = e(\tau_0)\cdot (\tau_0/\tau)^{1+c_s^2}[1+\mathcal 
O(\epsilon)]$.  

We will consider 
the evolution of linearized fluctuations on top of this background.  
The effect of these fluctuations on the background evolution
can then be included as a correction after the two point functions 
are known, i.e.
\begin{equation}
\frac{d \dlangle T^{\tau\tau} \drangle}{d\tau} = -\frac{ \dlangle 
T^{\tau\tau}\drangle  +  \dlangle \tau^2T^{\eta\eta} \drangle}{\tau},\ \ \
\end{equation}
where  the constitutive  relations take the form
\begin{align}
    \label{eave}
    \dlangle T^{\tau\tau} \drangle = e  +  \frac{\dlangle \vec{G}^2 \drangle}{e 
    +p} , \\
    \label{pLave}
     \dlangle \tau^2T^{\eta\eta} \drangle = c_s^2 e - \frac{4\eta}{3\tau}  + 
     \frac{\langle\!\langle (G^{\hat z})^2 \rangle\!\rangle}{e +p}.
\end{align}
Here and below $e(\tau)$ is the average rest frame energy density\footnote{
    $e(\tau)$ notates the average \emph{rest frame} energy density 
    and does not fluctuate; $\dlangle T^{\tau\tau}\drangle$ is
    the average energy density.
In general, the rest frame energy density $e + \delta e$ in a finite volume 
would be estimated from sample estimate of $T^{\tau\tau}$ and
$\vec{G}$  through the (ideal) constitutive
equations, $e + \delta e \simeq T^{\tau\tau} -  \frac{\vec{G}^2}{(1 + c_s^2) 
T^{\tau\tau}}$. Thus $e$ is given by \Eq{eave}, and $\delta e \simeq \delta 
T^{\tau\tau} - \delta (\vec{G^2}/T^{\tau\tau})/(1+c_s^2) \simeq \delta 
T^{\tau\tau}$.};
$\vec{G}$ is
the momentum density $\vec{G} = (T^{\tau x}, T^{\tau y}, \tau T^{\tau \eta})$, 
and all quantities are renormalized as explained more completely below.

There are two sorts of fluctuations to consider: fluctuations in 
the initial conditions (which are long range in rapidity), and hydrodynamic
fluctuations stemming from thermal noise (which are short range in rapidity). 
The
average over the initial conditions and noise  are 
denoted with $\llangle\ldots \rrangle_{\tau_0}$ and $\llangle\ldots \rrangle$ 
respectively,
while the average over both fluctuations is denoted with the double 
brackets $\dlangle \ldots \drangle$.
Since  the transverse momentum per
rapidity is conserved for boost invariant fields, approximately boost invariant 
initial fluctuations in $\tau T^{\tau i}$ remain important at late times.  In 
\Sect{icflucts} we study initial transverse momentum 
fluctuations, while in remainder of the paper we complete our study of
thermal fluctuations during a  Bjorken expansion.

\subsection{A  Bjorken expansion with initial transverse momentum 
fluctuations}
\label{icflucts}

After the initial passage of two large nuclei in a specific event,
each rapidity interval contains a finite amount of transverse momentum, 
although the event-averaged transverse momentum per rapidity is zero.
This initial transverse momentum  is spread
over a large rapidity range by the subsequent re-scatterings in the initial
state. 
Ultimately, this dynamical process can be described by (transverse) momentum 
diffusion in rapidity, and can be modeled with hydrodynamics and noise -- see 
\Sect{sketch}.
Here we will determine how long-range transverse momentum 
fluctuations in the initial state influence the evolution of the background 
energy density at late times.

As a model for the initial conditions in the $x,y$ plane,
we take Gaussian statistics for 
the initial transverse momentum fluctuations
\begin{align}
    \label{gabic}
    \llangle \tau_0 g^{i}_\perp(\tau_0,\xp)  \; \tau_0 g^{j}_\perp(\tau_0, \yp) 
    \rrangle_{\tau_0} =   \chit
    \delta^{ij} \delta^{2}(\xp  - \yp), 
\end{align}
where $g^{i}_\perp(\tau, \xp) \equiv T^{\tau i}$ is approximately independent 
of 
rapidity, so that each (large) rapidity interval is approximately boost 
invariant. 
Integrating over the transverse area $\A$, the total transverse momentum  per 
rapidity,
\begin{equation}
\frac{d p^{x}}{d\eta}  \equiv  \int_{\A} d^2x_\perp  \; \tau_0 
g^{x}_\perp(\tau_0, \xp) ,
\end{equation}
fluctuates from event to event with a scaled variance 
of
\begin{equation}
\label{chi0pp}
\chit \equiv \Big\langle \frac{1}{\A} \left(\frac{d p^{x}}{d\eta} \right)^2 
\Big\rangle_{\tau_0} .
\end{equation}

To find out how this fluctuating initial condition changes the evolution of
the system, we linearize the equations of motion of viscous hydrodynamics 
and Fourier transform with respect to the transverse coordinates
\begin{equation}
\vec{g}_\perp(\tau, \kp ) \equiv \int d^2x_\perp e^{i \kp \cdot \xp }  \, 
\vec{g}_\perp(\tau, \xp ).
\end{equation}
The full equations of motion are given in the next section, see \Eq{bjpert}.
Decomposing the transverse momentum fluctuation into longitudinal and 
transverse pieces
\begin{equation}
g_\perp^i(\tau, \kp) =  g_{L}^i(\kp)  \; + \; g_{T}^{i}(\kp), 
\end{equation}
with $\hat k^i_{\perp} g_{T}^{i} = 0$ and $g_{L}^i = \hat k^i_\perp \hat 
k^j_\perp g_\perp^j$,
we find that
the transverse piece obeys a two dimensional diffusion equation
\begin{equation}
\partial_{\tau} (\tau g_T^i) + \gamma_{\eta} k^2_\perp (\tau g_T^i) = 0,
\end{equation}
with initial conditions  specified by \Eq{gabic}
\begin{multline}
\label{T1T1correlatorIC}
\llangle \tau_0 g_T^i(\tau_0,\kp) \, \tau_0 g_T^j(\tau_0,-\kp') 
\rrangle_{\tau_0}= 
\chit \, (\delta^{ij} - \hat k^i \hat k^j)\, (2\pi)^2 \delta^{2}(\kp - \kp').
\end{multline}
Solving the diffusion equation with a time dependent 
diffusion constant $\gamma_\eta \propto \tau^{c_s^2}$, we see that the
variance at a specified space time point due to the fluctuating initial 
conditions is\footnote{Here we are neglecting the longitudinal contribution, 
$\llangle g_{L} g_{L} \rrangle$, which decreases more rapidly than $1/\tau$ at 
late times. }
\begin{align}
    \llangle \tau g^i_\perp (\tau,\xp) \tau g^j_\perp(\tau,\xp) 
    \rrangle_{\tau_0}  &=  
     \delta^{ij} \frac{\chit}{12\pi\gamma_\eta \tau}  \, .
\end{align}
Thus, we see that a fluctuating initial conditions contributes
quadratically to the average stress tensor 
\begin{subequations}
\begin{align}
    \frac{\llangle \tau^2 T^{\eta\eta} \rrangle_{\tau_0}}{e + p} &= \frac{p}{e 
    +p} -  \frac{4 \gamma_\eta}{3 \tau}, \\
    \frac{\llangle T^{xx} \rrangle_{\tau_0}}{e + p} &= \frac{p}{e +p} + 
    \frac{2\gamma_\eta}{3\tau}  + \left[
\frac{\chit}{\tau^2 (e+p)^2 } \right]
    \frac{1}{12 \pi \gamma_\eta \tau}, \\
    \llangle T^{yy} \rrangle_{\tau_0}  &= \llangle T^{xx} \rrangle_{\tau_0} , \\
    \llangle T^{\tau\tau} \rrangle_{\tau_0}  &= \llangle T^{xx} 
    \rrangle_{\tau_0} + \llangle T^{yy} \rrangle_{\tau_0} + \llangle \tau^2 
    T^{\eta\eta} \rrangle_{\tau_0},
\end{align}
\end{subequations}
where $p = c_s^2 e$.

\subsection{Kinetic equations of hydrodynamic fluctuations}
\label{bjkinetic}

To derive the kinetic equations we will follow the 
strategy of \Sect{grav1}, and expand all fluctuations in Fourier modes 
conjugate to transverse coordinates and rapidity,  e.g.
\begin{equation}
\delta e(\tau, \k) \equiv \int d\eta\, d^2 x_\perp \, e^{i \vec{k}_{\perp} 
\cdot \vec{x}_{\perp} + i \kappa \eta } \, \delta e(\tau, x_\perp, \eta).
\end{equation}
The linearized equations of motion of all hydrodynamic fields around the 
Bjorken background read
\begin{subequations}
\label{bjpert}
\begin{align}
    0&=\left(\frac{\partial}{\partial \tau} + 
    \frac{1+c_{s}^2}{\tau}\right)\delta e
+ i\vec k_{\perp}\cdot\vec g_{\perp} + i\kappa g^{\eta} +\xi^{\tau},\\
\vec 0_{\perp}&=\left(\frac{\partial}{\partial \tau} + \frac{1}{\tau}\right) 
\vec g_{\perp}
+ c_{s}^2 i\vec k_{\perp}\delta e + \gamma_{\eta}\left(k_{\perp}^2 +  
\frac{\kappa^2}{\tau^2}\right) \vec g_{\perp}  \nonumber \\
&+\frac{1}{3}\gamma_{\eta}\vec k_{\perp}\left(\vec k_{\perp}\cdot \vec 
g_{\perp} + \kappa g^{\eta}\right) +\vec\xi_{\perp},\\
0&=\left(\frac{\partial}{\partial \tau} + \frac{3}{\tau}\right) g^{\eta}
+ \frac{c_{s}^2 i \kappa}{\tau^2}\delta e + 
\gamma_{\eta}\left(k_{\perp}^2 + \frac{\kappa^2}{\tau^2}\right) 
g^{\eta}\nonumber \\
&+ \frac{1}{3\tau^2}\gamma_{\eta}\kappa
\left(\vec k_{\perp}\cdot\vec g_{\perp} + \kappa g^{\eta}\right) + \xi^{\eta}.
\end{align}
\end{subequations}
where $(g_\perp^x, g_\perp^y, g^{\eta}) = (T^{\tau x}, T^{\tau y}, T^{\tau 
\eta})$.
As in \Sect{grav1} and \ref{grav2} the hydrodynamic parameters in these 
equations (such as $\gamma_{\eta}$) are constructed from the bare parameters, 
$e_0(\Lambda)$, $p_0(\Lambda)$, $\eta_0(\Lambda)$ and
evolve according to ideal hydrodynamics, $e_0(\tau) = e_0(\tau_0) 
(\tau_0/\tau)^{1+ c_s^2}$.
We also neglected variation in viscosity $\delta \eta/\tau \ll \delta 
p,\delta e$, which is smaller by a factor
$\epsilon=\eta_0/((e_0+p_0)c_s^2\tau)\ll 1$ for conformal fluid.
Note also that the temporal noise component $\xi^{\tau}$ is smaller than 
$\xi^{i_{\perp}}$ 
and $\tau\xi^{\eta}$ by a factor $1/(k_*\tau)\sim \epsilon^{1/2}$ and the 
former can be neglected.

Following the procedure outlined in \Sect{gravitysec} we rewrite \Eqs{bjpert} 
in a compact matrix notation. We define $\vec G=(G^{\hat x},G^{\hat y},G^{\hat 
z})\equiv(\vec g_{\perp}, \tau g^{\eta})$ and $\vec K =(K_{\hat x},K_{\hat 
y},K_{\hat z})\equiv (\vec 
k_{\perp}, \kappa/\tau)$, so that equation of motion for $\phi_a\equiv 
(c_{s}\delta e,\vec G)$ is

\begin{equation}
\label{eq:fluct_ev}
\partial_\tau\phi_a(\tau, \k)
=-i\mathcal{L}_{ab}\phi_b - \mathcal{D}_{ab}\phi_b - \xi_a - 
\mathcal{P}_{ab}\phi_b,\\
\end{equation}
\begin{align}
\mathcal{L} &=
\begin{pmatrix}
0 & c_s\vec K\\
c_s\vec K & 0
\end{pmatrix},\quad
\mathcal{D} =\gamma_\eta \begin{pmatrix}
0 & 0\\
0 & K^2\delta_{\hat \imath \hat \jmath}
+ \frac{1}{3}K_{\hat \imath}K_{\hat \jmath}
\end{pmatrix}, \\
\mathcal{P} &= \frac{1}{\tau}
\begin{pmatrix}
1+c_s^2 & & &\\
 & 1 & & \\
 & & 1 & \\
 & & & 2
\end{pmatrix},
\end{align}
with noise correlator
\begin{multline}
\langle \xi_a(\tau,\k)\xi_b(\tau',-\k) \rangle =
\frac{2T(e_0+p_0)}{\tau} \mathcal{D}_{ab}(2\pi)^3\delta^3(\k-\k') \delta(\tau - 
\tau').
\end{multline}
Here $\delta^3(\k - \k') \equiv \delta^2(\vec{k}_\perp - \vec{k}'_{\perp}) 
\delta(\kappa - \kappa')$ and 
the factor of $1/\tau$ stems from the Jacobian of the coordinate
system  $\delta^4(x - x') /\sqrt{g(x)}$.

The kinetic equation for the two-point functions
\begin{equation}
\left<\phi_a(\tau,\k) 
\phi_b(\tau,-\k')\right>\equiv 
N_{ab}(\tau,\k)(2\pi)^3\delta^3(\k-\k'),
\end{equation}
is obtained similarly to \Sect{gravitysec}
\begin{equation}
\partial_\tau N(\tau,\k)
= -i[\mathcal L, N] - \{\mathcal D, N\} + \frac{2T(e_0+p_0)}{\tau}\mathcal D
-\{\mathcal P, N\}.
\end{equation}
The eigenvectors of $\mathcal{L}$ are of the same form as before, \Eq{vectors}, 
\begin{gather}
    (e_\pm)_a {=} \frac{1}{\sqrt{2}}
    \begin{pmatrix}
1\\
\pm\hat K
\end{pmatrix},\ \
(e_{T_1})_a {=} 
    \begin{pmatrix}
0\\
\vec T_1
\end{pmatrix},\ \
(e_{T_2})_a = 
    \begin{pmatrix}
0\\
\vec T_2
\end{pmatrix}.
\end{gather}
However, now the wavenumber vector $\vec{K}$
 is time dependent
\begin{subequations}
\begin{align}
    \hat{K}\equiv&  \frac{ (\vec{k}_\perp, \kappa/\tau)}{\sqrt{k_\perp^2 + 
    (\kappa/\tau)^2 } }  \equiv 
    (\sin\theta_K\cos\varphi_K,\sin\theta_K\sin\varphi_K,\cos\theta_K)  
    \, ,
\end{align}
\end{subequations}
The azimuthal angle $\varphi_K$ is independent of time due to 
 the residual rotational symmetry of the background in $xy$\nobreakdash-plane.
Following the same arguments as in \Sect{gravitysec}, we arrive at the kinetic 
equations for diagonal components
\begin{subequations}
\label{eq:kin_bj}
\begin{align}
\label{eq:kin_N++}
\partial_\tau N_{\pm\pm}
=& -\frac{4}{3}\gamma_\eta K^2
\left[N_{\pm\pm} - \frac{T(e_0+p_0)}{\tau} \right] 
-\frac{1}{\tau}\left(2+c_{s}^2+\cos^2\theta_K \right)N_{\pm\pm},\\
\label{eq:kin_N11}
\partial_\tau N_{T_1T_1}
=& -2\gamma_\eta K^2 
\left[N_{T_1T_1} - \frac{T(e_0+p_0)}{\tau}\right]
-\frac{2}{\tau}N_{T_1T_1},\\
\label{eq:kin_N22}
\partial_\tau N_{T_2T_2}
=& -2\gamma_\eta K^2
\left[N_{T_2T_2} - \frac{T(e_0+p_0)}{\tau}\right] 
-\frac{2}{\tau}\left(1+\sin^2\theta_K\right)N_{T_2T_2}.
\end{align}
\end{subequations}
The first terms on the right hand side describe relaxation of $N_{AA}$ toward  
local equilibrium $T(e_0+p_0)/\tau$, and the second terms drive $N_{AA}$ out of 
equilibrium through the interaction with the background flow.

We derived these equations relying on the scale separation given in
\Eq{scalesep2}.  The off-diagonal components between gapped modes (such as 
between 
the $\pm$ and $T_1$ and $T_2$ modes) are ignored because they rapidly rotate as 
discussed in \Sect{grav1}.
Note that the transverse mode $\phi_{T_1}$ is so chosen that it does not mix 
with the other modes.
This is possible because of the residual rotational symmetry in the $xy$-plane 
in the Bjorken expansion.
Therefore the kinetic equation for $N_{T_1T_1}$, \Eq{eq:kin_N11}, holds without 
the scale separation in \Eq{scalesep2} and is applicable for all wavenumbers 
$k$ from to zero to $1/\ell_{\rm mfp}$.

\subsection{Nonlinear fluctuations in the energy momentum tensor}
\label{bjstress}

\begin{figure}
\centering
\includegraphics{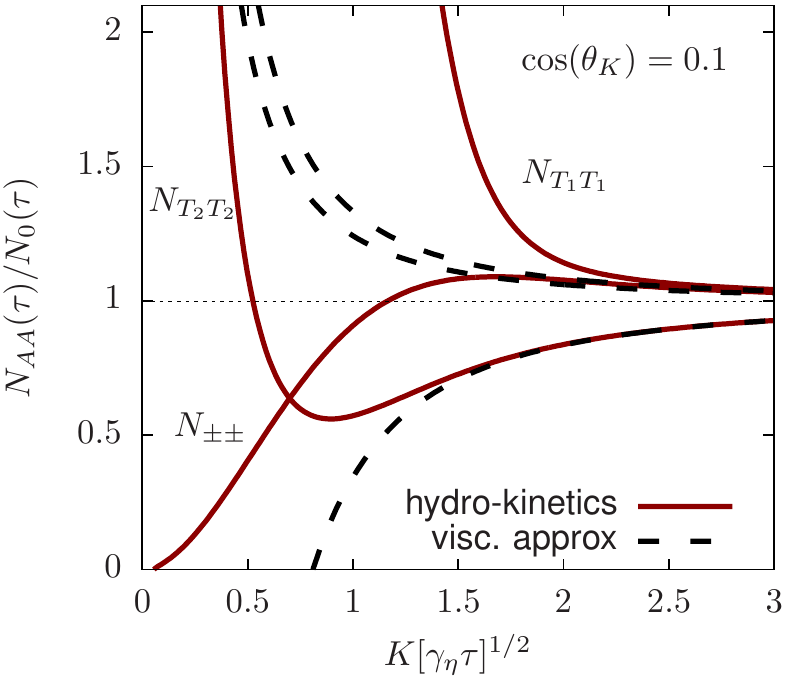}
\caption{Steady state solutions of \Eq{eq:kin_bj} for the two point 
energy-momentum correlation functions during a Bjorken 
expansion at late times, $\tau\gg \tau_0$. The correlations are plotted 
as a function of $K[\gamma_\eta \tau]^{1/2}$
for final time angle
$\cos\theta_K=0.1$.
For comparison leading order viscous solutions in $1/(\gamma_\eta K^2 \tau)$
are also shown, \Eq{eq:NAB_asmpt}. The differences of the steady state 
solutions from their asymptotic forms 
induces finite corrections 
to energy-momentum tensor, \Eq{eq:Tmunu_final}. }
\label{fig:NAA}
\end{figure}

Now let us investigate the solution of the kinetic equations close to the 
cutoff and isolate the UV divergent contribution. Solving \Eq{eq:kin_bj} in 
series of $1/(\gamma_\eta K^2\tau)$ we obtain an asymptotic solution for large 
$K/k_*$
\begin{subequations}
\label{eq:NAB_asmpt}
\begin{align}
\label{eq:N++_asmpt}
\frac{N_{\pm\pm}(\tau,\k)}{T(e_0+p_0)/\tau}&= 
1+ \frac{c_s^2 - \cos^2\theta_K}{\frac{4}{3}\gamma_{\eta} K^2\tau} 
+\ldots,\\
\label{eq:N11_asmpt}
\frac{N_{T_1T_1}(\tau,\k)}{T(e_0+p_0)/\tau}&=
1+ \frac{c_s^2}{\gamma_{\eta} K^2\tau}  +\ldots,\\
\label{eq:N22_asmpt}
\frac{N_{T_2T_2}(\tau,\k)}{T(e_0+p_0)/\tau}&=
1+ \frac{c_s^2 - \sin^2\theta_K}{\gamma_{\eta} K^2\tau} 
+ \ldots,
\end{align}
\end{subequations}
where we used $\partial_{\tau}[T(e_0+p_0)]\simeq 
-(1+2c_{s}^2)[T(e_0+p_0)]/\tau$ 
which is adequate for the desired accuracy of the present analysis.
For a given $K^2\gamma_\eta \tau = (K/k_*)^2$ and $\theta_K$ at final time 
$\tau$, we can solve \Eq{eq:kin_bj} numerically and find a steady state 
solution at late time 
$\tau \gg \tau_0$. We compare this steady state solution to the 
asymptotic form \Eq{eq:NAB_asmpt} in \Fig{fig:NAA}.

\Eq{eq:NAB_asmpt} is analogous to the ideal and first  viscous correction to 
the thermal distribution function, $f_0 + \delta f$, which are used in heavy ion
phenomenology and in determining the shear viscosity~\cite[see for example:][]{Teaney:2009qa}. At
large $K/k_*$ the  distribution $N_{AA}$ attains its equilibrium value,
$T(e_0 + p_0)/\tau$, up to viscous corrections of order $\tau_R/\tau$, where 
$\tau_{R}$ is a typical relaxation time for a mode of momentum $K$, $\tau_R 
\sim  1/\gamma_\eta K^2$.

The energy-momentum tensor averaged over fluctuations is given by
\begin{subequations}
\label{eq:Tmunu_tot}
\begin{align}
\label{eq:Ttt_tot}
\langle T^{\tau\tau}\rangle
&= e_0 + \frac{\langle \vec G^2\rangle}{e_0+p_0}, \\
\label{eq:Txx_tot}
\langle T^{xx}\rangle
&= p_0 + \frac{2\eta_0}{3\tau} + \frac{\langle (G^{\hat x})^2\rangle}{e_0+p_0}, 
\\
\label{eq:Tyy_tot}
\langle T^{yy}\rangle
&= p_0 + \frac{2\eta_0}{3\tau} + \frac{\langle (G^{\hat 
y})^2\rangle}{e_0+p_0},  \\
\label{eq:Tzz_tot}
\langle \tau^2 T^{\eta\eta}\rangle
&=p_0-\frac{4\eta_0}{3\tau} + \frac{\langle (G^{\hat z})^2\rangle}{e_0+p_0}.
\end{align}
\end{subequations}
Calculating $N_{ab}(\tau,\k) = \left[(e_A)_a N_{AB} (e_B)_b\right]$ from 
the kinetic theory, we determine 
$\langle\phi_a(\tau,\k)\phi_b(\tau,-\k)\rangle$ with 
$\phi_a = \left(c_s\delta e, \vec G\right)$ in Fourier 
space, yielding
\begin{align}
\langle\phi_a(x)\phi_b(x)\rangle
&={\int} \frac{d^2 k_{\perp}d\kappa}{(2\pi)^3} N_{ab}(\tau,\k), \nonumber \\
&= {\tau}{\int} \frac{K^2 dK d\cos \theta_K d\varphi_K}{(2\pi)^3}  
(e_A)_a N_{AB}(\tau,\k) (e_B)_{b} \label{eq:NLfluct}\, .
\end{align}
Note that the momentum integral is done in final time variables, 
$\vec{K}(\tau)$.
As shown below, the integration in Fourier space is divergent in the 
ultraviolet.
Therefore, we  regulate the integral by introducing a cutoff at $|\vec K|\sim 
\Lambda$.
In turn, the background quantities such as $e_0$ and $\eta_0$ must be 
renormalized and depend on $\Lambda$ so that the total result is independent of 
$\Lambda$.
The choice of $\Lambda$ is arbitrary as long as $k_{*} \ll \Lambda \ll 
1/\ell_{\rm  mfp}$ so that the non-linear contribution with  $|\vec K|\sim 
\Lambda$ is independent of the background flow.

The integration in \Eq{eq:NLfluct} includes the soft fluctuations for which the 
kinetic equation may not be applicable.
However, this contribution is suppressed by phase space and
the kinetic result can be extrapolated into this regime with negligible errors.

Combining Eqs.~\eqref{eq:NAB_asmpt}, \eqref{eq:Tmunu_tot} and 
\eqref{eq:NLfluct},  the energy momentum tensor is 
obtained as 
\begin{subequations}
\begin{align}
\langle T^{\tau\tau}\rangle  &= e_0 + 3T \int_{0}^{\Lambda}\frac{K^2dK}{2\pi^2} 
+ \Delta T^{\tau\tau}, \\
\langle T^{xx}\rangle  &= p_0 + \frac{2\eta_0}{3\tau}
+ T 
\int_{0}^{\Lambda}\frac{K^2dK}{2\pi^2}+\frac{17}{90}\frac{T(e_0+p_0)}{\eta_0\tau}
\int_{0}^{\Lambda}\frac{dK}{2\pi^2}  + \Delta T^{xx},  \\
\langle T^{yy}\rangle  &= p_0 + \frac{2\eta_0}{3\tau}
+ T \int_{0}^{\Lambda}\frac{K^2dK}{2\pi^2} 
+\frac{17}{90}\frac{T(e_0+p_0)}{\eta_0\tau}
\int_{0}^{\Lambda}\frac{dK}{2\pi^2}  + \Delta T^{yy},  \\
\langle \tau^2 T^{\eta\eta}\rangle  &= p_0 -\frac{4\eta_0}{3\tau}
+T\int_0^{\Lambda}\frac{K^2dK}{2\pi^2} 
-\frac{17}{45}\frac{T(e_0+p_0)}{\eta_0\tau}
\int_0^{\Lambda}\frac{dK}{2\pi^2} + \tau^2 \Delta T^{\eta\eta},
\end{align}
\end{subequations}
where the finite contributions $\Delta T^{\tau\tau}$, $\Delta T^{xx}$, $\Delta 
T^{yy}$, and $\tau^2\Delta T^{\eta\eta}$ are discussed in the next section.
By comparing terms with the same explicit $\tau$ dependence, the ultraviolet 
divergences are absorbed into the renormalized hydrodynamic variables
\begin{subequations}
\label{eq:renormalization}
\begin{align}
e &= e_0(\Lambda) + \frac{T\Lambda^3}{2\pi^2}, \\
p &= p_0(\Lambda) + \frac{T\Lambda^3}{6\pi^2}, \\
\eta &= \eta_0(\Lambda) + 
\frac{17\Lambda}{120\pi^2}\frac{T(e_0(\Lambda)+p_0(\Lambda))}{\eta_0(\Lambda)}.
\end{align}
\end{subequations}
Note that we do not assign a cut-off dependence to the temperature. 
The coefficients of the cubic and linear renormalizations of the pressure
and shear viscosity are independent of the background expansion, and match 
the static fluid results of \Sect{grav2}.
Here $e$, $p$, and $\eta$ are physical quantities at a given temperature $T$ in 
an infinite volume.
Using the physical quantities, the energy-momentum tensor is given as
\begin{subequations}
\label{renTij}
\begin{align}
\langle T^{\tau\tau}(\tau)\rangle &=e + \Delta T^{\tau\tau},\\
\langle T^{xx}(\tau)\rangle &=p + \frac{2\eta}{3\tau} + \Delta T^{xx},\\
\langle T^{yy}(\tau)\rangle &=p + \frac{2\eta}{3\tau} + \Delta T^{yy},\\
\langle \tau^2T^{\eta\eta}(\tau)\rangle  &=p - \frac{4\eta}{3\tau} + 
\tau^2\Delta T^{\eta\eta}.
\end{align}
\end{subequations}
If the two-point functions of the fluctuations were completely determined by 
the first two terms in \Eq{eq:NAB_asmpt}, their contributions would be 
completely absorbed by 
the renormalization of the background flow parameters such as $p_0(\Lambda)$ 
and $\eta_0(\Lambda)$.
However, the kinetic equations yield residual contributions, since the full 
solution deviates from its asymptotic form for $K \sim k_*$ as seen from 
\Fig{fig:NAA}. The purpose of hydrodynamics with noise is to capture this 
contribution.

Physically, the parameters $e_0(\Lambda)$, $p_0(\Lambda)$, and $\eta_0(\Lambda)$
in fluctuating hydrodynamics 
 reflect the equilibrium properties of modes above a cutoff $\Lambda$, which 
 have been already integrated out.
Equivalently, these parameters are determined by modes contained in a cell of 
size $a\sim 2\pi/\Lambda$.
For example, $p_0(\Lambda)$ is the partial pressure from equilibrated modes 
above the
cutoff (inside a cell), while the partial pressure from the modes below the 
cutoff
(larger than a cell size) is determined dynamically with fluctuating
hydrodynamics.
The second terms on the right hand sides of \Eq{eq:renormalization} are the 
contributions to each quantity from the modes below the cutoff, when all
of these long wavelength
modes are in perfect equilibrium in infinite volume.

\subsection{Out of equilibrium noise contributions to energy momentum tensor}
\label{sketch}
In this section we determine the residual contributions to the energy momentum 
tensor, $\Delta T^{\mu\nu}$, in \Eq{renTij} after the hydrodynamic parameters 
have been renormalized. We evaluate the precise numerical factors of the 
long-time tail terms for a 
Bjorken expansion (which is the main result of this paper), and identify 
additional contributions from  the noise at early times.
The mathematical procedure is somewhat involved, so here we outline the
calculation and present results, delegating the technical details to the 
Appendix \ref{app:finite}.

To find the full out of equilibrium correlators  we need to 
solve  \Eq{eq:kin_bj}, which can be written in the following general form
\begin{equation}\label{eq:formal}
\partial_\tau N_{AA}(\tau,\k) = f(\tau,\k) N_{AA}(\tau,\k) + g(\tau,\k),
\end{equation}
where  $f(\tau,\k)$ has contributions from both the dissipative and external 
forcing terms, and $g(\tau,\k)$ is the inhomogeneous term coming from 
the equilibrium correlation functions.
A formal solution of \Eq{eq:formal} is given by
\begin{align}
\label{eq:formalsol}
N_{AA}(\tau,\k) &= N_{AA}(\tau_0,\k) e^{\int_{\tau_0}^\tau d\tau' 
f(\tau',\k)}+\int_{\tau_0}^\tau d\tau'' g(\tau'',\k) e^{\int_{\tau_0}^{\tau''} 
d\tau' 
f(\tau',\k)}.
\end{align}
The first term describes the evolution of the initial correlation density
matrix $N_{AA}(\tau_0,\k)$ to final time $\tau$. 
The second term in \Eq{eq:formalsol} is  
the contribution from thermal fluctuations.
As we will see, only the $N_{T_1T_1}$ contribution is sensitive 
to the initial conditions and the thermal fluctuations at early times.
For the $T_1T_1$ correlator we will take the initial conditions described by 
\Eq{T1T1correlatorIC} in \Sect{icflucts},  $\tau_0^2 N_{T_1,T_1}(\tau_0,\k) = 
\chit \,2\pi \delta(\kappa)$.

Substituting the formal solution for $N_{AA}$ in \Eq{eq:Tmunu_tot} and 
\Eq{eq:NLfluct} 
we can determine the stress tensor at time $\tau \gg \tau_0$. 
 The integral $\int d^3k$ 
in \Eq{eq:NLfluct} diverges, but after 
subtracting $\Lambda^3$ and $\Lambda$ divergences  
discussed
in the previous section, we finally obtain 
the finite correction to energy momentum tensor $\Delta T^{\mu\nu}$. Writing 
\Eq{renTij} in full

\begin{subequations}
\label{eq:Tmunu_final}
\begin{align}
\label{eq:Tzz_finite}
\frac{\dlangle \tau^2 T^{\eta\eta}(\tau) \drangle}{e+p}
&= \frac{p}{e + p} - \frac{4 \gamma_\eta}{3\tau} 
  + \frac{1.08318}{s\, (4\pi \gamma_{\eta}\tau)^{3/2} }, \\
\label{eq:Txxyy_finite}
\frac{\dlangle T^{xx}(\tau) \drangle}{e+p}
&=  \frac{p}{e + p } + \frac{2 \gamma_\eta}{3\tau} + 
\left[ \frac{\chit{+}\delta \chit}{\tau^2 (e+p)^2} \right]  \frac{1}{(12\pi 
\gamma_\eta \tau)} 
-\,  \frac{0.273836}{s \; (4\pi \gamma_\eta \tau)^{3/2} },  \\
 \dlangle T^{yy} \drangle &= \dlangle T^{xx} \drangle, \\
\label{eq:Ttt_finite}
\dlangle T^{\tau\tau}\drangle &= \dlangle T^{xx} \drangle + \dlangle T^{yy} 
\drangle + \dlangle \tau^2 
T^{\eta\eta} \drangle.
\end{align}
\end{subequations}
The coefficients $1.08318$ and $-0.273836$ of the long-time tails, 
$1/(\gamma_\eta 
\tau)^{3/2}$ are obtained by numerical integration as explained in 
\app{app:finite}. The term 
$\chit+\delta \chit$ records the initial variance
in transverse momentum in a given rapidity slice (see \Eq{gabic} and 
\Eq{chi0pp}) 
together with 
the thermal contribution
\begin{equation}
\label{chi0ppcorrected}
\chit {+} \delta \chit = \llangle \frac{1}{\A} \left( 
\frac{dp^x}{d\eta}\right)^{\!2} \rrangle_{\!\!\tau_0} +  \left(\frac{T (e + p) 
\tau_0}{\sqrt{12\pi\gamma_\eta/\tau_0}} \right)_{\!\tau_0}  ,
\end{equation}
where the brackets $\left(\ldots \right)_{\tau_0}$ indicate
that all contained quantities are to be evaluated at the initial time, $\tau_0$.
We will provide an intuitive discussion of the result in the
next section.
\subsection{Qualitative discussion of \Eq{eq:Tmunu_final}}
\label{qualitative}
\subsubsection{Long time tails: $1/(\gamma_\eta \tau)^{3/2}$}
\label{qual_long_time}

Examining \Eq{eq:Tmunu_final} we see two groups of terms.  
The first group is proportional
to $1/(\gamma_\eta \tau)^{3/2}$ and  is
independent of initial conditions.  By contrast, the second group is 
proportional to
$1/(\gamma_\eta \tau)$, and depends on the  initial transverse momentum
fluctuations through the parameter $\chit + \delta \chit$ (see 
\Sect{icflucts}).  We will  first describe the terms
proportional to the fractional power $1/(\gamma_\eta \tau)^{3/2}$,  known 
as the long-time tails.

Squared fluctuations in equilibrium are of order $\langle\delta e(\vec 
x)\delta e(\vec y)\rangle_{\rm eq}/e^2 \sim \langle v^i(\vec x) v^j(\vec 
y)\rangle_{\rm eq}\sim  s^{-1}\delta(\vec x-\vec y)$, where $s$ is 
the entropy density (see \Eq{etcorrelators}).
Thus a fluctuation with wavenumber $k$ is suppressed by $\sqrt{k^3/s}$.
The suppression factor $k^3/s$ is roughly the inverse of the degrees of 
freedom inside a box of volume $ \Delta V  \sim (1/k)^3$, which must be a huge 
number for local thermodynamics to apply.
This is why the linear analysis of the hydrodynamic fluctuations is justified.

The energy momentum tensor in  viscous hydrodynamics is expanded  in powers of
gradients, leading to corrections in powers of $\epsilon\equiv
\eta/(e+p)\tau\ll1$. In addition, as discussed in \Sect{sec:kstar} the  
fluctuations with 
wavenumber of
order
$|\vec K|\sim k_*\sim 1/(\gamma_\eta\tau)^{1/2}$ dominate the nonlinear noise 
correction to the stress 
tensor, which 
is  suppressed by $s \Delta V \equiv
s/k_*^3\gg 1$.  
This correction to the longitudinal pressure reflects the equipartition of 
energy, with $\half T$ of energy per mode, and the number of non-equilibrium 
modes per volume $\sim k_{*}^3$.
To summarize, the reasoning in this paragraph leads to the following parametric 
estimate for the longitudinal stress  
\begin{eqnarray}
\frac{\langle\tau^2T^{\eta\eta}\rangle}{e+p}\sim 
\left[\frac{1}{4}+ \frac{\eta}{(e +p) \tau}  + \frac{1}{s (\gamma_\eta 
\tau)^{3/2}} 
+\cdots\right],
\end{eqnarray}
which is reflected  by  \Eq{eq:Tmunu_final}.

\subsubsection{ Transverse momentum diffusion in rapidity: $1/\gamma_{\eta} 
\tau$}
\label{diffusion_rapidity}

Additional corrections to the stress in \Eq{eq:Tmunu_final} decrease as 
$1/\gamma_\eta \tau$, in contrast to the long time tails.
As described in \Sect{icflucts}, long range (in rapidity) initial transverse 
momentum fluctuations 
correct the mean transverse pressures, $T^{xx}$ and $T^{yy}$,
by a term proportional to  $\chit/\gamma_\eta \tau$ (see \Eq{eq:Txxyy_finite}).
Hydrodynamic noise in the initial state adjusts this correction by adding to
the long range fluctuations of transverse momentum (see \Eqs{eq:Txxyy_finite} 
and \eq{chi0ppcorrected}).  The 
goal of this section is to explain this process qualitatively, and to 
quantitatively
explain the adjustment, $\chit \rightarrow \chit + \delta\chit$.

Formally, the $N_{T_1T_1}$ correlation function is 
sensitive to the noise at the initial time $\tau_0$, which arises
from a 
restricted region  of $\vec{K}$-space integration, $k_{\perp}\sim k_*$ 
and $ \kappa/\tau \sim k_{*}(\tau_0) (\tau_0/\tau) \sim k_*
(\tau_0/\tau)^{1/3}\ll k_*$. In this region the longitudinal momentum 
$\kappa/\tau_0$
reflects the dissipative scale $k_{*}(\tau_0)$ at the initial time $\tau_0$, 
while
the transverse momenta reflect the dissipative scale at final time $\tau$.

The dynamics in this phase space region is the following. During the initial
moments, thermal fluctuations lead to a local fluctuation of  transverse 
momentum in a given rapidity slice for each cell in the
transverse plane  
\begin{equation}
\llangle (\tau_0 \Delta g^x_\perp)^2 \rrangle \sim  \left(\frac{T\, (e +p) 
\,\tau_0 }{\Delta\eta \, (\Delta x_\perp)^2 } \right)_{\tau_0}.
\end{equation}
Here (as before) the brackets $(\ldots)_{\tau_0}$ indicate that all contained 
quantities should be evaluated at $\tau_0$.
During an initial time of order $\tau_0$, the momentum per rapidity diffuses to 
a finite 
longitudinal width~\cite{Gavin:2006xd} (see below)
\begin{equation}
\Delta \eta   \rightarrow   \sigma_{\eta}(\tau_0) \equiv 
\sqrt{6\gamma_\eta(\tau_0)/\tau_0} .
\end{equation}
The process is diffusive because the transverse momentum per rapidity is 
conserved.
The rapidity width is finite because the longitudinal expansion 
shuts off the diffusion process.
$\sigma_\eta(\tau_0)$ is broader than the rapidity width of subsequent 
interest, which is of order $\sigma_\eta(\tau)$. Thus, 
after an initial transient, the transverse momentum 
per rapidity may be considered approximately constant in time and rapidity, 
though localized in this  transverse plane
\begin{equation}
\llangle (\tau \Delta g^x_\perp)^2  \rrangle \sim   \left(\frac{T\, (e +p) 
\,\tau_0 }{\sqrt{\gamma_\eta/\tau_0} } \right)_{\tau_0} 
\frac{1}{(\Delta x_\perp)^2 }. 
\end{equation}
At much later times these transverse momentum fluctuations diffuse transversely 
(as described in \Sect{icflucts}) leading to a correction of order
\begin{equation}
\frac{ \llangle T^{xx} \rrangle }{e + p } \sim 
\frac{1}{\tau^2 (e + p)^2 } \left( \frac{T (e +p) \tau_0 
}{\sqrt{\gamma_\eta/\tau_0} } \right)_{\tau_0} \frac{1}{\gamma_\eta \tau} ,
\end{equation}
which qualitatively reproduces the correction in \Eq{eq:Txxyy_finite}.

Now we will briefly sketch this reasoning with equations.
At the early time moments $\tau \sim \tau_0$, the wave vector is predominantly 
longitudinal 
$\vec{K} \simeq (\vec{0}_\perp, \kappa/\tau)$ and the 
transverse momentum correlator
\begin{equation}
\llangle g^i_{\perp}(\tau,\k) g^{j}_{\perp}(\tau,-\k')\rrangle 
\equiv  N^{ij}(\k,\tau) (2\pi)^3 \delta^3(\k - \k') ,
\end{equation}
can be reconstructed from $N_{T_1T_1}$ and $N_{T_2T_2}$
\begin{equation}
N^{ij}(\tau,\k) = \sum_{A \in T_1, T_2} e_A^{i} e_A^{j} N_{AA}(\tau,\k),
\end{equation}
since $\vec{T}_1$ and $\vec{T}_2$ form a basis for the transverse plane.
In this limit, the equations of motion for $N_{T_1T_1}$ and $N_{T_2T_2}$ (see 
\Eq{eq:kin_N11} and \Eq{eq:kin_N22}) are the same,
and $N^{ij}$ satisfies a one dimensional  diffusion equation with a source at
early times
\begin{equation}
\label{longdiffuse}
\left( \partial_{\tau} +  2\gamma_{\eta} 
\left(\frac{\kappa}{\tau}\right)^2\right) (\tau^2 N^{ij} ) = 2\gamma_\eta 
\left(\frac{\kappa}{\tau}\right)^2 \, T(e+p)\tau \delta^{ij} \, .
\end{equation}
The lhs of \Eq{longdiffuse} represents the diffusion 
of transverse momentum
in rapidity,
 while the rhs represents the thermal transverse momentum fluctuations at
the earliest moments, which act as a source.
The source for the fluctuations, $2 T \eta \, (\kappa/\tau)^2$, 
is a rapidly decreasing function of time, and is dominant for times of order 
$\tau_0$.

The Green function 
propagating data from $\tau'$ to $\tau$ 
for the lhs of \Eq{longdiffuse} 
is  
\begin{align}
    G^{ij}(\tau\eta\xp|\tau'\eta'\xp') {=}
\frac{
e^{- (\eta - \eta')^2/(12 \gamma_\eta(\tau')/\tau')  }  
}{\sqrt{12\pi \gamma_\eta(\tau')/\tau'}} 
\delta^{ij} \delta^2(\xp - \xp'),
\end{align}
for $\tau \gg \tau'$.
Thus, a fluctuation localized in rapidity at time $\tau_0$ will diffuse to a 
finite rapidity width  of $\sigma_{\eta}(\tau_0)=\sqrt{6 
\gamma_\eta(\tau_0)/\tau_0}$  at late times\footnote{In 
\Refs{Gavin:2006xd,Gavin:2016hmv} the authors consider an initial 
distribution 
which is Gaussian in rapidity of width $\sigma_0$. During 
the expansion the width
is broadened by the diffusion process
\begin{equation}
\label{gavinresult}
\sigma^2_0 \rightarrow \sigma_0^2 + 6  \frac{\gamma_\eta(\tau_0) }{\tau_0}\,.
\end{equation}
These authors considered constant $\eta/(e + p)$ 
and found a factor of $4$ rather than $6$ in \Eq{gavinresult}.
}~\cite{Gavin:2006xd,Gavin:2016hmv}.
This 
is a small rapidity width in absolute units (since 
$\gamma_{\eta}(\tau_0)/\tau_0 \ll 1$ when hydrodynamics is a good 
approximation), but much broader than the rapidity
width of interest at the final time, $\gamma_\eta(\tau_0)/\tau_0 \gg 
\gamma_\eta(\tau)/\tau$.

Returning to \Eq{longdiffuse}, we solve the equation,
and determine the transverse momentum correlation function 
(in
the same rapidity slice)  
at an intermediate time $\tau'$ which 
is large compared to $\tau_0$ but much  
much less than the final time $\tau$, $\tau_0 \ll \tau' \ll \tau$
\begin{multline}
\tau'^2 \llangle g^{i}_\perp(\tau',\eta, \xp) g^{j}_\perp(\tau',\eta, \yp) 
\rrangle 
= \int  \frac{d\kappa d^2k_\perp}{(2\pi)^3 } e^{i\vec{k}_\perp \cdot (\xp - 
\yp)  
}
     \; \tau'^2 N^{ij}(\tau',\kappa) \, .
\end{multline}
Implementing these steps we find
\begin{multline}
\tau'^2 \llangle g^{i}_\perp(\tau',\eta,\xp) g^{j}_\perp(\tau',\eta, \yp) 
\rrangle =  
\left(\frac{T(e +p))\tau_0}{\sqrt{ 12\pi \gamma_\eta/\tau_0 
}}\right)_{\!\!\tau_0}\!\! \delta^{ij} \delta^2(\xp - \yp).
\end{multline}
This has the same form as the initial conditions described in \Sect{icflucts},
and fluctuations at the earliest moments simply increase the variance of
long range transverse momentum fluctuations by a constant amount
\begin{equation}
\delta \chit =  \left(\frac{T (e + p) \tau_0}{\sqrt{12\pi\gamma_\eta/\tau_0}} 
\right)_{\!\tau_0},
\end{equation}
reproducing \Eq{chi0ppcorrected}.
In a sense, this constant shift simply finalizes the thermalization process
described at the start of \Sect{icflucts}.  The correction $\delta \chit$ 
scales as
$\tau_0^{-1/3}$ and is therefore small compared to the first
term in \Eq{chi0ppcorrected} if $\tau_0$ is large compared to a typical 
thermalization time.

\section{Results and Discussion}
\label{results}
In this paper we determined a set of kinetic equations which describe
the evolution of hydrodynamic fluctuations during a Bjorken expansion.
We used these equations to
find the first fractional power correction 
to the longitudinal pressure, $\propto 1/(\tau T)^{3/2}$, at late times.
The evolution equations can be extended to much more general flows, 
and ultimately coupled to existing hydrodynamic codes.

The kinetic equations for hydrodynamic fluctuations 
are a WKB (or rotating wave) type approximation of the full stochastic 
hydrodynamic evolution
equations. This approximation
is justified because the relevant hydrodynamic modes have
wavenumbers of order
\begin{equation}
k_{*}  \sim \sqrt{ \frac{e  + p}{\eta \tau } } \, ,
\end{equation}
which is large compared to the inverse expansion rate, $1/\tau$. 
For example, the kinetic equation for the sound mode 
with wavenumber $\vec{K} = (\vec{k}_\perp , \kappa/\tau)$
interacting
with the Bjorken background takes the form of 
a relaxation type equation 
\begin{multline}
    \label{soundpp}
    \partial_\tau N_{++}(\tau,\k)
= -\frac{4}{3}\gamma_\eta K^2
\left[N_{++} - \frac{T(e_0+p_0)}{\tau} \right]  
-\frac{1}{\tau}\left(2+c_{s}^2+ \cos\theta_K \right)N_{++} \, .
\end{multline}
$N_{++}(\tau,\k)$ are  short wavelength (symmetrized) two point
functions of conserved stress tensor components, $\phi_+ \equiv (c_s \delta e + 
\hat{K} 
\cdot \vec{G})/\sqrt{2}$ in an evolving Bjorken hydrodynamic background (see 
\Sect{bjorken} and \Eq{eq:kin_bj} for the remaining modes).
At high wavenumbers $K \gg k_{*}$, the distribution function $N_{++}$ reaches
its equilibrium form $T (e_0+ p_0)/\tau$, up to first viscous
corrections  which may be found by solving \Eq{soundpp} order by order at large
$K/k_*$ (see \Eq{eq:N++_asmpt}).   This asymptotic 
form is responsible for the renormalization of the pressure and shear viscosity.
For wavenumbers of order $k_{*}$ the hydrodynamic
fluctuations are not in equilibrium at all, but reach a non-equilibrium steady
state at late times.  A graph of this non-equilibrium steady state is given in 
\Fig{fig:NAA}.  

The deviation of hydrodynamic fluctuations from equilibrium has
consequences for the evolution of the system. Indeed, the 
longitudinal pressure $\tau^2 T^{\eta\eta}$ receives a correction
from the unequilibrated modes 
\begin{multline}
    \label{tzzrepeat2}
\frac{\llangle \tau^2T^{\eta\eta}\rrangle }{e + p} =   \Big[ \frac{p}{e+p} -  
    \frac{4}{3} \frac{\gamma_\eta}{\tau}  
+     \frac{1.08318}{s \, (4\pi \gamma_\eta \tau)^{3/2} }   +   
\frac{(\lambda_1 - \eta\tau_\pi)}{e+p} \frac{8}{9\tau^2} \Big] \, ,
\end{multline}
where we have repeated \Eq{eq:Tzz_finite} for convenience.  The 
correction to the pressure ${\sim}T/(\gamma_\eta \tau)^{3/2}$ is of order 
${\sim}T k_{*}^3$, reflecting
the number of modes of order $k_{*}$ and the energy 
per mode, $\half T$.
In contrast to all previous analyses of long-time 
tails~\cite{Kovtun:2003vj,Kovtun:2011np},
the hydrodynamic fluctuations in the expanding case are not close to 
equilibrium, and a one loop expansion around equilibrium is not an 
appropriate approximation scheme. Our kinetic description effectively resums 
all diagrams contributing at the same order in the presence of 
expansion~\cite{Jeon:1995zm}.

Formally, the noise correction  is lower order than the correction due 
to second order hydrodynamics, which is proportional to a particular 
combination of second order parameters, $\lambda_1 - \eta \tau_\pi$. To 
quantify the importance 
of thermal fluctuations in practice,
we take representative numbers 
for the entropy from
the lattice~\cite{Borsanyi:2013bia,Bazavov:2014pvz}, estimates 
for the second order hydrodynamic coefficients
based on weakly and strongly coupled 
plasmas~\cite{York:2008rr,Bhattacharyya:2008jc,Baier:2007ix},
and an estimate for $\tau T$ at $\tau\sim 3.5\,{\rm fm}$ based on hydrodynamic
simulations\footnote{We take an estimate for the (approximately constant)
average entropy in the transverse plane from a recent LHC simulation for PbPb
collisions at $\sqrt{s}=2.76\,{\rm TeV}/{\rm nucleon}$, $\llangle \tau_{o}
s(\tau_o) \rrangle \simeq 4.0\,{\rm GeV}^2$~\cite{Mazeliauskas:2015vea}.  We 
take a time of $\tau \sim 3.5\,{\rm fm}$ (which is the time at which the 
elliptic flow develops~\cite[see
for example:][]{Teaney:2009qa}), where $T\simeq 250\,{\rm MeV}$. } 
\begin{subequations}
\begin{align}
    \frac{T^3}{s} \simeq&  \frac{1}{13.5}  \,, \\
    \frac{(\lambda_1 - \eta\tau_\pi)}{e+p} \simeq& -0.8 \left(\frac{\eta}{e+p} 
    \right)^2  \,, \\
     \tau T  \simeq 4.5 \, .
\end{align}
\end{subequations}
Then, for $\eta/s \simeq 1/4\pi$, \Eq{tzzrepeat2} evaluates to
\begin{multline}
    \label{oneby4pi}
\frac{ \llangle \tau^2 T^{\eta\eta} \rrangle }{e+ p} = \frac{1}{4} \Big[
    1.-0.092\,\left(\frac{4.5}{\tau T}\right)+0.034\,\left(\frac{4.5}{\tau 
    T}\right)^{3/2}
-0.00085 \left(\frac{4.5}{\tau T}\right)^{2} \Big],
\end{multline}
while for $\eta/s = 2/4\pi$, we find
\begin{multline}
    \label{twoby4pi}
\frac{ \llangle \tau^2 T^{\eta\eta} \rrangle }{e+ p} = \frac{1}{4} \Big[
    1.-0.185 \left(\frac{4.5}{\tau T}\right) + 0.013\left(\frac{4.5}{\tau 
    T}\right)^{3/2} 
   -0.0034\left(\frac{4.5}{\tau T}\right)^{2} \Big].
\end{multline}
For the smaller shear viscosity, \Eq{oneby4pi}, 
the 
nonlinear noise contribution completely dominates over
the second order hydro contribution.    For the larger shear
viscosity, \Eq{twoby4pi}, the noise remains three times larger
than  second order hydro, but this contribution is 
 only a $\sim 10\%$ of the first order viscous term. Finally, for $\eta/s \sim 
 3/4\pi$ the noise and second order
 hydro contributions become comparable.

The evolution of the average energy density of the system obeys
\begin{equation}
\frac{d \dlangle T^{\tau\tau} \drangle}{d\tau} = -\frac{ \dlangle 
T^{\tau\tau}\drangle  +  \dlangle \tau^2T^{\eta\eta} \drangle}{\tau},\ \ \
\end{equation}
where the double brackets notate an average over (long range in rapidity)
initial conditions and thermal noise\footnote{The
    longitudinal pressure in \Eq{tzzrepeat2} is independent of fluctuations in
    the initial conditions at late times. Thus, only 
    the average over the noise is relevant in this case,
$\dlangle \tau^2 T^{\eta\eta} \drangle = \llangle \tau^2 T^{\eta\eta} \rrangle$ 
}. To close the system of equations, the relationship between average energy 
density $\dlangle T^{\tau\tau} \drangle$ and the average rest frame energy  
density $e(\tau)$ must be specified, and this relation is given in 
\Eq{eq:Tmunu_final}.  
$T^{\tau\tau}$, $T^{xx}$,  and $T^{yy}$  are sensitive
to hydrodynamic noise at the earliest moments in addition
to the long-time tails.
In these cases thermal noise in the initial state
adds to the long-range  rapidity correlation functions of transverse momentum, 
which are already present without noise. 
This result is encapsulated by \Eq{chi0ppcorrected} and is discussed in
\Sect{icflucts} and \Sect{diffusion_rapidity}.

Although the analysis of hydrodynamic fluctuations in this paper was limited to 
conformal neutral fluids and a Bjorken expansion, the techniques developed here 
can be  applied to much 
more general flows. A next step is to generalize the kinetic equations in 
\Eq{eq:kin_bj} to an arbitrary expansion, and to couple such generalized 
equations to existing second order hydrodynamic codes.
In addition, it will be phenomenologically important to extend this work to 
non-conformal systems with net baryon number. Near the QCD critical point the 
noise will continue to grow without
bound, leading  to a critical renormalization of the  bulk
viscosity. In an expanding system these fluctuations will not be fully
equilibrated. We believe the formalism set up in this paper provides the first
steps towards quantitatively analyzing this rich dynamical regime.

\begin{subappendices}

\section{Computation of finite residual contributions}
\label{app:finite}
In this appendix we provide the details of the computation sketched in 
\Sect{sketch} 
for the residual out of equilibrium noise contribution  to the energy momentum 
tensor 
for a Bjorken background.
Let us scale the correlation density matrix  by the equilibrium 
value:
\begin{align}
R_{AA}(\tau,\k)\equiv \frac{N_{AA}(\tau,\k)}{T(e_0+p_0)/\tau}.
\end{align}
The kinetic equations of motion \Eq{eq:kin_bj} written for relative density 
matrix $R_{AA}$ are
\begin{align}
\partial_{\tau}R_{\pm\pm}
&= -\frac{4}{3}\gamma_{\eta}K^2
\left(R_{\pm\pm} -1\right)
+\frac{c_s^2-\cos^2\theta_K}{\tau}R_{\pm\pm},\\
\partial_{\tau}R_{T_1T_1}
&= -2\gamma_{\eta}K^2 \left(R_{T_1T_1} - 1\right)
+\frac{2c_s^2}{\tau}R_{T_1T_1},\\
\partial_{\tau}R_{T_2T_2}
&= -2\gamma_{\eta}K^2 \left(R_{T_2T_2} - 1\right)
+\frac{2(c_s^2-\sin^2\theta_K)}{\tau} R_{T_2T_2}.
\end{align}
Using dimensionless variables $t\equiv \tau'/\tau$ and $\vec r\equiv \vec 
K/k_*$ with $\vec K=(\vec k_{\perp},\kappa/\tau)$ and 
$k_*=1/(\gamma_\eta\tau)^{1/2}$ defined at $\tau$, the Green functions for 
the homogeneous parts are
\begin{align}
\label{eq:green_sound}
G_{\pm\pm}(\tau',\tau;\k)
&=\frac{1}{t^{c_s^2}}\frac{1}{\sqrt{A(t,\theta_K)}}\exp\left[-\frac{4}{3}r^2 
B(t,\theta_K)\right], \\
\label{eq:green_t1}
G_{T_1T_1}(\tau',\tau;\k)
&=\frac{1}{t^{2c_s^2}}\exp\left[
-2r^2 B(t,\theta_K)\right],\\
\label{eq:green_t2}
G_{T_2T_2}(\tau',\tau;\k)
&= t^{2-2c_s^2}A(t,\theta_K)
\exp\left[-2r^2B(t,\theta_K)\right],
\end{align}
where
\begin{align}
A(t,\theta_K)&\equiv\sin^2\theta_K + \frac{\cos^2\theta_K}{t^2}, \\
B(t,\theta_K)&\equiv
\frac{\sin^2\theta_K}{1+c_s^2}
\left(1-t^{1+c_s^2}\right)
+\frac{\cos^2\theta_K}{1-c_s^2}
\left(\frac{1}{t^{1-c_s^2}} -1\right).
\end{align}
With these Green functions, $R_{AA}$ due to thermal fluctuations (in contrast 
to initial fluctuations discussed in \Sect{icflucts}) is given by
\begin{align}
\label{RppGreen}
R_{++}(\tau,\k)&=
 \int_{\tau_0}^{\tau}d\tau' \frac{4}{3}\gamma_{\eta}(\tau')
\left(k_{\perp}^2+\frac{\kappa^2}{\tau'^2}\right)G_{++}(\tau',\tau;\k),
\end{align}
and similarly for the other modes (change 4/3 to 2 for the 
transverse modes). 
Since the asymptotic solution of $R_{AA}$ for large $K$ is known, we define the 
remainder of $R_{AA}$ as:
\begin{align}
\label{eq:rest_sound}
R^{(\rm r)}_{\pm\pm}(\tau,\k) &\equiv R_{\pm\pm}(\tau,\k) - \left(1 + 
\frac{c_s^2 - \cos^2\theta_K}{\frac{4}{3}\gamma_{\eta} K^2\tau}\right), \\
\label{eq:rest_t1}
R^{(\rm r)}_{T_1T_1}(\tau,\k) &\equiv R_{T_1T_1}(\tau,\k) - \left(1+ 
\frac{c_s^2}{\gamma_{\eta} K^2\tau}\right), \\
\label{eq:rest_t2}
R^{(\rm r)}_{T_2T_2}(\tau,\k) &\equiv R_{T_2T_2}(\tau,\k) - \left(1+ 
\frac{c_s^2 - \sin^2\theta_K}{\gamma_{\eta} K^2\tau}\right).
\end{align}

Using $R_{AA}^{(\rm r)}$ the residual contribution to the energy-momentum 
tensor is calculated from \Eq{eq:Ttt_tot} and \Eq{eq:NLfluct} as 
\begin{align}
\label{eq:DeltaTxx}
\Delta T^{xx} &= T
\int\frac{d^3K}{(2\pi)^3}\left[
\begin{aligned}
&\frac{R^{(\rm r)}_{++}+R^{(\rm r)}_{--}}{2} \sin^2\theta_K\cos^2\varphi_K \\
&+ R^{(\rm r)}_{T_1T_1}\sin^2\varphi_K \\
&+ R^{(\rm r)}_{T_2T_2}\cos^2\theta_K\cos^2\varphi_K 
\end{aligned}
\right],
\end{align}
\begin{align}
\label{eq:DeltaTyy}
\Delta T^{yy} &= T
\int\frac{d^3K}{(2\pi)^3}\left[
\begin{aligned}
&\frac{R^{(\rm r)}_{++}+R^{(\rm r)}_{--}}{2} \sin^2\theta_K\sin^2\varphi_K \\
&+ R^{(\rm r)}_{T_1T_1}\cos^2\varphi_K \\
&+ R^{(\rm r)}_{T_2T_2}\cos^2\theta_K\sin^2\varphi_K 
\end{aligned}
\right],
\end{align}
\begin{align}
\label{eq:DeltaTzz}
\tau^2 \Delta T^{\eta\eta} &= T
\int\frac{d^3K}{(2\pi)^3}\left[
\begin{aligned}
&\frac{R^{(\rm r)}_{++}+R^{(\rm r)}_{--}}{2} \cos^2\theta_K \\
&+ R^{(\rm r)}_{T_2T_2}\sin^2\theta_K
\end{aligned}
\right],
\end{align}
\begin{align}
\label{eq:DeltaTtt}
\Delta T^{\tau\tau} &= \Delta T^{xx} + \Delta T^{yy} + \tau^2\Delta 
T^{\eta\eta}.
\end{align}
Substituting the subtracted solution $R^{(\rm r)}_{AA}$ into 
\eqref{eq:DeltaTxx}-\eqref{eq:DeltaTtt} and performing $r$ integration with a 
Gaussian cutoff  $\exp[-r^2k_*^2/\Lambda^2]$, we get
\begin{align}
&\frac{\left[\tau^2\Delta T^{\eta\eta}(\tau)\right]}{T(\tau) k_*^3}
= 
\frac{3\sqrt{\pi}}{8}\int_{-1}^1\frac{d(\cos\theta_K)}{4\pi^2}\int_{\tau_0/\tau\to
 0}^1 dt \nonumber \\
& \ \ \ \times\left(
\frac{\frac{4}{3}\cos^2\theta_K\sqrt{A(t,\theta_K)}}
{\left[\frac{4}{3}B(t,\theta_K) + k_*^2/\Lambda^2\right]^{5/2}}
+ \frac{2t^{2-c_{s0}^2}\sin^2\theta_K  A(t,\theta_K)^2}
{\left[2B(t,\theta_K) + k_*^2/\Lambda^2\right]^{5/2}}
\right) \nonumber \\
& \ \ \ -\left[\mathcal O(\Lambda^3) + \mathcal O(\Lambda)\right],
\end{align}
\begin{align}
&\frac{\left[\Delta T^{xx}(\tau)+\Delta T^{yy}(\tau)\right]}{T(\tau) k_*^3}
= 
\frac{3\sqrt{\pi}}{8}\int_{-1}^1\frac{d(\cos\theta_K)}{4\pi^2}\int_{\tau_0/\tau}^1
 dt \nonumber \\
& \ \ \ \ \ \ \times\left(
\frac{\frac{4}{3}\sin^2\theta_K\sqrt{A(t,\theta_K)}}
{\left[\frac{4}{3}B(t,\theta_K) + k_*^2/\Lambda^2\right]^{5/2}}\right. 
\nonumber \\
& \ \ \ \ \ \ \ \ \ \ \ \left. 
+ \ \frac{2t^{2-c_s^2}\cos^2\theta_K A(t,\theta_K)^2
+ 2t^{-c_s^2}A(t,\theta_K)}
{\left[2B(t,\theta_K) + k_*^2/\Lambda^2\right]^{5/2}}
\right) \nonumber \\
& \ \ \ \ \ -\left[\mathcal O(\Lambda^3) + \mathcal O(\Lambda)\right] 
\end{align}
The ultraviolet divergent terms $\mathcal O(\Lambda^3,\Lambda)$ are from the 
asymptotic form of $R_{AA}$ at large $K$ in 
\eqref{eq:rest_sound}-\eqref{eq:rest_t2}.
Near $t=1$,  $B(t,\theta_K)\simeq 1-t$ and the cutoff $\Lambda$ regulates the 
divergence in time integral.
To isolate the divergences, we perform the partial integration twice and pick 
up cubic and linear divergences from the surface terms at $t=1$.
The resultant divergences are precisely canceled by $\mathcal 
O(\Lambda^3,\Lambda)$ terms.

After subtracting the ultraviolet divergences at $t=1$ and doing $\cos 
\theta_K$ integral analytically, the remaining time integration has to be done 
numerically.  $R^{(\rm r)}_{T_1T_1}$ mode contribution to 
$T^{xx}$ and $T^{yy}$ is divergent in the limit $\tau \gg \tau_0$.
Since the analytic behavior of the integrand around $t\sim 0$ is known, we can 
explicitly subtract the part sensitive to early times from the integrand to 
extract remaining 
finite pieces for $R^{(\rm r)}_{T_1T_1}$ mode. Numerical integration results 
necessary to find 
finite 
stress tensor corrections in \Eqs{eq:DeltaTxx}-\eq{eq:DeltaTzz} are summarized 
in 
Table \ref{tab1}. 
Summing contributions 
from the different modes to the longitudinal and transverse components of 
energy 
momentum tensor gives the numerical coefficients $1.08318$ and $-0.273836$ as 
seen 
in \Eq{eq:Tmunu_final}.

\begin{table}[t]
\def\arraystretch{1.5}
$
\begin{array}{|c|c|c|}
\hline
R^{(\rm r)}_{AA} & (4\pi)^{-3/2}\int d^3r R^{(\rm r)}_{AA} & (4\pi)^{-3/2}\int 
d^3r \cos^2 
\theta_K R^{(\rm r)}_{AA}\\
\hline
R^{(\rm r)}_{\pm\pm} & -0.439511 & 0.021281 \\ 
R^{(\rm r)}_{T_1T_1} & -\frac{\pi}{3\sqrt{6}}\approx-0.427517 & -0.467513 \\ 
R^{(\rm r)}_{T_2T_2} & 1.402539 & 0.340636\\
\hline
\end{array}
$
\caption{\label{tab1} Numerical values of finite pieces of regularized $R^{(\rm 
r)}_{AA}$ integrals for energy momentum tensor corrections. For the special 
case of $\int d^3r R^{(\rm r)}_{T_1T_1}$ the remaining one dimensional time 
integral can be done 
analytically.} 
\end{table}
\end{subappendices}


\chapter{Conclusion}
\label{chap:conclusion}

Fluctuations in heavy ion collisions are the key experimental observables, and
therefore a sophisticated understanding of perturbations is 
necessary for the precision extraction of the bulk properties of the
Quark Gluon Plasma (QGP). 
In this dissertation I discussed my 
work~\cite{Mazeliauskas:2015vea,Mazeliauskas:2015efa,Keegan:2016cpi,Akamatsu:2016llw}
 on the evolution of initial, 
pre-equilibrium, and 
thermal fluctuations in heavy ion collisions.

In \Chap{chap:pca} I used  Principal Component Analysis (PCA) to 
study the initial fluctuations in 
hydrodynamic  
simulations of heavy ion collisions at the CERN Large Hadron Collider 
(LHC)
($\snn=2.76\,\text{TeV}$). In particular, I explicitly
showed that factorization breaking of momentum dependent flow harmonics 
$v_n(p_T)$ (where $n=0\text{--}5$) can be 
quantitatively characterized by the interplay of subleading flows, which have 
an  
intuitive physical interpretation. For  the $n=0,1,3$ harmonics the subleading 
flow is a response to the radial excitation of the $n$th order eccentricity 
$\varepsilon_n$. In contrast, for  the $n=2,4,5$ harmonics a nonlinear mixing 
plays 
an important 
part in generating  subleading components. For example, in the case of the 
$n=2$ 
harmonic 
in peripheral collisions, the large leading elliptic flow $v_2$ mixes with  the 
radial  flow $v_0$ fluctuations and produces a new subleading elliptic flow. 
Correctly 
identifying different sources of flow helps  finding the optimal geometric and 
nonlinear 
predictors for the  flow harmonics.
 The main message of this work is that the momentum dependent harmonic 
flow $v_n(p_T)$ should be thought of not as a single 
response to  a
coarse grained initial state quantity, like $\varepsilon_n$,  but a 
superposition of 
responses to various aspects of the initial state geometry. The tools described 
in \Chap{chap:pca} give the most systematic way to date of unraveling the map 
between initial state fluctuations and the final state flow.

The PCA is a systematic 
and 
efficient way of decomposing flow into its most significant components. 
Importantly, studying the two-particle 
correlation functions in 
terms of dominant principal components utilizes all 
correlation data. 
Because  
subleading flows are more sensitive to the finer radial structure of the 
initial 
geometry, the subleading flows are an additional way of constraining 
the heavy ion collision models with already measured 
data. Subleading flows could also help interpreting some flow features in 
the ultra-central Pb+Pb
collisions, which are not yet reproduced by simulations~\cite{Jia:2014jca}.
  The PCA is a universal statistical method and can be  used for disentangling  
  correlations 
in rapidity $\eta$, where our understanding of fluctuations is still 
developing~\cite{Bhalerao:2014mua}.  
Finally, because the full two-particle correlation matrix
 is very well described by just two or three principle components,
the correct physical interpretation and quantitative reproduction of 
the leading principal components is a thorough test of our understanding 
of the 
QGP expansion.

In \Chap{chap:ekt} I looked at the equilibration of initial state fluctuations 
at the early stages of heavy ion collisions using the framework of weakly 
coupled 
kinetic theory~\cite{Baier:2000sb,Arnold:2002zm}.  Kinetic 
theory at realistic values of the coupling constant 
$\alpha_s\approx0.26$ (which 
is outside the strict 
weak coupling regime) predicts phenomenologically 
feasible (i.e. short enough) equilibration time of a uniform 
background~\cite{Kurkela:2015qoa}. In 
this work, I demonstrated  that transverse perturbations in the initial 
state also approach hydrodynamics  at the same time. Furthermore, 
the entire 
kinetic pre-thermal evolution can be conveniently  expressed in 
Green functions which 
map transverse perturbations in the initial state to energy momentum tensor at 
the time when hydrodynamics becomes applicable. Importantly, because the 
pre-thermal evolution time is short, it is enough to study equilibration within 
a causal region which is much smaller than the total transverse nuclear 
geometry.

Effective 
kinetic theory provides a first principle QCD description of the QGP 
equilibration and naturally connects the weak coupling initial 
state models like IP-Glasma to hydrodynamics---a connection 
which was so far missing. However, at the relevant 
energy scales in heavy ion 
collisions the coupling constant $\alpha_s\sim 0.3$ is not in the 
asymptotically weak coupling regime $\alpha_s\ll1$ and the extrapolated kinetic 
theory results should be taken cautiously.
 The kinetic theory domain of validity can be extended by 
non-perturbative treatment of soft momentum physics, e.g. 
non-perturbative anisotropic 
screening~\cite{Kajantie:2000iz,Braaten:1995jr,Kajantie:2002wa}. In a 
tangential 
development, the current work can be 
straightforwardly 
extended to study the chemical equilibration of quarks and gluons, and 
the 
production of out-of-equilibrium 
thermal photons.

The rapid equilibration in 
heavy ion collisions is a poorly understood problem,  which is even 
more acute for small collision systems with very short lifetimes.
Although the exact nature of equilibration 
is yet to be seen, the short evolution time and causality constrain the 
possible forms of response functions. Therefore an effective 
description can perhaps be 
inferred from 
the interpolation between the weak coupling theory and models 
based on holography that describe equilibration at infinitely strong coupling 
in QCD-like theories~\cite{Keegan:2015avk}.
 In fact, out-of-equilibrium quantum 
systems demonstrate surprisingly 
universal behavior in very different 
settings~\cite{Berges:2014bba,Berges:2015kfa,Heller:2016rtz}. The 
kinetic theory response functions computed in this work represents a practical 
implementation of the ``bottom-up'' thermalization scenario and is the first 
work directly connecting ``bottom-up'' to current hydrodynamic models of heavy 
ion collisions.

Finally, in \Chap{chap:noise}  I addressed the physics of thermal fluctuations 
in the hydrodynamically expanding QGP by constructing the 
effective hydro\nobreakdash-kinetic equations for the two-point 
correlation function.
The rapid Bjorken expansion of a medium
drives the hydrodynamic fluctuations out of equilibrium prescribed by the 
fluctuation-dissipation theorem. The competition between the 
expansion and dissipation leads to a steady state solution  of hydro-kinetic 
equations.
The difference between the equilibrium correlations and the steady state 
solution determines the so called ``long time tail" 
corrections to 
the energy momentum tensor, however the 
exact numerical factors for Bjorken expanding background were calculated here 
for 
the first time. The separation of scales between the long wavelength background 
physics and the dominant out-of-equilibrium hydrodynamic fluctuations
gives an 
intuitive picture for the 
nonlinear noise contributions.

The estimated nonlinear 
noise corrections in heavy ion collisions are potentially as large or larger 
than second order terms 
used in the modern hydrodynamic models and the presented framework of 
hydro-kinetic 
equations is an alternative to numerically challenging stochastic hydrodynamic
simulations. More importantly, the hydro-kinetic equations  are a leading 
order effective kinetic theory of noise and can be used to give a perturbative 
description of noise away from the QCD critical point. The 
framework 
can be straightforwardly applied to general background flows and non-conformal 
systems with non-vanishing bulk viscosity.

In summary, in this dissertation I discussed and applied several 
methods of 
separating and 
propagating initial and thermal fluctuations in heavy ion collisions: 
Principal 
Component Analysis, kinetic theory equilibration, and hydro-kinetic 
equations 
of 
noise.
The high energy Pb+Pb and Au+Au collisions at LHC 
and RHIC are likely to remain the most reliable grounds for determining QCD 
medium properties, but collisions in small systems (e.g. p+Pb or high 
multiplicity p+p) exhibit surprisingly similar signals of 
collectivity.  Principal Component Analysis and kinetic theory 
equilibration 
are promising tools to study initial fluctuations in these systems. On the 
other hand, the Beam Energy 
Scan will search for the QCD critical point in the low energy  and high baryon 
chemical potential region of the QCD phase diagram.  Since one of the 
signatures of 
a 
critical point is large stochastic fluctuations,
coupling hydro-kinetic equations with the advanced framework of viscous 
hydrodynamics is an 
important stepping stone for the correct modeling of the QGP matter near the 
critical point. 
Fluctuations will remain an important topic 
in heavy ion physics, and  the  methods developed in this dissertation 
will be of high relevance to the heavy ion research into the many body physics 
of 
QCD for the foreseeable future. 

\nocite{apsrev41Control}
\bibliography{master}

\begin{thebibliography}{237}%
\makeatletter
\providecommand \@ifxundefined [1]{%
 \@ifx{#1\undefined}
}%
\providecommand \@ifnum [1]{%
 \ifnum #1\expandafter \@firstoftwo
 \else \expandafter \@secondoftwo
 \fi
}%
\providecommand \@ifx [1]{%
 \ifx #1\expandafter \@firstoftwo
 \else \expandafter \@secondoftwo
 \fi
}%
\providecommand \natexlab [1]{#1}%
\providecommand \enquote  [1]{``#1''}%
\providecommand \bibnamefont  [1]{#1}%
\providecommand \bibfnamefont [1]{#1}%
\providecommand \citenamefont [1]{#1}%
\providecommand \href@noop [0]{\@secondoftwo}%
\providecommand \href [0]{\begingroup \@sanitize@url \@href}%
\providecommand \@href[1]{\@@startlink{#1}\@@href}%
\providecommand \@@href[1]{\endgroup#1\@@endlink}%
\providecommand \@sanitize@url [0]{\catcode `\\12\catcode `\$12\catcode
  `\&12\catcode `\#12\catcode `\^12\catcode `\_12\catcode `\%12\relax}%
\providecommand \@@startlink[1]{}%
\providecommand \@@endlink[0]{}%
\providecommand \url  [0]{\begingroup\@sanitize@url \@url }%
\providecommand \@url [1]{\endgroup\@href {#1}{\urlprefix }}%
\providecommand \urlprefix  [0]{URL }%
\providecommand \Eprint [0]{\href }%
\providecommand \doibase [0]{http://dx.doi.org/}%
\providecommand \selectlanguage [0]{\@gobble}%
\providecommand \bibinfo  [0]{\@secondoftwo}%
\providecommand \bibfield  [0]{\@secondoftwo}%
\providecommand \translation [1]{[#1]}%
\providecommand \BibitemOpen [0]{}%
\providecommand \bibitemStop [0]{}%
\providecommand \bibitemNoStop [0]{.\EOS\space}%
\providecommand \EOS [0]{\spacefactor3000\relax}%
\providecommand \BibitemShut  [1]{\csname bibitem#1\endcsname}%
\let\auto@bib@innerbib\@empty
\bibitem [{\citenamefont {Mazeliauskas}\ and\ \citenamefont
  {Teaney}(2015)}]{Mazeliauskas:2015vea}%
  \BibitemOpen
  \bibfield  {author} {\bibinfo {author} {\bibfnamefont {A.}~\bibnamefont
  {Mazeliauskas}}\ and\ \bibinfo {author} {\bibfnamefont {D.}~\bibnamefont
  {Teaney}},\ }\bibfield  {title} {\enquote {\bibinfo {title} {{Subleading
  harmonic flows in hydrodynamic simulations of heavy ion collisions}},}\
  }\href {\doibase 10.1103/PhysRevC.91.044902} {\bibfield  {journal} {\bibinfo
  {journal} {Phys. Rev.}\ }\textbf {\bibinfo {volume} {C91}},\ \bibinfo {pages}
  {044902} (\bibinfo {year} {2015})},\ \Eprint
  {http://arxiv.org/abs/1501.03138} {arXiv:1501.03138 [nucl-th]} \BibitemShut
  {NoStop}%
\bibitem [{\citenamefont {Mazeliauskas}\ and\ \citenamefont
  {Teaney}(2016)}]{Mazeliauskas:2015efa}%
  \BibitemOpen
  \bibfield  {author} {\bibinfo {author} {\bibfnamefont {A.}~\bibnamefont
  {Mazeliauskas}}\ and\ \bibinfo {author} {\bibfnamefont {D.}~\bibnamefont
  {Teaney}},\ }\bibfield  {title} {\enquote {\bibinfo {title} {{Fluctuations of
  harmonic and radial flow in heavy ion collisions with principal
  components}},}\ }\href {\doibase 10.1103/PhysRevC.93.024913} {\bibfield
  {journal} {\bibinfo  {journal} {Phys. Rev.}\ }\textbf {\bibinfo {volume}
  {C93}},\ \bibinfo {pages} {024913} (\bibinfo {year} {2016})},\ \Eprint
  {http://arxiv.org/abs/1509.07492} {arXiv:1509.07492 [nucl-th]} \BibitemShut
  {NoStop}%
\bibitem [{\citenamefont {Keegan}\ \emph
  {et~al.}(2016{\natexlab{a}})\citenamefont {Keegan}, \citenamefont {Kurkela},
  \citenamefont {Mazeliauskas},\ and\ \citenamefont {Teaney}}]{Keegan:2016cpi}%
  \BibitemOpen
  \bibfield  {author} {\bibinfo {author} {\bibfnamefont {L.}~\bibnamefont
  {Keegan}}, \bibinfo {author} {\bibfnamefont {A.}~\bibnamefont {Kurkela}},
  \bibinfo {author} {\bibfnamefont {A.}~\bibnamefont {Mazeliauskas}}, \ and\
  \bibinfo {author} {\bibfnamefont {D.}~\bibnamefont {Teaney}},\ }\bibfield
  {title} {\enquote {\bibinfo {title} {{Initial conditions for hydrodynamics
  from weakly coupled pre-equilibrium evolution}},}\ }\href@noop {} {\bibfield
  {journal} {\bibinfo  {journal} {JHEP}\ }\textbf {\bibinfo {volume} {08}},\
  \bibinfo {pages} {171} (\bibinfo {year} {2016}{\natexlab{a}})},\ \Eprint
  {http://arxiv.org/abs/1605.04287} {arXiv:1605.04287 [hep-ph]} \BibitemShut
  {NoStop}%
\bibitem [{\citenamefont {Akamatsu}\ \emph {et~al.}(2017)\citenamefont
  {Akamatsu}, \citenamefont {Mazeliauskas},\ and\ \citenamefont
  {Teaney}}]{Akamatsu:2016llw}%
  \BibitemOpen
  \bibfield  {author} {\bibinfo {author} {\bibfnamefont {Y.}~\bibnamefont
  {Akamatsu}}, \bibinfo {author} {\bibfnamefont {A.}~\bibnamefont
  {Mazeliauskas}}, \ and\ \bibinfo {author} {\bibfnamefont {D.}~\bibnamefont
  {Teaney}},\ }\bibfield  {title} {\enquote {\bibinfo {title} {{A kinetic
  regime of hydrodynamic fluctuations and long time tails for a Bjorken
  expansion}},}\ }\href {\doibase 110.1103/PhysRevC.95.014909} {\bibfield
  {journal} {\bibinfo  {journal} {Phys. Rev.}\ }\textbf {\bibinfo {volume}
  {C95}},\ \bibinfo {pages} {014909} (\bibinfo {year} {2017})},\ \Eprint
  {http://arxiv.org/abs/1606.07742} {arXiv:1606.07742 [nucl-th]} \BibitemShut
  {NoStop}%
\bibitem [{\citenamefont {Heisenberg}(1958)}]{heisenberg1958physics}%
  \BibitemOpen
  \bibfield  {author} {\bibinfo {author} {\bibfnamefont {W.}~\bibnamefont
  {Heisenberg}},\ }\href {https://books.google.com/books?id=Jz8LAQAAIAAJ}
  {\emph {\bibinfo {title} {Physics and philosophy: the revolution in modern
  science}}},\ World perspectives\ (\bibinfo  {publisher} {Harper},\ \bibinfo
  {year} {1958})\BibitemShut {NoStop}%
\bibitem [{\citenamefont {Fermi}(1934)}]{fermi}%
  \BibitemOpen
  \bibfield  {author} {\bibinfo {author} {\bibfnamefont {E.}~\bibnamefont
  {Fermi}},\ }\bibfield  {title} {\enquote {\bibinfo {title} {Versuch einer
  theorie der $\beta$-strahlen.{I}},}\ }\href {\doibase 10.1007/BF01351864}
  {\bibfield  {journal} {\bibinfo  {journal} {Zeitschrift f{\"u}r Physik}\
  }\textbf {\bibinfo {volume} {88}},\ \bibinfo {pages} {161--177} (\bibinfo
  {year} {1934})}\BibitemShut {NoStop}%
\bibitem [{\citenamefont {Yukawa}(1935)}]{yukawa}%
  \BibitemOpen
  \bibfield  {author} {\bibinfo {author} {\bibfnamefont {H.}~\bibnamefont
  {Yukawa}},\ }\bibfield  {title} {\enquote {\bibinfo {title} {On the
  interaction of elementary particles. {I}},}\ }\href
  {https://www.jstage.jst.go.jp/article/ppmsj1919/17/0/17_0_48/_article}
  {\bibfield  {journal} {\bibinfo  {journal} {Proceedings of the
  Physico-Mathematical Society of Japan. 3rd Series}\ }\textbf {\bibinfo
  {volume} {17}},\ \bibinfo {pages} {48--57} (\bibinfo {year}
  {1935})}\BibitemShut {NoStop}%
\bibitem [{\citenamefont {Glashow}(1961)}]{Glashow:1961tr}%
  \BibitemOpen
  \bibfield  {author} {\bibinfo {author} {\bibfnamefont {S.~L.}\ \bibnamefont
  {Glashow}},\ }\bibfield  {title} {\enquote {\bibinfo {title} {{Partial
  Symmetries of Weak Interactions}},}\ }\href {\doibase
  10.1016/0029-5582(61)90469-2} {\bibfield  {journal} {\bibinfo  {journal}
  {Nucl. Phys.}\ }\textbf {\bibinfo {volume} {22}},\ \bibinfo {pages}
  {579--588} (\bibinfo {year} {1961})}\BibitemShut {NoStop}%
\bibitem [{\citenamefont {Salam}(1968)}]{Salam:1968rm}%
  \BibitemOpen
  \bibfield  {author} {\bibinfo {author} {\bibfnamefont {A.}~\bibnamefont
  {Salam}},\ }\bibfield  {title} {\enquote {\bibinfo {title} {{Weak and
  Electromagnetic Interactions}},}\ }\bibfield  {booktitle} {\emph {\bibinfo
  {booktitle} {{8th Nobel Symposium Lerum, Sweden, May 19-25, 1968}}},\
  }\href@noop {} {\bibfield  {journal} {\bibinfo  {journal} {Conf. Proc.}\
  }\textbf {\bibinfo {volume} {C680519}},\ \bibinfo {pages} {367--377}
  (\bibinfo {year} {1968})}\BibitemShut {NoStop}%
\bibitem [{\citenamefont {Weinberg}(1967)}]{Weinberg:1967tq}%
  \BibitemOpen
  \bibfield  {author} {\bibinfo {author} {\bibfnamefont {S.}~\bibnamefont
  {Weinberg}},\ }\bibfield  {title} {\enquote {\bibinfo {title} {{A Model of
  Leptons}},}\ }\href {\doibase 10.1103/PhysRevLett.19.1264} {\bibfield
  {journal} {\bibinfo  {journal} {Phys. Rev. Lett.}\ }\textbf {\bibinfo
  {volume} {19}},\ \bibinfo {pages} {1264--1266} (\bibinfo {year}
  {1967})}\BibitemShut {NoStop}%
\bibitem [{\citenamefont {Gell-Mann}(1964)}]{GellMann:1964nj}%
  \BibitemOpen
  \bibfield  {author} {\bibinfo {author} {\bibfnamefont {M.}~\bibnamefont
  {Gell-Mann}},\ }\bibfield  {title} {\enquote {\bibinfo {title} {{A Schematic
  Model of Baryons and Mesons}},}\ }\href {\doibase
  10.1016/S0031-9163(64)92001-3} {\bibfield  {journal} {\bibinfo  {journal}
  {Phys. Lett.}\ }\textbf {\bibinfo {volume} {8}},\ \bibinfo {pages} {214--215}
  (\bibinfo {year} {1964})}\BibitemShut {NoStop}%
\bibitem [{\citenamefont {Zweig}(1964{\natexlab{a}})}]{Zweig:1981pd}%
  \BibitemOpen
  \bibfield  {author} {\bibinfo {author} {\bibfnamefont {G.}~\bibnamefont
  {Zweig}},\ }\bibfield  {title} {\enquote {\bibinfo {title} {{An SU(3) model
  for strong interaction symmetry and its breaking. Version 1}},}\ }\href@noop
  {} {\  (\bibinfo {year} {1964}{\natexlab{a}})}\BibitemShut {NoStop}%
\bibitem [{\citenamefont {Zweig}(1964{\natexlab{b}})}]{Zweig:1964jf}%
  \BibitemOpen
  \bibfield  {author} {\bibinfo {author} {\bibfnamefont {G.}~\bibnamefont
  {Zweig}},\ }\bibfield  {title} {\enquote {\bibinfo {title} {{An SU(3) model
  for strong interaction symmetry and its breaking. Version 2}},}\ }in\ \href
  {http://inspirehep.net/record/4674/files/cern-th-412.pdf} {\emph {\bibinfo
  {booktitle} {DEVELOPMENTS IN THE QUARK THEORY OF HADRONS. VOL. 1. 1964 -
  1978}}},\ \bibinfo {editor} {edited by\ \bibinfo {editor} {\bibfnamefont
  {D.}~\bibnamefont {Lichtenberg}}\ and\ \bibinfo {editor} {\bibfnamefont
  {S.~P.}\ \bibnamefont {Rosen}}}\ (\bibinfo {year} {1964})\ pp.\ \bibinfo
  {pages} {22--101}\BibitemShut {NoStop}%
\bibitem [{\citenamefont {Burgess}\ and\ \citenamefont
  {Moore}(2007)}]{Burgess}%
  \BibitemOpen
  \bibfield  {author} {\bibinfo {author} {\bibfnamefont {C.~P.}\ \bibnamefont
  {Burgess}}\ and\ \bibinfo {author} {\bibfnamefont {G.~D.}\ \bibnamefont
  {Moore}},\ }\href
  {http://proxy.library.stonybrook.edu/login?url=http://search.ebscohost.com/login.aspx?direct=true&db=cat03000a&AN=STB.001650650&site=eds-live&scope=site}
  {\emph {\bibinfo {title} {The standard model : a primer}}}\ (\bibinfo
  {publisher} {Cambridge : Cambridge University Press},\ \bibinfo {year}
  {2007})\BibitemShut {NoStop}%
\bibitem [{\citenamefont {Langacker}(2010)}]{Langacker}%
  \BibitemOpen
  \bibfield  {author} {\bibinfo {author} {\bibfnamefont {P.}~\bibnamefont
  {Langacker}},\ }\href
  {http://proxy.library.stonybrook.edu/login?url=http://search.ebscohost.com/login.aspx?direct=true&db=cat03000a&AN=STB.001650649&site=eds-live&scope=site}
  {\emph {\bibinfo {title} {The standard model and beyond}}},\ Series in high
  energy physics, cosmology, and gravitation\ (\bibinfo  {publisher} {Boca
  Raton, FL : CRC Press},\ \bibinfo {year} {2010})\BibitemShut {NoStop}%
\bibitem [{\citenamefont {Patrignani}(2016)}]{Olive:2016xmw}%
  \BibitemOpen
  \bibfield  {author} {\bibinfo {author} {\bibfnamefont {C.}~\bibnamefont
  {Patrignani}},\ }\bibfield  {title} {\enquote {\bibinfo {title} {{Review of
  Particle Physics}},}\ }\href {\doibase 10.1088/1674-1137/40/10/100001}
  {\bibfield  {journal} {\bibinfo  {journal} {Chin. Phys.}\ }\textbf {\bibinfo
  {volume} {C40}},\ \bibinfo {pages} {100001} (\bibinfo {year}
  {2016})}\BibitemShut {NoStop}%
\bibitem [{\citenamefont {Peskin}\ and\ \citenamefont
  {Schroeder}(1995)}]{Peskin}%
  \BibitemOpen
  \bibfield  {author} {\bibinfo {author} {\bibfnamefont {M.~E.}\ \bibnamefont
  {Peskin}}\ and\ \bibinfo {author} {\bibfnamefont {D.~V.}\ \bibnamefont
  {Schroeder}},\ }\href
  {http://proxy.library.stonybrook.edu/login?url=http://search.ebscohost.com/login.aspx?direct=true&db=cat03000a&AN=STB.000926000&site=eds-live&scope=site}
  {\emph {\bibinfo {title} {Introduction to quantum field theory}}}\ (\bibinfo
  {publisher} {Reading, Mass. : Addison-Wesley Pub. Co., c1995.},\ \bibinfo
  {year} {1995})\BibitemShut {NoStop}%
\bibitem [{\citenamefont {Woodard}(2009)}]{Woodard:2009ns}%
  \BibitemOpen
  \bibfield  {author} {\bibinfo {author} {\bibfnamefont {R.~P.}\ \bibnamefont
  {Woodard}},\ }\bibfield  {title} {\enquote {\bibinfo {title} {{How Far Are We
  from the Quantum Theory of Gravity?}}}\ }\href {\doibase
  10.1088/0034-4885/72/12/126002} {\bibfield  {journal} {\bibinfo  {journal}
  {Rept. Prog. Phys.}\ }\textbf {\bibinfo {volume} {72}},\ \bibinfo {pages}
  {126002} (\bibinfo {year} {2009})},\ \Eprint {http://arxiv.org/abs/0907.4238}
  {arXiv:0907.4238 [gr-qc]} \BibitemShut {NoStop}%
\bibitem [{\citenamefont {Yang}\ and\ \citenamefont
  {Mills}(1954)}]{Yang:1954ek}%
  \BibitemOpen
  \bibfield  {author} {\bibinfo {author} {\bibfnamefont {C.-N.}\ \bibnamefont
  {Yang}}\ and\ \bibinfo {author} {\bibfnamefont {R.~L.}\ \bibnamefont
  {Mills}},\ }\bibfield  {title} {\enquote {\bibinfo {title} {{Conservation of
  Isotopic Spin and Isotopic Gauge Invariance}},}\ }\href {\doibase
  10.1103/PhysRev.96.191} {\bibfield  {journal} {\bibinfo  {journal} {Phys.
  Rev.}\ }\textbf {\bibinfo {volume} {96}},\ \bibinfo {pages} {191--195}
  (\bibinfo {year} {1954})}\BibitemShut {NoStop}%
\bibitem [{\citenamefont {Politzer}(1973)}]{Politzer:1973fx}%
  \BibitemOpen
  \bibfield  {author} {\bibinfo {author} {\bibfnamefont {H.~D.}\ \bibnamefont
  {Politzer}},\ }\bibfield  {title} {\enquote {\bibinfo {title} {{Reliable
  Perturbative Results for Strong Interactions?}}}\ }\href {\doibase
  10.1103/PhysRevLett.30.1346} {\bibfield  {journal} {\bibinfo  {journal}
  {Phys. Rev. Lett.}\ }\textbf {\bibinfo {volume} {30}},\ \bibinfo {pages}
  {1346--1349} (\bibinfo {year} {1973})}\BibitemShut {NoStop}%
\bibitem [{\citenamefont {Gross}\ and\ \citenamefont
  {Wilczek}(1973)}]{Gross:1973id}%
  \BibitemOpen
  \bibfield  {author} {\bibinfo {author} {\bibfnamefont {D.~J.}\ \bibnamefont
  {Gross}}\ and\ \bibinfo {author} {\bibfnamefont {F.}~\bibnamefont
  {Wilczek}},\ }\bibfield  {title} {\enquote {\bibinfo {title} {{Ultraviolet
  Behavior of Nonabelian Gauge Theories}},}\ }\href {\doibase
  10.1103/PhysRevLett.30.1343} {\bibfield  {journal} {\bibinfo  {journal}
  {Phys. Rev. Lett.}\ }\textbf {\bibinfo {volume} {30}},\ \bibinfo {pages}
  {1343--1346} (\bibinfo {year} {1973})}\BibitemShut {NoStop}%
\bibitem [{\citenamefont {Wilson}(1974)}]{Wilson:1974sk}%
  \BibitemOpen
  \bibfield  {author} {\bibinfo {author} {\bibfnamefont {K.~G.}\ \bibnamefont
  {Wilson}},\ }\bibfield  {title} {\enquote {\bibinfo {title} {{Confinement of
  Quarks}},}\ }\href {\doibase 10.1103/PhysRevD.10.2445} {\bibfield  {journal}
  {\bibinfo  {journal} {Phys. Rev.}\ }\textbf {\bibinfo {volume} {D10}},\
  \bibinfo {pages} {2445--2459} (\bibinfo {year} {1974})}\BibitemShut {NoStop}%
\bibitem [{\citenamefont {Mandelstam}(1976)}]{Mandelstam:1974pi}%
  \BibitemOpen
  \bibfield  {author} {\bibinfo {author} {\bibfnamefont {S.}~\bibnamefont
  {Mandelstam}},\ }\bibfield  {title} {\enquote {\bibinfo {title} {{Vortices
  and Quark Confinement in Nonabelian Gauge Theories}},}\ }\bibfield
  {booktitle} {\emph {\bibinfo {booktitle} {{Phys. Rep. 23 (1976) 245-249, In
  *Gervais, J.L. (Ed.), Jacob, M. (Ed.): Non-linear and Collective Phenomena In
  Quantum Physics*, 12-16}}},\ }\href {\doibase 10.1016/0370-1573(76)90043-0}
  {\bibfield  {journal} {\bibinfo  {journal} {Phys. Rept.}\ }\textbf {\bibinfo
  {volume} {23}},\ \bibinfo {pages} {245--249} (\bibinfo {year}
  {1976})}\BibitemShut {NoStop}%
\bibitem [{\citenamefont {'t~Hooft}(1975)}]{tHooft:1975krp}%
  \BibitemOpen
  \bibfield  {author} {\bibinfo {author} {\bibfnamefont {G.}~\bibnamefont
  {'t~Hooft}},\ }\bibfield  {title} {\enquote {\bibinfo {title} {{Gauge Fields
  with Unified Weak, Electromagnetic, and Strong Interactions}},}\ }in\
  \href@noop {} {\emph {\bibinfo {booktitle} {{High-Energy Physics:
  Proceedings, EPS International Conference, Palermo, Italy, 23-28 June
  1975.}}}}\ (\bibinfo {year} {1975})\ p.\ \bibinfo {pages} {1225}\BibitemShut
  {NoStop}%
\bibitem [{\citenamefont {Jaffe}\ and\ \citenamefont
  {Witten}(2000)}]{Jaffe:2000ne}%
  \BibitemOpen
  \bibfield  {author} {\bibinfo {author} {\bibfnamefont {A.~M.}\ \bibnamefont
  {Jaffe}}\ and\ \bibinfo {author} {\bibfnamefont {E.}~\bibnamefont {Witten}},\
  }\bibfield  {title} {\enquote {\bibinfo {title} {{Quantum Yang-Mills
  theory}},}\ }\href
  {{http://www.claymath.org/sites/default/files/yangmills.pdf}} {\bibfield
  {journal} {\bibinfo  {journal} {Millennium Prize Problems}\ } (\bibinfo
  {year} {2000})}\BibitemShut {NoStop}%
\bibitem [{\citenamefont {Anderson}(1972)}]{Anderson393}%
  \BibitemOpen
  \bibfield  {author} {\bibinfo {author} {\bibfnamefont {P.~W.}\ \bibnamefont
  {Anderson}},\ }\bibfield  {title} {\enquote {\bibinfo {title} {More is
  different},}\ }\href {\doibase 10.1126/science.177.4047.393} {\bibfield
  {journal} {\bibinfo  {journal} {Science}\ }\textbf {\bibinfo {volume}
  {177}},\ \bibinfo {pages} {393--396} (\bibinfo {year} {1972})}\BibitemShut
  {NoStop}%
\bibitem [{\citenamefont {Shuryak}(1988)}]{Shuryak:1988ck}%
  \BibitemOpen
  \bibfield  {author} {\bibinfo {author} {\bibfnamefont {E.~V.}\ \bibnamefont
  {Shuryak}},\ }\href {\doibase 10.1142/9789812799425_fmatter} {\emph {\bibinfo
  {title} {{The QCD vacuum, hadrons and the superdense matter}}}},\ \bibinfo
  {series} {World Sci. Lect. Notes Phys.}, Vol.~\bibinfo {volume} {8}\
  (\bibinfo  {publisher} {Singapore ; New Jersey : World Scientific},\ \bibinfo
  {year} {1988})\BibitemShut {NoStop}%
\bibitem [{\citenamefont {Fodor}\ and\ \citenamefont
  {Hoelbling}(2012)}]{Fodor:2012gf}%
  \BibitemOpen
  \bibfield  {author} {\bibinfo {author} {\bibfnamefont {Z.}~\bibnamefont
  {Fodor}}\ and\ \bibinfo {author} {\bibfnamefont {C.}~\bibnamefont
  {Hoelbling}},\ }\bibfield  {title} {\enquote {\bibinfo {title} {{Light Hadron
  Masses from Lattice QCD}},}\ }\href {\doibase 10.1103/RevModPhys.84.449}
  {\bibfield  {journal} {\bibinfo  {journal} {Rev. Mod. Phys.}\ }\textbf
  {\bibinfo {volume} {84}},\ \bibinfo {pages} {449} (\bibinfo {year} {2012})},\
  \Eprint {http://arxiv.org/abs/1203.4789} {arXiv:1203.4789 [hep-lat]}
  \BibitemShut {NoStop}%
\bibitem [{\citenamefont {Ding}\ \emph {et~al.}(2015)\citenamefont {Ding},
  \citenamefont {Karsch},\ and\ \citenamefont {Mukherjee}}]{Ding:2015ona}%
  \BibitemOpen
  \bibfield  {author} {\bibinfo {author} {\bibfnamefont {H.-T.}\ \bibnamefont
  {Ding}}, \bibinfo {author} {\bibfnamefont {F.}~\bibnamefont {Karsch}}, \ and\
  \bibinfo {author} {\bibfnamefont {S.}~\bibnamefont {Mukherjee}},\ }\bibfield
  {title} {\enquote {\bibinfo {title} {{Thermodynamics of strong-interaction
  matter from Lattice QCD}},}\ }\href {\doibase 10.1142/S0218301315300076}
  {\bibfield  {journal} {\bibinfo  {journal} {Int. J. Mod. Phys.}\ }\textbf
  {\bibinfo {volume} {E24}},\ \bibinfo {pages} {1530007} (\bibinfo {year}
  {2015})},\ \Eprint {http://arxiv.org/abs/1504.05274} {arXiv:1504.05274
  [hep-lat]} \BibitemShut {NoStop}%
\bibitem [{\citenamefont {Veneziano}(1968)}]{Veneziano1968}%
  \BibitemOpen
  \bibfield  {author} {\bibinfo {author} {\bibfnamefont {G.}~\bibnamefont
  {Veneziano}},\ }\bibfield  {title} {\enquote {\bibinfo {title} {Construction
  of a crossing-simmetric, regge-behaved amplitude for linearly rising
  trajectories},}\ }\href {\doibase 10.1007/BF02824451} {\bibfield  {journal}
  {\bibinfo  {journal} {Il Nuovo Cimento A (1965-1970)}\ }\textbf {\bibinfo
  {volume} {57}},\ \bibinfo {pages} {190--197} (\bibinfo {year}
  {1968})}\BibitemShut {NoStop}%
\bibitem [{\citenamefont {Maldacena}(1999)}]{Maldacena:1997re}%
  \BibitemOpen
  \bibfield  {author} {\bibinfo {author} {\bibfnamefont {J.~M.}\ \bibnamefont
  {Maldacena}},\ }\bibfield  {title} {\enquote {\bibinfo {title} {{The Large N
  limit of superconformal field theories and supergravity}},}\ }\href {\doibase
  10.1023/A:1026654312961} {\bibfield  {journal} {\bibinfo  {journal} {Int. J.
  Theor. Phys.}\ }\textbf {\bibinfo {volume} {38}},\ \bibinfo {pages}
  {1113--1133} (\bibinfo {year} {1999})},\ \bibinfo {note} {[Adv. Theor. Math.
  Phys.2,231(1998)]},\ \Eprint {http://arxiv.org/abs/hep-th/9711200}
  {arXiv:hep-th/9711200 [hep-th]} \BibitemShut {NoStop}%
\bibitem [{\citenamefont {Chesler}\ and\ \citenamefont {van~der
  Schee}(2016)}]{Chesler:2016vft}%
  \BibitemOpen
  \bibfield  {author} {\bibinfo {author} {\bibfnamefont {P.~M.}\ \bibnamefont
  {Chesler}}\ and\ \bibinfo {author} {\bibfnamefont {W.}~\bibnamefont {van~der
  Schee}},\ }\bibfield  {title} {\enquote {\bibinfo {title} {{Early
  Thermalization, Hydrodynamics and Energy Loss in AdS/CFT}},}\ }in\ \href
  {\doibase 10.1142/9789814663717_0004} {\emph {\bibinfo {booktitle}
  {Quark-Gluon Plasma 5}}},\ \bibinfo {editor} {edited by\ \bibinfo {editor}
  {\bibfnamefont {X.-N.}\ \bibnamefont {Wang}}}\ (\bibinfo {year} {2016})\ pp.\
  \bibinfo {pages} {189--216}\BibitemShut {NoStop}%
\bibitem [{\citenamefont {Policastro}\ \emph {et~al.}(2002)\citenamefont
  {Policastro}, \citenamefont {Son},\ and\ \citenamefont
  {Starinets}}]{Policastro:2002se}%
  \BibitemOpen
  \bibfield  {author} {\bibinfo {author} {\bibfnamefont {G.}~\bibnamefont
  {Policastro}}, \bibinfo {author} {\bibfnamefont {D.~T.}\ \bibnamefont {Son}},
  \ and\ \bibinfo {author} {\bibfnamefont {A.~O.}\ \bibnamefont {Starinets}},\
  }\bibfield  {title} {\enquote {\bibinfo {title} {{From AdS / CFT
  correspondence to hydrodynamics}},}\ }\href {\doibase
  10.1088/1126-6708/2002/09/043} {\bibfield  {journal} {\bibinfo  {journal}
  {JHEP}\ }\textbf {\bibinfo {volume} {09}},\ \bibinfo {pages} {043} (\bibinfo
  {year} {2002})},\ \Eprint {http://arxiv.org/abs/hep-th/0205052}
  {arXiv:hep-th/0205052 [hep-th]} \BibitemShut {NoStop}%
\bibitem [{\citenamefont {Kovtun}\ \emph {et~al.}(2005)\citenamefont {Kovtun},
  \citenamefont {Son},\ and\ \citenamefont {Starinets}}]{Kovtun:2004de}%
  \BibitemOpen
  \bibfield  {author} {\bibinfo {author} {\bibfnamefont {P.}~\bibnamefont
  {Kovtun}}, \bibinfo {author} {\bibfnamefont {D.~T.}\ \bibnamefont {Son}}, \
  and\ \bibinfo {author} {\bibfnamefont {A.~O.}\ \bibnamefont {Starinets}},\
  }\bibfield  {title} {\enquote {\bibinfo {title} {{Viscosity in strongly
  interacting quantum field theories from black hole physics}},}\ }\href
  {\doibase 10.1103/PhysRevLett.94.111601} {\bibfield  {journal} {\bibinfo
  {journal} {Phys. Rev. Lett.}\ }\textbf {\bibinfo {volume} {94}},\ \bibinfo
  {pages} {111601} (\bibinfo {year} {2005})},\ \Eprint
  {http://arxiv.org/abs/hep-th/0405231} {arXiv:hep-th/0405231 [hep-th]}
  \BibitemShut {NoStop}%
\bibitem [{\citenamefont {Gyulassy}\ and\ \citenamefont
  {McLerran}(2005)}]{Gyulassy:2004zy}%
  \BibitemOpen
  \bibfield  {author} {\bibinfo {author} {\bibfnamefont {M.}~\bibnamefont
  {Gyulassy}}\ and\ \bibinfo {author} {\bibfnamefont {L.}~\bibnamefont
  {McLerran}},\ }\bibfield  {title} {\enquote {\bibinfo {title} {{New forms of
  QCD matter discovered at RHIC}},}\ }\bibfield  {booktitle} {\emph {\bibinfo
  {booktitle} {{Quark gluon plasma. New discoveries at RHIC: A case of strongly
  interacting quark gluon plasma. Proceedings, RBRC Workshop, Brookhaven,
  Upton, USA, May 14-15, 2004}}},\ }\href {\doibase
  10.1016/j.nuclphysa.2004.10.034} {\bibfield  {journal} {\bibinfo  {journal}
  {Nucl. Phys.}\ }\textbf {\bibinfo {volume} {A750}},\ \bibinfo {pages}
  {30--63} (\bibinfo {year} {2005})},\ \Eprint
  {http://arxiv.org/abs/nucl-th/0405013} {arXiv:nucl-th/0405013 [nucl-th]}
  \BibitemShut {NoStop}%
\bibitem [{\citenamefont {Shuryak}(1978)}]{Shuryak:1977ut}%
  \BibitemOpen
  \bibfield  {author} {\bibinfo {author} {\bibfnamefont {E.~V.}\ \bibnamefont
  {Shuryak}},\ }\bibfield  {title} {\enquote {\bibinfo {title} {{Theory of
  Hadronic Plasma}},}\ }\href
  {{http://www.jetp.ac.ru/cgi-bin/dn/e_047_02_0212.pdf}} {\bibfield  {journal}
  {\bibinfo  {journal} {Sov. Phys. JETP}\ }\textbf {\bibinfo {volume} {47}},\
  \bibinfo {pages} {212--219} (\bibinfo {year} {1978})},\ \bibinfo {note} {[Zh.
  Eksp. Teor. Fiz.74,408(1978)]}\BibitemShut {NoStop}%
\bibitem [{\citenamefont {Lattimer}(2012)}]{Lattimer:2012nd}%
  \BibitemOpen
  \bibfield  {author} {\bibinfo {author} {\bibfnamefont {J.~M.}\ \bibnamefont
  {Lattimer}},\ }\bibfield  {title} {\enquote {\bibinfo {title} {{The nuclear
  equation of state and neutron star masses}},}\ }\href {\doibase
  10.1146/annurev-nucl-102711-095018} {\bibfield  {journal} {\bibinfo
  {journal} {Ann. Rev. Nucl. Part. Sci.}\ }\textbf {\bibinfo {volume} {62}},\
  \bibinfo {pages} {485--515} (\bibinfo {year} {2012})},\ \Eprint
  {http://arxiv.org/abs/1305.3510} {arXiv:1305.3510 [nucl-th]} \BibitemShut
  {NoStop}%
\bibitem [{\citenamefont {Weinberg}(1977)}]{Weinberg}%
  \BibitemOpen
  \bibfield  {author} {\bibinfo {author} {\bibfnamefont {S.}~\bibnamefont
  {Weinberg}},\ }\href
  {http://proxy.library.stonybrook.edu/login?url=http://search.ebscohost.com/login.aspx?direct=true&db=cat03000a&AN=STB.000205669&site=eds-live&scope=site}
  {\emph {\bibinfo {title} {The first three minutes : a modern view of the
  origin of the universe / Steven Weinberg.}}}\ (\bibinfo  {publisher} {New
  York : Basic Books, c1977.},\ \bibinfo {year} {1977})\BibitemShut {NoStop}%
\bibitem [{\citenamefont {Wei-Qin}(1994)}]{Wei-Qin1994}%
  \BibitemOpen
  \bibfield  {author} {\bibinfo {author} {\bibfnamefont {C.}~\bibnamefont
  {Wei-Qin}},\ }\enquote {\bibinfo {title} {Some physical analyses in
  relativistic heavy ion collisions},}\ in\ \href {\doibase
  10.1007/978-94-011-1060-0_5} {\emph {\bibinfo {booktitle} {Cosmology and
  Particle Physics}}},\ \bibinfo {editor} {edited by\ \bibinfo {editor}
  {\bibfnamefont {V.}~\bibnamefont {de~Sabbata}}\ and\ \bibinfo {editor}
  {\bibfnamefont {H.}~\bibnamefont {Tso-Hsiu}}}\ (\bibinfo  {publisher}
  {Springer Netherlands},\ \bibinfo {address} {Dordrecht},\ \bibinfo {year}
  {1994})\ pp.\ \bibinfo {pages} {49--65}\BibitemShut {NoStop}%
\bibitem [{\citenamefont {Fischer}\ and\ \citenamefont
  {Jowett}(2014)}]{Fischer:2014wfa}%
  \BibitemOpen
  \bibfield  {author} {\bibinfo {author} {\bibfnamefont {W.}~\bibnamefont
  {Fischer}}\ and\ \bibinfo {author} {\bibfnamefont {J.~M.}\ \bibnamefont
  {Jowett}},\ }\bibfield  {title} {\enquote {\bibinfo {title} {{Ion
  Colliders}},}\ }\href {\doibase 10.1142/S1793626814300047} {\bibfield
  {journal} {\bibinfo  {journal} {Rev. Accel. Sci. Tech.}\ }\textbf {\bibinfo
  {volume} {7}},\ \bibinfo {pages} {49--76} (\bibinfo {year}
  {2014})}\BibitemShut {NoStop}%
\bibitem [{\citenamefont {Wang}(2016)}]{Wang:2016opj}%
  \BibitemOpen
  \bibinfo {editor} {\bibfnamefont {X.-N.}\ \bibnamefont {Wang}},\ ed.,\ \href
  {\doibase 10.1142/9533} {\emph {\bibinfo {title} {{Quark-Gluon Plasma 5}}}}\
  (\bibinfo  {publisher} {World Scientific},\ \bibinfo {address} {New Jersey},\
  \bibinfo {year} {2016})\BibitemShut {NoStop}%
\bibitem [{\citenamefont {Baym}(2002)}]{Baym:2001in}%
  \BibitemOpen
  \bibfield  {author} {\bibinfo {author} {\bibfnamefont {G.}~\bibnamefont
  {Baym}},\ }\bibfield  {title} {\enquote {\bibinfo {title} {{RHIC: From dreams
  to beams in two decades}},}\ }\bibfield  {booktitle} {\emph {\bibinfo
  {booktitle} {{Quark matter 2001. Proceedings, 15th International Conference
  on Ultrarelativistic nucleus nucleus collisions, QM 2001, Stony Brook, USA,
  January 15-20, 2001}}},\ }\href {\doibase 10.1016/S0375-9474(01)01342-2}
  {\bibfield  {journal} {\bibinfo  {journal} {Nucl. Phys.}\ }\textbf {\bibinfo
  {volume} {A698}},\ \bibinfo {pages} {XXIII--XXXII} (\bibinfo {year}
  {2002})},\ \Eprint {http://arxiv.org/abs/hep-ph/0104138}
  {arXiv:hep-ph/0104138 [hep-ph]} \BibitemShut {NoStop}%
\bibitem [{\citenamefont {Fischer}(2016)}]{RHIC_Runs}%
  \BibitemOpen
  \bibfield  {author} {\bibinfo {author} {\bibfnamefont {W.}~\bibnamefont
  {Fischer}},\ }\href@noop {} {\enquote {\bibinfo {title} {Run overview of the
  relativistic heavy ion collider},}\ }\bibinfo {howpublished}
  {\url{http://www.agsrhichome.bnl.gov/RHIC/Runs/index.html}} (\bibinfo {year}
  {2016})\BibitemShut {NoStop}%
\bibitem [{\citenamefont {Luo}(2016)}]{Luo:2015doi}%
  \BibitemOpen
  \bibfield  {author} {\bibinfo {author} {\bibfnamefont {X.}~\bibnamefont
  {Luo}},\ }\bibfield  {title} {\enquote {\bibinfo {title} {{Exploring the QCD
  Phase Structure with Beam Energy Scan in Heavy-ion Collisions}},}\ }\bibfield
   {booktitle} {\emph {\bibinfo {booktitle} {{Proceedings, 25th International
  Conference on Ultra-Relativistic Nucleus-Nucleus Collisions (Quark Matter
  2015): Kobe, Japan, September 27-October 3, 2015}}},\ }\href {\doibase
  10.1016/j.nuclphysa.2016.03.025} {\bibfield  {journal} {\bibinfo  {journal}
  {Nucl. Phys.}\ }\textbf {\bibinfo {volume} {A956}},\ \bibinfo {pages}
  {75--82} (\bibinfo {year} {2016})},\ \Eprint
  {http://arxiv.org/abs/1512.09215} {arXiv:1512.09215 [nucl-ex]} \BibitemShut
  {NoStop}%
\bibitem [{\citenamefont {Muller}\ \emph {et~al.}(2012)\citenamefont {Muller},
  \citenamefont {Schukraft},\ and\ \citenamefont {Wyslouch}}]{Muller:2012zq}%
  \BibitemOpen
  \bibfield  {author} {\bibinfo {author} {\bibfnamefont {B.}~\bibnamefont
  {Muller}}, \bibinfo {author} {\bibfnamefont {J.}~\bibnamefont {Schukraft}}, \
  and\ \bibinfo {author} {\bibfnamefont {B.}~\bibnamefont {Wyslouch}},\
  }\bibfield  {title} {\enquote {\bibinfo {title} {{First Results from Pb+Pb
  collisions at the LHC}},}\ }\href {\doibase
  10.1146/annurev-nucl-102711-094910} {\bibfield  {journal} {\bibinfo
  {journal} {Ann. Rev. Nucl. Part. Sci.}\ }\textbf {\bibinfo {volume} {62}},\
  \bibinfo {pages} {361--386} (\bibinfo {year} {2012})},\ \Eprint
  {http://arxiv.org/abs/1202.3233} {arXiv:1202.3233 [hep-ex]} \BibitemShut
  {NoStop}%
\bibitem [{\citenamefont {Chatrchyan}\ \emph {et~al.}(2013)\citenamefont
  {Chatrchyan} \emph {et~al.}}]{CMS:2012qk}%
  \BibitemOpen
  \bibfield  {author} {\bibinfo {author} {\bibfnamefont {S.}~\bibnamefont
  {Chatrchyan}} \emph {et~al.} (\bibinfo {collaboration} {CMS}),\ }\bibfield
  {title} {\enquote {\bibinfo {title} {{Observation of long-range near-side
  angular correlations in proton-lead collisions at the LHC}},}\ }\href
  {\doibase 10.1016/j.physletb.2012.11.025} {\bibfield  {journal} {\bibinfo
  {journal} {Phys. Lett.}\ }\textbf {\bibinfo {volume} {B718}},\ \bibinfo
  {pages} {795--814} (\bibinfo {year} {2013})},\ \Eprint
  {http://arxiv.org/abs/1210.5482} {arXiv:1210.5482 [nucl-ex]} \BibitemShut
  {NoStop}%
\bibitem [{\citenamefont {Aaboud}\ \emph {et~al.}(2016)\citenamefont {Aaboud}
  \emph {et~al.}}]{Aaboud:2016yar}%
  \BibitemOpen
  \bibfield  {author} {\bibinfo {author} {\bibfnamefont {M.}~\bibnamefont
  {Aaboud}} \emph {et~al.} (\bibinfo {collaboration} {ATLAS}),\ }\bibfield
  {title} {\enquote {\bibinfo {title} {{Measurements of long-range azimuthal
  anisotropies and associated Fourier coefficients for $pp$ collisions at
  $\sqrt{s}=5.02$ and $13$ TeV and $p$+Pb collisions at
  $\sqrt{s_{\mathrm{NN}}}=5.02$ TeV with the ATLAS detector}},}\ }\href@noop {}
  {\  (\bibinfo {year} {2016})},\ \Eprint {http://arxiv.org/abs/1609.06213}
  {arXiv:1609.06213 [nucl-ex]} \BibitemShut {NoStop}%
\bibitem [{\citenamefont {Aad}\ \emph {et~al.}(2013)\citenamefont {Aad} \emph
  {et~al.}}]{Aad:2012gla}%
  \BibitemOpen
  \bibfield  {author} {\bibinfo {author} {\bibfnamefont {G.}~\bibnamefont
  {Aad}} \emph {et~al.} (\bibinfo {collaboration} {ATLAS}),\ }\bibfield
  {title} {\enquote {\bibinfo {title} {{Observation of Associated Near-Side and
  Away-Side Long-Range Correlations in $\sqrt{s_{NN}}$=5.02  TeV
  Proton-Lead Collisions with the ATLAS Detector}},}\ }\href {\doibase
  10.1103/PhysRevLett.110.182302} {\bibfield  {journal} {\bibinfo  {journal}
  {Phys. Rev. Lett.}\ }\textbf {\bibinfo {volume} {110}},\ \bibinfo {pages}
  {182302} (\bibinfo {year} {2013})},\ \Eprint {http://arxiv.org/abs/1212.5198}
  {arXiv:1212.5198 [hep-ex]} \BibitemShut {NoStop}%
\bibitem [{\citenamefont {{Community White Paper}}(2012)}]{NSAC}%
  \BibitemOpen
  \bibfield  {author} {\bibinfo {author} {\bibnamefont {{Community White
  Paper}}},\ }\href@noop {} {\enquote {\bibinfo {title} {{Hot and Dense QCD
  Matter}},}\ }\bibinfo {howpublished}
  {\url{https://www.bnl.gov/npp/docs/Bass_RHI_WP_final.pdf}} (\bibinfo {year}
  {2012})\BibitemShut {NoStop}%
\bibitem [{\citenamefont {Ackermann}\ \emph {et~al.}(2001)\citenamefont
  {Ackermann} \emph {et~al.}}]{Ackermann:2000tr}%
  \BibitemOpen
  \bibfield  {author} {\bibinfo {author} {\bibfnamefont {K.~H.}\ \bibnamefont
  {Ackermann}} \emph {et~al.} (\bibinfo {collaboration} {STAR}),\ }\bibfield
  {title} {\enquote {\bibinfo {title} {{Elliptic flow in Au + Au collisions at
  (S(NN))**(1/2) = 130 GeV}},}\ }\href {\doibase 10.1103/PhysRevLett.86.402}
  {\bibfield  {journal} {\bibinfo  {journal} {Phys. Rev. Lett.}\ }\textbf
  {\bibinfo {volume} {86}},\ \bibinfo {pages} {402--407} (\bibinfo {year}
  {2001})},\ \Eprint {http://arxiv.org/abs/nucl-ex/0009011}
  {arXiv:nucl-ex/0009011 [nucl-ex]} \BibitemShut {NoStop}%
\bibitem [{\citenamefont {Teaney}(2003)}]{Teaney:2003kp}%
  \BibitemOpen
  \bibfield  {author} {\bibinfo {author} {\bibfnamefont {D.}~\bibnamefont
  {Teaney}},\ }\bibfield  {title} {\enquote {\bibinfo {title} {{The Effects of
  viscosity on spectra, elliptic flow, and HBT radii}},}\ }\href {\doibase
  10.1103/PhysRevC.68.034913} {\bibfield  {journal} {\bibinfo  {journal} {Phys.
  Rev.}\ }\textbf {\bibinfo {volume} {C68}},\ \bibinfo {pages} {034913}
  (\bibinfo {year} {2003})},\ \Eprint {http://arxiv.org/abs/nucl-th/0301099}
  {arXiv:nucl-th/0301099 [nucl-th]} \BibitemShut {NoStop}%
\bibitem [{\citenamefont {Romatschke}\ and\ \citenamefont
  {Romatschke}(2007)}]{Romatschke:2007mq}%
  \BibitemOpen
  \bibfield  {author} {\bibinfo {author} {\bibfnamefont {P.}~\bibnamefont
  {Romatschke}}\ and\ \bibinfo {author} {\bibfnamefont {U.}~\bibnamefont
  {Romatschke}},\ }\bibfield  {title} {\enquote {\bibinfo {title} {{Viscosity
  Information from Relativistic Nuclear Collisions: How Perfect is the Fluid
  Observed at RHIC?}}}\ }\href {\doibase 10.1103/PhysRevLett.99.172301}
  {\bibfield  {journal} {\bibinfo  {journal} {Phys. Rev. Lett.}\ }\textbf
  {\bibinfo {volume} {99}},\ \bibinfo {pages} {172301} (\bibinfo {year}
  {2007})},\ \Eprint {http://arxiv.org/abs/0706.1522} {arXiv:0706.1522
  [nucl-th]} \BibitemShut {NoStop}%
\bibitem [{\citenamefont {Danielewicz}\ and\ \citenamefont
  {Gyulassy}(1985)}]{Danielewicz:1984ww}%
  \BibitemOpen
  \bibfield  {author} {\bibinfo {author} {\bibfnamefont {P.}~\bibnamefont
  {Danielewicz}}\ and\ \bibinfo {author} {\bibfnamefont {M.}~\bibnamefont
  {Gyulassy}},\ }\bibfield  {title} {\enquote {\bibinfo {title} {{Dissipative
  Phenomena in Quark Gluon Plasmas}},}\ }\href {\doibase
  10.1103/PhysRevD.31.53} {\bibfield  {journal} {\bibinfo  {journal} {Phys.
  Rev.}\ }\textbf {\bibinfo {volume} {D31}},\ \bibinfo {pages} {53--62}
  (\bibinfo {year} {1985})}\BibitemShut {NoStop}%
\bibitem [{\citenamefont {Antinori}\ \emph {et~al.}(2016)\citenamefont
  {Antinori} \emph {et~al.}}]{Antinori:2016zxe}%
  \BibitemOpen
  \bibfield  {author} {\bibinfo {author} {\bibfnamefont {F.}~\bibnamefont
  {Antinori}} \emph {et~al.},\ }\bibfield  {title} {\enquote {\bibinfo {title}
  {{Thoughts on heavy-ion physics in the high luminosity era: the soft
  sector}},}\ }\href@noop {} {\  (\bibinfo {year} {2016})},\ \Eprint
  {http://arxiv.org/abs/1604.03310} {arXiv:1604.03310 [hep-ph]} \BibitemShut
  {NoStop}%
\bibitem [{\citenamefont {Shen}\ and\ \citenamefont
  {Heinz}(2015)}]{Shen:2015msa}%
  \BibitemOpen
  \bibfield  {author} {\bibinfo {author} {\bibfnamefont {C.}~\bibnamefont
  {Shen}}\ and\ \bibinfo {author} {\bibfnamefont {U.}~\bibnamefont {Heinz}},\
  }\bibfield  {title} {\enquote {\bibinfo {title} {{The road to precision:
  Extraction of the specific shear viscosity of the quark-gluon plasma}},}\
  }\href {\doibase 10.1080/10619127.2015.1006502} {\bibfield  {journal}
  {\bibinfo  {journal} {Nucl. Phys. News}\ }\textbf {\bibinfo {volume} {25}},\
  \bibinfo {pages} {6--11} (\bibinfo {year} {2015})},\ \Eprint
  {http://arxiv.org/abs/1507.01558} {arXiv:1507.01558 [nucl-th]} \BibitemShut
  {NoStop}%
\bibitem [{\citenamefont {Alver}\ and\ \citenamefont
  {Roland}(2010)}]{Alver:2010gr}%
  \BibitemOpen
  \bibfield  {author} {\bibinfo {author} {\bibfnamefont {B.}~\bibnamefont
  {Alver}}\ and\ \bibinfo {author} {\bibfnamefont {G.}~\bibnamefont {Roland}},\
  }\bibfield  {title} {\enquote {\bibinfo {title} {{Collision geometry
  fluctuations and triangular flow in heavy-ion collisions}},}\ }\href
  {\doibase 10.1103/PhysRevC.82.039903, 10.1103/PhysRevC.81.054905} {\bibfield
  {journal} {\bibinfo  {journal} {Phys. Rev.}\ }\textbf {\bibinfo {volume}
  {C81}},\ \bibinfo {pages} {054905} (\bibinfo {year} {2010})},\ \bibinfo
  {note} {[Erratum: Phys. Rev.C82,039903(2010)]},\ \Eprint
  {http://arxiv.org/abs/1003.0194} {arXiv:1003.0194 [nucl-th]} \BibitemShut
  {NoStop}%
\bibitem [{\citenamefont {Staig}\ and\ \citenamefont
  {Shuryak}(2011)}]{Staig:2011wj}%
  \BibitemOpen
  \bibfield  {author} {\bibinfo {author} {\bibfnamefont {P.}~\bibnamefont
  {Staig}}\ and\ \bibinfo {author} {\bibfnamefont {E.}~\bibnamefont
  {Shuryak}},\ }\bibfield  {title} {\enquote {\bibinfo {title} {{The Fate of
  the Initial State Fluctuations in Heavy Ion Collisions. III The Second Act of
  Hydrodynamics}},}\ }\href {\doibase 10.1103/PhysRevC.84.044912} {\bibfield
  {journal} {\bibinfo  {journal} {Phys. Rev.}\ }\textbf {\bibinfo {volume}
  {C84}},\ \bibinfo {pages} {044912} (\bibinfo {year} {2011})},\ \Eprint
  {http://arxiv.org/abs/1105.0676} {arXiv:1105.0676 [nucl-th]} \BibitemShut
  {NoStop}%
\bibitem [{\citenamefont {Schenke}\ \emph
  {et~al.}(2012{\natexlab{a}})\citenamefont {Schenke}, \citenamefont {Jeon},\
  and\ \citenamefont {Gale}}]{Schenke:2011bn}%
  \BibitemOpen
  \bibfield  {author} {\bibinfo {author} {\bibfnamefont {B.}~\bibnamefont
  {Schenke}}, \bibinfo {author} {\bibfnamefont {S.}~\bibnamefont {Jeon}}, \
  and\ \bibinfo {author} {\bibfnamefont {C.}~\bibnamefont {Gale}},\ }\bibfield
  {title} {\enquote {\bibinfo {title} {{Higher flow harmonics from (3+1)D
  event-by-event viscous hydrodynamics}},}\ }\href {\doibase
  10.1103/PhysRevC.85.024901} {\bibfield  {journal} {\bibinfo  {journal} {Phys.
  Rev.}\ }\textbf {\bibinfo {volume} {C85}},\ \bibinfo {pages} {024901}
  (\bibinfo {year} {2012}{\natexlab{a}})},\ \Eprint
  {http://arxiv.org/abs/1109.6289} {arXiv:1109.6289 [hep-ph]} \BibitemShut
  {NoStop}%
\bibitem [{\citenamefont {Chatrchyan}\ \emph
  {et~al.}(2014{\natexlab{a}})\citenamefont {Chatrchyan} \emph
  {et~al.}}]{Chatrchyan:2013kba}%
  \BibitemOpen
  \bibfield  {author} {\bibinfo {author} {\bibfnamefont {S.}~\bibnamefont
  {Chatrchyan}} \emph {et~al.} (\bibinfo {collaboration} {CMS}),\ }\bibfield
  {title} {\enquote {\bibinfo {title} {{Measurement of higher-order harmonic
  azimuthal anisotropy in PbPb collisions at $\sqrt{s_{NN}}$ = 2.76 TeV}},}\
  }\href {\doibase 10.1103/PhysRevC.89.044906} {\bibfield  {journal} {\bibinfo
  {journal} {Phys. Rev.}\ }\textbf {\bibinfo {volume} {C89}},\ \bibinfo {pages}
  {044906} (\bibinfo {year} {2014}{\natexlab{a}})},\ \Eprint
  {http://arxiv.org/abs/1310.8651} {arXiv:1310.8651 [nucl-ex]} \BibitemShut
  {NoStop}%
\bibitem [{\citenamefont {Aad}\ \emph {et~al.}(2012{\natexlab{a}})\citenamefont
  {Aad} \emph {et~al.}}]{ATLAS:2012at}%
  \BibitemOpen
  \bibfield  {author} {\bibinfo {author} {\bibfnamefont {G.}~\bibnamefont
  {Aad}} \emph {et~al.} (\bibinfo {collaboration} {ATLAS}),\ }\bibfield
  {title} {\enquote {\bibinfo {title} {{Measurement of the azimuthal anisotropy
  for charged particle production in $\sqrt{s_{NN}}=2.76$ TeV lead-lead
  collisions with the ATLAS detector}},}\ }\href {\doibase
  10.1103/PhysRevC.86.014907} {\bibfield  {journal} {\bibinfo  {journal} {Phys.
  Rev.}\ }\textbf {\bibinfo {volume} {C86}},\ \bibinfo {pages} {014907}
  (\bibinfo {year} {2012}{\natexlab{a}})},\ \Eprint
  {http://arxiv.org/abs/1203.3087} {arXiv:1203.3087 [hep-ex]} \BibitemShut
  {NoStop}%
\bibitem [{\citenamefont {Aamodt}\ \emph
  {et~al.}(2011{\natexlab{a}})\citenamefont {Aamodt} \emph
  {et~al.}}]{ALICE:2011ab}%
  \BibitemOpen
  \bibfield  {author} {\bibinfo {author} {\bibfnamefont {K.}~\bibnamefont
  {Aamodt}} \emph {et~al.} (\bibinfo {collaboration} {ALICE}),\ }\bibfield
  {title} {\enquote {\bibinfo {title} {{Higher harmonic anisotropic flow
  measurements of charged particles in Pb-Pb collisions at $\sqrt{s_{NN}}$=2.76
  TeV}},}\ }\href {\doibase 10.1103/PhysRevLett.107.032301} {\bibfield
  {journal} {\bibinfo  {journal} {Phys. Rev. Lett.}\ }\textbf {\bibinfo
  {volume} {107}},\ \bibinfo {pages} {032301} (\bibinfo {year}
  {2011}{\natexlab{a}})},\ \Eprint {http://arxiv.org/abs/1105.3865}
  {arXiv:1105.3865 [nucl-ex]} \BibitemShut {NoStop}%
\bibitem [{\citenamefont {Adare}\ \emph {et~al.}(2011)\citenamefont {Adare}
  \emph {et~al.}}]{Adare:2011tg}%
  \BibitemOpen
  \bibfield  {author} {\bibinfo {author} {\bibfnamefont {A.}~\bibnamefont
  {Adare}} \emph {et~al.} (\bibinfo {collaboration} {PHENIX}),\ }\bibfield
  {title} {\enquote {\bibinfo {title} {{Measurements of Higher-Order Flow
  Harmonics in Au+Au Collisions at $\sqrt{s_{NN}} = 200$ GeV}},}\ }\href
  {\doibase 10.1103/PhysRevLett.107.252301} {\bibfield  {journal} {\bibinfo
  {journal} {Phys. Rev. Lett.}\ }\textbf {\bibinfo {volume} {107}},\ \bibinfo
  {pages} {252301} (\bibinfo {year} {2011})},\ \Eprint
  {http://arxiv.org/abs/1105.3928} {arXiv:1105.3928 [nucl-ex]} \BibitemShut
  {NoStop}%
\bibitem [{\citenamefont {Adams}\ \emph {et~al.}(2004)\citenamefont {Adams}
  \emph {et~al.}}]{Adams:2003zg}%
  \BibitemOpen
  \bibfield  {author} {\bibinfo {author} {\bibfnamefont {J.}~\bibnamefont
  {Adams}} \emph {et~al.} (\bibinfo {collaboration} {STAR}),\ }\bibfield
  {title} {\enquote {\bibinfo {title} {{Azimuthal anisotropy at RHIC: The First
  and fourth harmonics}},}\ }\href {\doibase 10.1103/PhysRevLett.92.062301}
  {\bibfield  {journal} {\bibinfo  {journal} {Phys. Rev. Lett.}\ }\textbf
  {\bibinfo {volume} {92}},\ \bibinfo {pages} {062301} (\bibinfo {year}
  {2004})},\ \Eprint {http://arxiv.org/abs/nucl-ex/0310029}
  {arXiv:nucl-ex/0310029 [nucl-ex]} \BibitemShut {NoStop}%
\bibitem [{\citenamefont {Adamczyk}\ \emph {et~al.}(2013)\citenamefont
  {Adamczyk} \emph {et~al.}}]{Adamczyk:2013waa}%
  \BibitemOpen
  \bibfield  {author} {\bibinfo {author} {\bibfnamefont {L.}~\bibnamefont
  {Adamczyk}} \emph {et~al.} (\bibinfo {collaboration} {STAR}),\ }\bibfield
  {title} {\enquote {\bibinfo {title} {{Third Harmonic Flow of Charged
  Particles in Au+Au Collisions at sqrtsNN = 200 GeV}},}\ }\href {\doibase
  10.1103/PhysRevC.88.014904} {\bibfield  {journal} {\bibinfo  {journal} {Phys.
  Rev.}\ }\textbf {\bibinfo {volume} {C88}},\ \bibinfo {pages} {014904}
  (\bibinfo {year} {2013})},\ \Eprint {http://arxiv.org/abs/1301.2187}
  {arXiv:1301.2187 [nucl-ex]} \BibitemShut {NoStop}%
\bibitem [{\citenamefont {Gale}\ \emph
  {et~al.}(2013{\natexlab{a}})\citenamefont {Gale}, \citenamefont {Jeon},
  \citenamefont {Schenke}, \citenamefont {Tribedy},\ and\ \citenamefont
  {Venugopalan}}]{Gale:2012rq}%
  \BibitemOpen
  \bibfield  {author} {\bibinfo {author} {\bibfnamefont {C.}~\bibnamefont
  {Gale}}, \bibinfo {author} {\bibfnamefont {S.}~\bibnamefont {Jeon}}, \bibinfo
  {author} {\bibfnamefont {B.}~\bibnamefont {Schenke}}, \bibinfo {author}
  {\bibfnamefont {P.}~\bibnamefont {Tribedy}}, \ and\ \bibinfo {author}
  {\bibfnamefont {R.}~\bibnamefont {Venugopalan}},\ }\bibfield  {title}
  {\enquote {\bibinfo {title} {{Event-by-event anisotropic flow in heavy-ion
  collisions from combined Yang-Mills and viscous fluid dynamics}},}\ }\href
  {\doibase 10.1103/PhysRevLett.110.012302} {\bibfield  {journal} {\bibinfo
  {journal} {Phys. Rev. Lett.}\ }\textbf {\bibinfo {volume} {110}},\ \bibinfo
  {pages} {012302} (\bibinfo {year} {2013}{\natexlab{a}})},\ \Eprint
  {http://arxiv.org/abs/1209.6330} {arXiv:1209.6330 [nucl-th]} \BibitemShut
  {NoStop}%
\bibitem [{\citenamefont {Heinz}\ and\ \citenamefont
  {Snellings}(2013)}]{Heinz:2013th}%
  \BibitemOpen
  \bibfield  {author} {\bibinfo {author} {\bibfnamefont {U.}~\bibnamefont
  {Heinz}}\ and\ \bibinfo {author} {\bibfnamefont {R.}~\bibnamefont
  {Snellings}},\ }\bibfield  {title} {\enquote {\bibinfo {title} {{Collective
  flow and viscosity in relativistic heavy-ion collisions}},}\ }\href {\doibase
  10.1146/annurev-nucl-102212-170540} {\bibfield  {journal} {\bibinfo
  {journal} {Ann. Rev. Nucl. Part. Sci.}\ }\textbf {\bibinfo {volume} {63}},\
  \bibinfo {pages} {123--151} (\bibinfo {year} {2013})},\ \Eprint
  {http://arxiv.org/abs/1301.2826} {arXiv:1301.2826 [nucl-th]} \BibitemShut
  {NoStop}%
\bibitem [{\citenamefont {Luzum}\ and\ \citenamefont
  {Petersen}(2014)}]{Luzum:2013yya}%
  \BibitemOpen
  \bibfield  {author} {\bibinfo {author} {\bibfnamefont {M.}~\bibnamefont
  {Luzum}}\ and\ \bibinfo {author} {\bibfnamefont {H.}~\bibnamefont
  {Petersen}},\ }\bibfield  {title} {\enquote {\bibinfo {title} {{Initial State
  Fluctuations and Final State Correlations in Relativistic Heavy-Ion
  Collisions}},}\ }\href {\doibase 10.1088/0954-3899/41/6/063102} {\bibfield
  {journal} {\bibinfo  {journal} {J. Phys.}\ }\textbf {\bibinfo {volume}
  {G41}},\ \bibinfo {pages} {063102} (\bibinfo {year} {2014})},\ \Eprint
  {http://arxiv.org/abs/1312.5503} {arXiv:1312.5503 [nucl-th]} \BibitemShut
  {NoStop}%
\bibitem [{\citenamefont {Gale}\ \emph
  {et~al.}(2013{\natexlab{b}})\citenamefont {Gale}, \citenamefont {Jeon},\ and\
  \citenamefont {Schenke}}]{Gale:2013da}%
  \BibitemOpen
  \bibfield  {author} {\bibinfo {author} {\bibfnamefont {C.}~\bibnamefont
  {Gale}}, \bibinfo {author} {\bibfnamefont {S.}~\bibnamefont {Jeon}}, \ and\
  \bibinfo {author} {\bibfnamefont {B.}~\bibnamefont {Schenke}},\ }\bibfield
  {title} {\enquote {\bibinfo {title} {{Hydrodynamic Modeling of Heavy-Ion
  Collisions}},}\ }\href {\doibase 10.1142/S0217751X13400113} {\bibfield
  {journal} {\bibinfo  {journal} {Int. J. Mod. Phys.}\ }\textbf {\bibinfo
  {volume} {A28}},\ \bibinfo {pages} {1340011} (\bibinfo {year}
  {2013}{\natexlab{b}})},\ \Eprint {http://arxiv.org/abs/1301.5893}
  {arXiv:1301.5893 [nucl-th]} \BibitemShut {NoStop}%
\bibitem [{\citenamefont {Romatschke}(2017)}]{Romatschke:2016hle}%
  \BibitemOpen
  \bibfield  {author} {\bibinfo {author} {\bibfnamefont {P.}~\bibnamefont
  {Romatschke}},\ }\bibfield  {title} {\enquote {\bibinfo {title} {{Do nuclear
  collisions create a locally equilibrated quark–gluon plasma?}}}\ }\href
  {\doibase 10.1140/epjc/s10052-016-4567-x} {\bibfield  {journal} {\bibinfo
  {journal} {Eur. Phys. J.}\ }\textbf {\bibinfo {volume} {C77}},\ \bibinfo
  {pages} {21} (\bibinfo {year} {2017})},\ \Eprint
  {http://arxiv.org/abs/1609.02820} {arXiv:1609.02820 [nucl-th]} \BibitemShut
  {NoStop}%
\bibitem [{\citenamefont {Kurkela}(2016)}]{Kurkela:2016vts}%
  \BibitemOpen
  \bibfield  {author} {\bibinfo {author} {\bibfnamefont {A.}~\bibnamefont
  {Kurkela}},\ }\bibfield  {title} {\enquote {\bibinfo {title} {{Initial state
  of Heavy-Ion Collisions: Isotropization and thermalization}},}\ }\bibfield
  {booktitle} {\emph {\bibinfo {booktitle} {{Proceedings, 25th International
  Conference on Ultra-Relativistic Nucleus-Nucleus Collisions (Quark Matter
  2015): Kobe, Japan, September 27-October 3, 2015}}},\ }\href {\doibase
  10.1016/j.nuclphysa.2016.01.069} {\bibfield  {journal} {\bibinfo  {journal}
  {Nucl. Phys.}\ }\textbf {\bibinfo {volume} {A956}},\ \bibinfo {pages}
  {136--143} (\bibinfo {year} {2016})},\ \Eprint
  {http://arxiv.org/abs/1601.03283} {arXiv:1601.03283 [hep-ph]} \BibitemShut
  {NoStop}%
\bibitem [{\citenamefont {Gelis}(2016)}]{Gelis:2016rnt}%
  \BibitemOpen
  \bibfield  {author} {\bibinfo {author} {\bibfnamefont {F.}~\bibnamefont
  {Gelis}},\ }\bibfield  {title} {\enquote {\bibinfo {title} {{Initial State
  and Thermalization in the Color Glass Condensate Framework}},}\ }in\ \href
  {\doibase 10.1142/9789814663717_0002} {\emph {\bibinfo {booktitle}
  {Quark-Gluon Plasma 5}}},\ \bibinfo {editor} {edited by\ \bibinfo {editor}
  {\bibfnamefont {X.-N.}\ \bibnamefont {Wang}}}\ (\bibinfo {year} {2016})\ pp.\
  \bibinfo {pages} {67--129}\BibitemShut {NoStop}%
\bibitem [{\citenamefont {Loizides}(2016)}]{Loizides:2016tew}%
  \BibitemOpen
  \bibfield  {author} {\bibinfo {author} {\bibfnamefont {C.}~\bibnamefont
  {Loizides}},\ }\bibfield  {title} {\enquote {\bibinfo {title} {{Experimental
  overview on small collision systems at the LHC}},}\ }\bibfield  {booktitle}
  {\emph {\bibinfo {booktitle} {{Proceedings, 25th International Conference on
  Ultra-Relativistic Nucleus-Nucleus Collisions (Quark Matter 2015): Kobe,
  Japan, September 27-October 3, 2015}}},\ }\href {\doibase
  10.1016/j.nuclphysa.2016.04.022} {\bibfield  {journal} {\bibinfo  {journal}
  {Nucl. Phys.}\ }\textbf {\bibinfo {volume} {A956}},\ \bibinfo {pages}
  {200--207} (\bibinfo {year} {2016})},\ \Eprint
  {http://arxiv.org/abs/1602.09138} {arXiv:1602.09138 [nucl-ex]} \BibitemShut
  {NoStop}%
\bibitem [{\citenamefont {Stankus}(2016)}]{Stankus:2016usz}%
  \BibitemOpen
  \bibfield  {author} {\bibinfo {author} {\bibfnamefont {P.}~\bibnamefont
  {Stankus}},\ }\bibfield  {title} {\enquote {\bibinfo {title} {{Experimental
  overview on small colliding systems at RHIC}},}\ }\bibfield  {booktitle}
  {\emph {\bibinfo {booktitle} {{Proceedings, 25th International Conference on
  Ultra-Relativistic Nucleus-Nucleus Collisions (Quark Matter 2015): Kobe,
  Japan, September 27-October 3, 2015}}},\ }\href {\doibase
  10.1016/j.nuclphysa.2016.05.016} {\bibfield  {journal} {\bibinfo  {journal}
  {Nucl. Phys.}\ }\textbf {\bibinfo {volume} {A956}},\ \bibinfo {pages}
  {192--199} (\bibinfo {year} {2016})}\BibitemShut {NoStop}%
\bibitem [{\citenamefont {Stephanov}\ \emph {et~al.}(1998)\citenamefont
  {Stephanov}, \citenamefont {Rajagopal},\ and\ \citenamefont
  {Shuryak}}]{Stephanov:1998dy}%
  \BibitemOpen
  \bibfield  {author} {\bibinfo {author} {\bibfnamefont {M.~A.}\ \bibnamefont
  {Stephanov}}, \bibinfo {author} {\bibfnamefont {K.}~\bibnamefont
  {Rajagopal}}, \ and\ \bibinfo {author} {\bibfnamefont {E.~V.}\ \bibnamefont
  {Shuryak}},\ }\bibfield  {title} {\enquote {\bibinfo {title} {{Signatures of
  the tricritical point in QCD}},}\ }\href {\doibase
  10.1103/PhysRevLett.81.4816} {\bibfield  {journal} {\bibinfo  {journal}
  {Phys. Rev. Lett.}\ }\textbf {\bibinfo {volume} {81}},\ \bibinfo {pages}
  {4816--4819} (\bibinfo {year} {1998})},\ \Eprint
  {http://arxiv.org/abs/hep-ph/9806219} {arXiv:hep-ph/9806219 [hep-ph]}
  \BibitemShut {NoStop}%
\bibitem [{\citenamefont {Teaney}(2010)}]{Teaney:2009qa}%
  \BibitemOpen
  \bibfield  {author} {\bibinfo {author} {\bibfnamefont {D.~A.}\ \bibnamefont
  {Teaney}},\ }\bibfield  {title} {\enquote {\bibinfo {title} {{Viscous
  Hydrodynamics and the Quark Gluon Plasma}},}\ }in\ \href {\doibase
  10.1142/9789814293297_0004} {\emph {\bibinfo {booktitle} {Quark-Gluon Plasma
  4}}}\ (\bibinfo  {publisher} {World Scientific Pub Co Pte Lt},\ \bibinfo
  {year} {2010})\ pp.\ \bibinfo {pages} {207--266},\ \Eprint
  {http://arxiv.org/abs/0905.2433} {arXiv:0905.2433 [nucl-th]} \BibitemShut
  {NoStop}%
\bibitem [{\citenamefont {Romatschke}(2010)}]{Romatschke:2009im}%
  \BibitemOpen
  \bibfield  {author} {\bibinfo {author} {\bibfnamefont {P.}~\bibnamefont
  {Romatschke}},\ }\bibfield  {title} {\enquote {\bibinfo {title} {{New
  Developments in Relativistic Viscous Hydrodynamics}},}\ }\href {\doibase
  10.1142/S0218301310014613} {\bibfield  {journal} {\bibinfo  {journal} {Int.
  J. Mod. Phys.}\ }\textbf {\bibinfo {volume} {E19}},\ \bibinfo {pages} {1--53}
  (\bibinfo {year} {2010})},\ \Eprint {http://arxiv.org/abs/0902.3663}
  {arXiv:0902.3663 [hep-ph]} \BibitemShut {NoStop}%
\bibitem [{\citenamefont {Bjorken}(1983)}]{Bjorken:1982qr}%
  \BibitemOpen
  \bibfield  {author} {\bibinfo {author} {\bibfnamefont {J.~D.}\ \bibnamefont
  {Bjorken}},\ }\bibfield  {title} {\enquote {\bibinfo {title} {{Highly
  Relativistic Nucleus-Nucleus Collisions: The Central Rapidity Region}},}\
  }\href {\doibase 10.1103/PhysRevD.27.140} {\bibfield  {journal} {\bibinfo
  {journal} {Phys. Rev.}\ }\textbf {\bibinfo {volume} {D27}},\ \bibinfo {pages}
  {140--151} (\bibinfo {year} {1983})}\BibitemShut {NoStop}%
\bibitem [{\citenamefont {Aad}\ \emph {et~al.}(2012{\natexlab{b}})\citenamefont
  {Aad} \emph {et~al.}}]{ATLAS:2011ag}%
  \BibitemOpen
  \bibfield  {author} {\bibinfo {author} {\bibfnamefont {G.}~\bibnamefont
  {Aad}} \emph {et~al.} (\bibinfo {collaboration} {ATLAS}),\ }\bibfield
  {title} {\enquote {\bibinfo {title} {{Measurement of the centrality
  dependence of the charged particle pseudorapidity distribution in lead-lead
  collisions at $\sqrt{s_{NN}}=2.76$ TeV with the ATLAS detector}},}\ }\href
  {\doibase 10.1016/j.physletb.2012.02.045} {\bibfield  {journal} {\bibinfo
  {journal} {Phys. Lett.}\ }\textbf {\bibinfo {volume} {B710}},\ \bibinfo
  {pages} {363--382} (\bibinfo {year} {2012}{\natexlab{b}})},\ \Eprint
  {http://arxiv.org/abs/1108.6027} {arXiv:1108.6027 [hep-ex]} \BibitemShut
  {NoStop}%
\bibitem [{\citenamefont {Chatrchyan}\ \emph {et~al.}(2011)\citenamefont
  {Chatrchyan} \emph {et~al.}}]{Chatrchyan:2011pb}%
  \BibitemOpen
  \bibfield  {author} {\bibinfo {author} {\bibfnamefont {S.}~\bibnamefont
  {Chatrchyan}} \emph {et~al.} (\bibinfo {collaboration} {CMS}),\ }\bibfield
  {title} {\enquote {\bibinfo {title} {{Dependence on pseudorapidity and
  centrality of charged hadron production in PbPb collisions at a
  nucleon-nucleon centre-of-mass energy of 2.76 TeV}},}\ }\href {\doibase
  10.1007/JHEP08(2011)141} {\bibfield  {journal} {\bibinfo  {journal} {JHEP}\
  }\textbf {\bibinfo {volume} {08}},\ \bibinfo {pages} {141} (\bibinfo {year}
  {2011})},\ \Eprint {http://arxiv.org/abs/1107.4800} {arXiv:1107.4800
  [nucl-ex]} \BibitemShut {NoStop}%
\bibitem [{\citenamefont {McLerran}(2002)}]{McLerran:2001sr}%
  \BibitemOpen
  \bibfield  {author} {\bibinfo {author} {\bibfnamefont {L.~D.}\ \bibnamefont
  {McLerran}},\ }\bibfield  {title} {\enquote {\bibinfo {title} {{The Color
  glass condensate and small x physics: Four lectures}},}\ }\bibfield
  {booktitle} {\emph {\bibinfo {booktitle} {{Lectures on quark matter.
  Proceedings, 40. International Universitätswochen for theoretical physics,
  40th Winter School, IUKT 40: Schladming, Austria, March 3-10, 2001}}},\
  }\href {\doibase 10.1007/3-540-45792-5_8} {\bibfield  {journal} {\bibinfo
  {journal} {Lect. Notes Phys.}\ }\textbf {\bibinfo {volume} {583}},\ \bibinfo
  {pages} {291--334} (\bibinfo {year} {2002})},\ \Eprint
  {http://arxiv.org/abs/hep-ph/0104285} {arXiv:hep-ph/0104285 [hep-ph]}
  \BibitemShut {NoStop}%
\bibitem [{\citenamefont {Huovinen}(2003)}]{Huovinen:2002rn}%
  \BibitemOpen
  \bibfield  {author} {\bibinfo {author} {\bibfnamefont {P.}~\bibnamefont
  {Huovinen}},\ }\bibfield  {title} {\enquote {\bibinfo {title} {{Results of
  the hydrodynamics approach to heavy ion collisions}},}\ }\bibfield
  {booktitle} {\emph {\bibinfo {booktitle} {{Proceedings, 16th International
  Conference on Ultra-Relativistic nucleus nucleus collisions (Quark Matter
  2012): Nantes, France, July 18-24, 2002}}},\ }\href {\doibase
  10.1016/S0375-9474(02)01439-2} {\bibfield  {journal} {\bibinfo  {journal}
  {Nucl. Phys.}\ }\textbf {\bibinfo {volume} {A715}},\ \bibinfo {pages}
  {299--308} (\bibinfo {year} {2003})},\ \Eprint
  {http://arxiv.org/abs/nucl-th/0210024} {arXiv:nucl-th/0210024 [nucl-th]}
  \BibitemShut {NoStop}%
\bibitem [{\citenamefont {Lappi}(2016)}]{Lappi:2015jka}%
  \BibitemOpen
  \bibfield  {author} {\bibinfo {author} {\bibfnamefont {T.}~\bibnamefont
  {Lappi}},\ }\bibfield  {title} {\enquote {\bibinfo {title} {{Initial state in
  heavy ion collisions}},}\ }in\ \href {\doibase
  10.1016/j.nuclphysbps.2016.05.007} {\emph {\bibinfo {booktitle}
  {{Proceedings, 7th International Conference on Hard and Electromagnetic
  Probes of High-Energy Nuclear Collisions (Hard Probes 2015): Montréal,
  Québec, Canada, June 29-July 3, 2015}}}}\ (\bibinfo {year} {2016})\ \Eprint
  {http://arxiv.org/abs/1509.04503} {arXiv:1509.04503 [hep-ph]} \BibitemShut
  {NoStop}%
\bibitem [{\citenamefont {Angeli}\ and\ \citenamefont
  {Marinova}(2013)}]{Angeli201369}%
  \BibitemOpen
  \bibfield  {author} {\bibinfo {author} {\bibfnamefont {I.}~\bibnamefont
  {Angeli}}\ and\ \bibinfo {author} {\bibfnamefont {K.}~\bibnamefont
  {Marinova}},\ }\bibfield  {title} {\enquote {\bibinfo {title} {Table of
  experimental nuclear ground state charge radii: An update},}\ }\href
  {\doibase http://dx.doi.org/10.1016/j.adt.2011.12.006} {\bibfield  {journal}
  {\bibinfo  {journal} {Atomic Data and Nuclear Data Tables}\ }\textbf
  {\bibinfo {volume} {99}},\ \bibinfo {pages} {69 -- 95} (\bibinfo {year}
  {2013})}\BibitemShut {NoStop}%
\bibitem [{\citenamefont {Miller}\ \emph {et~al.}(2007)\citenamefont {Miller},
  \citenamefont {Reygers}, \citenamefont {Sanders},\ and\ \citenamefont
  {Steinberg}}]{Miller:2007ri}%
  \BibitemOpen
  \bibfield  {author} {\bibinfo {author} {\bibfnamefont {M.~L.}\ \bibnamefont
  {Miller}}, \bibinfo {author} {\bibfnamefont {K.}~\bibnamefont {Reygers}},
  \bibinfo {author} {\bibfnamefont {S.~J.}\ \bibnamefont {Sanders}}, \ and\
  \bibinfo {author} {\bibfnamefont {P.}~\bibnamefont {Steinberg}},\ }\bibfield
  {title} {\enquote {\bibinfo {title} {{Glauber modeling in high energy nuclear
  collisions}},}\ }\href {\doibase 10.1146/annurev.nucl.57.090506.123020}
  {\bibfield  {journal} {\bibinfo  {journal} {Ann. Rev. Nucl. Part. Sci.}\
  }\textbf {\bibinfo {volume} {57}},\ \bibinfo {pages} {205--243} (\bibinfo
  {year} {2007})},\ \Eprint {http://arxiv.org/abs/nucl-ex/0701025}
  {arXiv:nucl-ex/0701025 [nucl-ex]} \BibitemShut {NoStop}%
\bibitem [{\citenamefont {Blaizot}\ \emph {et~al.}(2014)\citenamefont
  {Blaizot}, \citenamefont {Broniowski},\ and\ \citenamefont
  {Ollitrault}}]{Blaizot:2014wba}%
  \BibitemOpen
  \bibfield  {author} {\bibinfo {author} {\bibfnamefont {J.-P.}\ \bibnamefont
  {Blaizot}}, \bibinfo {author} {\bibfnamefont {W.}~\bibnamefont {Broniowski}},
  \ and\ \bibinfo {author} {\bibfnamefont {J.-Y.}\ \bibnamefont {Ollitrault}},\
  }\bibfield  {title} {\enquote {\bibinfo {title} {{Correlations in the Monte
  Carlo Glauber model}},}\ }\href {\doibase 10.1103/PhysRevC.90.034906}
  {\bibfield  {journal} {\bibinfo  {journal} {Phys. Rev.}\ }\textbf {\bibinfo
  {volume} {C90}},\ \bibinfo {pages} {034906} (\bibinfo {year} {2014})},\
  \Eprint {http://arxiv.org/abs/1405.3274} {arXiv:1405.3274 [nucl-th]}
  \BibitemShut {NoStop}%
\bibitem [{\citenamefont {Alver}\ \emph {et~al.}(2008)\citenamefont {Alver},
  \citenamefont {Baker}, \citenamefont {Loizides},\ and\ \citenamefont
  {Steinberg}}]{Alver:2008aq}%
  \BibitemOpen
  \bibfield  {author} {\bibinfo {author} {\bibfnamefont {B.}~\bibnamefont
  {Alver}}, \bibinfo {author} {\bibfnamefont {M.}~\bibnamefont {Baker}},
  \bibinfo {author} {\bibfnamefont {C.}~\bibnamefont {Loizides}}, \ and\
  \bibinfo {author} {\bibfnamefont {P.}~\bibnamefont {Steinberg}},\ }\bibfield
  {title} {\enquote {\bibinfo {title} {{The PHOBOS Glauber Monte Carlo}},}\
  }\href@noop {} {\  (\bibinfo {year} {2008})},\ \Eprint
  {http://arxiv.org/abs/0805.4411} {arXiv:0805.4411 [nucl-ex]} \BibitemShut
  {NoStop}%
\bibitem [{\citenamefont {Hirano}\ and\ \citenamefont
  {Nara}(2009)}]{Hirano:2009ah}%
  \BibitemOpen
  \bibfield  {author} {\bibinfo {author} {\bibfnamefont {T.}~\bibnamefont
  {Hirano}}\ and\ \bibinfo {author} {\bibfnamefont {Y.}~\bibnamefont {Nara}},\
  }\bibfield  {title} {\enquote {\bibinfo {title} {{Eccentricity fluctuation
  effects on elliptic flow in relativistic heavy ion collisions}},}\ }\href
  {\doibase 10.1103/PhysRevC.79.064904} {\bibfield  {journal} {\bibinfo
  {journal} {Phys. Rev.}\ }\textbf {\bibinfo {volume} {C79}},\ \bibinfo {pages}
  {064904} (\bibinfo {year} {2009})},\ \Eprint {http://arxiv.org/abs/0904.4080}
  {arXiv:0904.4080 [nucl-th]} \BibitemShut {NoStop}%
\bibitem [{\citenamefont {Qiu}(2013)}]{Qiu:2013wca}%
  \BibitemOpen
  \bibfield  {author} {\bibinfo {author} {\bibfnamefont {Z.}~\bibnamefont
  {Qiu}},\ }\emph {\bibinfo {title} {{Event-by-event Hydrodynamic Simulations
  for Relativistic Heavy-ion Collisions}}},\ \href
  {https://inspirehep.net/record/1247294/files/arXiv:1308.2182.pdf} {Ph.D.
  thesis},\ \bibinfo  {school} {Ohio State U.} (\bibinfo {year} {2013}),\
  \Eprint {http://arxiv.org/abs/1308.2182} {arXiv:1308.2182 [nucl-th]}
  \BibitemShut {NoStop}%
\bibitem [{\citenamefont {Qiu}\ and\ \citenamefont {Heinz}(2011)}]{Qiu:2011iv}%
  \BibitemOpen
  \bibfield  {author} {\bibinfo {author} {\bibfnamefont {Z.}~\bibnamefont
  {Qiu}}\ and\ \bibinfo {author} {\bibfnamefont {U.~W.}\ \bibnamefont
  {Heinz}},\ }\bibfield  {title} {\enquote {\bibinfo {title} {{Event-by-event
  shape and flow fluctuations of relativistic heavy-ion collision
  fireballs}},}\ }\href {\doibase 10.1103/PhysRevC.84.024911} {\bibfield
  {journal} {\bibinfo  {journal} {Phys. Rev.}\ }\textbf {\bibinfo {volume}
  {C84}},\ \bibinfo {pages} {024911} (\bibinfo {year} {2011})},\ \Eprint
  {http://arxiv.org/abs/1104.0650} {arXiv:1104.0650 [nucl-th]} \BibitemShut
  {NoStop}%
\bibitem [{\citenamefont {Iancu}\ \emph {et~al.}(2002)\citenamefont {Iancu},
  \citenamefont {Leonidov},\ and\ \citenamefont {McLerran}}]{Iancu:2002xk}%
  \BibitemOpen
  \bibfield  {author} {\bibinfo {author} {\bibfnamefont {E.}~\bibnamefont
  {Iancu}}, \bibinfo {author} {\bibfnamefont {A.}~\bibnamefont {Leonidov}}, \
  and\ \bibinfo {author} {\bibfnamefont {L.}~\bibnamefont {McLerran}},\
  }\bibfield  {title} {\enquote {\bibinfo {title} {{The Color glass condensate:
  An Introduction}},}\ }in\ \href
  {http://alice.cern.ch/format/showfull?sysnb=2297268} {\emph {\bibinfo
  {booktitle} {{QCD perspectives on hot and dense matter. Proceedings, NATO
  Advanced Study Institute, Summer School, Cargese, France, August 6-18,
  2001}}}}\ (\bibinfo {year} {2002})\ pp.\ \bibinfo {pages} {73--145},\ \Eprint
  {http://arxiv.org/abs/hep-ph/0202270} {arXiv:hep-ph/0202270 [hep-ph]}
  \BibitemShut {NoStop}%
\bibitem [{\citenamefont {Iancu}\ and\ \citenamefont
  {Venugopalan}(2003)}]{Iancu:2003xm}%
  \BibitemOpen
  \bibfield  {author} {\bibinfo {author} {\bibfnamefont {E.}~\bibnamefont
  {Iancu}}\ and\ \bibinfo {author} {\bibfnamefont {R.}~\bibnamefont
  {Venugopalan}},\ }\bibfield  {title} {\enquote {\bibinfo {title} {{The Color
  glass condensate and high-energy scattering in QCD}},}\ }in\ \href@noop {}
  {\emph {\bibinfo {booktitle} {{In *Hwa, R.C. (ed.) et al.: Quark gluon
  plasma* 249-3363}}}}\ (\bibinfo {year} {2003})\ \Eprint
  {http://arxiv.org/abs/hep-ph/0303204} {arXiv:hep-ph/0303204 [hep-ph]}
  \BibitemShut {NoStop}%
\bibitem [{\citenamefont {Gelis}\ \emph {et~al.}(2010)\citenamefont {Gelis},
  \citenamefont {Iancu}, \citenamefont {Jalilian-Marian},\ and\ \citenamefont
  {Venugopalan}}]{Gelis:2010nm}%
  \BibitemOpen
  \bibfield  {author} {\bibinfo {author} {\bibfnamefont {F.}~\bibnamefont
  {Gelis}}, \bibinfo {author} {\bibfnamefont {E.}~\bibnamefont {Iancu}},
  \bibinfo {author} {\bibfnamefont {J.}~\bibnamefont {Jalilian-Marian}}, \ and\
  \bibinfo {author} {\bibfnamefont {R.}~\bibnamefont {Venugopalan}},\
  }\bibfield  {title} {\enquote {\bibinfo {title} {{The Color Glass
  Condensate}},}\ }\href {\doibase 10.1146/annurev.nucl.010909.083629}
  {\bibfield  {journal} {\bibinfo  {journal} {Ann. Rev. Nucl. Part. Sci.}\
  }\textbf {\bibinfo {volume} {60}},\ \bibinfo {pages} {463--489} (\bibinfo
  {year} {2010})},\ \Eprint {http://arxiv.org/abs/1002.0333} {arXiv:1002.0333
  [hep-ph]} \BibitemShut {NoStop}%
\bibitem [{\citenamefont {Gelis}\ \emph {et~al.}(2007)\citenamefont {Gelis},
  \citenamefont {Lappi},\ and\ \citenamefont {Venugopalan}}]{Gelis:2007kn}%
  \BibitemOpen
  \bibfield  {author} {\bibinfo {author} {\bibfnamefont {F.}~\bibnamefont
  {Gelis}}, \bibinfo {author} {\bibfnamefont {T.}~\bibnamefont {Lappi}}, \ and\
  \bibinfo {author} {\bibfnamefont {R.}~\bibnamefont {Venugopalan}},\
  }\bibfield  {title} {\enquote {\bibinfo {title} {{High energy scattering in
  Quantum Chromodynamics}},}\ }\bibfield  {booktitle} {\emph {\bibinfo
  {booktitle} {{Hadron physics. Proceedings, 10th International Workshop,
  Florianopolis, Brazil, April 26-31, 2007}}},\ }\href {\doibase
  10.1142/S0218301307008331} {\bibfield  {journal} {\bibinfo  {journal} {Int.
  J. Mod. Phys.}\ }\textbf {\bibinfo {volume} {E16}},\ \bibinfo {pages}
  {2595--2637} (\bibinfo {year} {2007})},\ \Eprint
  {http://arxiv.org/abs/0708.0047} {arXiv:0708.0047 [hep-ph]} \BibitemShut
  {NoStop}%
\bibitem [{\citenamefont {Lappi}(2011)}]{Lappi:2011ju}%
  \BibitemOpen
  \bibfield  {author} {\bibinfo {author} {\bibfnamefont {T.}~\bibnamefont
  {Lappi}},\ }\bibfield  {title} {\enquote {\bibinfo {title} {{Gluon spectrum
  in the glasma from JIMWLK evolution}},}\ }\href {\doibase
  10.1016/j.physletb.2011.08.011} {\bibfield  {journal} {\bibinfo  {journal}
  {Phys. Lett.}\ }\textbf {\bibinfo {volume} {B703}},\ \bibinfo {pages}
  {325--330} (\bibinfo {year} {2011})},\ \Eprint
  {http://arxiv.org/abs/1105.5511} {arXiv:1105.5511 [hep-ph]} \BibitemShut
  {NoStop}%
\bibitem [{\citenamefont {Albacete}\ and\ \citenamefont
  {Marquet}(2014)}]{Albacete:2014fwa}%
  \BibitemOpen
  \bibfield  {author} {\bibinfo {author} {\bibfnamefont {J.~L.}\ \bibnamefont
  {Albacete}}\ and\ \bibinfo {author} {\bibfnamefont {C.}~\bibnamefont
  {Marquet}},\ }\bibfield  {title} {\enquote {\bibinfo {title} {{Gluon
  saturation and initial conditions for relativistic heavy ion collisions}},}\
  }\href {\doibase 10.1016/j.ppnp.2014.01.004} {\bibfield  {journal} {\bibinfo
  {journal} {Prog. Part. Nucl. Phys.}\ }\textbf {\bibinfo {volume} {76}},\
  \bibinfo {pages} {1--42} (\bibinfo {year} {2014})},\ \Eprint
  {http://arxiv.org/abs/1401.4866} {arXiv:1401.4866 [hep-ph]} \BibitemShut
  {NoStop}%
\bibitem [{\citenamefont {Aaron}\ \emph {et~al.}(2010)\citenamefont {Aaron}
  \emph {et~al.}}]{Aaron:2009aa}%
  \BibitemOpen
  \bibfield  {author} {\bibinfo {author} {\bibfnamefont {F.~D.}\ \bibnamefont
  {Aaron}} \emph {et~al.} (\bibinfo {collaboration} {ZEUS, H1}),\ }\bibfield
  {title} {\enquote {\bibinfo {title} {{Combined Measurement and QCD Analysis
  of the Inclusive e+- p Scattering Cross Sections at HERA}},}\ }\href
  {\doibase 10.1007/JHEP01(2010)109} {\bibfield  {journal} {\bibinfo  {journal}
  {JHEP}\ }\textbf {\bibinfo {volume} {01}},\ \bibinfo {pages} {109} (\bibinfo
  {year} {2010})},\ \Eprint {http://arxiv.org/abs/0911.0884} {arXiv:0911.0884
  [hep-ex]} \BibitemShut {NoStop}%
\bibitem [{\citenamefont {Krasnitz}\ and\ \citenamefont
  {Venugopalan}(2000)}]{Krasnitz:1999wc}%
  \BibitemOpen
  \bibfield  {author} {\bibinfo {author} {\bibfnamefont {A.}~\bibnamefont
  {Krasnitz}}\ and\ \bibinfo {author} {\bibfnamefont {R.}~\bibnamefont
  {Venugopalan}},\ }\bibfield  {title} {\enquote {\bibinfo {title} {{The
  Initial energy density of gluons produced in very high-energy nuclear
  collisions}},}\ }\href {\doibase 10.1103/PhysRevLett.84.4309} {\bibfield
  {journal} {\bibinfo  {journal} {Phys. Rev. Lett.}\ }\textbf {\bibinfo
  {volume} {84}},\ \bibinfo {pages} {4309--4312} (\bibinfo {year} {2000})},\
  \Eprint {http://arxiv.org/abs/hep-ph/9909203} {arXiv:hep-ph/9909203 [hep-ph]}
  \BibitemShut {NoStop}%
\bibitem [{\citenamefont {Berges}\ \emph
  {et~al.}(2014{\natexlab{a}})\citenamefont {Berges}, \citenamefont
  {Boguslavski}, \citenamefont {Schlichting},\ and\ \citenamefont
  {Venugopalan}}]{Berges:2013eia}%
  \BibitemOpen
  \bibfield  {author} {\bibinfo {author} {\bibfnamefont {J.}~\bibnamefont
  {Berges}}, \bibinfo {author} {\bibfnamefont {K.}~\bibnamefont {Boguslavski}},
  \bibinfo {author} {\bibfnamefont {S.}~\bibnamefont {Schlichting}}, \ and\
  \bibinfo {author} {\bibfnamefont {R.}~\bibnamefont {Venugopalan}},\
  }\bibfield  {title} {\enquote {\bibinfo {title} {{Turbulent thermalization
  process in heavy-ion collisions at ultrarelativistic energies}},}\ }\href
  {\doibase 10.1103/PhysRevD.89.074011} {\bibfield  {journal} {\bibinfo
  {journal} {Phys. Rev.}\ }\textbf {\bibinfo {volume} {D89}},\ \bibinfo {pages}
  {074011} (\bibinfo {year} {2014}{\natexlab{a}})},\ \Eprint
  {http://arxiv.org/abs/1303.5650} {arXiv:1303.5650 [hep-ph]} \BibitemShut
  {NoStop}%
\bibitem [{\citenamefont {Berges}\ \emph
  {et~al.}(2014{\natexlab{b}})\citenamefont {Berges}, \citenamefont
  {Boguslavski}, \citenamefont {Schlichting},\ and\ \citenamefont
  {Venugopalan}}]{Berges:2013fga}%
  \BibitemOpen
  \bibfield  {author} {\bibinfo {author} {\bibfnamefont {J.}~\bibnamefont
  {Berges}}, \bibinfo {author} {\bibfnamefont {K.}~\bibnamefont {Boguslavski}},
  \bibinfo {author} {\bibfnamefont {S.}~\bibnamefont {Schlichting}}, \ and\
  \bibinfo {author} {\bibfnamefont {R.}~\bibnamefont {Venugopalan}},\
  }\bibfield  {title} {\enquote {\bibinfo {title} {{Universal attractor in a
  highly occupied non-Abelian plasma}},}\ }\href {\doibase
  10.1103/PhysRevD.89.114007} {\bibfield  {journal} {\bibinfo  {journal} {Phys.
  Rev.}\ }\textbf {\bibinfo {volume} {D89}},\ \bibinfo {pages} {114007}
  (\bibinfo {year} {2014}{\natexlab{b}})},\ \Eprint
  {http://arxiv.org/abs/1311.3005} {arXiv:1311.3005 [hep-ph]} \BibitemShut
  {NoStop}%
\bibitem [{\citenamefont {Baier}\ \emph {et~al.}(2001)\citenamefont {Baier},
  \citenamefont {Mueller}, \citenamefont {Schiff},\ and\ \citenamefont
  {Son}}]{Baier:2000sb}%
  \BibitemOpen
  \bibfield  {author} {\bibinfo {author} {\bibfnamefont {R.}~\bibnamefont
  {Baier}}, \bibinfo {author} {\bibfnamefont {A.~H.}\ \bibnamefont {Mueller}},
  \bibinfo {author} {\bibfnamefont {D.}~\bibnamefont {Schiff}}, \ and\ \bibinfo
  {author} {\bibfnamefont {D.~T.}\ \bibnamefont {Son}},\ }\bibfield  {title}
  {\enquote {\bibinfo {title} {{'Bottom up' thermalization in heavy ion
  collisions}},}\ }\href {\doibase 10.1016/S0370-2693(01)00191-5} {\bibfield
  {journal} {\bibinfo  {journal} {Phys. Lett.}\ }\textbf {\bibinfo {volume}
  {B502}},\ \bibinfo {pages} {51--58} (\bibinfo {year} {2001})},\ \Eprint
  {http://arxiv.org/abs/hep-ph/0009237} {arXiv:hep-ph/0009237 [hep-ph]}
  \BibitemShut {NoStop}%
\bibitem [{\citenamefont {Aamodt}\ \emph
  {et~al.}(2011{\natexlab{b}})\citenamefont {Aamodt} \emph
  {et~al.}}]{Aamodt:2010cz}%
  \BibitemOpen
  \bibfield  {author} {\bibinfo {author} {\bibfnamefont {K.}~\bibnamefont
  {Aamodt}} \emph {et~al.} (\bibinfo {collaboration} {ALICE}),\ }\bibfield
  {title} {\enquote {\bibinfo {title} {{Centrality dependence of the
  charged-particle multiplicity density at mid-rapidity in Pb-Pb collisions at
  $\sqrt{s_{NN}}=2.76$ TeV}},}\ }\href {\doibase
  10.1103/PhysRevLett.106.032301} {\bibfield  {journal} {\bibinfo  {journal}
  {Phys. Rev. Lett.}\ }\textbf {\bibinfo {volume} {106}},\ \bibinfo {pages}
  {032301} (\bibinfo {year} {2011}{\natexlab{b}})},\ \Eprint
  {http://arxiv.org/abs/1012.1657} {arXiv:1012.1657 [nucl-ex]} \BibitemShut
  {NoStop}%
\bibitem [{\citenamefont {Ollitrault}(1992)}]{Ollitrault:1992bk}%
  \BibitemOpen
  \bibfield  {author} {\bibinfo {author} {\bibfnamefont {J.-Y.}\ \bibnamefont
  {Ollitrault}},\ }\bibfield  {title} {\enquote {\bibinfo {title} {{Anisotropy
  as a signature of transverse collective flow}},}\ }\href {\doibase
  10.1103/PhysRevD.46.229} {\bibfield  {journal} {\bibinfo  {journal} {Phys.
  Rev.}\ }\textbf {\bibinfo {volume} {D46}},\ \bibinfo {pages} {229--245}
  (\bibinfo {year} {1992})}\BibitemShut {NoStop}%
\bibitem [{\citenamefont {Landau}\ and\ \citenamefont
  {Lifshit︠s︡}(1959)}]{LandauFluids}%
  \BibitemOpen
  \bibfield  {author} {\bibinfo {author} {\bibfnamefont {L.~D.}\ \bibnamefont
  {Landau}}\ and\ \bibinfo {author} {\bibfnamefont {E.~M.}\ \bibnamefont
  {Lifshit︠s︡}},\ }\href
  {http://proxy.library.stonybrook.edu/login?url=http://search.ebscohost.com/login.aspx?direct=true&db=cat03000a&AN=STB.000516093&site=eds-live&scope=site}
  {\emph {\bibinfo {title} {Fluid mechanics}}},\ \bibinfo {series} {Course of
  theoretical physics}, Vol.~\bibinfo {volume} {6}\ (\bibinfo  {publisher}
  {London, Pergamon Press; Reading, Mass., Addison-Wesley Pub. Co., 1959.},\
  \bibinfo {year} {1959})\BibitemShut {NoStop}%
\bibitem [{\citenamefont {Kovtun}(2012)}]{Kovtun:2012rj}%
  \BibitemOpen
  \bibfield  {author} {\bibinfo {author} {\bibfnamefont {P.}~\bibnamefont
  {Kovtun}},\ }\bibfield  {title} {\enquote {\bibinfo {title} {{Lectures on
  hydrodynamic fluctuations in relativistic theories}},}\ }\bibfield
  {booktitle} {\emph {\bibinfo {booktitle} {{INT Summer School on Applications
  of String Theory Seattle, Washington, USA, July 18-29, 2011}}},\ }\href
  {\doibase 10.1088/1751-8113/45/47/473001} {\bibfield  {journal} {\bibinfo
  {journal} {J. Phys.}\ }\textbf {\bibinfo {volume} {A45}},\ \bibinfo {pages}
  {473001} (\bibinfo {year} {2012})},\ \Eprint {http://arxiv.org/abs/1205.5040}
  {arXiv:1205.5040 [hep-th]} \BibitemShut {NoStop}%
\bibitem [{\citenamefont {Luzum}\ and\ \citenamefont
  {Ollitrault}(2013)}]{Luzum:2012wu}%
  \BibitemOpen
  \bibfield  {author} {\bibinfo {author} {\bibfnamefont {M.}~\bibnamefont
  {Luzum}}\ and\ \bibinfo {author} {\bibfnamefont {J.-Y.}\ \bibnamefont
  {Ollitrault}},\ }\bibfield  {title} {\enquote {\bibinfo {title} {{Extracting
  the shear viscosity of the quark-gluon plasma from flow in ultra-central
  heavy-ion collisions}},}\ }\bibfield  {booktitle} {\emph {\bibinfo
  {booktitle} {{Proceedings, 23rd International Conference on Ultrarelativistic
  Nucleus-Nucleus Collisions : Quark Matter 2012 (QM 2012): Washington, DC,
  USA, August 13-18, 2012}}},\ }\href {\doibase
  10.1016/j.nuclphysa.2013.02.028} {\bibfield  {journal} {\bibinfo  {journal}
  {Nucl. Phys.}\ }\textbf {\bibinfo {volume} {A904-905}},\ \bibinfo {pages}
  {377c--380c} (\bibinfo {year} {2013})},\ \Eprint
  {http://arxiv.org/abs/1210.6010} {arXiv:1210.6010 [nucl-th]} \BibitemShut
  {NoStop}%
\bibitem [{\citenamefont {Song}\ \emph {et~al.}(2011)\citenamefont {Song},
  \citenamefont {Bass}, \citenamefont {Heinz}, \citenamefont {Hirano},\ and\
  \citenamefont {Shen}}]{Song:2010mg}%
  \BibitemOpen
  \bibfield  {author} {\bibinfo {author} {\bibfnamefont {H.}~\bibnamefont
  {Song}}, \bibinfo {author} {\bibfnamefont {S.~A.}\ \bibnamefont {Bass}},
  \bibinfo {author} {\bibfnamefont {U.}~\bibnamefont {Heinz}}, \bibinfo
  {author} {\bibfnamefont {T.}~\bibnamefont {Hirano}}, \ and\ \bibinfo {author}
  {\bibfnamefont {C.}~\bibnamefont {Shen}},\ }\bibfield  {title} {\enquote
  {\bibinfo {title} {{200 A GeV Au+Au collisions serve a nearly perfect
  quark-gluon liquid}},}\ }\href {\doibase 10.1103/PhysRevLett.106.192301,
  10.1103/PhysRevLett.109.139904} {\bibfield  {journal} {\bibinfo  {journal}
  {Phys. Rev. Lett.}\ }\textbf {\bibinfo {volume} {106}},\ \bibinfo {pages}
  {192301} (\bibinfo {year} {2011})},\ \bibinfo {note} {[Erratum: Phys. Rev.
  Lett.109,139904(2012)]},\ \Eprint {http://arxiv.org/abs/1011.2783}
  {arXiv:1011.2783 [nucl-th]} \BibitemShut {NoStop}%
\bibitem [{\citenamefont {Bernhard}\ \emph {et~al.}(2016)\citenamefont
  {Bernhard}, \citenamefont {Moreland}, \citenamefont {Bass}, \citenamefont
  {Liu},\ and\ \citenamefont {Heinz}}]{Bernhard:2016tnd}%
  \BibitemOpen
  \bibfield  {author} {\bibinfo {author} {\bibfnamefont {J.~E.}\ \bibnamefont
  {Bernhard}}, \bibinfo {author} {\bibfnamefont {J.~S.}\ \bibnamefont
  {Moreland}}, \bibinfo {author} {\bibfnamefont {S.~A.}\ \bibnamefont {Bass}},
  \bibinfo {author} {\bibfnamefont {J.}~\bibnamefont {Liu}}, \ and\ \bibinfo
  {author} {\bibfnamefont {U.}~\bibnamefont {Heinz}},\ }\bibfield  {title}
  {\enquote {\bibinfo {title} {{Applying Bayesian parameter estimation to
  relativistic heavy-ion collisions: simultaneous characterization of the
  initial state and quark-gluon plasma medium}},}\ }\href {\doibase
  10.1103/PhysRevC.94.024907} {\bibfield  {journal} {\bibinfo  {journal} {Phys.
  Rev.}\ }\textbf {\bibinfo {volume} {C94}},\ \bibinfo {pages} {024907}
  (\bibinfo {year} {2016})},\ \Eprint {http://arxiv.org/abs/1605.03954}
  {arXiv:1605.03954 [nucl-th]} \BibitemShut {NoStop}%
\bibitem [{\citenamefont {Karsch}\ \emph {et~al.}(2008)\citenamefont {Karsch},
  \citenamefont {Kharzeev},\ and\ \citenamefont {Tuchin}}]{Karsch:2007jc}%
  \BibitemOpen
  \bibfield  {author} {\bibinfo {author} {\bibfnamefont {F.}~\bibnamefont
  {Karsch}}, \bibinfo {author} {\bibfnamefont {D.}~\bibnamefont {Kharzeev}}, \
  and\ \bibinfo {author} {\bibfnamefont {K.}~\bibnamefont {Tuchin}},\
  }\bibfield  {title} {\enquote {\bibinfo {title} {{Universal properties of
  bulk viscosity near the QCD phase transition}},}\ }\href {\doibase
  10.1016/j.physletb.2008.01.080} {\bibfield  {journal} {\bibinfo  {journal}
  {Phys. Lett.}\ }\textbf {\bibinfo {volume} {B663}},\ \bibinfo {pages}
  {217--221} (\bibinfo {year} {2008})},\ \Eprint
  {http://arxiv.org/abs/0711.0914} {arXiv:0711.0914 [hep-ph]} \BibitemShut
  {NoStop}%
\bibitem [{\citenamefont {Arnold}\ \emph {et~al.}(2006)\citenamefont {Arnold},
  \citenamefont {Dogan},\ and\ \citenamefont {Moore}}]{Arnold:2006fz}%
  \BibitemOpen
  \bibfield  {author} {\bibinfo {author} {\bibfnamefont {P.~B.}\ \bibnamefont
  {Arnold}}, \bibinfo {author} {\bibfnamefont {C.}~\bibnamefont {Dogan}}, \
  and\ \bibinfo {author} {\bibfnamefont {G.~D.}\ \bibnamefont {Moore}},\
  }\bibfield  {title} {\enquote {\bibinfo {title} {{The Bulk Viscosity of
  High-Temperature QCD}},}\ }\href {\doibase 10.1103/PhysRevD.74.085021}
  {\bibfield  {journal} {\bibinfo  {journal} {Phys. Rev.}\ }\textbf {\bibinfo
  {volume} {D74}},\ \bibinfo {pages} {085021} (\bibinfo {year} {2006})},\
  \Eprint {http://arxiv.org/abs/hep-ph/0608012} {arXiv:hep-ph/0608012 [hep-ph]}
  \BibitemShut {NoStop}%
\bibitem [{\citenamefont {Meyer}(2008)}]{Meyer:2007dy}%
  \BibitemOpen
  \bibfield  {author} {\bibinfo {author} {\bibfnamefont {H.~B.}\ \bibnamefont
  {Meyer}},\ }\bibfield  {title} {\enquote {\bibinfo {title} {{A Calculation of
  the bulk viscosity in SU(3) gluodynamics}},}\ }\href {\doibase
  10.1103/PhysRevLett.100.162001} {\bibfield  {journal} {\bibinfo  {journal}
  {Phys. Rev. Lett.}\ }\textbf {\bibinfo {volume} {100}},\ \bibinfo {pages}
  {162001} (\bibinfo {year} {2008})},\ \Eprint {http://arxiv.org/abs/0710.3717}
  {arXiv:0710.3717 [hep-lat]} \BibitemShut {NoStop}%
\bibitem [{\citenamefont {Ryu}\ \emph {et~al.}(2015)\citenamefont {Ryu},
  \citenamefont {Paquet}, \citenamefont {Shen}, \citenamefont {Denicol},
  \citenamefont {Schenke}, \citenamefont {Jeon},\ and\ \citenamefont
  {Gale}}]{Ryu:2015vwa}%
  \BibitemOpen
  \bibfield  {author} {\bibinfo {author} {\bibfnamefont {S.}~\bibnamefont
  {Ryu}}, \bibinfo {author} {\bibfnamefont {J.~F.}\ \bibnamefont {Paquet}},
  \bibinfo {author} {\bibfnamefont {C.}~\bibnamefont {Shen}}, \bibinfo {author}
  {\bibfnamefont {G.~S.}\ \bibnamefont {Denicol}}, \bibinfo {author}
  {\bibfnamefont {B.}~\bibnamefont {Schenke}}, \bibinfo {author} {\bibfnamefont
  {S.}~\bibnamefont {Jeon}}, \ and\ \bibinfo {author} {\bibfnamefont
  {C.}~\bibnamefont {Gale}},\ }\bibfield  {title} {\enquote {\bibinfo {title}
  {{Importance of the Bulk Viscosity of QCD in Ultrarelativistic Heavy-Ion
  Collisions}},}\ }\href {\doibase 10.1103/PhysRevLett.115.132301} {\bibfield
  {journal} {\bibinfo  {journal} {Phys. Rev. Lett.}\ }\textbf {\bibinfo
  {volume} {115}},\ \bibinfo {pages} {132301} (\bibinfo {year} {2015})},\
  \Eprint {http://arxiv.org/abs/1502.01675} {arXiv:1502.01675 [nucl-th]}
  \BibitemShut {NoStop}%
\bibitem [{\citenamefont {Hiscock}\ and\ \citenamefont
  {Lindblom}(1983)}]{Hiscock:1983zz}%
  \BibitemOpen
  \bibfield  {author} {\bibinfo {author} {\bibfnamefont {W.~A.}\ \bibnamefont
  {Hiscock}}\ and\ \bibinfo {author} {\bibfnamefont {L.}~\bibnamefont
  {Lindblom}},\ }\bibfield  {title} {\enquote {\bibinfo {title} {{Stability and
  causality in dissipative relativistic fluids}},}\ }\href {\doibase
  10.1016/0003-4916(83)90288-9} {\bibfield  {journal} {\bibinfo  {journal}
  {Annals Phys.}\ }\textbf {\bibinfo {volume} {151}},\ \bibinfo {pages}
  {466--496} (\bibinfo {year} {1983})}\BibitemShut {NoStop}%
\bibitem [{\citenamefont {Israel}(1976)}]{Israel:1976tn}%
  \BibitemOpen
  \bibfield  {author} {\bibinfo {author} {\bibfnamefont {W.}~\bibnamefont
  {Israel}},\ }\bibfield  {title} {\enquote {\bibinfo {title} {{Nonstationary
  irreversible thermodynamics: A Causal relativistic theory}},}\ }\href
  {\doibase 10.1016/0003-4916(76)90064-6} {\bibfield  {journal} {\bibinfo
  {journal} {Annals Phys.}\ }\textbf {\bibinfo {volume} {100}},\ \bibinfo
  {pages} {310--331} (\bibinfo {year} {1976})}\BibitemShut {NoStop}%
\bibitem [{\citenamefont {Baier}\ \emph {et~al.}(2008)\citenamefont {Baier},
  \citenamefont {Romatschke}, \citenamefont {Son}, \citenamefont {Starinets},\
  and\ \citenamefont {Stephanov}}]{Baier:2007ix}%
  \BibitemOpen
  \bibfield  {author} {\bibinfo {author} {\bibfnamefont {R.}~\bibnamefont
  {Baier}}, \bibinfo {author} {\bibfnamefont {P.}~\bibnamefont {Romatschke}},
  \bibinfo {author} {\bibfnamefont {D.~T.}\ \bibnamefont {Son}}, \bibinfo
  {author} {\bibfnamefont {A.~O.}\ \bibnamefont {Starinets}}, \ and\ \bibinfo
  {author} {\bibfnamefont {M.~A.}\ \bibnamefont {Stephanov}},\ }\bibfield
  {title} {\enquote {\bibinfo {title} {{Relativistic viscous hydrodynamics,
  conformal invariance, and holography}},}\ }\href {\doibase
  10.1088/1126-6708/2008/04/100} {\bibfield  {journal} {\bibinfo  {journal}
  {JHEP}\ }\textbf {\bibinfo {volume} {04}},\ \bibinfo {pages} {100} (\bibinfo
  {year} {2008})},\ \Eprint {http://arxiv.org/abs/0712.2451} {arXiv:0712.2451
  [hep-th]} \BibitemShut {NoStop}%
\bibitem [{\citenamefont {Bhattacharyya}\ \emph {et~al.}(2008)\citenamefont
  {Bhattacharyya}, \citenamefont {Hubeny}, \citenamefont {Minwalla},\ and\
  \citenamefont {Rangamani}}]{Bhattacharyya:2008jc}%
  \BibitemOpen
  \bibfield  {author} {\bibinfo {author} {\bibfnamefont {S.}~\bibnamefont
  {Bhattacharyya}}, \bibinfo {author} {\bibfnamefont {V.~E.}\ \bibnamefont
  {Hubeny}}, \bibinfo {author} {\bibfnamefont {S.}~\bibnamefont {Minwalla}}, \
  and\ \bibinfo {author} {\bibfnamefont {M.}~\bibnamefont {Rangamani}},\
  }\bibfield  {title} {\enquote {\bibinfo {title} {{Nonlinear Fluid Dynamics
  from Gravity}},}\ }\href {\doibase 10.1088/1126-6708/2008/02/045} {\bibfield
  {journal} {\bibinfo  {journal} {JHEP}\ }\textbf {\bibinfo {volume} {02}},\
  \bibinfo {pages} {045} (\bibinfo {year} {2008})},\ \Eprint
  {http://arxiv.org/abs/0712.2456} {arXiv:0712.2456 [hep-th]} \BibitemShut
  {NoStop}%
\bibitem [{\citenamefont {Bazavov}\ \emph {et~al.}(2014)\citenamefont {Bazavov}
  \emph {et~al.}}]{Bazavov:2014pvz}%
  \BibitemOpen
  \bibfield  {author} {\bibinfo {author} {\bibfnamefont {A.}~\bibnamefont
  {Bazavov}} \emph {et~al.} (\bibinfo {collaboration} {HotQCD}),\ }\bibfield
  {title} {\enquote {\bibinfo {title} {{Equation of state in ( 2+1 )-flavor
  QCD}},}\ }\href {\doibase 10.1103/PhysRevD.90.094503} {\bibfield  {journal}
  {\bibinfo  {journal} {Phys. Rev.}\ }\textbf {\bibinfo {volume} {D90}},\
  \bibinfo {pages} {094503} (\bibinfo {year} {2014})},\ \Eprint
  {http://arxiv.org/abs/1407.6387} {arXiv:1407.6387 [hep-lat]} \BibitemShut
  {NoStop}%
\bibitem [{\citenamefont {Aoki}\ \emph {et~al.}(2006)\citenamefont {Aoki},
  \citenamefont {Endrodi}, \citenamefont {Fodor}, \citenamefont {Katz},\ and\
  \citenamefont {Szabo}}]{Aoki:2006we}%
  \BibitemOpen
  \bibfield  {author} {\bibinfo {author} {\bibfnamefont {Y.}~\bibnamefont
  {Aoki}}, \bibinfo {author} {\bibfnamefont {G.}~\bibnamefont {Endrodi}},
  \bibinfo {author} {\bibfnamefont {Z.}~\bibnamefont {Fodor}}, \bibinfo
  {author} {\bibfnamefont {S.~D.}\ \bibnamefont {Katz}}, \ and\ \bibinfo
  {author} {\bibfnamefont {K.~K.}\ \bibnamefont {Szabo}},\ }\bibfield  {title}
  {\enquote {\bibinfo {title} {{The Order of the quantum chromodynamics
  transition predicted by the standard model of particle physics}},}\ }\href
  {\doibase 10.1038/nature05120} {\bibfield  {journal} {\bibinfo  {journal}
  {Nature}\ }\textbf {\bibinfo {volume} {443}},\ \bibinfo {pages} {675--678}
  (\bibinfo {year} {2006})},\ \Eprint {http://arxiv.org/abs/hep-lat/0611014}
  {arXiv:hep-lat/0611014 [hep-lat]} \BibitemShut {NoStop}%
\bibitem [{\citenamefont {Stephanov}(2004)}]{Stephanov:2004wx}%
  \BibitemOpen
  \bibfield  {author} {\bibinfo {author} {\bibfnamefont {M.~A.}\ \bibnamefont
  {Stephanov}},\ }\bibfield  {title} {\enquote {\bibinfo {title} {{QCD phase
  diagram and the critical point}},}\ }\bibfield  {booktitle} {\emph {\bibinfo
  {booktitle} {{Non-perturbative quantum chromodynamics. Proceedings, 8th
  Workshop, Paris, France, June 7-11, 2004}}},\ }\href {\doibase
  10.1142/S0217751X05027965} {\bibfield  {journal} {\bibinfo  {journal} {Prog.
  Theor. Phys. Suppl.}\ }\textbf {\bibinfo {volume} {153}},\ \bibinfo {pages}
  {139--156} (\bibinfo {year} {2004})},\ \bibinfo {note} {[Int. J. Mod.
  Phys.A20,4387(2005)]},\ \Eprint {http://arxiv.org/abs/hep-ph/0402115}
  {arXiv:hep-ph/0402115 [hep-ph]} \BibitemShut {NoStop}%
\bibitem [{\citenamefont {{STAR Collaboration}}(2014)}]{bes2star}%
  \BibitemOpen
  \bibfield  {author} {\bibinfo {author} {\bibnamefont {{STAR
  Collaboration}}},\ }\href@noop {} {\enquote {\bibinfo {title} {{Studying the
  Phase Diagram of QCD Matter at RHIC}},}\ }\bibinfo {howpublished}
  {\url{https://drupal.star.bnl.gov/STAR/starnotes/public/sn0598}} (\bibinfo
  {year} {2014})\BibitemShut {NoStop}%
\bibitem [{\citenamefont {{PHENIX Collaboration}}(2014)}]{bes2phenix}%
  \BibitemOpen
  \bibfield  {author} {\bibinfo {author} {\bibnamefont {{PHENIX
  Collaboration}}},\ }\href@noop {} {\enquote {\bibinfo {title} {{Beam Energy
  Scan II (2018--2019)}},}\ }\bibinfo {howpublished}
  {\url{http://www.phenix.bnl.gov/phenix/WWW/publish/dave/sPHENIX/BES_II_whitepaper.pdf}}
  (\bibinfo {year} {2014})\BibitemShut {NoStop}%
\bibitem [{\citenamefont {Huovinen}\ and\ \citenamefont
  {Petreczky}(2010)}]{Huovinen:2009yb}%
  \BibitemOpen
  \bibfield  {author} {\bibinfo {author} {\bibfnamefont {P.}~\bibnamefont
  {Huovinen}}\ and\ \bibinfo {author} {\bibfnamefont {P.}~\bibnamefont
  {Petreczky}},\ }\bibfield  {title} {\enquote {\bibinfo {title} {{QCD Equation
  of State and Hadron Resonance Gas}},}\ }\href {\doibase
  10.1016/j.nuclphysa.2010.02.015} {\bibfield  {journal} {\bibinfo  {journal}
  {Nucl. Phys.}\ }\textbf {\bibinfo {volume} {A837}},\ \bibinfo {pages}
  {26--53} (\bibinfo {year} {2010})},\ \Eprint {http://arxiv.org/abs/0912.2541}
  {arXiv:0912.2541 [hep-ph]} \BibitemShut {NoStop}%
\bibitem [{\citenamefont {Abelev}\ \emph {et~al.}(2012)\citenamefont {Abelev}
  \emph {et~al.}}]{Abelev:2012wca}%
  \BibitemOpen
  \bibfield  {author} {\bibinfo {author} {\bibfnamefont {B.}~\bibnamefont
  {Abelev}} \emph {et~al.} (\bibinfo {collaboration} {ALICE}),\ }\bibfield
  {title} {\enquote {\bibinfo {title} {{Pion, Kaon, and Proton Production in
  Central Pb--Pb Collisions at $\sqrt{s_{NN}} = 2.76$ TeV}},}\ }\href {\doibase
  10.1103/PhysRevLett.109.252301} {\bibfield  {journal} {\bibinfo  {journal}
  {Phys. Rev. Lett.}\ }\textbf {\bibinfo {volume} {109}},\ \bibinfo {pages}
  {252301} (\bibinfo {year} {2012})},\ \Eprint {http://arxiv.org/abs/1208.1974}
  {arXiv:1208.1974 [hep-ex]} \BibitemShut {NoStop}%
\bibitem [{\citenamefont {Cooper}\ and\ \citenamefont
  {Frye}(1974)}]{Cooper:1974mv}%
  \BibitemOpen
  \bibfield  {author} {\bibinfo {author} {\bibfnamefont {F.}~\bibnamefont
  {Cooper}}\ and\ \bibinfo {author} {\bibfnamefont {G.}~\bibnamefont {Frye}},\
  }\bibfield  {title} {\enquote {\bibinfo {title} {{Comment on the Single
  Particle Distribution in the Hydrodynamic and Statistical Thermodynamic
  Models of Multiparticle Production}},}\ }\href {\doibase
  10.1103/PhysRevD.10.186} {\bibfield  {journal} {\bibinfo  {journal} {Phys.
  Rev.}\ }\textbf {\bibinfo {volume} {D10}},\ \bibinfo {pages} {186} (\bibinfo
  {year} {1974})}\BibitemShut {NoStop}%
\bibitem [{\citenamefont {Dusling}\ \emph {et~al.}(2010)\citenamefont
  {Dusling}, \citenamefont {Moore},\ and\ \citenamefont
  {Teaney}}]{Dusling:2009df}%
  \BibitemOpen
  \bibfield  {author} {\bibinfo {author} {\bibfnamefont {K.}~\bibnamefont
  {Dusling}}, \bibinfo {author} {\bibfnamefont {G.~D.}\ \bibnamefont {Moore}},
  \ and\ \bibinfo {author} {\bibfnamefont {D.}~\bibnamefont {Teaney}},\
  }\bibfield  {title} {\enquote {\bibinfo {title} {{Radiative energy loss and
  v(2) spectra for viscous hydrodynamics}},}\ }\href {\doibase
  10.1103/PhysRevC.81.034907} {\bibfield  {journal} {\bibinfo  {journal} {Phys.
  Rev.}\ }\textbf {\bibinfo {volume} {C81}},\ \bibinfo {pages} {034907}
  (\bibinfo {year} {2010})},\ \Eprint {http://arxiv.org/abs/0909.0754}
  {arXiv:0909.0754 [nucl-th]} \BibitemShut {NoStop}%
\bibitem [{\citenamefont {Petersen}\ \emph {et~al.}(2008)\citenamefont
  {Petersen}, \citenamefont {Steinheimer}, \citenamefont {Burau}, \citenamefont
  {Bleicher},\ and\ \citenamefont {Stocker}}]{Petersen:2008dd}%
  \BibitemOpen
  \bibfield  {author} {\bibinfo {author} {\bibfnamefont {H.}~\bibnamefont
  {Petersen}}, \bibinfo {author} {\bibfnamefont {J.}~\bibnamefont
  {Steinheimer}}, \bibinfo {author} {\bibfnamefont {G.}~\bibnamefont {Burau}},
  \bibinfo {author} {\bibfnamefont {M.}~\bibnamefont {Bleicher}}, \ and\
  \bibinfo {author} {\bibfnamefont {H.}~\bibnamefont {Stocker}},\ }\bibfield
  {title} {\enquote {\bibinfo {title} {{A Fully Integrated Transport Approach
  to Heavy Ion Reactions with an Intermediate Hydrodynamic Stage}},}\ }\href
  {\doibase 10.1103/PhysRevC.78.044901} {\bibfield  {journal} {\bibinfo
  {journal} {Phys. Rev.}\ }\textbf {\bibinfo {volume} {C78}},\ \bibinfo {pages}
  {044901} (\bibinfo {year} {2008})},\ \Eprint {http://arxiv.org/abs/0806.1695}
  {arXiv:0806.1695 [nucl-th]} \BibitemShut {NoStop}%
\bibitem [{\citenamefont {Dusling}\ and\ \citenamefont
  {Teaney}(2008)}]{Dusling:2007gi}%
  \BibitemOpen
  \bibfield  {author} {\bibinfo {author} {\bibfnamefont {K.}~\bibnamefont
  {Dusling}}\ and\ \bibinfo {author} {\bibfnamefont {D.}~\bibnamefont
  {Teaney}},\ }\bibfield  {title} {\enquote {\bibinfo {title} {{Simulating
  elliptic flow with viscous hydrodynamics}},}\ }\href {\doibase
  10.1103/PhysRevC.77.034905} {\bibfield  {journal} {\bibinfo  {journal} {Phys.
  Rev.}\ }\textbf {\bibinfo {volume} {C77}},\ \bibinfo {pages} {034905}
  (\bibinfo {year} {2008})},\ \Eprint {http://arxiv.org/abs/0710.5932}
  {arXiv:0710.5932 [nucl-th]} \BibitemShut {NoStop}%
\bibitem [{\citenamefont {Molnar}\ and\ \citenamefont
  {Huovinen}(2008)}]{Molnar:2008xj}%
  \BibitemOpen
  \bibfield  {author} {\bibinfo {author} {\bibfnamefont {D.}~\bibnamefont
  {Molnar}}\ and\ \bibinfo {author} {\bibfnamefont {P.}~\bibnamefont
  {Huovinen}},\ }\bibfield  {title} {\enquote {\bibinfo {title} {{Dissipative
  effects from transport and viscous hydrodynamics}},}\ }\bibfield  {booktitle}
  {\emph {\bibinfo {booktitle} {{Proceedings, 20th International Conference on
  Ultra-Relativistic Nucleus-Nucleus Collisions (QM 2008): Jaipur, India,
  February 4-10, 2008}}},\ }\href {\doibase 10.1088/0954-3899/35/10/104125}
  {\bibfield  {journal} {\bibinfo  {journal} {J. Phys.}\ }\textbf {\bibinfo
  {volume} {G35}},\ \bibinfo {pages} {104125} (\bibinfo {year} {2008})},\
  \Eprint {http://arxiv.org/abs/0806.1367} {arXiv:0806.1367 [nucl-th]}
  \BibitemShut {NoStop}%
\bibitem [{\citenamefont {Casalderrey-Solana}\ and\ \citenamefont
  {Salgado}(2007)}]{CasalderreySolana:2007pr}%
  \BibitemOpen
  \bibfield  {author} {\bibinfo {author} {\bibfnamefont {J.}~\bibnamefont
  {Casalderrey-Solana}}\ and\ \bibinfo {author} {\bibfnamefont {C.~A.}\
  \bibnamefont {Salgado}},\ }\bibfield  {title} {\enquote {\bibinfo {title}
  {{Introductory lectures on jet quenching in heavy ion collisions}},}\
  }\bibfield  {booktitle} {\emph {\bibinfo {booktitle} {{Theoretical physics.
  Proceedings, 47th Cracow School, Zakopane, Poland, June 14-22, 2007}}},\
  }\href@noop {} {\bibfield  {journal} {\bibinfo  {journal} {Acta Phys.
  Polon.}\ }\textbf {\bibinfo {volume} {B38}},\ \bibinfo {pages} {3731--3794}
  (\bibinfo {year} {2007})},\ \Eprint {http://arxiv.org/abs/0712.3443}
  {arXiv:0712.3443 [hep-ph]} \BibitemShut {NoStop}%
\bibitem [{\citenamefont {d'Enterria}(2010)}]{dEnterria:2009xfs}%
  \BibitemOpen
  \bibfield  {author} {\bibinfo {author} {\bibfnamefont {D.}~\bibnamefont
  {d'Enterria}},\ }\bibfield  {title} {\enquote {\bibinfo {title} {{Jet
  quenching}},}\ }\href {\doibase 10.1007/978-3-642-01539-7_16} {\bibfield
  {journal} {\bibinfo  {journal} {Landolt-Bornstein}\ }\textbf {\bibinfo
  {volume} {23}},\ \bibinfo {pages} {471} (\bibinfo {year} {2010})},\ \Eprint
  {http://arxiv.org/abs/0902.2011} {arXiv:0902.2011 [nucl-ex]} \BibitemShut
  {NoStop}%
\bibitem [{\citenamefont {Voloshin}\ \emph {et~al.}(2008)\citenamefont
  {Voloshin}, \citenamefont {Poskanzer},\ and\ \citenamefont
  {Snellings}}]{Voloshin:2008dg}%
  \BibitemOpen
  \bibfield  {author} {\bibinfo {author} {\bibfnamefont {S.~A.}\ \bibnamefont
  {Voloshin}}, \bibinfo {author} {\bibfnamefont {A.~M.}\ \bibnamefont
  {Poskanzer}}, \ and\ \bibinfo {author} {\bibfnamefont {R.}~\bibnamefont
  {Snellings}},\ }\bibfield  {title} {\enquote {\bibinfo {title} {{Collective
  phenomena in non-central nuclear collisions}},}\ }\href@noop {} {\  (\bibinfo
  {year} {2008})},\ \Eprint {http://arxiv.org/abs/0809.2949} {arXiv:0809.2949
  [nucl-ex]} \BibitemShut {NoStop}%
\bibitem [{\citenamefont {Voloshin}\ and\ \citenamefont
  {Zhang}(1996)}]{Voloshin:1994mz}%
  \BibitemOpen
  \bibfield  {author} {\bibinfo {author} {\bibfnamefont {S.}~\bibnamefont
  {Voloshin}}\ and\ \bibinfo {author} {\bibfnamefont {Y.}~\bibnamefont
  {Zhang}},\ }\bibfield  {title} {\enquote {\bibinfo {title} {{Flow study in
  relativistic nuclear collisions by Fourier expansion of Azimuthal particle
  distributions}},}\ }\href {\doibase 10.1007/s002880050141} {\bibfield
  {journal} {\bibinfo  {journal} {Z. Phys.}\ }\textbf {\bibinfo {volume}
  {C70}},\ \bibinfo {pages} {665--672} (\bibinfo {year} {1996})},\ \Eprint
  {http://arxiv.org/abs/hep-ph/9407282} {arXiv:hep-ph/9407282 [hep-ph]}
  \BibitemShut {NoStop}%
\bibitem [{\citenamefont {Khachatryan}\ \emph {et~al.}(2015)\citenamefont
  {Khachatryan} \emph {et~al.}}]{Khachatryan:2015oea}%
  \BibitemOpen
  \bibfield  {author} {\bibinfo {author} {\bibfnamefont {V.}~\bibnamefont
  {Khachatryan}} \emph {et~al.} (\bibinfo {collaboration} {CMS}),\ }\bibfield
  {title} {\enquote {\bibinfo {title} {{Evidence for transverse momentum and
  pseudorapidity dependent event plane fluctuations in PbPb and pPb
  collisions}},}\ }\href {\doibase 10.1103/PhysRevC.92.034911} {\bibfield
  {journal} {\bibinfo  {journal} {Phys. Rev.}\ }\textbf {\bibinfo {volume}
  {C92}},\ \bibinfo {pages} {034911} (\bibinfo {year} {2015})},\ \Eprint
  {http://arxiv.org/abs/1503.01692} {arXiv:1503.01692 [nucl-ex]} \BibitemShut
  {NoStop}%
\bibitem [{\citenamefont {Chatrchyan}\ \emph
  {et~al.}(2014{\natexlab{b}})\citenamefont {Chatrchyan} \emph
  {et~al.}}]{CMS:2013bza}%
  \BibitemOpen
  \bibfield  {author} {\bibinfo {author} {\bibfnamefont {S.}~\bibnamefont
  {Chatrchyan}} \emph {et~al.} (\bibinfo {collaboration} {CMS}),\ }\bibfield
  {title} {\enquote {\bibinfo {title} {{Studies of azimuthal dihadron
  correlations in ultra-central PbPb collisions at $\sqrt{s_{NN}} =$ 2.76
  TeV}},}\ }\href {\doibase 10.1007/JHEP02(2014)088} {\bibfield  {journal}
  {\bibinfo  {journal} {JHEP}\ }\textbf {\bibinfo {volume} {02}},\ \bibinfo
  {pages} {088} (\bibinfo {year} {2014}{\natexlab{b}})},\ \Eprint
  {http://arxiv.org/abs/1312.1845} {arXiv:1312.1845 [nucl-ex]} \BibitemShut
  {NoStop}%
\bibitem [{\citenamefont {Aamodt}\ \emph {et~al.}(2012)\citenamefont {Aamodt}
  \emph {et~al.}}]{Aamodt:2011by}%
  \BibitemOpen
  \bibfield  {author} {\bibinfo {author} {\bibfnamefont {K.}~\bibnamefont
  {Aamodt}} \emph {et~al.} (\bibinfo {collaboration} {ALICE}),\ }\bibfield
  {title} {\enquote {\bibinfo {title} {{Harmonic decomposition of two-particle
  angular correlations in Pb-Pb collisions at $\sqrt{s_{NN}}=$ 2.76 TeV}},}\
  }\href {\doibase 10.1016/j.physletb.2012.01.060} {\bibfield  {journal}
  {\bibinfo  {journal} {Phys. Lett.}\ }\textbf {\bibinfo {volume} {B708}},\
  \bibinfo {pages} {249--264} (\bibinfo {year} {2012})},\ \Eprint
  {http://arxiv.org/abs/1109.2501} {arXiv:1109.2501 [nucl-ex]} \BibitemShut
  {NoStop}%
\bibitem [{\citenamefont {Gardim}\ \emph {et~al.}(2013)\citenamefont {Gardim},
  \citenamefont {Grassi}, \citenamefont {Luzum},\ and\ \citenamefont
  {Ollitrault}}]{Gardim:2012im}%
  \BibitemOpen
  \bibfield  {author} {\bibinfo {author} {\bibfnamefont {F.~G.}\ \bibnamefont
  {Gardim}}, \bibinfo {author} {\bibfnamefont {F.}~\bibnamefont {Grassi}},
  \bibinfo {author} {\bibfnamefont {M.}~\bibnamefont {Luzum}}, \ and\ \bibinfo
  {author} {\bibfnamefont {J.-Y.}\ \bibnamefont {Ollitrault}},\ }\bibfield
  {title} {\enquote {\bibinfo {title} {{Breaking of factorization of
  two-particle correlations in hydrodynamics}},}\ }\href {\doibase
  10.1103/PhysRevC.87.031901} {\bibfield  {journal} {\bibinfo  {journal} {Phys.
  Rev.}\ }\textbf {\bibinfo {volume} {C87}},\ \bibinfo {pages} {031901}
  (\bibinfo {year} {2013})},\ \Eprint {http://arxiv.org/abs/1211.0989}
  {arXiv:1211.0989 [nucl-th]} \BibitemShut {NoStop}%
\bibitem [{\citenamefont {Ellis}\ \emph {et~al.}(1996)\citenamefont {Ellis},
  \citenamefont {Stirling},\ and\ \citenamefont {Webber}}]{Ellis}%
  \BibitemOpen
  \bibfield  {author} {\bibinfo {author} {\bibfnamefont {R.~K.}\ \bibnamefont
  {Ellis}}, \bibinfo {author} {\bibfnamefont {W.~J.}\ \bibnamefont {Stirling}},
  \ and\ \bibinfo {author} {\bibfnamefont {B.~R.}\ \bibnamefont {Webber}},\
  }\href
  {http://proxy.library.stonybrook.edu/login?url=http://search.ebscohost.com/login.aspx?direct=true&db=cat03000a&AN=STB.000962540&site=eds-live&scope=site}
  {\emph {\bibinfo {title} {QCD and collider physics}}},\ Cambridge monographs
  on particle physics, nuclear physics, and cosmology: 8\ (\bibinfo
  {publisher} {Cambridge ; New York : Cambridge University Press, 1996.},\
  \bibinfo {year} {1996})\BibitemShut {NoStop}%
\bibitem [{\citenamefont {Adam}\ \emph
  {et~al.}(2016{\natexlab{a}})\citenamefont {Adam} \emph
  {et~al.}}]{Adam:2015lda}%
  \BibitemOpen
  \bibfield  {author} {\bibinfo {author} {\bibfnamefont {J.}~\bibnamefont
  {Adam}} \emph {et~al.} (\bibinfo {collaboration} {ALICE}),\ }\bibfield
  {title} {\enquote {\bibinfo {title} {{Direct photon production in Pb-Pb
  collisions at $\sqrt{s_{\rm{NN}}} =$ 2.76 TeV}},}\ }\href {\doibase
  10.1016/j.physletb.2016.01.020} {\bibfield  {journal} {\bibinfo  {journal}
  {Phys. Lett.}\ }\textbf {\bibinfo {volume} {B754}},\ \bibinfo {pages}
  {235--248} (\bibinfo {year} {2016}{\natexlab{a}})},\ \Eprint
  {http://arxiv.org/abs/1509.07324} {arXiv:1509.07324 [nucl-ex]} \BibitemShut
  {NoStop}%
\bibitem [{\citenamefont {Blaizot}\ \emph {et~al.}(2003)\citenamefont
  {Blaizot}, \citenamefont {Iancu},\ and\ \citenamefont
  {Rebhan}}]{Blaizot:2003tw}%
  \BibitemOpen
  \bibfield  {author} {\bibinfo {author} {\bibfnamefont {J.-P.}\ \bibnamefont
  {Blaizot}}, \bibinfo {author} {\bibfnamefont {E.}~\bibnamefont {Iancu}}, \
  and\ \bibinfo {author} {\bibfnamefont {A.}~\bibnamefont {Rebhan}},\
  }\bibfield  {title} {\enquote {\bibinfo {title} {{Thermodynamics of the high
  temperature quark gluon plasma}},}\ }in\ \href {\doibase
  10.1142/9789812795533_0002} {\emph {\bibinfo {booktitle} {{In *Hwa, R.C.
  (ed.) et al.: Quark gluon plasma* 60-122}}}}\ (\bibinfo {year} {2003})\
  \Eprint {http://arxiv.org/abs/hep-ph/0303185} {arXiv:hep-ph/0303185 [hep-ph]}
  \BibitemShut {NoStop}%
\bibitem [{\citenamefont {Braaten}\ and\ \citenamefont
  {Pisarski}(1990)}]{Braaten:1989mz}%
  \BibitemOpen
  \bibfield  {author} {\bibinfo {author} {\bibfnamefont {E.}~\bibnamefont
  {Braaten}}\ and\ \bibinfo {author} {\bibfnamefont {R.~D.}\ \bibnamefont
  {Pisarski}},\ }\bibfield  {title} {\enquote {\bibinfo {title} {{Soft
  Amplitudes in Hot Gauge Theories: A General Analysis}},}\ }\href {\doibase
  10.1016/0550-3213(90)90508-B} {\bibfield  {journal} {\bibinfo  {journal}
  {Nucl. Phys.}\ }\textbf {\bibinfo {volume} {B337}},\ \bibinfo {pages}
  {569--634} (\bibinfo {year} {1990})}\BibitemShut {NoStop}%
\bibitem [{\citenamefont {Arnold}\ \emph
  {et~al.}(2003{\natexlab{a}})\citenamefont {Arnold}, \citenamefont {Moore},\
  and\ \citenamefont {Yaffe}}]{Arnold:2002zm}%
  \BibitemOpen
  \bibfield  {author} {\bibinfo {author} {\bibfnamefont {P.~B.}\ \bibnamefont
  {Arnold}}, \bibinfo {author} {\bibfnamefont {G.~D.}\ \bibnamefont {Moore}}, \
  and\ \bibinfo {author} {\bibfnamefont {L.~G.}\ \bibnamefont {Yaffe}},\
  }\bibfield  {title} {\enquote {\bibinfo {title} {{Effective kinetic theory
  for high temperature gauge theories}},}\ }\href {\doibase
  10.1088/1126-6708/2003/01/030} {\bibfield  {journal} {\bibinfo  {journal}
  {JHEP}\ }\textbf {\bibinfo {volume} {01}},\ \bibinfo {pages} {030} (\bibinfo
  {year} {2003}{\natexlab{a}})},\ \Eprint {http://arxiv.org/abs/hep-ph/0209353}
  {arXiv:hep-ph/0209353 [hep-ph]} \BibitemShut {NoStop}%
\bibitem [{\citenamefont {Mueller}(2000)}]{Mueller:1999fp}%
  \BibitemOpen
  \bibfield  {author} {\bibinfo {author} {\bibfnamefont {A.~H.}\ \bibnamefont
  {Mueller}},\ }\bibfield  {title} {\enquote {\bibinfo {title} {{Toward
  equilibration in the early stages after a high-energy heavy ion
  collision}},}\ }\href {\doibase 10.1016/S0550-3213(99)00502-7} {\bibfield
  {journal} {\bibinfo  {journal} {Nucl. Phys.}\ }\textbf {\bibinfo {volume}
  {B572}},\ \bibinfo {pages} {227--240} (\bibinfo {year} {2000})},\ \Eprint
  {http://arxiv.org/abs/hep-ph/9906322} {arXiv:hep-ph/9906322 [hep-ph]}
  \BibitemShut {NoStop}%
\bibitem [{\citenamefont {Arnold}\ \emph {et~al.}(2002)\citenamefont {Arnold},
  \citenamefont {Moore},\ and\ \citenamefont {Yaffe}}]{Arnold:2002ja}%
  \BibitemOpen
  \bibfield  {author} {\bibinfo {author} {\bibfnamefont {P.~B.}\ \bibnamefont
  {Arnold}}, \bibinfo {author} {\bibfnamefont {G.~D.}\ \bibnamefont {Moore}}, \
  and\ \bibinfo {author} {\bibfnamefont {L.~G.}\ \bibnamefont {Yaffe}},\
  }\bibfield  {title} {\enquote {\bibinfo {title} {{Photon and gluon emission
  in relativistic plasmas}},}\ }\href {\doibase 10.1088/1126-6708/2002/06/030}
  {\bibfield  {journal} {\bibinfo  {journal} {JHEP}\ }\textbf {\bibinfo
  {volume} {06}},\ \bibinfo {pages} {030} (\bibinfo {year} {2002})},\ \Eprint
  {http://arxiv.org/abs/hep-ph/0204343} {arXiv:hep-ph/0204343 [hep-ph]}
  \BibitemShut {NoStop}%
\bibitem [{\citenamefont {Arnold}\ \emph
  {et~al.}(2003{\natexlab{b}})\citenamefont {Arnold}, \citenamefont {Moore},\
  and\ \citenamefont {Yaffe}}]{Arnold:2003zc}%
  \BibitemOpen
  \bibfield  {author} {\bibinfo {author} {\bibfnamefont {P.~B.}\ \bibnamefont
  {Arnold}}, \bibinfo {author} {\bibfnamefont {G.~D.}\ \bibnamefont {Moore}}, \
  and\ \bibinfo {author} {\bibfnamefont {L.~G.}\ \bibnamefont {Yaffe}},\
  }\bibfield  {title} {\enquote {\bibinfo {title} {{Transport coefficients in
  high temperature gauge theories. 2. Beyond leading log}},}\ }\href {\doibase
  10.1088/1126-6708/2003/05/051} {\bibfield  {journal} {\bibinfo  {journal}
  {JHEP}\ }\textbf {\bibinfo {volume} {05}},\ \bibinfo {pages} {051} (\bibinfo
  {year} {2003}{\natexlab{b}})},\ \Eprint {http://arxiv.org/abs/hep-ph/0302165}
  {arXiv:hep-ph/0302165 [hep-ph]} \BibitemShut {NoStop}%
\bibitem [{\citenamefont {Braaten}\ and\ \citenamefont
  {Nieto}(1996{\natexlab{a}})}]{Braaten:1995ju}%
  \BibitemOpen
  \bibfield  {author} {\bibinfo {author} {\bibfnamefont {E.}~\bibnamefont
  {Braaten}}\ and\ \bibinfo {author} {\bibfnamefont {A.}~\bibnamefont
  {Nieto}},\ }\bibfield  {title} {\enquote {\bibinfo {title} {{On the
  convergence of perturbative QCD at high temperature}},}\ }\href {\doibase
  10.1103/PhysRevLett.76.1417} {\bibfield  {journal} {\bibinfo  {journal}
  {Phys. Rev. Lett.}\ }\textbf {\bibinfo {volume} {76}},\ \bibinfo {pages}
  {1417--1420} (\bibinfo {year} {1996}{\natexlab{a}})},\ \Eprint
  {http://arxiv.org/abs/hep-ph/9508406} {arXiv:hep-ph/9508406 [hep-ph]}
  \BibitemShut {NoStop}%
\bibitem [{\citenamefont {Andersen}\ \emph {et~al.}(2011)\citenamefont
  {Andersen}, \citenamefont {Leganger}, \citenamefont {Strickland},\ and\
  \citenamefont {Su}}]{Andersen:2011sf}%
  \BibitemOpen
  \bibfield  {author} {\bibinfo {author} {\bibfnamefont {J.~O.}\ \bibnamefont
  {Andersen}}, \bibinfo {author} {\bibfnamefont {L.~E.}\ \bibnamefont
  {Leganger}}, \bibinfo {author} {\bibfnamefont {M.}~\bibnamefont
  {Strickland}}, \ and\ \bibinfo {author} {\bibfnamefont {N.}~\bibnamefont
  {Su}},\ }\bibfield  {title} {\enquote {\bibinfo {title} {{Three-loop HTL QCD
  thermodynamics}},}\ }\href {\doibase 10.1007/JHEP08(2011)053} {\bibfield
  {journal} {\bibinfo  {journal} {JHEP}\ }\textbf {\bibinfo {volume} {08}},\
  \bibinfo {pages} {053} (\bibinfo {year} {2011})},\ \Eprint
  {http://arxiv.org/abs/1103.2528} {arXiv:1103.2528 [hep-ph]} \BibitemShut
  {NoStop}%
\bibitem [{\citenamefont {Haque}\ \emph {et~al.}(2014)\citenamefont {Haque},
  \citenamefont {Bandyopadhyay}, \citenamefont {Andersen}, \citenamefont
  {Mustafa}, \citenamefont {Strickland},\ and\ \citenamefont
  {Su}}]{Haque:2014rua}%
  \BibitemOpen
  \bibfield  {author} {\bibinfo {author} {\bibfnamefont {N.}~\bibnamefont
  {Haque}}, \bibinfo {author} {\bibfnamefont {A.}~\bibnamefont
  {Bandyopadhyay}}, \bibinfo {author} {\bibfnamefont {J.~O.}\ \bibnamefont
  {Andersen}}, \bibinfo {author} {\bibfnamefont {M.~G.}\ \bibnamefont
  {Mustafa}}, \bibinfo {author} {\bibfnamefont {M.}~\bibnamefont {Strickland}},
  \ and\ \bibinfo {author} {\bibfnamefont {N.}~\bibnamefont {Su}},\ }\bibfield
  {title} {\enquote {\bibinfo {title} {{Three-loop HTLpt thermodynamics at
  finite temperature and chemical potential}},}\ }\href {\doibase
  10.1007/JHEP05(2014)027} {\bibfield  {journal} {\bibinfo  {journal} {JHEP}\
  }\textbf {\bibinfo {volume} {05}},\ \bibinfo {pages} {027} (\bibinfo {year}
  {2014})},\ \Eprint {http://arxiv.org/abs/1402.6907} {arXiv:1402.6907
  [hep-ph]} \BibitemShut {NoStop}%
\bibitem [{\citenamefont {Ghiglieri}\ \emph {et~al.}(2016)\citenamefont
  {Ghiglieri}, \citenamefont {Moore},\ and\ \citenamefont
  {Teaney}}]{Ghiglieri:2015ala}%
  \BibitemOpen
  \bibfield  {author} {\bibinfo {author} {\bibfnamefont {J.}~\bibnamefont
  {Ghiglieri}}, \bibinfo {author} {\bibfnamefont {G.~D.}\ \bibnamefont
  {Moore}}, \ and\ \bibinfo {author} {\bibfnamefont {D.}~\bibnamefont
  {Teaney}},\ }\bibfield  {title} {\enquote {\bibinfo {title} {{Jet-Medium
  Interactions at NLO in a Weakly-Coupled Quark-Gluon Plasma}},}\ }\href
  {\doibase 10.1007/JHEP03(2016)095} {\bibfield  {journal} {\bibinfo  {journal}
  {JHEP}\ }\textbf {\bibinfo {volume} {03}},\ \bibinfo {pages} {095} (\bibinfo
  {year} {2016})},\ \Eprint {http://arxiv.org/abs/1509.07773} {arXiv:1509.07773
  [hep-ph]} \BibitemShut {NoStop}%
\bibitem [{\citenamefont {Fukushima}(2017)}]{Fukushima:2016xgg}%
  \BibitemOpen
  \bibfield  {author} {\bibinfo {author} {\bibfnamefont {K.}~\bibnamefont
  {Fukushima}},\ }\bibfield  {title} {\enquote {\bibinfo {title} {{Evolution to
  the Quark-Gluon Plasma}},}\ }\href {\doibase 10.1088/1361-6633/80/2/022301}
  {\bibfield  {journal} {\bibinfo  {journal} {Rept. Prog. Phys.}\ }\textbf
  {\bibinfo {volume} {80}},\ \bibinfo {pages} {022301} (\bibinfo {year}
  {2017})},\ \Eprint {http://arxiv.org/abs/1603.02340} {arXiv:1603.02340
  [nucl-th]} \BibitemShut {NoStop}%
\bibitem [{\citenamefont {Mrowczynski}\ \emph {et~al.}(2016)\citenamefont
  {Mrowczynski}, \citenamefont {Schenke},\ and\ \citenamefont
  {Strickland}}]{Mrowczynski:2016etf}%
  \BibitemOpen
  \bibfield  {author} {\bibinfo {author} {\bibfnamefont {S.}~\bibnamefont
  {Mrowczynski}}, \bibinfo {author} {\bibfnamefont {B.}~\bibnamefont
  {Schenke}}, \ and\ \bibinfo {author} {\bibfnamefont {M.}~\bibnamefont
  {Strickland}},\ }\bibfield  {title} {\enquote {\bibinfo {title} {{Color
  Instabilities in the Quark-Gluon Plasma}},}\ }\href@noop {} {\  (\bibinfo
  {year} {2016})},\ \Eprint {http://arxiv.org/abs/1603.08946} {arXiv:1603.08946
  [hep-ph]} \BibitemShut {NoStop}%
\bibitem [{\citenamefont {Arnold}\ \emph
  {et~al.}(2005{\natexlab{a}})\citenamefont {Arnold}, \citenamefont {Moore},\
  and\ \citenamefont {Yaffe}}]{Arnold:2005vb}%
  \BibitemOpen
  \bibfield  {author} {\bibinfo {author} {\bibfnamefont {P.~B.}\ \bibnamefont
  {Arnold}}, \bibinfo {author} {\bibfnamefont {G.~D.}\ \bibnamefont {Moore}}, \
  and\ \bibinfo {author} {\bibfnamefont {L.~G.}\ \bibnamefont {Yaffe}},\
  }\bibfield  {title} {\enquote {\bibinfo {title} {{The Fate of non-Abelian
  plasma instabilities in 3+1 dimensions}},}\ }\href {\doibase
  10.1103/PhysRevD.72.054003} {\bibfield  {journal} {\bibinfo  {journal} {Phys.
  Rev.}\ }\textbf {\bibinfo {volume} {D72}},\ \bibinfo {pages} {054003}
  (\bibinfo {year} {2005}{\natexlab{a}})},\ \Eprint
  {http://arxiv.org/abs/hep-ph/0505212} {arXiv:hep-ph/0505212 [hep-ph]}
  \BibitemShut {NoStop}%
\bibitem [{\citenamefont {Berges}\ \emph
  {et~al.}(2014{\natexlab{c}})\citenamefont {Berges}, \citenamefont
  {Boguslavski}, \citenamefont {Schlichting},\ and\ \citenamefont
  {Venugopalan}}]{Berges:2013lsa}%
  \BibitemOpen
  \bibfield  {author} {\bibinfo {author} {\bibfnamefont {J.}~\bibnamefont
  {Berges}}, \bibinfo {author} {\bibfnamefont {K.}~\bibnamefont {Boguslavski}},
  \bibinfo {author} {\bibfnamefont {S.}~\bibnamefont {Schlichting}}, \ and\
  \bibinfo {author} {\bibfnamefont {R.}~\bibnamefont {Venugopalan}},\
  }\bibfield  {title} {\enquote {\bibinfo {title} {{Basin of attraction for
  turbulent thermalization and the range of validity of classical-statistical
  simulations}},}\ }\href {\doibase 10.1007/JHEP05(2014)054} {\bibfield
  {journal} {\bibinfo  {journal} {JHEP}\ }\textbf {\bibinfo {volume} {05}},\
  \bibinfo {pages} {054} (\bibinfo {year} {2014}{\natexlab{c}})},\ \Eprint
  {http://arxiv.org/abs/1312.5216} {arXiv:1312.5216 [hep-ph]} \BibitemShut
  {NoStop}%
\bibitem [{\citenamefont {Berges}\ \emph {et~al.}(2015)\citenamefont {Berges},
  \citenamefont {Boguslavski}, \citenamefont {Schlichting},\ and\ \citenamefont
  {Venugopalan}}]{Berges:2014bba}%
  \BibitemOpen
  \bibfield  {author} {\bibinfo {author} {\bibfnamefont {J.}~\bibnamefont
  {Berges}}, \bibinfo {author} {\bibfnamefont {K.}~\bibnamefont {Boguslavski}},
  \bibinfo {author} {\bibfnamefont {S.}~\bibnamefont {Schlichting}}, \ and\
  \bibinfo {author} {\bibfnamefont {R.}~\bibnamefont {Venugopalan}},\
  }\bibfield  {title} {\enquote {\bibinfo {title} {{Universality far from
  equilibrium: From superfluid Bose gases to heavy-ion collisions}},}\ }\href
  {\doibase 10.1103/PhysRevLett.114.061601} {\bibfield  {journal} {\bibinfo
  {journal} {Phys. Rev. Lett.}\ }\textbf {\bibinfo {volume} {114}},\ \bibinfo
  {pages} {061601} (\bibinfo {year} {2015})},\ \Eprint
  {http://arxiv.org/abs/1408.1670} {arXiv:1408.1670 [hep-ph]} \BibitemShut
  {NoStop}%
\bibitem [{\citenamefont {Kajantie}\ \emph {et~al.}(2001)\citenamefont
  {Kajantie}, \citenamefont {Laine}, \citenamefont {Rummukainen},\ and\
  \citenamefont {Schroder}}]{Kajantie:2000iz}%
  \BibitemOpen
  \bibfield  {author} {\bibinfo {author} {\bibfnamefont {K.}~\bibnamefont
  {Kajantie}}, \bibinfo {author} {\bibfnamefont {M.}~\bibnamefont {Laine}},
  \bibinfo {author} {\bibfnamefont {K.}~\bibnamefont {Rummukainen}}, \ and\
  \bibinfo {author} {\bibfnamefont {Y.}~\bibnamefont {Schroder}},\ }\bibfield
  {title} {\enquote {\bibinfo {title} {{How to resum long distance
  contributions to the QCD pressure?}}}\ }\href {\doibase
  10.1103/PhysRevLett.86.10} {\bibfield  {journal} {\bibinfo  {journal} {Phys.
  Rev. Lett.}\ }\textbf {\bibinfo {volume} {86}},\ \bibinfo {pages} {10--13}
  (\bibinfo {year} {2001})},\ \Eprint {http://arxiv.org/abs/hep-ph/0007109}
  {arXiv:hep-ph/0007109 [hep-ph]} \BibitemShut {NoStop}%
\bibitem [{\citenamefont {Braaten}\ and\ \citenamefont
  {Nieto}(1996{\natexlab{b}})}]{Braaten:1995jr}%
  \BibitemOpen
  \bibfield  {author} {\bibinfo {author} {\bibfnamefont {E.}~\bibnamefont
  {Braaten}}\ and\ \bibinfo {author} {\bibfnamefont {A.}~\bibnamefont
  {Nieto}},\ }\bibfield  {title} {\enquote {\bibinfo {title} {{Free energy of
  QCD at high temperature}},}\ }\href {\doibase 10.1103/PhysRevD.53.3421}
  {\bibfield  {journal} {\bibinfo  {journal} {Phys. Rev.}\ }\textbf {\bibinfo
  {volume} {D53}},\ \bibinfo {pages} {3421--3437} (\bibinfo {year}
  {1996}{\natexlab{b}})},\ \Eprint {http://arxiv.org/abs/hep-ph/9510408}
  {arXiv:hep-ph/9510408 [hep-ph]} \BibitemShut {NoStop}%
\bibitem [{\citenamefont {Kajantie}\ \emph {et~al.}(2003)\citenamefont
  {Kajantie}, \citenamefont {Laine}, \citenamefont {Rummukainen},\ and\
  \citenamefont {Schroder}}]{Kajantie:2002wa}%
  \BibitemOpen
  \bibfield  {author} {\bibinfo {author} {\bibfnamefont {K.}~\bibnamefont
  {Kajantie}}, \bibinfo {author} {\bibfnamefont {M.}~\bibnamefont {Laine}},
  \bibinfo {author} {\bibfnamefont {K.}~\bibnamefont {Rummukainen}}, \ and\
  \bibinfo {author} {\bibfnamefont {Y.}~\bibnamefont {Schroder}},\ }\bibfield
  {title} {\enquote {\bibinfo {title} {{The Pressure of hot QCD up to g6
  ln(1/g)}},}\ }\href {\doibase 10.1103/PhysRevD.67.105008} {\bibfield
  {journal} {\bibinfo  {journal} {Phys. Rev.}\ }\textbf {\bibinfo {volume}
  {D67}},\ \bibinfo {pages} {105008} (\bibinfo {year} {2003})},\ \Eprint
  {http://arxiv.org/abs/hep-ph/0211321} {arXiv:hep-ph/0211321 [hep-ph]}
  \BibitemShut {NoStop}%
\bibitem [{\citenamefont {Heinz}\ \emph {et~al.}(2013)\citenamefont {Heinz},
  \citenamefont {Qiu},\ and\ \citenamefont {Shen}}]{Heinz:2013bua}%
  \BibitemOpen
  \bibfield  {author} {\bibinfo {author} {\bibfnamefont {U.}~\bibnamefont
  {Heinz}}, \bibinfo {author} {\bibfnamefont {Z.}~\bibnamefont {Qiu}}, \ and\
  \bibinfo {author} {\bibfnamefont {C.}~\bibnamefont {Shen}},\ }\bibfield
  {title} {\enquote {\bibinfo {title} {{Fluctuating flow angles and anisotropic
  flow measurements}},}\ }\href {\doibase 10.1103/PhysRevC.87.034913}
  {\bibfield  {journal} {\bibinfo  {journal} {Phys. Rev.}\ }\textbf {\bibinfo
  {volume} {C87}},\ \bibinfo {pages} {034913} (\bibinfo {year} {2013})},\
  \Eprint {http://arxiv.org/abs/1302.3535} {arXiv:1302.3535 [nucl-th]}
  \BibitemShut {NoStop}%
\bibitem [{\citenamefont {Kozlov}\ \emph {et~al.}(2014)\citenamefont {Kozlov},
  \citenamefont {Luzum}, \citenamefont {Denicol}, \citenamefont {Jeon},\ and\
  \citenamefont {Gale}}]{Kozlov:2014fqa}%
  \BibitemOpen
  \bibfield  {author} {\bibinfo {author} {\bibfnamefont {I.}~\bibnamefont
  {Kozlov}}, \bibinfo {author} {\bibfnamefont {M.}~\bibnamefont {Luzum}},
  \bibinfo {author} {\bibfnamefont {G.}~\bibnamefont {Denicol}}, \bibinfo
  {author} {\bibfnamefont {S.}~\bibnamefont {Jeon}}, \ and\ \bibinfo {author}
  {\bibfnamefont {C.}~\bibnamefont {Gale}},\ }\bibfield  {title} {\enquote
  {\bibinfo {title} {{Transverse momentum structure of pair correlations as a
  signature of collective behavior in small collision systems}},}\ }\href@noop
  {} {\  (\bibinfo {year} {2014})},\ \Eprint {http://arxiv.org/abs/1405.3976}
  {arXiv:1405.3976 [nucl-th]} \BibitemShut {NoStop}%
\bibitem [{\citenamefont {Jolliffe}(2002)}]{citeulike:11071912}%
  \BibitemOpen
  \bibfield  {author} {\bibinfo {author} {\bibfnamefont {I.~T.}\ \bibnamefont
  {Jolliffe}},\ }\href {\doibase 10.1007/b98835} {\emph {\bibinfo {title}
  {{Principal Component Analysis}}}},\ Springer Series in Statistics\ (\bibinfo
   {publisher} {Springer-Verlag},\ \bibinfo {year} {2002})\BibitemShut
  {NoStop}%
\bibitem [{\citenamefont {Borga}\ \emph {et~al.}(1997)\citenamefont {Borga},
  \citenamefont {Landelius},\ and\ \citenamefont {Knutsson}}]{diva2:288565}%
  \BibitemOpen
  \bibfield  {author} {\bibinfo {author} {\bibfnamefont {M.}~\bibnamefont
  {Borga}}, \bibinfo {author} {\bibfnamefont {T.}~\bibnamefont {Landelius}}, \
  and\ \bibinfo {author} {\bibfnamefont {H.}~\bibnamefont {Knutsson}},\
  }\href@noop {} {\emph {\bibinfo {title} {{A Unified Approach to PCA, PLS, MLR
  and CCA}}}},\ \bibinfo {type} {Other academic}\ \bibinfo {number}
  {LiTH-ISY-R, 1992}\ (\bibinfo  {institution} {Linköping University,
  Department of Electrical Engineering},\ \bibinfo {address} {Sweden},\
  \bibinfo {year} {1997})\BibitemShut {NoStop}%
\bibitem [{\citenamefont {Bhalerao}\ \emph {et~al.}(2015)\citenamefont
  {Bhalerao}, \citenamefont {Ollitrault}, \citenamefont {Pal},\ and\
  \citenamefont {Teaney}}]{Bhalerao:2014mua}%
  \BibitemOpen
  \bibfield  {author} {\bibinfo {author} {\bibfnamefont {R.~S.}\ \bibnamefont
  {Bhalerao}}, \bibinfo {author} {\bibfnamefont {J.-Y.}\ \bibnamefont
  {Ollitrault}}, \bibinfo {author} {\bibfnamefont {S.}~\bibnamefont {Pal}}, \
  and\ \bibinfo {author} {\bibfnamefont {D.}~\bibnamefont {Teaney}},\
  }\bibfield  {title} {\enquote {\bibinfo {title} {{Principal component
  analysis of event-by-event fluctuations}},}\ }\href {\doibase
  10.1103/PhysRevLett.114.152301} {\bibfield  {journal} {\bibinfo  {journal}
  {Phys. Rev. Lett.}\ }\textbf {\bibinfo {volume} {114}},\ \bibinfo {pages}
  {152301} (\bibinfo {year} {2015})},\ \Eprint {http://arxiv.org/abs/1410.7739}
  {arXiv:1410.7739 [nucl-th]} \BibitemShut {NoStop}%
\bibitem [{\citenamefont {Milosevic}(2016)}]{Milosevic:2016tiw}%
  \BibitemOpen
  \bibfield  {author} {\bibinfo {author} {\bibfnamefont {J.}~\bibnamefont
  {Milosevic}} (\bibinfo {collaboration} {CMS}),\ }\bibfield  {title} {\enquote
  {\bibinfo {title} {{Principle Component Analysis of two-particle correlations
  in PbPb and pPb collisions at CMS}},}\ }\bibfield  {booktitle} {\emph
  {\bibinfo {booktitle} {{Proceedings, 25th International Conference on
  Ultra-Relativistic Nucleus-Nucleus Collisions (Quark Matter 2015): Kobe,
  Japan, September 27-October 3, 2015}}},\ }\href {\doibase
  10.1016/j.nuclphysa.2016.03.022} {\bibfield  {journal} {\bibinfo  {journal}
  {Nucl. Phys.}\ }\textbf {\bibinfo {volume} {A956}},\ \bibinfo {pages}
  {308--311} (\bibinfo {year} {2016})}\BibitemShut {NoStop}%
\bibitem [{\citenamefont {Aad}\ \emph {et~al.}(2014{\natexlab{a}})\citenamefont
  {Aad} \emph {et~al.}}]{Aad:2014lta}%
  \BibitemOpen
  \bibfield  {author} {\bibinfo {author} {\bibfnamefont {G.}~\bibnamefont
  {Aad}} \emph {et~al.} (\bibinfo {collaboration} {ATLAS}),\ }\bibfield
  {title} {\enquote {\bibinfo {title} {{Measurement of long-range
  pseudorapidity correlations and azimuthal harmonics in $\sqrt{s_{NN}}=5.02$
  TeV proton-lead collisions with the ATLAS detector}},}\ }\href {\doibase
  10.1103/PhysRevC.90.044906} {\bibfield  {journal} {\bibinfo  {journal} {Phys.
  Rev.}\ }\textbf {\bibinfo {volume} {C90}},\ \bibinfo {pages} {044906}
  (\bibinfo {year} {2014}{\natexlab{a}})},\ \Eprint
  {http://arxiv.org/abs/1409.1792} {arXiv:1409.1792 [hep-ex]} \BibitemShut
  {NoStop}%
\bibitem [{\citenamefont {Bozek}\ \emph {et~al.}(2011)\citenamefont {Bozek},
  \citenamefont {Broniowski},\ and\ \citenamefont {Moreira}}]{Bozek:2010vz}%
  \BibitemOpen
  \bibfield  {author} {\bibinfo {author} {\bibfnamefont {P.}~\bibnamefont
  {Bozek}}, \bibinfo {author} {\bibfnamefont {W.}~\bibnamefont {Broniowski}}, \
  and\ \bibinfo {author} {\bibfnamefont {J.}~\bibnamefont {Moreira}},\
  }\bibfield  {title} {\enquote {\bibinfo {title} {\{Torqued fireballs in
  relativistic heavy-ion collisions\}},}\ }\href {\doibase
  10.1103/PhysRevC.83.034911} {\bibfield  {journal} {\bibinfo  {journal} {Phys.
  Rev.}\ }\textbf {\bibinfo {volume} {C83}},\ \bibinfo {pages} {034911}
  (\bibinfo {year} {2011})},\ \Eprint {http://arxiv.org/abs/1011.3354}
  {arXiv:1011.3354 [nucl-th]} \BibitemShut {NoStop}%
\%\%CITATION = ARXIV:1011.3354;\%\%
\bibitem [{\citenamefont {Bzdak}\ and\ \citenamefont
  {Teaney}(2013)}]{Bzdak:2012tp}%
  \BibitemOpen
  \bibfield  {author} {\bibinfo {author} {\bibfnamefont {A.}~\bibnamefont
  {Bzdak}}\ and\ \bibinfo {author} {\bibfnamefont {D.}~\bibnamefont {Teaney}},\
  }\bibfield  {title} {\enquote {\bibinfo {title} {{Longitudinal fluctuations
  of the fireball density in heavy-ion collisions}},}\ }\href {\doibase
  10.1103/PhysRevC.87.024906} {\bibfield  {journal} {\bibinfo  {journal} {Phys.
  Rev.}\ }\textbf {\bibinfo {volume} {C87}},\ \bibinfo {pages} {024906}
  (\bibinfo {year} {2013})},\ \Eprint {http://arxiv.org/abs/1210.1965}
  {arXiv:1210.1965 [nucl-th]} \BibitemShut {NoStop}%
\bibitem [{\citenamefont {Jia}\ and\ \citenamefont {Huo}(2014)}]{Jia:2014ysa}%
  \BibitemOpen
  \bibfield  {author} {\bibinfo {author} {\bibfnamefont {J.}~\bibnamefont
  {Jia}}\ and\ \bibinfo {author} {\bibfnamefont {P.}~\bibnamefont {Huo}},\
  }\bibfield  {title} {\enquote {\bibinfo {title} {{Forward-backward
  eccentricity and participant-plane angle fluctuations and their influences on
  longitudinal dynamics of collective flow}},}\ }\href {\doibase
  10.1103/PhysRevC.90.034915} {\bibfield  {journal} {\bibinfo  {journal} {Phys.
  Rev.}\ }\textbf {\bibinfo {volume} {C90}},\ \bibinfo {pages} {034915}
  (\bibinfo {year} {2014})},\ \Eprint {http://arxiv.org/abs/1403.6077}
  {arXiv:1403.6077 [nucl-th]} \BibitemShut {NoStop}%
\bibitem [{\citenamefont {Jia}\ \emph {et~al.}(2016)\citenamefont {Jia},
  \citenamefont {Radhakrishnan},\ and\ \citenamefont {Zhou}}]{Jia:2015jga}%
  \BibitemOpen
  \bibfield  {author} {\bibinfo {author} {\bibfnamefont {J.}~\bibnamefont
  {Jia}}, \bibinfo {author} {\bibfnamefont {S.}~\bibnamefont {Radhakrishnan}},
  \ and\ \bibinfo {author} {\bibfnamefont {M.}~\bibnamefont {Zhou}},\
  }\bibfield  {title} {\enquote {\bibinfo {title} {{Forward-backward
  multiplicity fluctuation and longitudinal harmonics in high-energy nuclear
  collisions}},}\ }\href {\doibase 10.1103/PhysRevC.93.044905} {\bibfield
  {journal} {\bibinfo  {journal} {Phys. Rev.}\ }\textbf {\bibinfo {volume}
  {C93}},\ \bibinfo {pages} {044905} (\bibinfo {year} {2016})},\ \Eprint
  {http://arxiv.org/abs/1506.03496} {arXiv:1506.03496 [nucl-th]} \BibitemShut
  {NoStop}%
\bibitem [{\citenamefont {Arnold}\ \emph
  {et~al.}(2005{\natexlab{b}})\citenamefont {Arnold}, \citenamefont {Lenaghan},
  \citenamefont {Moore},\ and\ \citenamefont {Yaffe}}]{Arnold:2004ti}%
  \BibitemOpen
  \bibfield  {author} {\bibinfo {author} {\bibfnamefont {P.~B.}\ \bibnamefont
  {Arnold}}, \bibinfo {author} {\bibfnamefont {J.}~\bibnamefont {Lenaghan}},
  \bibinfo {author} {\bibfnamefont {G.~D.}\ \bibnamefont {Moore}}, \ and\
  \bibinfo {author} {\bibfnamefont {L.~G.}\ \bibnamefont {Yaffe}},\ }\bibfield
  {title} {\enquote {\bibinfo {title} {{Apparent thermalization due to plasma
  instabilities in quark-gluon plasma}},}\ }\href {\doibase
  10.1103/PhysRevLett.94.072302} {\bibfield  {journal} {\bibinfo  {journal}
  {Phys. Rev. Lett.}\ }\textbf {\bibinfo {volume} {94}},\ \bibinfo {pages}
  {072302} (\bibinfo {year} {2005}{\natexlab{b}})},\ \Eprint
  {http://arxiv.org/abs/nucl-th/0409068} {arXiv:nucl-th/0409068 [nucl-th]}
  \BibitemShut {NoStop}%
\bibitem [{\citenamefont {Niemi}\ \emph {et~al.}(2016)\citenamefont {Niemi},
  \citenamefont {Eskola},\ and\ \citenamefont {Paatelainen}}]{Niemi:2015qia}%
  \BibitemOpen
  \bibfield  {author} {\bibinfo {author} {\bibfnamefont {H.}~\bibnamefont
  {Niemi}}, \bibinfo {author} {\bibfnamefont {K.~J.}\ \bibnamefont {Eskola}}, \
  and\ \bibinfo {author} {\bibfnamefont {R.}~\bibnamefont {Paatelainen}},\
  }\bibfield  {title} {\enquote {\bibinfo {title} {{Event-by-event fluctuations
  in a perturbative QCD + saturation + hydrodynamics model: Determining QCD
  matter shear viscosity in ultrarelativistic heavy-ion collisions}},}\ }\href
  {\doibase 10.1103/PhysRevC.93.024907} {\bibfield  {journal} {\bibinfo
  {journal} {Phys. Rev.}\ }\textbf {\bibinfo {volume} {C93}},\ \bibinfo {pages}
  {024907} (\bibinfo {year} {2016})},\ \Eprint
  {http://arxiv.org/abs/1505.02677} {arXiv:1505.02677 [hep-ph]} \BibitemShut
  {NoStop}%
\bibitem [{\citenamefont {Kurkela}\ and\ \citenamefont
  {Zhu}(2015)}]{Kurkela:2015qoa}%
  \BibitemOpen
  \bibfield  {author} {\bibinfo {author} {\bibfnamefont {A.}~\bibnamefont
  {Kurkela}}\ and\ \bibinfo {author} {\bibfnamefont {Y.}~\bibnamefont {Zhu}},\
  }\bibfield  {title} {\enquote {\bibinfo {title} {{Isotropization and
  hydrodynamization in weakly coupled heavy-ion collisions}},}\ }\href
  {\doibase 10.1103/PhysRevLett.115.182301} {\bibfield  {journal} {\bibinfo
  {journal} {Phys. Rev. Lett.}\ }\textbf {\bibinfo {volume} {115}},\ \bibinfo
  {pages} {182301} (\bibinfo {year} {2015})},\ \Eprint
  {http://arxiv.org/abs/1506.06647} {arXiv:1506.06647 [hep-ph]} \BibitemShut
  {NoStop}%
\bibitem [{\citenamefont {Keegan}\ \emph
  {et~al.}(2016{\natexlab{b}})\citenamefont {Keegan}, \citenamefont {Kurkela},
  \citenamefont {Romatschke}, \citenamefont {van~der Schee},\ and\
  \citenamefont {Zhu}}]{Keegan:2015avk}%
  \BibitemOpen
  \bibfield  {author} {\bibinfo {author} {\bibfnamefont {L.}~\bibnamefont
  {Keegan}}, \bibinfo {author} {\bibfnamefont {A.}~\bibnamefont {Kurkela}},
  \bibinfo {author} {\bibfnamefont {P.}~\bibnamefont {Romatschke}}, \bibinfo
  {author} {\bibfnamefont {W.}~\bibnamefont {van~der Schee}}, \ and\ \bibinfo
  {author} {\bibfnamefont {Y.}~\bibnamefont {Zhu}},\ }\bibfield  {title}
  {\enquote {\bibinfo {title} {{Weak and strong coupling equilibration in
  nonabelian gauge theories}},}\ }\href {\doibase 10.1007/JHEP04(2016)031}
  {\bibfield  {journal} {\bibinfo  {journal} {JHEP}\ }\textbf {\bibinfo
  {volume} {04}},\ \bibinfo {pages} {031} (\bibinfo {year}
  {2016}{\natexlab{b}})},\ \Eprint {http://arxiv.org/abs/1512.05347}
  {arXiv:1512.05347 [hep-th]} \BibitemShut {NoStop}%
\bibitem [{\citenamefont {Schenke}\ \emph
  {et~al.}(2012{\natexlab{b}})\citenamefont {Schenke}, \citenamefont
  {Tribedy},\ and\ \citenamefont {Venugopalan}}]{Schenke:2012wb}%
  \BibitemOpen
  \bibfield  {author} {\bibinfo {author} {\bibfnamefont {B.}~\bibnamefont
  {Schenke}}, \bibinfo {author} {\bibfnamefont {P.}~\bibnamefont {Tribedy}}, \
  and\ \bibinfo {author} {\bibfnamefont {R.}~\bibnamefont {Venugopalan}},\
  }\bibfield  {title} {\enquote {\bibinfo {title} {{Fluctuating Glasma initial
  conditions and flow in heavy ion collisions}},}\ }\href {\doibase
  10.1103/PhysRevLett.108.252301} {\bibfield  {journal} {\bibinfo  {journal}
  {Phys. Rev. Lett.}\ }\textbf {\bibinfo {volume} {108}},\ \bibinfo {pages}
  {252301} (\bibinfo {year} {2012}{\natexlab{b}})},\ \Eprint
  {http://arxiv.org/abs/1202.6646} {arXiv:1202.6646 [nucl-th]} \BibitemShut
  {NoStop}%
\bibitem [{\citenamefont {Broniowski}\ \emph
  {et~al.}(2009{\natexlab{a}})\citenamefont {Broniowski}, \citenamefont
  {Florkowski}, \citenamefont {Chojnacki},\ and\ \citenamefont
  {Kisiel}}]{Broniowski:2008qk}%
  \BibitemOpen
  \bibfield  {author} {\bibinfo {author} {\bibfnamefont {W.}~\bibnamefont
  {Broniowski}}, \bibinfo {author} {\bibfnamefont {W.}~\bibnamefont
  {Florkowski}}, \bibinfo {author} {\bibfnamefont {M.}~\bibnamefont
  {Chojnacki}}, \ and\ \bibinfo {author} {\bibfnamefont {A.}~\bibnamefont
  {Kisiel}},\ }\bibfield  {title} {\enquote {\bibinfo {title} {{Free-streaming
  approximation in early dynamics of relativistic heavy-ion collisions}},}\
  }\href {\doibase 10.1103/PhysRevC.80.034902} {\bibfield  {journal} {\bibinfo
  {journal} {Phys. Rev.}\ }\textbf {\bibinfo {volume} {C80}},\ \bibinfo {pages}
  {034902} (\bibinfo {year} {2009}{\natexlab{a}})},\ \Eprint
  {http://arxiv.org/abs/0812.3393} {arXiv:0812.3393 [nucl-th]} \BibitemShut
  {NoStop}%
\bibitem [{\citenamefont {Liu}\ \emph {et~al.}(2015)\citenamefont {Liu},
  \citenamefont {Shen},\ and\ \citenamefont {Heinz}}]{Liu:2015nwa}%
  \BibitemOpen
  \bibfield  {author} {\bibinfo {author} {\bibfnamefont {J.}~\bibnamefont
  {Liu}}, \bibinfo {author} {\bibfnamefont {C.}~\bibnamefont {Shen}}, \ and\
  \bibinfo {author} {\bibfnamefont {U.}~\bibnamefont {Heinz}},\ }\bibfield
  {title} {\enquote {\bibinfo {title} {{Pre-equilibrium evolution effects on
  heavy-ion collision observables}},}\ }\href {\doibase
  10.1103/PhysRevC.92.049904, 10.1103/PhysRevC.91.064906} {\bibfield  {journal}
  {\bibinfo  {journal} {Phys. Rev.}\ }\textbf {\bibinfo {volume} {C91}},\
  \bibinfo {pages} {064906} (\bibinfo {year} {2015})},\ \bibinfo {note}
  {[Erratum: Phys. Rev.C92,no.4,049904(2015)]},\ \Eprint
  {http://arxiv.org/abs/1504.02160} {arXiv:1504.02160 [nucl-th]} \BibitemShut
  {NoStop}%
\bibitem [{\citenamefont {van~der Schee}\ \emph {et~al.}(2013)\citenamefont
  {van~der Schee}, \citenamefont {Romatschke},\ and\ \citenamefont
  {Pratt}}]{vanderSchee:2013pia}%
  \BibitemOpen
  \bibfield  {author} {\bibinfo {author} {\bibfnamefont {W.}~\bibnamefont
  {van~der Schee}}, \bibinfo {author} {\bibfnamefont {P.}~\bibnamefont
  {Romatschke}}, \ and\ \bibinfo {author} {\bibfnamefont {S.}~\bibnamefont
  {Pratt}},\ }\bibfield  {title} {\enquote {\bibinfo {title} {{Fully Dynamical
  Simulation of Central Nuclear Collisions}},}\ }\href {\doibase
  10.1103/PhysRevLett.111.222302} {\bibfield  {journal} {\bibinfo  {journal}
  {Phys. Rev. Lett.}\ }\textbf {\bibinfo {volume} {111}},\ \bibinfo {pages}
  {222302} (\bibinfo {year} {2013})},\ \Eprint {http://arxiv.org/abs/1307.2539}
  {arXiv:1307.2539} \BibitemShut {NoStop}%
\bibitem [{\citenamefont {Romatschke}(2015)}]{Romatschke:2015gxa}%
  \BibitemOpen
  \bibfield  {author} {\bibinfo {author} {\bibfnamefont {P.}~\bibnamefont
  {Romatschke}},\ }\bibfield  {title} {\enquote {\bibinfo {title} {{Light-Heavy
  Ion Collisions: A window into pre-equilibrium QCD dynamics?}}}\ }\href
  {\doibase 10.1140/epjc/s10052-015-3509-3} {\bibfield  {journal} {\bibinfo
  {journal} {Eur. Phys. J.}\ }\textbf {\bibinfo {volume} {C75}},\ \bibinfo
  {pages} {305} (\bibinfo {year} {2015})},\ \Eprint
  {http://arxiv.org/abs/1502.04745} {arXiv:1502.04745 [nucl-th]} \BibitemShut
  {NoStop}%
\bibitem [{\citenamefont {Martinez}\ \emph {et~al.}(2012)\citenamefont
  {Martinez}, \citenamefont {Ryblewski},\ and\ \citenamefont
  {Strickland}}]{Martinez:2012tu}%
  \BibitemOpen
  \bibfield  {author} {\bibinfo {author} {\bibfnamefont {M.}~\bibnamefont
  {Martinez}}, \bibinfo {author} {\bibfnamefont {R.}~\bibnamefont {Ryblewski}},
  \ and\ \bibinfo {author} {\bibfnamefont {M.}~\bibnamefont {Strickland}},\
  }\bibfield  {title} {\enquote {\bibinfo {title} {{Boost-Invariant
  (2+1)-dimensional Anisotropic Hydrodynamics}},}\ }\href {\doibase
  10.1103/PhysRevC.85.064913} {\bibfield  {journal} {\bibinfo  {journal} {Phys.
  Rev.}\ }\textbf {\bibinfo {volume} {C85}},\ \bibinfo {pages} {064913}
  (\bibinfo {year} {2012})},\ \Eprint {http://arxiv.org/abs/1204.1473}
  {arXiv:1204.1473 [nucl-th]} \BibitemShut {NoStop}%
\bibitem [{\citenamefont {Landau}\ and\ \citenamefont
  {Lifshits}(1958)}]{LandauStatPart1}%
  \BibitemOpen
  \bibfield  {author} {\bibinfo {author} {\bibfnamefont {L.~D.}\ \bibnamefont
  {Landau}}\ and\ \bibinfo {author} {\bibfnamefont {E.~M.}\ \bibnamefont
  {Lifshits}},\ }\href
  {http://proxy.library.stonybrook.edu/login?url=http://search.ebscohost.com/login.aspx?direct=true&db=cat03000a&AN=STB.000516107&site=eds-live&scope=site}
  {\emph {\bibinfo {title} {Statistical physics: Part I}}},\ \bibinfo {series}
  {Course of theoretical physics}, Vol.~\bibinfo {volume} {5}\ (\bibinfo
  {publisher} {London, Pergamon Press; Reading, Mass., Addison-Wesley Pub. Co.,
  1958.},\ \bibinfo {year} {1958})\BibitemShut {NoStop}%
\bibitem [{\citenamefont {Landau}\ \emph {et~al.}(1978)\citenamefont {Landau},
  \citenamefont {Lifshits},\ and\ \citenamefont
  {Pitaevskii}}]{LandauStatPart2}%
  \BibitemOpen
  \bibfield  {author} {\bibinfo {author} {\bibfnamefont {L.~D.}\ \bibnamefont
  {Landau}}, \bibinfo {author} {\bibfnamefont {E.~M.}\ \bibnamefont
  {Lifshits}}, \ and\ \bibinfo {author} {\bibfnamefont {L.~P.}\ \bibnamefont
  {Pitaevskii}},\ }\href
  {http://proxy.library.stonybrook.edu/login?url=http://search.ebscohost.com/login.aspx?direct=true&db=cat03000a&AN=STB.000442124&site=eds-live&scope=site}
  {\emph {\bibinfo {title} {Statistical physics: Part II}}},\ \bibinfo {series}
  {Course of theoretical physics}, Vol.~\bibinfo {volume} {9}\ (\bibinfo
  {publisher} {Oxford ; New York : Pergamon Press, 1980.},\ \bibinfo {year}
  {1978})\BibitemShut {NoStop}%
\bibitem [{\citenamefont {Alder}\ and\ \citenamefont
  {Wainwright}(1967)}]{velocity_auto}%
  \BibitemOpen
  \bibfield  {author} {\bibinfo {author} {\bibfnamefont {B.~J.}\ \bibnamefont
  {Alder}}\ and\ \bibinfo {author} {\bibfnamefont {T.~E.}\ \bibnamefont
  {Wainwright}},\ }\bibfield  {title} {\enquote {\bibinfo {title} {Velocity
  autocorrelations for hard spheres},}\ }\href {\doibase
  10.1103/PhysRevLett.18.988} {\bibfield  {journal} {\bibinfo  {journal} {Phys.
  Rev. Lett.}\ }\textbf {\bibinfo {volume} {18}},\ \bibinfo {pages} {988--990}
  (\bibinfo {year} {1967})}\BibitemShut {NoStop}%
\bibitem [{\citenamefont {Alder}\ and\ \citenamefont
  {Wainwright}(1970)}]{wainwright2}%
  \BibitemOpen
  \bibfield  {author} {\bibinfo {author} {\bibfnamefont {B.~J.}\ \bibnamefont
  {Alder}}\ and\ \bibinfo {author} {\bibfnamefont {T.~E.}\ \bibnamefont
  {Wainwright}},\ }\bibfield  {title} {\enquote {\bibinfo {title} {Decay of the
  velocity autocorrelation function},}\ }\href {\doibase 10.1103/PhysRevA.1.18}
  {\bibfield  {journal} {\bibinfo  {journal} {Phys. Rev. A}\ }\textbf {\bibinfo
  {volume} {1}},\ \bibinfo {pages} {18--21} (\bibinfo {year}
  {1970})}\BibitemShut {NoStop}%
\bibitem [{\citenamefont {Zwanzig}\ and\ \citenamefont
  {Bixon}(1970)}]{bixon_zwanzig}%
  \BibitemOpen
  \bibfield  {author} {\bibinfo {author} {\bibfnamefont {R.}~\bibnamefont
  {Zwanzig}}\ and\ \bibinfo {author} {\bibfnamefont {M.}~\bibnamefont
  {Bixon}},\ }\bibfield  {title} {\enquote {\bibinfo {title} {Hydrodynamic
  theory of the velocity correlation function},}\ }\href {\doibase
  10.1103/PhysRevA.2.2005} {\bibfield  {journal} {\bibinfo  {journal} {Phys.
  Rev. A}\ }\textbf {\bibinfo {volume} {2}},\ \bibinfo {pages} {2005--2012}
  (\bibinfo {year} {1970})}\BibitemShut {NoStop}%
\bibitem [{\citenamefont {Kovtun}\ \emph {et~al.}(2011)\citenamefont {Kovtun},
  \citenamefont {Moore},\ and\ \citenamefont {Romatschke}}]{Kovtun:2011np}%
  \BibitemOpen
  \bibfield  {author} {\bibinfo {author} {\bibfnamefont {P.}~\bibnamefont
  {Kovtun}}, \bibinfo {author} {\bibfnamefont {G.~D.}\ \bibnamefont {Moore}}, \
  and\ \bibinfo {author} {\bibfnamefont {P.}~\bibnamefont {Romatschke}},\
  }\bibfield  {title} {\enquote {\bibinfo {title} {{The stickiness of sound: An
  absolute lower limit on viscosity and the breakdown of second order
  relativistic hydrodynamics}},}\ }\href {\doibase 10.1103/PhysRevD.84.025006}
  {\bibfield  {journal} {\bibinfo  {journal} {Phys. Rev.}\ }\textbf {\bibinfo
  {volume} {D84}},\ \bibinfo {pages} {025006} (\bibinfo {year} {2011})},\
  \Eprint {http://arxiv.org/abs/1104.1586} {arXiv:1104.1586 [hep-ph]}
  \BibitemShut {NoStop}%
\bibitem [{\citenamefont {Kovtun}\ and\ \citenamefont
  {Yaffe}(2003)}]{Kovtun:2003vj}%
  \BibitemOpen
  \bibfield  {author} {\bibinfo {author} {\bibfnamefont {P.}~\bibnamefont
  {Kovtun}}\ and\ \bibinfo {author} {\bibfnamefont {L.~G.}\ \bibnamefont
  {Yaffe}},\ }\bibfield  {title} {\enquote {\bibinfo {title} {{Hydrodynamic
  fluctuations, long time tails, and supersymmetry}},}\ }\href {\doibase
  10.1103/PhysRevD.68.025007} {\bibfield  {journal} {\bibinfo  {journal} {Phys.
  Rev.}\ }\textbf {\bibinfo {volume} {D68}},\ \bibinfo {pages} {025007}
  (\bibinfo {year} {2003})},\ \Eprint {http://arxiv.org/abs/hep-th/0303010}
  {arXiv:hep-th/0303010 [hep-th]} \BibitemShut {NoStop}%
\bibitem [{\citenamefont {Gavin}\ and\ \citenamefont
  {Abdel-Aziz}(2006)}]{Gavin:2006xd}%
  \BibitemOpen
  \bibfield  {author} {\bibinfo {author} {\bibfnamefont {S.}~\bibnamefont
  {Gavin}}\ and\ \bibinfo {author} {\bibfnamefont {M.}~\bibnamefont
  {Abdel-Aziz}},\ }\bibfield  {title} {\enquote {\bibinfo {title} {{Measuring
  Shear Viscosity Using Transverse Momentum Correlations in Relativistic
  Nuclear Collisions}},}\ }\href {\doibase 10.1103/PhysRevLett.97.162302}
  {\bibfield  {journal} {\bibinfo  {journal} {Phys. Rev. Lett.}\ }\textbf
  {\bibinfo {volume} {97}},\ \bibinfo {pages} {162302} (\bibinfo {year}
  {2006})},\ \Eprint {http://arxiv.org/abs/nucl-th/0606061}
  {arXiv:nucl-th/0606061 [nucl-th]} \BibitemShut {NoStop}%
\bibitem [{\citenamefont {Kapusta}\ \emph {et~al.}(2012)\citenamefont
  {Kapusta}, \citenamefont {Muller},\ and\ \citenamefont
  {Stephanov}}]{Kapusta:2011gt}%
  \BibitemOpen
  \bibfield  {author} {\bibinfo {author} {\bibfnamefont {J.~I.}\ \bibnamefont
  {Kapusta}}, \bibinfo {author} {\bibfnamefont {B.}~\bibnamefont {Muller}}, \
  and\ \bibinfo {author} {\bibfnamefont {M.}~\bibnamefont {Stephanov}},\
  }\bibfield  {title} {\enquote {\bibinfo {title} {{Relativistic Theory of
  Hydrodynamic Fluctuations with Applications to Heavy Ion Collisions}},}\
  }\href {\doibase 10.1103/PhysRevC.85.054906} {\bibfield  {journal} {\bibinfo
  {journal} {Phys. Rev.}\ }\textbf {\bibinfo {volume} {C85}},\ \bibinfo {pages}
  {054906} (\bibinfo {year} {2012})},\ \Eprint {http://arxiv.org/abs/1112.6405}
  {arXiv:1112.6405 [nucl-th]} \BibitemShut {NoStop}%
\bibitem [{\citenamefont {Yan}\ and\ \citenamefont
  {Gr{\"o}nqvist}(2016)}]{Yan:2015lfa}%
  \BibitemOpen
  \bibfield  {author} {\bibinfo {author} {\bibfnamefont {L.}~\bibnamefont
  {Yan}}\ and\ \bibinfo {author} {\bibfnamefont {H.}~\bibnamefont
  {Gr{\"o}nqvist}},\ }\bibfield  {title} {\enquote {\bibinfo {title}
  {{Hydrodynamical noise and Gubser flow}},}\ }\href {\doibase
  10.1007/JHEP03(2016)121} {\bibfield  {journal} {\bibinfo  {journal} {JHEP}\
  }\textbf {\bibinfo {volume} {03}},\ \bibinfo {pages} {121} (\bibinfo {year}
  {2016})},\ \Eprint {http://arxiv.org/abs/1511.07198} {arXiv:1511.07198
  [nucl-th]} \BibitemShut {NoStop}%
\bibitem [{\citenamefont {Young}\ \emph {et~al.}(2015)\citenamefont {Young},
  \citenamefont {Kapusta}, \citenamefont {Gale}, \citenamefont {Jeon},\ and\
  \citenamefont {Schenke}}]{Young:2014pka}%
  \BibitemOpen
  \bibfield  {author} {\bibinfo {author} {\bibfnamefont {C.}~\bibnamefont
  {Young}}, \bibinfo {author} {\bibfnamefont {J.~I.}\ \bibnamefont {Kapusta}},
  \bibinfo {author} {\bibfnamefont {C.}~\bibnamefont {Gale}}, \bibinfo {author}
  {\bibfnamefont {S.}~\bibnamefont {Jeon}}, \ and\ \bibinfo {author}
  {\bibfnamefont {B.}~\bibnamefont {Schenke}},\ }\bibfield  {title} {\enquote
  {\bibinfo {title} {{Thermally Fluctuating Second-Order Viscous Hydrodynamics
  and Heavy-Ion Collisions}},}\ }\href {\doibase 10.1103/PhysRevC.91.044901}
  {\bibfield  {journal} {\bibinfo  {journal} {Phys. Rev.}\ }\textbf {\bibinfo
  {volume} {C91}},\ \bibinfo {pages} {044901} (\bibinfo {year} {2015})},\
  \Eprint {http://arxiv.org/abs/1407.1077} {arXiv:1407.1077 [nucl-th]}
  \BibitemShut {NoStop}%
\bibitem [{\citenamefont {Nagai}\ \emph {et~al.}(2016)\citenamefont {Nagai},
  \citenamefont {Kurita}, \citenamefont {Murase},\ and\ \citenamefont
  {Hirano}}]{Nagai:2016wyx}%
  \BibitemOpen
  \bibfield  {author} {\bibinfo {author} {\bibfnamefont {K.}~\bibnamefont
  {Nagai}}, \bibinfo {author} {\bibfnamefont {R.}~\bibnamefont {Kurita}},
  \bibinfo {author} {\bibfnamefont {K.}~\bibnamefont {Murase}}, \ and\ \bibinfo
  {author} {\bibfnamefont {T.}~\bibnamefont {Hirano}},\ }\bibfield  {title}
  {\enquote {\bibinfo {title} {{Causal hydrodynamic fluctuation in Bjorken
  expansion}},}\ }\bibfield  {booktitle} {\emph {\bibinfo {booktitle}
  {{Proceedings, 25th International Conference on Ultra-Relativistic
  Nucleus-Nucleus Collisions (Quark Matter 2015): Kobe, Japan, September
  27-October 3, 2015}}},\ }\href {\doibase 10.1016/j.nuclphysa.2016.02.007}
  {\bibfield  {journal} {\bibinfo  {journal} {Nucl. Phys.}\ }\textbf {\bibinfo
  {volume} {A956}},\ \bibinfo {pages} {781--784} (\bibinfo {year} {2016})},\
  \Eprint {http://arxiv.org/abs/1602.00794} {arXiv:1602.00794 [nucl-th]}
  \BibitemShut {NoStop}%
\bibitem [{\citenamefont {Murase}\ and\ \citenamefont
  {Hirano}(2016)}]{Murase:2016rhl}%
  \BibitemOpen
  \bibfield  {author} {\bibinfo {author} {\bibfnamefont {K.}~\bibnamefont
  {Murase}}\ and\ \bibinfo {author} {\bibfnamefont {T.}~\bibnamefont
  {Hirano}},\ }\bibfield  {title} {\enquote {\bibinfo {title} {{Hydrodynamic
  fluctuations and dissipation in an integrated dynamical model}},}\ }\bibfield
   {booktitle} {\emph {\bibinfo {booktitle} {{Proceedings, 25th International
  Conference on Ultra-Relativistic Nucleus-Nucleus Collisions (Quark Matter
  2015): Kobe, Japan, September 27-October 3, 2015}}},\ }\href {\doibase
  10.1016/j.nuclphysa.2016.01.011} {\bibfield  {journal} {\bibinfo  {journal}
  {Nucl. Phys.}\ }\textbf {\bibinfo {volume} {A956}},\ \bibinfo {pages}
  {276--279} (\bibinfo {year} {2016})},\ \Eprint
  {http://arxiv.org/abs/1601.02260} {arXiv:1601.02260 [nucl-th]} \BibitemShut
  {NoStop}%
\bibitem [{\citenamefont {Aad}\ \emph {et~al.}(2014{\natexlab{b}})\citenamefont
  {Aad} \emph {et~al.}}]{Aad:2014fla}%
  \BibitemOpen
  \bibfield  {author} {\bibinfo {author} {\bibfnamefont {G.}~\bibnamefont
  {Aad}} \emph {et~al.} (\bibinfo {collaboration} {ATLAS}),\ }\bibfield
  {title} {\enquote {\bibinfo {title} {{Measurement of event-plane correlations
  in $\sqrt{s_{NN}}=2.76$ TeV lead-lead collisions with the ATLAS detector}},}\
  }\href {\doibase 10.1103/PhysRevC.90.024905} {\bibfield  {journal} {\bibinfo
  {journal} {Phys. Rev.}\ }\textbf {\bibinfo {volume} {C90}},\ \bibinfo {pages}
  {024905} (\bibinfo {year} {2014}{\natexlab{b}})},\ \Eprint
  {http://arxiv.org/abs/1403.0489} {arXiv:1403.0489 [hep-ex]} \BibitemShut
  {NoStop}%
\bibitem [{\citenamefont {Aad}\ \emph {et~al.}(2015)\citenamefont {Aad} \emph
  {et~al.}}]{Aad:2015lwa}%
  \BibitemOpen
  \bibfield  {author} {\bibinfo {author} {\bibfnamefont {G.}~\bibnamefont
  {Aad}} \emph {et~al.} (\bibinfo {collaboration} {ATLAS}),\ }\bibfield
  {title} {\enquote {\bibinfo {title} {{Measurement of the correlation between
  flow harmonics of different order in lead-lead collisions at
  $\sqrt{s_{NN}}$=2.76 TeV with the ATLAS detector}},}\ }\href {\doibase
  10.1103/PhysRevC.92.034903} {\bibfield  {journal} {\bibinfo  {journal} {Phys.
  Rev.}\ }\textbf {\bibinfo {volume} {C92}},\ \bibinfo {pages} {034903}
  (\bibinfo {year} {2015})},\ \Eprint {http://arxiv.org/abs/1504.01289}
  {arXiv:1504.01289 [hep-ex]} \BibitemShut {NoStop}%
\bibitem [{\citenamefont {Adam}\ \emph
  {et~al.}(2016{\natexlab{b}})\citenamefont {Adam} \emph
  {et~al.}}]{Adam:2015eta}%
  \BibitemOpen
  \bibfield  {author} {\bibinfo {author} {\bibfnamefont {J.}~\bibnamefont
  {Adam}} \emph {et~al.} (\bibinfo {collaboration} {ALICE}),\ }\bibfield
  {title} {\enquote {\bibinfo {title} {{Event shape engineering for inclusive
  spectra and elliptic flow in Pb-Pb collisions at $\sqrt{s_{\rm{NN}}}=2.76$
  TeV}},}\ }\href {\doibase {10.1103/PhysRevC.93.034916}} {\bibfield  {journal}
  {\bibinfo  {journal} {Phys. Rev.}\ }\textbf {\bibinfo {volume} {C93}},\
  \bibinfo {pages} {034916} (\bibinfo {year} {2016}{\natexlab{b}})},\ \Eprint
  {http://arxiv.org/abs/1507.06194} {arXiv:1507.06194 [nucl-ex]} \BibitemShut
  {NoStop}%
\bibitem [{\citenamefont {Adamczyk}\ \emph {et~al.}(2015)\citenamefont
  {Adamczyk} \emph {et~al.}}]{Adamczyk:2015xjc}%
  \BibitemOpen
  \bibfield  {author} {\bibinfo {author} {\bibfnamefont {L.}~\bibnamefont
  {Adamczyk}} \emph {et~al.} (\bibinfo {collaboration} {STAR}),\ }\bibfield
  {title} {\enquote {\bibinfo {title} {{Long-range pseudorapidity dihadron
  correlations in $d$+Au collisions at $\sqrt{s_{\rm NN}}=200$ GeV}},}\ }\href
  {\doibase 10.1016/j.physletb.2015.05.075} {\bibfield  {journal} {\bibinfo
  {journal} {Phys. Lett.}\ }\textbf {\bibinfo {volume} {B747}},\ \bibinfo
  {pages} {265--271} (\bibinfo {year} {2015})},\ \Eprint
  {http://arxiv.org/abs/1502.07652} {arXiv:1502.07652 [nucl-ex]} \BibitemShut
  {NoStop}%
\bibitem [{\citenamefont {Adare}\ \emph {et~al.}(2015)\citenamefont {Adare}
  \emph {et~al.}}]{Adare:2015ctn}%
  \BibitemOpen
  \bibfield  {author} {\bibinfo {author} {\bibfnamefont {A.}~\bibnamefont
  {Adare}} \emph {et~al.} (\bibinfo {collaboration} {PHENIX}),\ }\bibfield
  {title} {\enquote {\bibinfo {title} {{Measurements of elliptic and triangular
  flow in high-multiplicity $^{3}$He$+$Au collisions at $\sqrt{s_{_{NN}}}=200$
  GeV}},}\ }\href {\doibase 10.1103/PhysRevLett.115.142301} {\bibfield
  {journal} {\bibinfo  {journal} {Phys. Rev. Lett.}\ }\textbf {\bibinfo
  {volume} {115}},\ \bibinfo {pages} {142301} (\bibinfo {year} {2015})},\
  \Eprint {http://arxiv.org/abs/1507.06273} {arXiv:1507.06273 [nucl-ex]}
  \BibitemShut {NoStop}%
\bibitem [{\citenamefont {Borghini}\ and\ \citenamefont
  {Ollitrault}(2006)}]{Borghini:2005kd}%
  \BibitemOpen
  \bibfield  {author} {\bibinfo {author} {\bibfnamefont {N.}~\bibnamefont
  {Borghini}}\ and\ \bibinfo {author} {\bibfnamefont {J.-Y.}\ \bibnamefont
  {Ollitrault}},\ }\bibfield  {title} {\enquote {\bibinfo {title} {{Momentum
  spectra, anisotropic flow, and ideal fluids}},}\ }\href {\doibase
  10.1016/j.physletb.2006.09.062} {\bibfield  {journal} {\bibinfo  {journal}
  {Phys. Lett.}\ }\textbf {\bibinfo {volume} {B642}},\ \bibinfo {pages}
  {227--231} (\bibinfo {year} {2006})},\ \Eprint
  {http://arxiv.org/abs/nucl-th/0506045} {arXiv:nucl-th/0506045 [nucl-th]}
  \BibitemShut {NoStop}%
\bibitem [{\citenamefont {Gardim}\ \emph {et~al.}(2012)\citenamefont {Gardim},
  \citenamefont {Grassi}, \citenamefont {Luzum},\ and\ \citenamefont
  {Ollitrault}}]{Gardim:2011xv}%
  \BibitemOpen
  \bibfield  {author} {\bibinfo {author} {\bibfnamefont {F.~G.}\ \bibnamefont
  {Gardim}}, \bibinfo {author} {\bibfnamefont {F.}~\bibnamefont {Grassi}},
  \bibinfo {author} {\bibfnamefont {M.}~\bibnamefont {Luzum}}, \ and\ \bibinfo
  {author} {\bibfnamefont {J.-Y.}\ \bibnamefont {Ollitrault}},\ }\bibfield
  {title} {\enquote {\bibinfo {title} {{Mapping the hydrodynamic response to
  the initial geometry in heavy-ion collisions}},}\ }\href {\doibase
  10.1103/PhysRevC.85.024908} {\bibfield  {journal} {\bibinfo  {journal} {Phys.
  Rev.}\ }\textbf {\bibinfo {volume} {C85}},\ \bibinfo {pages} {024908}
  (\bibinfo {year} {2012})},\ \Eprint {http://arxiv.org/abs/1111.6538}
  {arXiv:1111.6538 [nucl-th]} \BibitemShut {NoStop}%
\bibitem [{\citenamefont {Teaney}\ and\ \citenamefont
  {Yan}(2012)}]{Teaney:2012ke}%
  \BibitemOpen
  \bibfield  {author} {\bibinfo {author} {\bibfnamefont {D.}~\bibnamefont
  {Teaney}}\ and\ \bibinfo {author} {\bibfnamefont {L.}~\bibnamefont {Yan}},\
  }\bibfield  {title} {\enquote {\bibinfo {title} {{Non linearities in the
  harmonic spectrum of heavy ion collisions with ideal and viscous
  hydrodynamics}},}\ }\href {\doibase 10.1103/PhysRevC.86.044908} {\bibfield
  {journal} {\bibinfo  {journal} {Phys. Rev.}\ }\textbf {\bibinfo {volume}
  {C86}},\ \bibinfo {pages} {044908} (\bibinfo {year} {2012})},\ \Eprint
  {http://arxiv.org/abs/1206.1905} {arXiv:1206.1905 [nucl-th]} \BibitemShut
  {NoStop}%
\bibitem [{\citenamefont {Laine}\ and\ \citenamefont
  {Schroder}(2006)}]{Laine:2006cp}%
  \BibitemOpen
  \bibfield  {author} {\bibinfo {author} {\bibfnamefont {M.}~\bibnamefont
  {Laine}}\ and\ \bibinfo {author} {\bibfnamefont {Y.}~\bibnamefont
  {Schroder}},\ }\bibfield  {title} {\enquote {\bibinfo {title} {{Quark mass
  thresholds in QCD thermodynamics}},}\ }\href {\doibase
  10.1103/PhysRevD.73.085009} {\bibfield  {journal} {\bibinfo  {journal} {Phys.
  Rev.}\ }\textbf {\bibinfo {volume} {D73}},\ \bibinfo {pages} {085009}
  (\bibinfo {year} {2006})},\ \Eprint {http://arxiv.org/abs/hep-ph/0603048}
  {arXiv:hep-ph/0603048 [hep-ph]} \BibitemShut {NoStop}%
\bibitem [{\citenamefont {Luzum}\ and\ \citenamefont
  {Romatschke}(2008)}]{Luzum:2008cw}%
  \BibitemOpen
  \bibfield  {author} {\bibinfo {author} {\bibfnamefont {M.}~\bibnamefont
  {Luzum}}\ and\ \bibinfo {author} {\bibfnamefont {P.}~\bibnamefont
  {Romatschke}},\ }\bibfield  {title} {\enquote {\bibinfo {title} {{Conformal
  Relativistic Viscous Hydrodynamics: Applications to RHIC results at
  s(NN)**(1/2) = 200-GeV}},}\ }\href {\doibase 10.1103/PhysRevC.78.034915,
  10.1103/PhysRevC.79.039903} {\bibfield  {journal} {\bibinfo  {journal} {Phys.
  Rev.}\ }\textbf {\bibinfo {volume} {C78}},\ \bibinfo {pages} {034915}
  (\bibinfo {year} {2008})},\ \bibinfo {note} {[Erratum: Phys.
  Rev.C79,039903(2009)]},\ \Eprint {http://arxiv.org/abs/0804.4015}
  {arXiv:0804.4015 [nucl-th]} \BibitemShut {NoStop}%
\bibitem [{\citenamefont {Başar}\ and\ \citenamefont
  {Teaney}(2014)}]{Basar:2013hea}%
  \BibitemOpen
  \bibfield  {author} {\bibinfo {author} {\bibfnamefont {G.}~\bibnamefont
  {Başar}}\ and\ \bibinfo {author} {\bibfnamefont {D.}~\bibnamefont
  {Teaney}},\ }\bibfield  {title} {\enquote {\bibinfo {title} {{Scaling
  relation between pA and AA collisions}},}\ }\href {\doibase
  10.1103/PhysRevC.90.054903} {\bibfield  {journal} {\bibinfo  {journal} {Phys.
  Rev.}\ }\textbf {\bibinfo {volume} {C90}},\ \bibinfo {pages} {054903}
  (\bibinfo {year} {2014})},\ \Eprint {http://arxiv.org/abs/1312.6770}
  {arXiv:1312.6770 [nucl-th]} \BibitemShut {NoStop}%
\bibitem [{\citenamefont {Broniowski}\ \emph
  {et~al.}(2009{\natexlab{b}})\citenamefont {Broniowski}, \citenamefont
  {Chojnacki},\ and\ \citenamefont {Obara}}]{Broniowski:2009fm}%
  \BibitemOpen
  \bibfield  {author} {\bibinfo {author} {\bibfnamefont {W.}~\bibnamefont
  {Broniowski}}, \bibinfo {author} {\bibfnamefont {M.}~\bibnamefont
  {Chojnacki}}, \ and\ \bibinfo {author} {\bibfnamefont {L.}~\bibnamefont
  {Obara}},\ }\bibfield  {title} {\enquote {\bibinfo {title} {{Size
  fluctuations of the initial source and the event-by-event transverse momentum
  fluctuations in relativistic heavy-ion collisions}},}\ }\href {\doibase
  10.1103/PhysRevC.80.051902} {\bibfield  {journal} {\bibinfo  {journal} {Phys.
  Rev.}\ }\textbf {\bibinfo {volume} {C80}},\ \bibinfo {pages} {051902}
  (\bibinfo {year} {2009}{\natexlab{b}})},\ \Eprint
  {http://arxiv.org/abs/0907.3216} {arXiv:0907.3216 [nucl-th]} \BibitemShut
  {NoStop}%
\bibitem [{\citenamefont {Bozek}\ and\ \citenamefont
  {Broniowski}(2012)}]{Bozek:2012fw}%
  \BibitemOpen
  \bibfield  {author} {\bibinfo {author} {\bibfnamefont {P.}~\bibnamefont
  {Bozek}}\ and\ \bibinfo {author} {\bibfnamefont {W.}~\bibnamefont
  {Broniowski}},\ }\bibfield  {title} {\enquote {\bibinfo {title}
  {{Transverse-momentum fluctuations in relativistic heavy-ion collisions from
  event-by-event viscous hydrodynamics}},}\ }\href {\doibase
  10.1103/PhysRevC.85.044910} {\bibfield  {journal} {\bibinfo  {journal} {Phys.
  Rev.}\ }\textbf {\bibinfo {volume} {C85}},\ \bibinfo {pages} {044910}
  (\bibinfo {year} {2012})},\ \Eprint {http://arxiv.org/abs/1203.1810}
  {arXiv:1203.1810 [nucl-th]} \BibitemShut {NoStop}%
\bibitem [{\citenamefont {Qiu}\ and\ \citenamefont {Heinz}(2012)}]{Qiu:2012uy}%
  \BibitemOpen
  \bibfield  {author} {\bibinfo {author} {\bibfnamefont {Z.}~\bibnamefont
  {Qiu}}\ and\ \bibinfo {author} {\bibfnamefont {U.}~\bibnamefont {Heinz}},\
  }\bibfield  {title} {\enquote {\bibinfo {title} {{Hydrodynamic event-plane
  correlations in Pb+Pb collisions at $\sqrt{s}=2.76$ATeV}},}\ }\href {\doibase
  10.1016/j.physletb.2012.09.030} {\bibfield  {journal} {\bibinfo  {journal}
  {Phys. Lett.}\ }\textbf {\bibinfo {volume} {B717}},\ \bibinfo {pages}
  {261--265} (\bibinfo {year} {2012})},\ \Eprint
  {http://arxiv.org/abs/1208.1200} {arXiv:1208.1200 [nucl-th]} \BibitemShut
  {NoStop}%
\bibitem [{\citenamefont {Gardim}\ \emph {et~al.}(2015)\citenamefont {Gardim},
  \citenamefont {Noronha-Hostler}, \citenamefont {Luzum},\ and\ \citenamefont
  {Grassi}}]{Gardim:2014tya}%
  \BibitemOpen
  \bibfield  {author} {\bibinfo {author} {\bibfnamefont {F.~G.}\ \bibnamefont
  {Gardim}}, \bibinfo {author} {\bibfnamefont {J.}~\bibnamefont
  {Noronha-Hostler}}, \bibinfo {author} {\bibfnamefont {M.}~\bibnamefont
  {Luzum}}, \ and\ \bibinfo {author} {\bibfnamefont {F.}~\bibnamefont
  {Grassi}},\ }\bibfield  {title} {\enquote {\bibinfo {title} {{Effects of
  viscosity on the mapping of initial to final state in heavy ion
  collisions}},}\ }\href {\doibase 10.1103/PhysRevC.91.034902} {\bibfield
  {journal} {\bibinfo  {journal} {Phys. Rev.}\ }\textbf {\bibinfo {volume}
  {C91}},\ \bibinfo {pages} {034902} (\bibinfo {year} {2015})},\ \Eprint
  {http://arxiv.org/abs/1411.2574} {arXiv:1411.2574 [nucl-th]} \BibitemShut
  {NoStop}%
\bibitem [{\citenamefont {Bernhard}\ \emph {et~al.}(2015)\citenamefont
  {Bernhard}, \citenamefont {Marcy}, \citenamefont {Coleman-Smith},
  \citenamefont {Huzurbazar}, \citenamefont {Wolpert},\ and\ \citenamefont
  {Bass}}]{Bernhard:2015hxa}%
  \BibitemOpen
  \bibfield  {author} {\bibinfo {author} {\bibfnamefont {J.~E.}\ \bibnamefont
  {Bernhard}}, \bibinfo {author} {\bibfnamefont {P.~W.}\ \bibnamefont {Marcy}},
  \bibinfo {author} {\bibfnamefont {C.~E.}\ \bibnamefont {Coleman-Smith}},
  \bibinfo {author} {\bibfnamefont {S.}~\bibnamefont {Huzurbazar}}, \bibinfo
  {author} {\bibfnamefont {R.~L.}\ \bibnamefont {Wolpert}}, \ and\ \bibinfo
  {author} {\bibfnamefont {S.~A.}\ \bibnamefont {Bass}},\ }\bibfield  {title}
  {\enquote {\bibinfo {title} {{Quantifying properties of hot and dense QCD
  matter through systematic model-to-data comparison}},}\ }\href {\doibase
  10.1103/PhysRevC.91.054910} {\bibfield  {journal} {\bibinfo  {journal} {Phys.
  Rev.}\ }\textbf {\bibinfo {volume} {C91}},\ \bibinfo {pages} {054910}
  (\bibinfo {year} {2015})},\ \Eprint {http://arxiv.org/abs/1502.00339}
  {arXiv:1502.00339 [nucl-th]} \BibitemShut {NoStop}%
\bibitem [{\citenamefont {Niemi}\ and\ \citenamefont
  {Denicol}(2014)}]{Niemi:2014wta}%
  \BibitemOpen
  \bibfield  {author} {\bibinfo {author} {\bibfnamefont {H.}~\bibnamefont
  {Niemi}}\ and\ \bibinfo {author} {\bibfnamefont {G.~S.}\ \bibnamefont
  {Denicol}},\ }\bibfield  {title} {\enquote {\bibinfo {title} {{How large is
  the Knudsen number reached in fluid dynamical simulations of
  ultrarelativistic heavy ion collisions?}}}\ }\href@noop {} {\  (\bibinfo
  {year} {2014})},\ \Eprint {http://arxiv.org/abs/1404.7327} {arXiv:1404.7327
  [nucl-th]} \BibitemShut {NoStop}%
\bibitem [{\citenamefont {Noronha-Hostler}\ \emph {et~al.}(2016)\citenamefont
  {Noronha-Hostler}, \citenamefont {Noronha},\ and\ \citenamefont
  {Gyulassy}}]{Noronha-Hostler:2015coa}%
  \BibitemOpen
  \bibfield  {author} {\bibinfo {author} {\bibfnamefont {J.}~\bibnamefont
  {Noronha-Hostler}}, \bibinfo {author} {\bibfnamefont {J.}~\bibnamefont
  {Noronha}}, \ and\ \bibinfo {author} {\bibfnamefont {M.}~\bibnamefont
  {Gyulassy}},\ }\bibfield  {title} {\enquote {\bibinfo {title} {{Sensitivity
  of flow harmonics to subnucleon scale fluctuations in heavy ion
  collisions}},}\ }\href {\doibase 10.1103/PhysRevC.93.024909} {\bibfield
  {journal} {\bibinfo  {journal} {Phys. Rev.}\ }\textbf {\bibinfo {volume}
  {C93}},\ \bibinfo {pages} {024909} (\bibinfo {year} {2016})},\ \Eprint
  {http://arxiv.org/abs/1508.02455} {arXiv:1508.02455 [nucl-th]} \BibitemShut
  {NoStop}%
\bibitem [{\citenamefont {Vredevoogd}\ and\ \citenamefont
  {Pratt}(2009)}]{Vredevoogd:2008id}%
  \BibitemOpen
  \bibfield  {author} {\bibinfo {author} {\bibfnamefont {J.}~\bibnamefont
  {Vredevoogd}}\ and\ \bibinfo {author} {\bibfnamefont {S.}~\bibnamefont
  {Pratt}},\ }\bibfield  {title} {\enquote {\bibinfo {title} {{Universal Flow
  in the First Stage of Relativistic Heavy Ion Collisions}},}\ }\href {\doibase
  10.1103/PhysRevC.79.044915} {\bibfield  {journal} {\bibinfo  {journal} {Phys.
  Rev.}\ }\textbf {\bibinfo {volume} {C79}},\ \bibinfo {pages} {044915}
  (\bibinfo {year} {2009})},\ \Eprint {http://arxiv.org/abs/0810.4325}
  {arXiv:0810.4325 [nucl-th]} \BibitemShut {NoStop}%
\bibitem [{\citenamefont {York}\ and\ \citenamefont
  {Moore}(2009)}]{York:2008rr}%
  \BibitemOpen
  \bibfield  {author} {\bibinfo {author} {\bibfnamefont {M.~A.}\ \bibnamefont
  {York}}\ and\ \bibinfo {author} {\bibfnamefont {G.~D.}\ \bibnamefont
  {Moore}},\ }\bibfield  {title} {\enquote {\bibinfo {title} {{Second order
  hydrodynamic coefficients from kinetic theory}},}\ }\href {\doibase
  10.1103/PhysRevD.79.054011} {\bibfield  {journal} {\bibinfo  {journal} {Phys.
  Rev.}\ }\textbf {\bibinfo {volume} {D79}},\ \bibinfo {pages} {054011}
  (\bibinfo {year} {2009})},\ \Eprint {http://arxiv.org/abs/0811.0729}
  {arXiv:0811.0729 [hep-ph]} \BibitemShut {NoStop}%
\bibitem [{\citenamefont {Kovner}\ \emph {et~al.}(1995)\citenamefont {Kovner},
  \citenamefont {McLerran},\ and\ \citenamefont {Weigert}}]{Kovner:1995ja}%
  \BibitemOpen
  \bibfield  {author} {\bibinfo {author} {\bibfnamefont {A.}~\bibnamefont
  {Kovner}}, \bibinfo {author} {\bibfnamefont {L.~D.}\ \bibnamefont
  {McLerran}}, \ and\ \bibinfo {author} {\bibfnamefont {H.}~\bibnamefont
  {Weigert}},\ }\bibfield  {title} {\enquote {\bibinfo {title} {{Gluon
  production from nonAbelian Weizsacker-Williams fields in nucleus-nucleus
  collisions}},}\ }\href {\doibase 10.1103/PhysRevD.52.6231} {\bibfield
  {journal} {\bibinfo  {journal} {Phys. Rev.}\ }\textbf {\bibinfo {volume}
  {D52}},\ \bibinfo {pages} {6231--6237} (\bibinfo {year} {1995})},\ \Eprint
  {http://arxiv.org/abs/hep-ph/9502289} {arXiv:hep-ph/9502289 [hep-ph]}
  \BibitemShut {NoStop}%
\bibitem [{\citenamefont {Lappi}\ and\ \citenamefont
  {McLerran}(2006)}]{Lappi:2006fp}%
  \BibitemOpen
  \bibfield  {author} {\bibinfo {author} {\bibfnamefont {T.}~\bibnamefont
  {Lappi}}\ and\ \bibinfo {author} {\bibfnamefont {L.}~\bibnamefont
  {McLerran}},\ }\bibfield  {title} {\enquote {\bibinfo {title} {{Some features
  of the glasma}},}\ }\href {\doibase 10.1016/j.nuclphysa.2006.04.001}
  {\bibfield  {journal} {\bibinfo  {journal} {Nucl. Phys.}\ }\textbf {\bibinfo
  {volume} {A772}},\ \bibinfo {pages} {200--212} (\bibinfo {year} {2006})},\
  \Eprint {http://arxiv.org/abs/hep-ph/0602189} {arXiv:hep-ph/0602189 [hep-ph]}
  \BibitemShut {NoStop}%
\bibitem [{\citenamefont {Epelbaum}\ and\ \citenamefont
  {Gelis}(2013)}]{Gelis:2013rba}%
  \BibitemOpen
  \bibfield  {author} {\bibinfo {author} {\bibfnamefont {T.}~\bibnamefont
  {Epelbaum}}\ and\ \bibinfo {author} {\bibfnamefont {F.}~\bibnamefont
  {Gelis}},\ }\bibfield  {title} {\enquote {\bibinfo {title} {{Pressure
  isotropization in high energy heavy ion collisions}},}\ }\href {\doibase
  10.1103/PhysRevLett.111.232301} {\bibfield  {journal} {\bibinfo  {journal}
  {Phys. Rev. Lett.}\ }\textbf {\bibinfo {volume} {111}},\ \bibinfo {pages}
  {232301} (\bibinfo {year} {2013})},\ \Eprint {http://arxiv.org/abs/1307.2214}
  {arXiv:1307.2214 [hep-ph]} \BibitemShut {NoStop}%
\bibitem [{\citenamefont {Chen}\ \emph {et~al.}(2015)\citenamefont {Chen},
  \citenamefont {Fries}, \citenamefont {Kapusta},\ and\ \citenamefont
  {Li}}]{Chen:2015wia}%
  \BibitemOpen
  \bibfield  {author} {\bibinfo {author} {\bibfnamefont {G.}~\bibnamefont
  {Chen}}, \bibinfo {author} {\bibfnamefont {R.~J.}\ \bibnamefont {Fries}},
  \bibinfo {author} {\bibfnamefont {J.~I.}\ \bibnamefont {Kapusta}}, \ and\
  \bibinfo {author} {\bibfnamefont {Y.}~\bibnamefont {Li}},\ }\bibfield
  {title} {\enquote {\bibinfo {title} {{Early Time Dynamics of Gluon Fields in
  High Energy Nuclear Collisions}},}\ }\href {\doibase
  10.1103/PhysRevC.92.064912} {\bibfield  {journal} {\bibinfo  {journal} {Phys.
  Rev.}\ }\textbf {\bibinfo {volume} {C92}},\ \bibinfo {pages} {064912}
  (\bibinfo {year} {2015})},\ \Eprint {http://arxiv.org/abs/1507.03524}
  {arXiv:1507.03524 [nucl-th]} \BibitemShut {NoStop}%
\bibitem [{\citenamefont {Li}\ and\ \citenamefont
  {Kapusta}(2016)}]{Li:2016eqr}%
  \BibitemOpen
  \bibfield  {author} {\bibinfo {author} {\bibfnamefont {M.}~\bibnamefont
  {Li}}\ and\ \bibinfo {author} {\bibfnamefont {J.~I.}\ \bibnamefont
  {Kapusta}},\ }\bibfield  {title} {\enquote {\bibinfo {title} {{Analytic
  calculation of the energy-momentum tensor in heavy ion collisions from color
  glass condensate}},}\ }\href {\doibase 10.1103/PhysRevC.94.024908} {\bibfield
   {journal} {\bibinfo  {journal} {Phys. Rev.}\ }\textbf {\bibinfo {volume}
  {C94}},\ \bibinfo {pages} {024908} (\bibinfo {year} {2016})},\ \Eprint
  {http://arxiv.org/abs/1602.09060} {arXiv:1602.09060 [nucl-th]} \BibitemShut
  {NoStop}%
\bibitem [{\citenamefont {Kurkela}\ and\ \citenamefont
  {Moore}(2011{\natexlab{a}})}]{Kurkela:2011ti}%
  \BibitemOpen
  \bibfield  {author} {\bibinfo {author} {\bibfnamefont {A.}~\bibnamefont
  {Kurkela}}\ and\ \bibinfo {author} {\bibfnamefont {G.~D.}\ \bibnamefont
  {Moore}},\ }\bibfield  {title} {\enquote {\bibinfo {title} {{Thermalization
  in Weakly Coupled Nonabelian Plasmas}},}\ }\href {\doibase
  10.1007/JHEP12(2011)044} {\bibfield  {journal} {\bibinfo  {journal} {JHEP}\
  }\textbf {\bibinfo {volume} {12}},\ \bibinfo {pages} {044} (\bibinfo {year}
  {2011}{\natexlab{a}})},\ \Eprint {http://arxiv.org/abs/1107.5050}
  {arXiv:1107.5050 [hep-ph]} \BibitemShut {NoStop}%
\bibitem [{\citenamefont {Kurkela}\ and\ \citenamefont
  {Moore}(2011{\natexlab{b}})}]{Kurkela:2011ub}%
  \BibitemOpen
  \bibfield  {author} {\bibinfo {author} {\bibfnamefont {A.}~\bibnamefont
  {Kurkela}}\ and\ \bibinfo {author} {\bibfnamefont {G.~D.}\ \bibnamefont
  {Moore}},\ }\bibfield  {title} {\enquote {\bibinfo {title} {{Bjorken Flow,
  Plasma Instabilities, and Thermalization}},}\ }\href {\doibase
  10.1007/JHEP11(2011)120} {\bibfield  {journal} {\bibinfo  {journal} {JHEP}\
  }\textbf {\bibinfo {volume} {11}},\ \bibinfo {pages} {120} (\bibinfo {year}
  {2011}{\natexlab{b}})},\ \Eprint {http://arxiv.org/abs/1108.4684}
  {arXiv:1108.4684 [hep-ph]} \BibitemShut {NoStop}%
\bibitem [{\citenamefont {Mueller}\ and\ \citenamefont
  {Son}(2004)}]{Mueller:2002gd}%
  \BibitemOpen
  \bibfield  {author} {\bibinfo {author} {\bibfnamefont {A.~H.}\ \bibnamefont
  {Mueller}}\ and\ \bibinfo {author} {\bibfnamefont {D.~T.}\ \bibnamefont
  {Son}},\ }\bibfield  {title} {\enquote {\bibinfo {title} {{On the Equivalence
  between the Boltzmann equation and classical field theory at large occupation
  numbers}},}\ }\href {\doibase 10.1016/j.physletb.2003.12.047} {\bibfield
  {journal} {\bibinfo  {journal} {Phys. Lett.}\ }\textbf {\bibinfo {volume}
  {B582}},\ \bibinfo {pages} {279--287} (\bibinfo {year} {2004})},\ \Eprint
  {http://arxiv.org/abs/hep-ph/0212198} {arXiv:hep-ph/0212198 [hep-ph]}
  \BibitemShut {NoStop}%
\bibitem [{\citenamefont {Jeon}(2005)}]{Jeon:2004dh}%
  \BibitemOpen
  \bibfield  {author} {\bibinfo {author} {\bibfnamefont {S.}~\bibnamefont
  {Jeon}},\ }\bibfield  {title} {\enquote {\bibinfo {title} {{The Boltzmann
  equation in classical and quantum field theory}},}\ }\href {\doibase
  10.1103/PhysRevC.72.014907} {\bibfield  {journal} {\bibinfo  {journal} {Phys.
  Rev.}\ }\textbf {\bibinfo {volume} {C72}},\ \bibinfo {pages} {014907}
  (\bibinfo {year} {2005})},\ \Eprint {http://arxiv.org/abs/hep-ph/0412121}
  {arXiv:hep-ph/0412121 [hep-ph]} \BibitemShut {NoStop}%
\bibitem [{\citenamefont {Abraao~York}\ \emph {et~al.}(2014)\citenamefont
  {Abraao~York}, \citenamefont {Kurkela}, \citenamefont {Lu},\ and\
  \citenamefont {Moore}}]{York:2014wja}%
  \BibitemOpen
  \bibfield  {author} {\bibinfo {author} {\bibfnamefont {M.~C.}\ \bibnamefont
  {Abraao~York}}, \bibinfo {author} {\bibfnamefont {A.}~\bibnamefont
  {Kurkela}}, \bibinfo {author} {\bibfnamefont {E.}~\bibnamefont {Lu}}, \ and\
  \bibinfo {author} {\bibfnamefont {G.~D.}\ \bibnamefont {Moore}},\ }\bibfield
  {title} {\enquote {\bibinfo {title} {{UV cascade in classical Yang-Mills
  theory via kinetic theory}},}\ }\href {\doibase 10.1103/PhysRevD.89.074036}
  {\bibfield  {journal} {\bibinfo  {journal} {Phys. Rev.}\ }\textbf {\bibinfo
  {volume} {D89}},\ \bibinfo {pages} {074036} (\bibinfo {year} {2014})},\
  \Eprint {http://arxiv.org/abs/1401.3751} {arXiv:1401.3751 [hep-ph]}
  \BibitemShut {NoStop}%
\bibitem [{\citenamefont {Romatschke}(2016)}]{Romatschke:2015gic}%
  \BibitemOpen
  \bibfield  {author} {\bibinfo {author} {\bibfnamefont {P.}~\bibnamefont
  {Romatschke}},\ }\bibfield  {title} {\enquote {\bibinfo {title} {{Retarded
  correlators in kinetic theory: branch cuts, poles and hydrodynamic onset
  transitions}},}\ }\href {\doibase 10.1140/epjc/s10052-016-4169-7} {\bibfield
  {journal} {\bibinfo  {journal} {Eur. Phys. J.}\ }\textbf {\bibinfo {volume}
  {C76}},\ \bibinfo {pages} {352} (\bibinfo {year} {2016})},\ \Eprint
  {http://arxiv.org/abs/1512.02641} {arXiv:1512.02641 [hep-th]} \BibitemShut
  {NoStop}%
\bibitem [{\citenamefont {Policastro}\ \emph {et~al.}(2001)\citenamefont
  {Policastro}, \citenamefont {Son},\ and\ \citenamefont
  {Starinets}}]{Policastro:2001yc}%
  \BibitemOpen
  \bibfield  {author} {\bibinfo {author} {\bibfnamefont {G.}~\bibnamefont
  {Policastro}}, \bibinfo {author} {\bibfnamefont {D.~T.}\ \bibnamefont {Son}},
  \ and\ \bibinfo {author} {\bibfnamefont {A.~O.}\ \bibnamefont {Starinets}},\
  }\bibfield  {title} {\enquote {\bibinfo {title} {{The Shear viscosity of
  strongly coupled N=4 supersymmetric Yang-Mills plasma}},}\ }\href {\doibase
  10.1103/PhysRevLett.87.081601} {\bibfield  {journal} {\bibinfo  {journal}
  {Phys. Rev. Lett.}\ }\textbf {\bibinfo {volume} {87}},\ \bibinfo {pages}
  {081601} (\bibinfo {year} {2001})},\ \Eprint
  {http://arxiv.org/abs/hep-th/0104066} {arXiv:hep-th/0104066 [hep-th]}
  \BibitemShut {NoStop}%
\bibitem [{\citenamefont {Kapusta}\ and\ \citenamefont
  {Torres-Rincon}(2012)}]{Kapusta:2012zb}%
  \BibitemOpen
  \bibfield  {author} {\bibinfo {author} {\bibfnamefont {J.~I.}\ \bibnamefont
  {Kapusta}}\ and\ \bibinfo {author} {\bibfnamefont {J.~M.}\ \bibnamefont
  {Torres-Rincon}},\ }\bibfield  {title} {\enquote {\bibinfo {title} {{Thermal
  Conductivity and Chiral Critical Point in Heavy Ion Collisions}},}\ }\href
  {\doibase 10.1103/PhysRevC.86.054911} {\bibfield  {journal} {\bibinfo
  {journal} {Phys. Rev.}\ }\textbf {\bibinfo {volume} {C86}},\ \bibinfo {pages}
  {054911} (\bibinfo {year} {2012})},\ \Eprint {http://arxiv.org/abs/1209.0675}
  {arXiv:1209.0675 [nucl-th]} \BibitemShut {NoStop}%
\bibitem [{\citenamefont {Bell}\ \emph {et~al.}(2007)\citenamefont {Bell},
  \citenamefont {Garcia},\ and\ \citenamefont {Williams}}]{bell2007numerical}%
  \BibitemOpen
  \bibfield  {author} {\bibinfo {author} {\bibfnamefont {J.~B.}\ \bibnamefont
  {Bell}}, \bibinfo {author} {\bibfnamefont {A.~L.}\ \bibnamefont {Garcia}}, \
  and\ \bibinfo {author} {\bibfnamefont {S.~A.}\ \bibnamefont {Williams}},\
  }\bibfield  {title} {\enquote {\bibinfo {title} {Numerical methods for the
  stochastic {Landau-Lifshitz Navier-Stokes} equations},}\ }\href {\doibase
  10.1103/PhysRevE.76.016708} {\bibfield  {journal} {\bibinfo  {journal} {Phys.
  Rev. E}\ }\textbf {\bibinfo {volume} {76}},\ \bibinfo {pages} {016708}
  (\bibinfo {year} {2007})}\BibitemShut {NoStop}%
\bibitem [{\citenamefont {Donev}\ \emph {et~al.}(2011)\citenamefont {Donev},
  \citenamefont {Bell}, \citenamefont {de~la Fuente},\ and\ \citenamefont
  {Garcia}}]{donev2011diffusive}%
  \BibitemOpen
  \bibfield  {author} {\bibinfo {author} {\bibfnamefont {A.}~\bibnamefont
  {Donev}}, \bibinfo {author} {\bibfnamefont {J.~B.}\ \bibnamefont {Bell}},
  \bibinfo {author} {\bibfnamefont {A.}~\bibnamefont {de~la Fuente}}, \ and\
  \bibinfo {author} {\bibfnamefont {A.~L.}\ \bibnamefont {Garcia}},\ }\bibfield
   {title} {\enquote {\bibinfo {title} {Diffusive transport by thermal velocity
  fluctuations},}\ }\href {\doibase 10.1103/PhysRevLett.106.204501} {\bibfield
  {journal} {\bibinfo  {journal} {Phys. Rev. Lett.}\ }\textbf {\bibinfo
  {volume} {106}},\ \bibinfo {pages} {204501} (\bibinfo {year}
  {2011})}\BibitemShut {NoStop}%
\bibitem [{\citenamefont {Usabiaga}\ \emph {et~al.}(2012)\citenamefont
  {Usabiaga}, \citenamefont {Bell}, \citenamefont {Delgado-Buscalioni},
  \citenamefont {Donev}, \citenamefont {Fai}, \citenamefont {Griffith},\ and\
  \citenamefont {Peskin}}]{balboa2012staggered}%
  \BibitemOpen
  \bibfield  {author} {\bibinfo {author} {\bibfnamefont {F.~B.}\ \bibnamefont
  {Usabiaga}}, \bibinfo {author} {\bibfnamefont {J.~B.}\ \bibnamefont {Bell}},
  \bibinfo {author} {\bibfnamefont {R.}~\bibnamefont {Delgado-Buscalioni}},
  \bibinfo {author} {\bibfnamefont {A.}~\bibnamefont {Donev}}, \bibinfo
  {author} {\bibfnamefont {T.~G.}\ \bibnamefont {Fai}}, \bibinfo {author}
  {\bibfnamefont {B.~E.}\ \bibnamefont {Griffith}}, \ and\ \bibinfo {author}
  {\bibfnamefont {C.~S.}\ \bibnamefont {Peskin}},\ }\bibfield  {title}
  {\enquote {\bibinfo {title} {Staggered schemes for fluctuating
  hydrodynamics},}\ }\href {\doibase 10.1137/120864520} {\bibfield  {journal}
  {\bibinfo  {journal} {Multiscale Modeling \& Simulation}\ }\textbf {\bibinfo
  {volume} {10}},\ \bibinfo {pages} {1369--1408} (\bibinfo {year}
  {2012})}\BibitemShut {NoStop}%
\bibitem [{\citenamefont {Arnold}\ and\ \citenamefont
  {Yaffe}(1998)}]{Arnold:1997gh}%
  \BibitemOpen
  \bibfield  {author} {\bibinfo {author} {\bibfnamefont {P.~B.}\ \bibnamefont
  {Arnold}}\ and\ \bibinfo {author} {\bibfnamefont {L.~G.}\ \bibnamefont
  {Yaffe}},\ }\bibfield  {title} {\enquote {\bibinfo {title} {{Effective
  theories for real time correlations in hot plasmas}},}\ }\href {\doibase
  10.1103/PhysRevD.57.1178} {\bibfield  {journal} {\bibinfo  {journal} {Phys.
  Rev.}\ }\textbf {\bibinfo {volume} {D57}},\ \bibinfo {pages} {1178--1192}
  (\bibinfo {year} {1998})},\ \Eprint {http://arxiv.org/abs/hep-ph/9709449}
  {arXiv:hep-ph/9709449 [hep-ph]} \BibitemShut {NoStop}%
\bibitem [{\citenamefont {Son}\ and\ \citenamefont
  {Starinets}(2007)}]{Son:2007vk}%
  \BibitemOpen
  \bibfield  {author} {\bibinfo {author} {\bibfnamefont {D.~T.}\ \bibnamefont
  {Son}}\ and\ \bibinfo {author} {\bibfnamefont {A.~O.}\ \bibnamefont
  {Starinets}},\ }\bibfield  {title} {\enquote {\bibinfo {title} {{Viscosity,
  Black Holes, and Quantum Field Theory}},}\ }\href {\doibase
  10.1146/annurev.nucl.57.090506.123120} {\bibfield  {journal} {\bibinfo
  {journal} {Ann. Rev. Nucl. Part. Sci.}\ }\textbf {\bibinfo {volume} {57}},\
  \bibinfo {pages} {95--118} (\bibinfo {year} {2007})},\ \Eprint
  {http://arxiv.org/abs/0704.0240} {arXiv:0704.0240 [hep-th]} \BibitemShut
  {NoStop}%
\bibitem [{\citenamefont {Hong}\ and\ \citenamefont
  {Teaney}(2010)}]{Hong:2010at}%
  \BibitemOpen
  \bibfield  {author} {\bibinfo {author} {\bibfnamefont {J.}~\bibnamefont
  {Hong}}\ and\ \bibinfo {author} {\bibfnamefont {D.}~\bibnamefont {Teaney}},\
  }\bibfield  {title} {\enquote {\bibinfo {title} {{Spectral densities for hot
  QCD plasmas in a leading log approximation}},}\ }\href {\doibase
  10.1103/PhysRevC.82.044908} {\bibfield  {journal} {\bibinfo  {journal} {Phys.
  Rev.}\ }\textbf {\bibinfo {volume} {C82}},\ \bibinfo {pages} {044908}
  (\bibinfo {year} {2010})},\ \Eprint {http://arxiv.org/abs/1003.0699}
  {arXiv:1003.0699 [nucl-th]} \BibitemShut {NoStop}%
\bibitem [{\citenamefont {Caron-Huot}\ and\ \citenamefont
  {Saremi}(2010)}]{CaronHuot:2009iq}%
  \BibitemOpen
  \bibfield  {author} {\bibinfo {author} {\bibfnamefont {S.}~\bibnamefont
  {Caron-Huot}}\ and\ \bibinfo {author} {\bibfnamefont {O.}~\bibnamefont
  {Saremi}},\ }\bibfield  {title} {\enquote {\bibinfo {title} {{Hydrodynamic
  Long-Time tails From Anti de Sitter Space}},}\ }\href {\doibase
  10.1007/JHEP11(2010)013} {\bibfield  {journal} {\bibinfo  {journal} {JHEP}\
  }\textbf {\bibinfo {volume} {11}},\ \bibinfo {pages} {013} (\bibinfo {year}
  {2010})},\ \Eprint {http://arxiv.org/abs/0909.4525} {arXiv:0909.4525
  [hep-th]} \BibitemShut {NoStop}%
\bibitem [{\citenamefont {Pomeau}\ and\ \citenamefont
  {Resibois}(1974)}]{Pomeau:1974hg}%
  \BibitemOpen
  \bibfield  {author} {\bibinfo {author} {\bibfnamefont {Y.}~\bibnamefont
  {Pomeau}}\ and\ \bibinfo {author} {\bibfnamefont {P.}~\bibnamefont
  {Resibois}},\ }\bibfield  {title} {\enquote {\bibinfo {title} {{Time
  Dependent Correlation Functions and Mode-Mode Coupling Theories}},}\
  }\href@noop {} {\bibfield  {journal} {\bibinfo  {journal} {Submitted to:
  Phys. Rept.}\ } (\bibinfo {year} {1974})}\BibitemShut {NoStop}%
\bibitem [{\citenamefont {Fox}(1978)}]{Fox1978}%
  \BibitemOpen
  \bibfield  {author} {\bibinfo {author} {\bibfnamefont {R.~F.}\ \bibnamefont
  {Fox}},\ }\bibfield  {title} {\enquote {\bibinfo {title} {{Gaussian
  stochastic processes in physics}},}\ }\href {\doibase
  10.1016/0370-1573(78)90145-X} {\bibfield  {journal} {\bibinfo  {journal}
  {Phys. Rept.}\ }\textbf {\bibinfo {volume} {48}},\ \bibinfo {pages}
  {179--283} (\bibinfo {year} {1978})}\BibitemShut {NoStop}%
\bibitem [{\citenamefont {Jeon}\ and\ \citenamefont
  {Yaffe}(1996)}]{Jeon:1995zm}%
  \BibitemOpen
  \bibfield  {author} {\bibinfo {author} {\bibfnamefont {S.}~\bibnamefont
  {Jeon}}\ and\ \bibinfo {author} {\bibfnamefont {L.~G.}\ \bibnamefont
  {Yaffe}},\ }\bibfield  {title} {\enquote {\bibinfo {title} {{From quantum
  field theory to hydrodynamics: Transport coefficients and effective kinetic
  theory}},}\ }\href {\doibase 10.1103/PhysRevD.53.5799} {\bibfield  {journal}
  {\bibinfo  {journal} {Phys. Rev.}\ }\textbf {\bibinfo {volume} {D53}},\
  \bibinfo {pages} {5799--5809} (\bibinfo {year} {1996})},\ \Eprint
  {http://arxiv.org/abs/hep-ph/9512263} {arXiv:hep-ph/9512263 [hep-ph]}
  \BibitemShut {NoStop}%
\bibitem [{\citenamefont {Gavin}\ \emph {et~al.}(2016)\citenamefont {Gavin},
  \citenamefont {Moschelli},\ and\ \citenamefont {Zin}}]{Gavin:2016hmv}%
  \BibitemOpen
  \bibfield  {author} {\bibinfo {author} {\bibfnamefont {S.}~\bibnamefont
  {Gavin}}, \bibinfo {author} {\bibfnamefont {G.}~\bibnamefont {Moschelli}}, \
  and\ \bibinfo {author} {\bibfnamefont {C.}~\bibnamefont {Zin}},\ }\bibfield
  {title} {\enquote {\bibinfo {title} {{Rapidity Correlation Structure in
  Nuclear Collisions}},}\ }\href {\doibase 10.1103/PhysRevC.94.024921}
  {\bibfield  {journal} {\bibinfo  {journal} {Phys. Rev.}\ }\textbf {\bibinfo
  {volume} {C94}},\ \bibinfo {pages} {024921} (\bibinfo {year} {2016})},\
  \Eprint {http://arxiv.org/abs/1606.02692} {arXiv:1606.02692 [nucl-th]}
  \BibitemShut {NoStop}%
\bibitem [{\citenamefont {Borsanyi}\ \emph {et~al.}(2014)\citenamefont
  {Borsanyi}, \citenamefont {Fodor}, \citenamefont {Hoelbling}, \citenamefont
  {Katz}, \citenamefont {Krieg},\ and\ \citenamefont
  {Szabo}}]{Borsanyi:2013bia}%
  \BibitemOpen
  \bibfield  {author} {\bibinfo {author} {\bibfnamefont {S.}~\bibnamefont
  {Borsanyi}}, \bibinfo {author} {\bibfnamefont {Z.}~\bibnamefont {Fodor}},
  \bibinfo {author} {\bibfnamefont {C.}~\bibnamefont {Hoelbling}}, \bibinfo
  {author} {\bibfnamefont {S.~D.}\ \bibnamefont {Katz}}, \bibinfo {author}
  {\bibfnamefont {S.}~\bibnamefont {Krieg}}, \ and\ \bibinfo {author}
  {\bibfnamefont {K.~K.}\ \bibnamefont {Szabo}},\ }\bibfield  {title} {\enquote
  {\bibinfo {title} {{Full result for the QCD equation of state with 2+1
  flavors}},}\ }\href {\doibase 10.1016/j.physletb.2014.01.007} {\bibfield
  {journal} {\bibinfo  {journal} {Phys. Lett.}\ }\textbf {\bibinfo {volume}
  {B730}},\ \bibinfo {pages} {99--104} (\bibinfo {year} {2014})},\ \Eprint
  {http://arxiv.org/abs/1309.5258} {arXiv:1309.5258 [hep-lat]} \BibitemShut
  {NoStop}%
\bibitem [{\citenamefont {Jia}(2014)}]{Jia:2014jca}%
  \BibitemOpen
  \bibfield  {author} {\bibinfo {author} {\bibfnamefont {J.}~\bibnamefont
  {Jia}},\ }\bibfield  {title} {\enquote {\bibinfo {title} {\{Event-shape
  fluctuations and flow correlations in ultra-relativistic heavy-ion
  collisions\}},}\ }\href {\doibase 10.1088/0954-3899/41/12/124003} {\bibfield
  {journal} {\bibinfo  {journal} {J. Phys.}\ }\textbf {\bibinfo {volume}
  {G41}},\ \bibinfo {pages} {124003} (\bibinfo {year} {2014})},\ \Eprint
  {http://arxiv.org/abs/1407.6057} {arXiv:1407.6057 [nucl-ex]} \BibitemShut
  {NoStop}%
\%\%CITATION = ARXIV:1407.6057;\%\%
\bibitem [{\citenamefont {Berges}(2015)}]{Berges:2015kfa}%
  \BibitemOpen
  \bibfield  {author} {\bibinfo {author} {\bibfnamefont {J.}~\bibnamefont
  {Berges}},\ }\bibfield  {title} {\enquote {\bibinfo {title} {{Nonequilibrium
  Quantum Fields: From Cold Atoms to Cosmology}},}\ }\href@noop {} {\
  (\bibinfo {year} {2015})},\ \Eprint {http://arxiv.org/abs/1503.02907}
  {arXiv:1503.02907 [hep-ph]} \BibitemShut {NoStop}%
\bibitem [{\citenamefont {Heller}\ \emph {et~al.}(2016)\citenamefont {Heller},
  \citenamefont {Kurkela},\ and\ \citenamefont {Spalinski}}]{Heller:2016rtz}%
  \BibitemOpen
  \bibfield  {author} {\bibinfo {author} {\bibfnamefont {M.~P.}\ \bibnamefont
  {Heller}}, \bibinfo {author} {\bibfnamefont {A.}~\bibnamefont {Kurkela}}, \
  and\ \bibinfo {author} {\bibfnamefont {M.}~\bibnamefont {Spalinski}},\
  }\bibfield  {title} {\enquote {\bibinfo {title} {{Hydrodynamization and
  transient modes of expanding plasma in kinetic theory}},}\ }\href@noop {} {\
  (\bibinfo {year} {2016})},\ \Eprint {http://arxiv.org/abs/1609.04803}
  {arXiv:1609.04803 [nucl-th]} \BibitemShut {NoStop}%
\end{thebibliography}%

\end{document}